\documentclass[11pt]{article}
\usepackage[margin=1in]{geometry}
\usepackage{graphicx} % Required for inserting images

\usepackage[backend=biber]{biblatex}
\usepackage{commands-tam}
\usepackage{hyperref}
\usepackage{subcaption}% \captionsetup{compatibility=false}
\usepackage{mathrsfs}
\usepackage{wrapfig}
\usepackage{enumitem}

\usepackage{amsmath}
\usepackage{amssymb}
\usepackage{amsthm}
\usepackage{thmtools, thm-restate}
\newtheorem{definition}{Definition}
\newtheorem{lemma}{Lemma}
\newtheorem{theorem}{Theorem}
\newtheorem{corollary}{Corollary}
\newtheorem{remark}{Remark}

\usepackage{multirow}
\usepackage[capitalize, noabbrev]{cleveref}
\usepackage[usenames, dvipsnames]{xcolor}
\usepackage{tikz}
\usetikzlibrary{matrix}
\usetikzlibrary{arrows.meta, angles, shapes.geometric}
\usetikzlibrary{calc, positioning}
\usetikzlibrary{tikzmark}
\usetikzlibrary{math, quotes}
\usetikzlibrary{patterns, fadings, patterns.meta, backgrounds}
\usetikzlibrary{shapes.callouts}

\usepackage{fractalify}

\tikzset{gate/.style={draw,circle}}

\newcommand\coldot[2][]{ \tikz{\node[yshift=-.2ex, circle, fill=#2] (#1) {};} }
\newcommand\tealdot[1][]{\coldot[#1]{teal}}
\newcommand\pinkdot[1][]{\coldot[#1]{pink}}
\newcommand\golddot[1][]{\coldot[#1]{Dandelion}}

\newcommand\gammashape{
\begin{tikzpicture}
  \node[draw] (O) at (0, 0) {};
  \node[draw, right=0 of O] {};
  \node[draw, above=0 of O] {};
  \node[fill=red, circle, scale=.2, anchor=center] at (O.center) {};
  \end{tikzpicture}
}

\newcommand\drawcacarpet[1]{ % Argument = profondeur
  \pgfmathsetmacro\level{int(#1)}
  \ifnum #1 = 0 \relax
  \filldraw[fill=gray!50] (0, 0) -- (0, 1) -- (1, 1) -- (1, 0) -- cycle;
  \else
  \pgfmathsetmacro\next{int(#1 - 1)}
  \begin{scope}[scale=.166666666]
    \foreach \x in {0, ..., 5}
    \foreach \y in {0, ..., 5}
    {
      \ifthenelse{\numexpr\(\x=3\and\y=3\)\OR\(\x=2\and\y=2\)\OR\(\x=3\AND\y=2\)\OR\(\x=2\AND\y=3\)}
      {

      }
      {
        \begin{scope}[shift={(\x, \y)}]
          \drawcacarpet{\next}
        \end{scope}
      }
    }
  \end{scope}
  \fi
}

\newcommand\drawinputsW[2][]{\draw[-{Stealth}] ($ #2 +(-.7, -.3) $) .. controls ($ #2 +(-.5, -.5) $) .. #2 node[pos=-.2] {#1};}

\newcommand\drawinputNe[2][]{\draw[-{Stealth}] ($ #2 +(.3, .7) $) .. controls ($ #2 +(.5, .5) $) .. #2 node[pos=-.2] {#1};}

\newcommand\drawinputEw[2][]{\draw[-{Stealth}{Stealth}] ($ #2 +(-.7, 0) $) -- #2 node[pos=-.4] {#1};}
\newcommand\drawinputWe[2][]{\draw[-{Stealth}{Stealth}] ($ #2 +(.7, 0) $) -- #2 node[pos=-.4] {#1};}

\newcommand\drawinputE[2][]{\draw[-{Stealth}] ($ #2 +(.7, 0) $) -- #2 node[pos=-.5] {#1};}
\newcommand\drawinputW[2][]{\draw[-{Stealth}] ($ #2 +(-.7, 0) $) -- #2 node[pos=-.5] {#1};}

\newcommand\drawinputEn[2][]{\draw[-{Stealth}] ($ #2 +(.7, .3) $) .. controls ($ #2 +(.5, .5) $) .. #2 node[pos=-.2] {#1};}

\newcommand\drawinputSw[2][]{\draw[-{Stealth}] ($ #2 +(-.3, -.7) $) .. controls ($ #2 +(-.5, -.5) $) .. #2 node[pos=-.2] {#1};}
\newcommand\drawinputWs[2][]{\draw[-{Stealth}] ($ #2 +(-.7, -.4) $) .. controls ($ #2 +(-.5, -.5) $) .. #2 node[pos=-.2] {#1};}

\newcommand\drawoutputE[2][]{\draw[-{Stealth}] #2 edge node [pos=1.2] {#1} +(.9, 0);}
\newcommand\drawoutputN[2][]{\draw[-{Stealth}] #2 edge node [pos=1.2] {#1} +(0, .9);}
\newcommand\drawoutputnE[2][]{\draw[-{Stealth}] #2 .. controls +(.5, .5) .. +(.9, .1) node[pos=1.1] {#1};}
\newcommand\drawoutputnW[2][]{\draw[-{Stealth}] #2 .. controls +(-.5, .5) .. +(-.9, .1) node [pos=1.1] {#1};}

\newcommand\drawoutputNe[2][]{\draw[-{Stealth}] #2 .. controls +(.5, .5) .. +(.1, .9) node [pos=1.1] {#1};}
\newcommand\drawoutputsE[2][]{\draw[-{Stealth}] #2 .. controls +(.5, -.5) .. +(.9, -.1) node [pos=1.1] {#1};}
\newcommand\drawoutputSe[2][]{\draw[-{Stealth}] #2 .. controls +(.5, -.5) .. +(.1, -.9) node [pos=1.1] {#1};}

\tikzset{
  singleGlue/.pic={
    \begin{scope}[scale=.2]
      \fill[fill=red] (-.4,0) -- (.4, 0) -- (.2, 1) -- (-.2, 1) -- cycle;
    \end{scope}
  }
}

\tikzset{
  doubleGlue/.pic={
    \begin{scope}[scale=.2]
      \fill[fill=black] (-.7, -1) rectangle (-.2, 0);
      \fill[fill=black] (.2, -1) rectangle (.7, 0);
    \end{scope}
  }
}

\tikzset{
  spaceinvader/.pic={
    \begin{scope}[scale=.08]
      \foreach \x/\y in {0/0, 0/1, 1/1, 1/2, 1/3, 2/0, 2/1, 2/3, 2/4, 3/1, 3/2, 3/3, 4/0, 4/1, 4/3, 4/4, 5/1, 5/2, 5/3, 6/0, 6/1}
      \filldraw[pic actions, shift={(\x, \y)}] (0, 0) rectangle (1, 1);
      \coordinate (-right-foot) at (6,0);
      \coordinate (-right-antenna) at (4,5);
      \coordinate (-left-foot) at (0,0);
    \end{scope}
  }
}

\tikzset{
  pics/posWithinK/.style args={#1/#2}{
    code={
      \begin{scope}[scale=.1, shift={(-3,-3)}]
        \draw[-] (0,0) grid (6,6);
        \fill[#2] (2,2) rectangle(4,4);
        \fill[black] #1 rectangle +(1,1);
        \coordinate(-east) at (6,3);
        \coordinate(-west) at (0,3);
      \end{scope}  
    }
  }
}

\newcommand{\neighcross}[4]{
  \begin{scope}[scale=.33]
    \ifthenelse{\equal{#1}{}}{}{\filldraw[fill=#1] (0, 1) rectangle +(1, 1);}
    \ifthenelse{\equal{#2}{}}{}{\filldraw[fill=#2] (1, 0) rectangle +(1, 1);}
    \ifthenelse{\equal{#3}{}}{}{\filldraw[fill=#3] (0, -1) rectangle +(1, 1);}
    \ifthenelse{\equal{#4}{}}{}{\filldraw[fill=#4] (-1, 0) rectangle +(1, 1);}
    \filldraw[fill=white] (0, 0) rectangle +(1, 1);
    \node at (.5, .5) {?};
  \end{scope}
}

\newcommand{\neighcrossnodes}[4]{
  \begin{scope}[scale=.5]
    \ifthenelse{\equal{#1}{}}{}{\draw (0, 1) rectangle +(1, 1) node[pos=.5] {#1};}
    \ifthenelse{\equal{#2}{}}{}{\draw (1, 0) rectangle +(1, 1) node[pos=.5] {#2};}
    \ifthenelse{\equal{#3}{}}{}{\draw (0, -1) rectangle +(1, 1) node[pos=.5] {#3};}
    \ifthenelse{\equal{#4}{}}{}{\draw (-1, 0) rectangle +(1, 1) node[pos=.5] {#4};}
    \filldraw[fill=white] (0, 0) rectangle +(1, 1);
    \node at (.5, .5) {?};
  \end{scope}
}

\newcommand{\neighcrosspics}[4]{
  \begin{scope}[scale=1.5]
    \ifthenelse{\equal{#1}{}}{}{\draw (0, 1) rectangle +(1, 1) pic[pos=.5, scale=.75] {#1};}
    \ifthenelse{\equal{#2}{}}{}{\draw (1, 0) rectangle +(1, 1) pic[pos=.5, scale=.75] {#2};}
    \ifthenelse{\equal{#3}{}}{}{\draw (0, -1) rectangle +(1, 1) pic[pos=.5, scale=.75] {#3};}
    \ifthenelse{\equal{#4}{}}{}{\draw (-1, 0) rectangle +(1, 1) pic[pos=.5, scale=.75] {#4};}
    \filldraw[fill=gray!10!white] (0, 0) rectangle +(1, 1);
    \node at (.5, .5) {*};
  \end{scope}
}

\usepackage{wasysym}
\usepackage{readarray}

\usepackage{FBattribute}
\newcommand\ZZ{\ensuremath{\mathbb{Z}^2}}
\newcommand\bigquot[2]{\ensuremath{\left\lfloor}\frac{#1}{#2}\right\rfloor}
\newcommand\quot[2]{\ensuremath{\lfloor{#1}/{#2}\rfloor}}

\newcommand\cacarpet{\ensuremath{K^{\infty}}}

\newcommand\tiles[1][]{\FBattribute[#1]{tile}}
\newcommand\Ss{\ensuremath{\mathcal{T}}}

\newcommand\FreeSeqs[1]{\ensuremath{\mathcal{H}^{\operatorname{Free}}[#1]}}
\newcommand\Assemblies[1]{\ensuremath{#1^{\subset\ZZ}}}

\newcommand\TerminalProds[1]{\ensuremath{\mathcal{A}_\square[#1]}}

\newcommand\Dir{\ensuremath{\{N, E, S, W\}}}
\newcommand\PDir{\ensuremath{\mathcal{P}(\Dir)}}

\newcommand\Wires{\operatorname{Wires}}
\newcommand\Wirings{\mathcal{W}}

\newcommand\winputset[1][]{\FBattribute[#1]{Inputs}}
\newcommand\winertset[1][]{\FBattribute[#1]{Inert}}
\newcommand\woutputset[1][]{\FBattribute[#1]{Outputs}}

\newcommand\woutputmap[1][]{\FBattribute[#1]{output-num}}
\newcommand\woutputwires[1][]{\FBattribute[#1]{outwires}}

\newcommand\gatefun[1][]{\FBattribute[#1]{func}}
\newcommand\gatewiring[1][]{\FBattribute[#1]{wiring}}
\newcommand\Gates[1]{\ensuremath{\operatorname{Gates}(#1)}}
\newcommand\gatecolor[1][]{
  \def\tempa{}%
  \def\tempb{#1}%
  \ifx\tempa\tempb{}
    \ensuremath\operatorname{label}
  \else
    \ensuremath{\operatorname{label}(#1)}
  \fi
}

\newcommand\innerevalfunc[1]{\ensuremath{\widetilde{#1}}}
\newcommand\evalfunc[1]{\ensuremath{\bar{#1}}}
\newcommand\decgate{\ensuremath{\operatorname{dec-gate}}}
\newcommand\pinputs{\ensuremath{\operatorname{pred-inputs}}}
\newcommand\psymb{\ensuremath{\operatorname{pred-symbols}}}

\newcommand*\finst[1]{\ensuremath{\operatorname{instantiate}_{#1}}}
\newcommand\gcolor[1]{\ensuremath{\mathfrak{#1}}}
\newcommand\wirepos[2]{#1 (\textrm{with wire #2})}
\newcommand\lmpos[1][]{\FBattribute[#1]{pos}}

\newcommand\lmpar[1][]{\ensuremath{#1\cdot\operatorname{parent}}}
\newcommand\lmpwiring[1][]{\FBattribute[#1]{parent-wiring}}
\newcommand\lmpcolor[1][]{\FBattribute[#1]{parent-label}}
\newcommand\fincr[1]{
  \def\tempa{}%
  \def\tempb{#1}%
  \ifx\tempa\tempb{}
    \ensuremath{\operatorname{incr}}
  \else
    \ensuremath{\operatorname{incr}[#1]}
  \fi
}
\newcommand\frepar[1]{
  \def\tempa{}%
  \def\tempb{#1}%
  \ifx\tempa\tempb{}
    \ensuremath{\operatorname{set-parent}}
  \else
    \ensuremath{\operatorname{set-parent}[#1]}
  \fi
}
\newcommand\decgatew{\ensuremath{\operatorname{dec-gate}_w}}
\newcommand\decgatel{\ensuremath{\operatorname{dec-gate}_l}}
\newcommand\mloc[1][]{\FBattribute[#1]{parent-label_2}}
\newcommand\mglob[1][]{\FBattribute[#1]{global}}
\newcommand\gmcolor[1][]{\FBattribute[#1]{label_{2}}}
\newcommand\gmanclocal[1][]{\FBattribute[#1]{ancestor-msg}}

\newcommand\extractfunc{\ensuremath{\operatorname{extract}}}
\newcommand\embedfunc{\ensuremath{\operatorname{embed}}}
\newcommand\fdecode[2][]{
  \def\tempa{}%
  \def\tempb{#1}%
  \ifx\tempa\tempb{}
  \ensuremath{\operatorname{decode}_{#2}}
  \else
  \ensuremath{\operatorname{decode}_{#2}[#1]}
  \fi
}
\newcommand\fgincr[2]{
  \def\tempa{}%
  \def\tempb{#1}%
  \ifx\tempa\tempb{}
  \ensuremath{\operatorname{incr}_{#2}}
  \else
  \ensuremath{\operatorname{incr}_{#1}[#2]}
  \fi
}

\newcommand\complete[1]{#1^\bullet}
\newcommand\tree{\ensuremath{\mathcal{TREE-DEC}}}
\newcommand\treedec[1]{\tree[\complete{#1}]}
\newcommand\ezEmbed[1][]{\ensuremath{\operatorname{squash}_{#1}}}
\newcommand\pd{\ensuremath{\mathring{-}}}
\newcommand\support[1][]{\FBattribute[#1]{support}}
\newcommand\FreeAssemblies[1]{\ensuremath{{#1}^{\operatorname{Free}}}}
\newcommand\fizziness{\operatorname{fz}}
\newcommand\moreFizzy{\ensuremath{>_{\fizziness}}}
\newcommand\moreFizzyEq{\ensuremath{\geq_{\fizziness}}}
\newcommand\glue[1][]{\FBattribute[#1]{glue}}
\newcommand\present[1][]{\FBattribute[#1]{present}}
\newcommand\paths[1][]{\FBattribute[#1]{successor}}
\newcommand\movie{\ensuremath{\mathcal{MOVIE}}}
\newcommand\far{\ensuremath{\operatorname{Far}}}
\newcommand\near{\ensuremath{\operatorname{Near}}}

\newcommand\Cause{\operatorname{Cause}}

\usepackage[textsize=scriptsize]{todonotes}

\title{Strict Self-Assembly of Discrete Self-Similar Fractals in the abstract Tile-Assembly Model}

\author{Florent Becker \and Daniel Hader \and Matthew J. Patitz}
\date{}

\begin{document}

\maketitle

\begin{abstract}
This paper answers a long-standing open question in tile-assembly theory, namely that it is possible to strictly assemble discrete self-similar fractals (DSSFs) in the abstract Tile-Assembly Model (aTAM). We prove this in 2 separate ways, each taking advantage of a novel set of tools. One of our constructions shows that specializing the notion of a \emph{quine}, a program which prints its own output, to the language of tile-assembly naturally induces a fractal structure. The other construction introduces \emph{self-describing circuits} as a means to abstractly represent the information flow through a tile-assembly construction and shows that such circuits may be constructed for a relative of the Sierpinski carpet, and indeed many other DSSFs, through a process of fixed-point iteration. This later result, or more specifically the machinery used in its construction, further enable us to provide a polynomial time procedure for deciding whether any given subset of $\mathbb{Z}^2$ will generate an aTAM producible DSSF. To this end, we also introduce the \emph{Tree Pump Theorem}, a result analogous to the important \emph{Window Movie Lemma}, but with requirements on the set of productions rather than on the self-assembling system itself.
\end{abstract}

\thispagestyle{empty}

\clearpage
\pagenumbering{arabic}

\section{Introduction}
\label{sec:introduction}

In tile-based self-assembly or \emph{tile-assembly}, primitive units called \emph{tiles} combine via local interactions to form complex structures. The study of tile-assembly is ostensibly largely motivated by work in DNA-nanotechnology where synthetic DNA strands are made to combine, through the mechanism of nucleotide base pairing dynamics, into structures that resemble microscopic square tiles with selectively sticky sides. With careful design, these DNA tiles, under well chosen conditions (temperature, concentration, salinity, etc.), then tend to combine with each other to form desired structures or even follow the execution of algorithms \cite{woods2019diverse,SeemanMaoLaBeanReif,Reif02,SchulmanRedundancy,rothemund2004algorithmic,evans2014crystals,SchulYurWinfEvolution,OrigamiSeed}. The abstract Tile-Assembly Model (aTAM) is a simplified, mathematical model of tile-assembly, and while the aTAM is often used to aid in the design of DNA-based self-assembling systems \cite{rothemund2004algorithmic,OrigamiSeed,evans2014crystals}, it has been known since its introduction that the aTAM exhibits rich computational dynamics and even Turing universality\cite{Winf98}. Since then, theoretical investigations into the aTAM (and variants thereof) have unveiled a complex tapestry of hierarchies, asymptotic bounds, and intricate algorithmic constructions which has blossomed into a sort of computability and complexity theory for these uniquely geometric computational models \cite{SolWin07,NeedForSeed,RotWin00,Versus,BeckerTimeOpt,DDDxModelSimICALP,jVersus1,SummersRectangles,DDDIU,Temp1PathsSTOC,TempOneNotIU,WoodsMeunierSTOC,WoodsIUSurvey,DirectedNotIU}. Even compared to other geometric computational models such as Cellular Automata \cite{gutowitz1991cellular,mitchell2005computation} or Tilings \cite{penrose1979pentaplexity,Culik1997,ConwayTilingGroups}, the aTAM has uniquely asynchronous dynamics, making this landscape of results quite a change of scenery.

One method for investigating the aTAM involves characterizing the shapes and patterns it is capable of assembling. Early results in this direction found, for instance, that any finite shape may be constructed with an asymptotically optimal number of unique tile types (proportional to the Kolmogorov complexity of the shape) if one is willing to scale the shape by some finite but typically unfortunately large factor\cite{SolWin07}. While the simulation of a Turing machine within the aTAM is relatively simple, used there (and in many other results) as a primitive \emph{tile gadget} within a more complex aTAM construction, this result in particular illustrates an important specificity of the aTAM compared to more conventional models of computation\footnote{with perhaps the exception of cellular automata}, namely the relevance of how the computation is embedded into the geometric construction process of shapes. This makes it pertinent to consider what shapes are difficult (or impossible) to self-assemble. As is the case in many areas of study focused on geometry, fractals provide an interesting challenge.

While no single definition suffices for the term ``fractal'', convenient for our purposes are \emph{discrete self-similar fractals} (DSSFs) \cite{jSSADST}, the limit points of self-substitution tilings of finite connected subsets of $\mathbb{Z}^2$. DSSFs are infinite and aperiodic which is enough in its own right to make a construction difficult, but further still DSSFs are sparse meaning that there is generally little contiguous space for tiles to simulate a Turing machine computation or something analogous. It's important to note here that this difficulty is only present in the context of \emph{strict} self-assembly where tiles are only allowed to occupy locations in the target shape. This is in contrast to \emph{weak} self-assembly wherein tiles may encroach into negative space, and the target shape is characterized by a specific subset of tile types. While it is trivial to weakly self-assemble a host of DSSFs, strict self-assembly of DSSFs has remained an open question. In \cite{jSSADST} it was shown that the DSSF analog of the Sierpiński triangle is impossible to strictly assemble, and such impossibility results have since grown to include subsets of DSSFs such as so-called \emph{pinch point} fractals, etc. \cite{jSADSSF,ScaledPierFractals,HFractals,TreeFractals}. These results depend fundamentally on the fact that each stage of such a fractal is connected by a bounded number of tiles to the previous stage. Interestingly, this seems to hold ``just barely'', and in \cite{LutzShutters12} and \cite{jSSADST} it was shown that by allotting only a logarithmically increasing amount of additional connectivity between fractal stages, using processes called \emph{lacing} and \emph{fibering} respectively, it becomes possible to strictly self-assemble these DSSFs. Another feature of fractal shapes is a notion of dimension which may generally fall between integer values. More precisely, this notion generally assigns some value to how the amount of space occupied by the shape scales with its radius. In the context of DSSFs in the aTAM, $\zeta$-dimension \cite{ZD} serves as a convenient and common choice for this metric. It may be tempting to presume that a non-integer $\zeta$-dimension plays a role in characterizing the assemble-able shapes. Yet, there have been aTAM constructions which can be configured to assemble shapes with any $\zeta$-dimension among a dense subset of the interval $(1, 2)$, though the resulting shapes are distinctly not DSSFs nor are they even sparse\cite{hader2023fractalNACO}.

In this paper, we close the question of strictly assembling DSSFs in the positive. We show not only that DSSFs can be assembled in the aTAM, we do so in 2 separate ways, and provide a polynomial time algorithm for deciding whether a DSSF generated by a given substitution rule can be assembled. While these results answer a long-standing open question in the field, more important are the novel and, we believe, delightful techniques developed for their construction. One of our DSSF constructions arises naturally from the question ``what does a quine look like in the aTAM?'' where a quine typically refers to a program which prints its own description. Here we substitute typical notions of \emph{description} in the context of a universal Turing machine with that of an \emph{intrinsically universal} (IU) tileset in the aTAM \cite{IUSA}. An IU tileset is one capable of simulating all other aTAM systems in some class, not just symbolically as with a Turing machine, but geometrically so that the full dynamics of tile attachment are captured at scale. Our quine therefore is a tile system which grows into a scaled version of itself. Our IU tileset, while restricted to simulating just a subset (albeit useful subset) of all possible aTAM systems, is also certainly the smallest IU tileset published that has been fully implemented in an aTAM simulator and can be downloaded \cite{QuineAssemblySoftware}. This construction can also be made to have essentially any $\zeta$-dimension desired between $1$ and $2$. Our other DSSF construction relies on a novel scheme for identifying self-assemble-able shapes by abstracting the flow of information through an aTAM system's tiles into what we call \emph{self-describing circuits}. These circuits translate naturally into the tile types necessary for assembly while allowing for precise reasoning about their functionality, and through fixed-point iteration we construct a self-describing circuit for a scaled version of a close relative of the Sierpiński Carpet. This latter construction generalizes to all generators (subsets of $\mathbb{Z}^2$ to be self-substitution-tiled) with sufficient \emph{bandwidth}, a measure of a shape's ability to transmit information between their opposite sides. Namely, this generalization relies on having at least 2 disjoint paths between the both the north and south sides and the east and west sides of the generator simultaneously.

Finally, and perhaps most importantly, we characterize the set of generators  that result in DSSFs which can be self-assembled in the aTAM. Specifically, we show that the assembly of a DSSF in the aTAM fundamentally relies on the bandwidth requirement of our circuit construction: its bandwidth requirement is actually a characterization up to a finite amount of iterations of the generator. Moreover, it may be determined for any generator shape in polynomial time using a maximum flow algorithm. This result is not only satisfyingly intuitive, especially given that the impossibility proof for constructing pinch point fractals depends on the same intuition, but additionally our proof involves a wide range of novel devices which we believe are of general interest to the tile-assembly community. This includes, among other things, the use of ordinal indexed assembly sequences to reason about infinite ones and an important theorem we call the \emph{Tree Pump Theorem} which serves a similar purpose to the incredibly useful \emph{Window Movie Lemma}\cite{meunier_intrinsic_2014}.
%, while being more readily applicable in situations the latter is not.
Especially striking is the fact that the Tree Pump Theorem's hypotheses are about the shape of the productions of a system rather than the system itself, allowing anyone to prove in one fell swoop that a given shape cannot be obtained by \emph{any} aTAM system.

\section{Overview of the Concepts}
\label{sec:definitions}

%\section{Preliminary definitions}\label{sec:prelims}

Please see Section \ref{sec:extended-defs} for more detailed definitions, and \cref{fig:concept_map} for the articulation of the concepts and results of the paper.

\subsection{The abstract Tile Assembly Model}\label{sec:aTAM}

The definitions we use for the aTAM are borrowed and adapted from \cite{DDDIU} and we note that \cite{RotWin00} and \cite{jSSADST} are good introductions to the model for unfamiliar readers.
The basic components of the aTAM are unrotatable 2-dimensional square \emph{tiles}. Each side of a tile can have a \emph{glue}, each of which is composed of a string \emph{label} and an integer \emph{strength}. All glues with strength 0 are assumed to have identical labels and are called \emph{null} glues. A \emph{tile type} is uniquely defined by an assignment of glues to its four sides, and a tile is an instance of a tile type. When two tiles are placed so that they have abutting sides, if those two sides share the same glue, they \emph{bind} with the strength of that shared glue. An \emph{assembly} is a configuration of tiles in $\mathbb{Z}^2$, and it is said to be $\tau$-stable if bonds whose strengths sum to at least $\tau$ must be broken in order to separate the assembly into two or more components. We say that assembly $\alpha$ is a \emph{subassembly} of $\beta$ (written $\alpha \sqsubseteq \beta$) if $\dom \alpha \subseteq \dom \beta$ and $\alpha$ and $\beta$ have the same tile types on their shared domain.

A \emph{tile-assembly system} (TAS) is a triple $\calT=(T,\sigma,\tau)$, where $T$ is a tileset, $\sigma$ is a finite $\tau$-stable assembly called the \emph{seed assembly}, and $\tau\in\mathbb{Z}^+$ is called the \emph{binding threshold} (a.k.a. \emph{temperature}). A tileset must be finite, but an infinite number of copies of each tile type are assumed to be present.
%If the seed $\sigma$ consists of a single tile, i.e. $|\sigma| = 1$, we say $\calT$ is \emph{singly-seeded}.
If an assembly $\alpha$ can be produced in $\calT$ by a series of single-tile $\tau$-stable additions, starting from the seed assembly $\sigma$, we say that $\alpha$ is a \emph{producible} assembly, and use $\prodasm{\calT}$ to denote the set of producible assemblies. If no additional tiles can $\tau$-stably bind to $\alpha$, we say that $\alpha$ is a \emph{terminal} assembly, and use $\termasm{\calT}$ to denote the set of terminal assemblies. If $|\termasm{\calT}| = 1$, i.e., there is exactly one terminal assembly, we say that $\calT$ is \emph{directed}. Note that both of our DSSF constructions are directed.
%When $\calT$ is clear from context, we may omit $\calT$ from notation.
We say $\calT$ \emph{strictly self-assembles shape $S$} if for any $\alpha \in \TerminalProds{\calT}$, $\dom \alpha = S$.

%Given a location $\vec{l} \in \mathbb{Z}^d$, let $\var{Color}(\alpha,\vec{l})$ define a function that takes as input an assembly and a location and returns the color of the tile at that location (and is undefined if $\vec{l} \not \in \dom(\alpha)$).
%
%If $\calT$ is directed and $\alpha$ is its unique terminal assembly\todo{Change this definition so it doesn't require directedness}, we say that $\calT$ \emph{weakly self-assembles pattern $P$} iff $\dom_{C_P}(\alpha) = P$ and $\forall (\vec{l},c) \in P, c = \var{Color}(\alpha,\vec{l})$.
%
%We say $\calT$ \emph{weakly self-assembles shape $P$} iff for all $\alpha \in \mathcal{A}_{\Box}[\mathcal{T}]$, $\dom_{C_P}(\alpha) = P$ and $\forall (\vec{l},c) \in P, c = \var{Color}(\alpha,\vec{l})$.
%

When a tile initially attaches to an assembly, we call the sides with glues that perform that initial bonding its \emph{input sides}, and any other sides with non-null glues, to which later attaching tiles bind, its \emph{output sides}. We define \emph{standard} aTAM systems as those which (1) are directed, (2) have for each tile type fixed and distinct sets of input sides and output sides, (3) produce no assemblies with an input side glue on its exterior, (4) have for each tile type the fact that its input sides are either a single side with a strength-2 glue or two diagonally adjacent sides with glues of strength-1, (5) have no glue mismatches in the single terminal assembly, and (6) never produce assemblies containing locations with cumulative inputs of strength 2 without a corresponding tile to attach in that location.
Since the aTAM is computationally universal \cite{Winf98}, and relatively straightforward algorithms exist for performing computational simulations of aTAM systems, there exists an aTAM tileset $S$ such that, for any arbitrary aTAM system $\calT$, a system using $S$ that is appropriately seeded can computationally simulate $\calT$. However, in \cite{IUSA} it was shown that a more natural notion geometry-preserving simulation is also possible. This notion of simulation, called \emph{intrinsic simulation}, occurs when, for some $c \in \mathbb{Z}^+$, each $c \times c$ block of tiles of a simulating system $\mathcal{S}$ are mapped to a single tile of the system $\calT$ being simulated. These $c \times c$ blocks of tiles are referred to as \emph{macrotiles} and they must be formed in orderings that match the orderings of tile attachments in $\calT$. (For technical details and definitions related to the aTAM and intrinsic simulation, please refer to Section \ref{sec:aTAM-extended}.)

\subsection{Discrete Self-Similar Fractal Shapes}

Fractals are usually defined in some continuous space such as $\mathbb{R}^2$ or $\mathbb{C}^2$. In the context of self-assembly, one needs to work with \emph{discrete} fractals which are defined as fixed points of some 2 dimensional substitution.
A substitution $\sigma_G(X):\mathcal{P}(\mathbb{Z}^2) \to \mathcal{P}(\mathbb{Z}^2)$ based on a finite pattern $G$, called the \emph{generator}, replaces each pixel in a pattern $X$ with a copy of $G$.

\begin{definition}[Dicrete Self-Similar Fractal]
  Let $G$ be a finite subset of $\mathbb{N}^2$ with $(0, 0) \in G$. The discrete self-similar fractal shape with generator $G$ is 
  \( G^{\infty} = \bigcup_{i \in \mathbb{N}} \sigma_G^i(\{(0,0)\}) \subset \mathbb{N}^2. \)
  The $i$-th step of the fractal is $G^i = \sigma_G^i(\{(0, 0)\}$.
\end{definition}

The \emph{Sierpinski's Cacarpet} is the self-similar fractal shape $\cacarpet$ with generator \(K = \{0,\ldots, 5\}^2 \setminus \{2, 3\}^2\), represented on \cref{fig:cacarpet_main}; $\kappa = \sigma_K$ is the associated substitution.

\begin{figure}
  \centering
  \begin{tikzpicture}
    \input{tikz/cacarpet}
  \end{tikzpicture}
  \caption{The substitution $\kappa$ of the Sierpinski's Cacarpet.}
  \label{fig:cacarpet_main}
\end{figure}

\section{Overview  of our Results}
\label{sec:tour_of_results}
\subsection{Fine Characterizations of the dynamics of the aTAM}

% Characterizing which fractals can self-assemble in the aTAM requires a fine understanding of the dynamics of this geometric and asynchronous model. 
Three general results about the aTAM form the foundation of our fractal results. Each of them advances the field of self-assembly and has general relevance within it.
% , as well as for the understanding of geometrical, asynchronous computation models. 

\paragraph{Standard aTAM systems are intrinsically universal and admit Quines}
\label{sec:quine-blurb}

\begin{theorem}\label{thm:standardIU}
The class of standard aTAM tile assembly systems is intrinsically universal.
\end{theorem}

Due to space constraints, the details of our construction can be found in Section \ref{sec:standard-IU}. Notably, we fully implemented the tile set using a relatively small number of tile types ($< 10,000)$.
In this section we describe the construction of an aTAM quine. In the broader context of computability theory, a quine refers to a program that, with no input, will output precisely its own description. Here, \emph{description} is generally defined with respect to a universal machine $U$. If simulating $U$ on the pair $(d,x)$ yields the same results as machine $M(x)$ for all $x$, then $d$ is a description of $M$ in the context of $U$. A quine is then simply a description $d$ of a machine $M_d(x) = U(d,x)$ which outputs $d$ for all $x$.
Here we describe a natural sense in which this definition can be sufficiently specialized using the aTAM as the model of computation and intrinsic universality in place of Turing universality. In this context, the universal machine is replaced by an intrinsically universal tile set $U$. The notion of \emph{description} here has to be modified to accommodate this change; for a description of a tile system $\mathcal{T}$ to be meaningful in the context of $U$ requires that $\mathcal{T}$ be encoded in some way that tiles in $U$ can use to intrinsically simulate $\mathcal{T}$. To this end, we introduce a notion of \emph{seeding} where a tile system $\calS$ is said to \emph{seed} a system $\calT$ with respect to $U$ if $\calS$ grows into the shape of a macrotile which can initiate an intrinsic simulation of $\calT$ using the tiles in $U$.

%For languages supporting format strings, Quines are readily constructable by feeding a well designed format string itself to be used as a sub-string. Even in languages without format strings, this behavior may be emulated supposing that there is a flexible enough means for a program to generate, find, and modify the source code of other programs and execute them as sub-routines. Our aTAM quine construction is more akin to the former approach with tile glues being used to encode symbols to be manipulated by tile attachments. Here we formally define an aTAM quine and demonstrate an approach to constructing one.

\begin{figure}
    \centering
    \subfloat[][]{
    \label{fig:quine-overview}
    \includegraphics[width=3.2in]{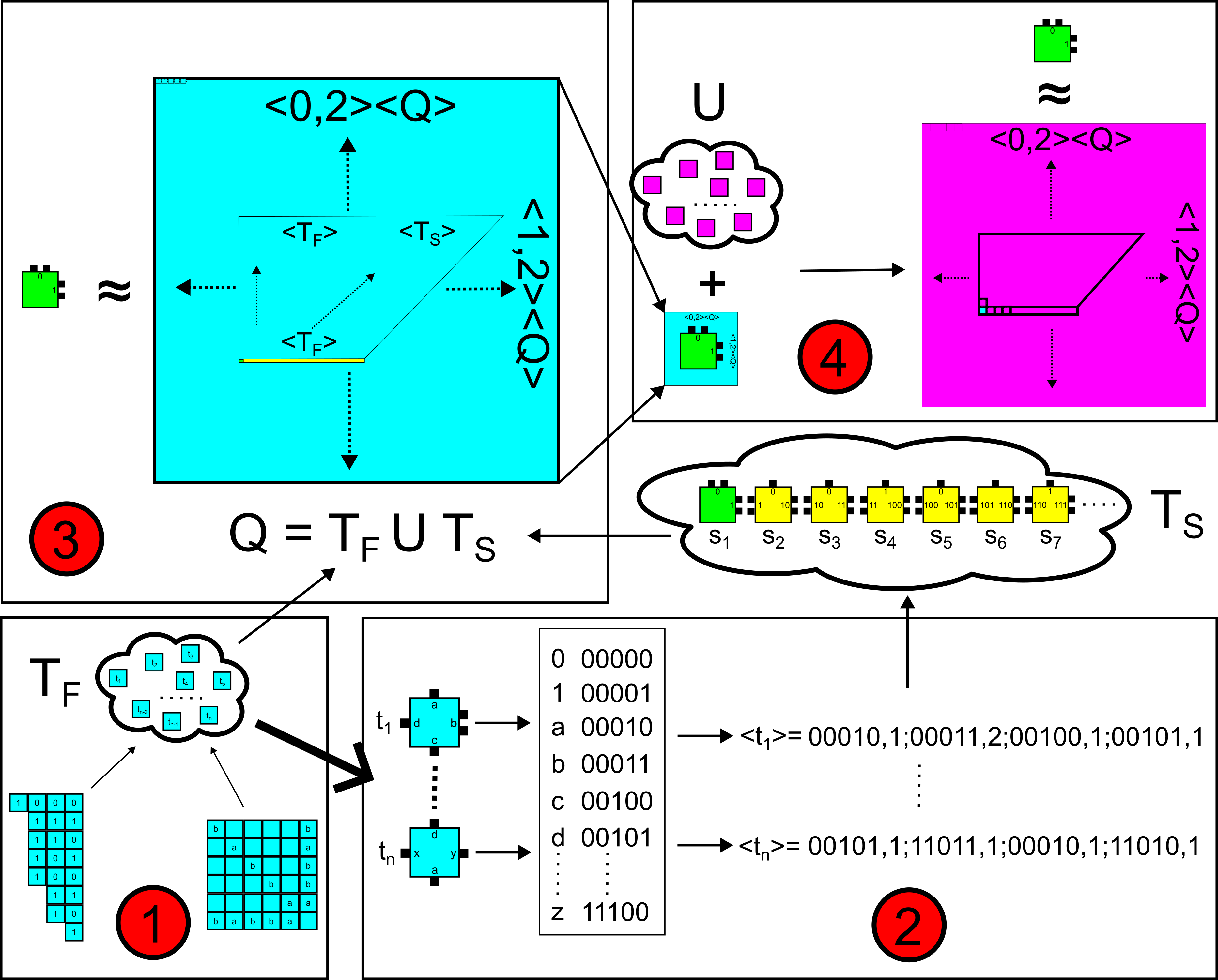}

    }
    \quad
    \subfloat[][]{
    \label{fig:squaring-quine-tour}
    \includegraphics[width=0.4\textwidth]{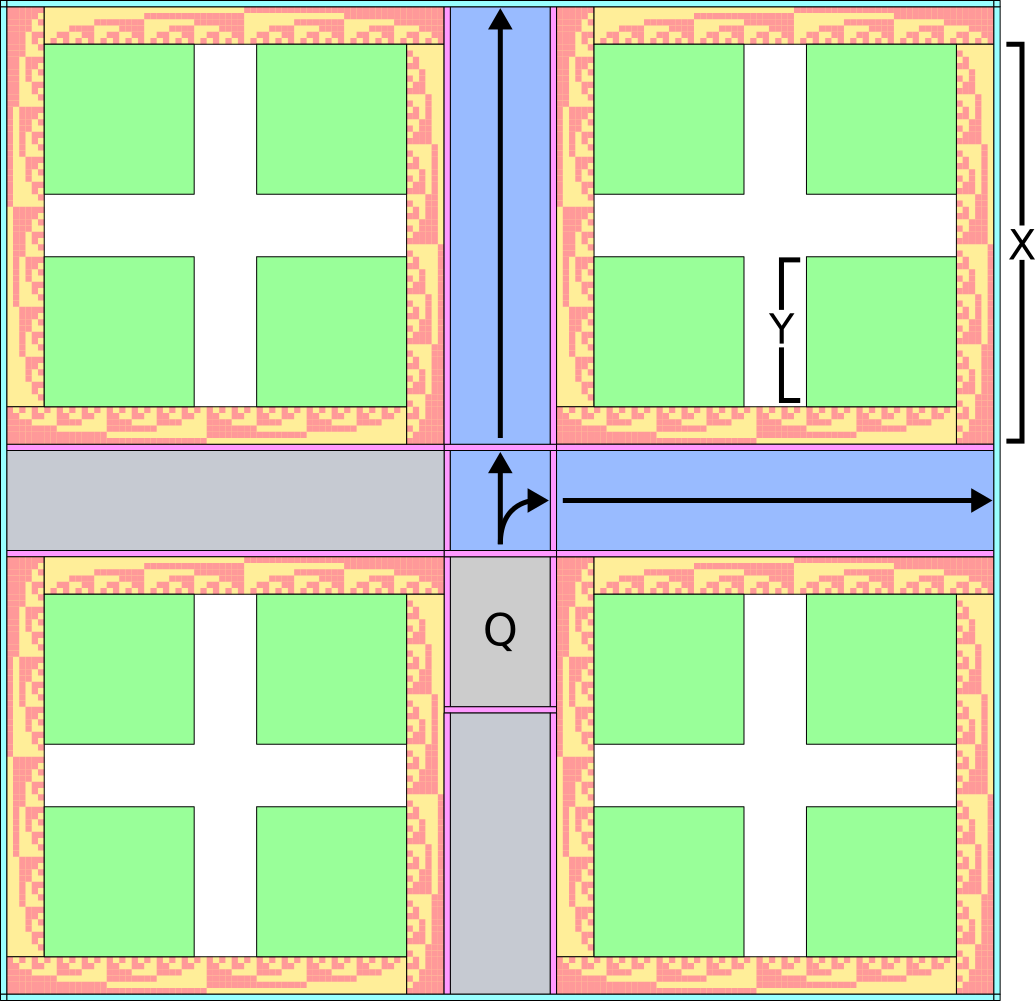}
    }
    \caption{(a) A schematic depiction of the construction of the quine system. (1) Subsets of tile types capable of various algorithmic functions (e.g. binary counting, rotation of values, etc.) make up the ``functional'' tileset $T_F$. (2) The glue labels and tile types of $T_F$ are encoded to generate the ``seed row'' tileset $T_S$ that self-assembles a row presenting an encoding of $T_F$ (plus some additional necessary information) via its northern glues. The seed tile is shown as green and the rest of the seed row as yellow. (3) Using the full tileset $Q =  T_F \cup T_S$, the system grows from the seed tile to form the seed row, then the tiles of $T_F$ cause upward growth that computes the definitions of the seed row tiles (via their northern glues and locations), appending those definitions to those provided by the seed row. This results in an assembly with the full definition of $Q$ encoded in its northern glues. Further growth by the tiles of $T_F$ causes that assembly to grow into a terminal square macrotile that is consistent with the definition of a seed assembly representing the quine system's seed tile for the IU tileset $U$. (4) If the terminal assembly from (3) were used as the seed assembly for a system including the tile types from $U$, that system would simulate the quine system. That is, each tile of $Q$ would be simulated by a macrotile composed of tiles from $U$, resulting in a terminal assembly that is a macro-macrotile representing the seed tile of $Q$. (b) Schematic depiction of the formation of a macrotile square by the quine construction adapted for DSSFs. The dimension $X$, counted by binary counters, can be set to an arbitrary value as long as it is greater than the height of the grey rectangle (created by the earlier stages of the quine assembly growth), and the dimension $Y$ can be set to be any value between $0$ and $X/2$. Careful setting of $X$ and $Y$ values can allow for arbitrary percentages of the area contained within the macrotile to be empty space.}
\end{figure}

\begin{definition}
    Fix a tileset $U$ which is IU for some class $\mathfrak{C}$ of TASs and let the corresponding representation and seed generation functions be $\mathcal{R}$ and $S$ respectively. We say that a TAS $\mathcal{Q}=(Q,\sigma_Q,\tau_Q)$ \emph{seeds} the TAS $\calT = (T, \sigma_T, \tau_T)$ with respect to $(U, \mathcal{R}, S)$ if:
(1) $\calT \in \mathfrak{C}$,
(2) $\left|\sigma_Q\right| = 1$,
(3) $\mathcal{Q}$ is directed, and
(4) the macrotile seed assembly $S(\sigma_T)$ corresponding to $\sigma_\calT$ has the same shape and glues presented on its exterior as the terminal assembly of $\mathcal{Q}$.
\end{definition}

In other words, for $\mathcal{Q}$ to seed $\mathcal{T}$ with respect to $U$ means that $\mathcal{Q}$ will grow into an assembly which behaves in all ways as a valid macrotile seed for the intrinsic simulation of $\mathcal{T}$ by tiles in $U$.

\begin{definition}
\label{def:quine}
    Given IU tileset $U$ (and its corresponding representation and seed generating functions $\mathcal{R}$ and $S$), an \emph{aTAM quine} is a TAS $\mathcal{Q} = (Q, \sigma, \tau)$ that seeds itself with respect to $(U, \mathcal{R}, S)$.
\end{definition}

\begin{restatable}{theorem}{thmQuine}
\label{thm:quine}
There exists an aTAM quine.
\end{restatable}

We prove Theorem \ref{thm:quine} by construction, demonstrating an aTAM quine $\mathcal{Q} = (Q, \sigma, 2)$ for the tileset $U$ that is IU for standard aTAM systems (and its corresponding representation and seed generating functions $R$ and $S$) given in the proof of Theorem \ref{thm:standardIU}. %We first describe our aTAM TAS $\mathcal{Q} = (Q,\sigma,2)$ at a high-level to give an overview of its functional components, then provide more detail about how each component works. 
A high-level, schematic depiction of $\mathcal{Q}$ can be seen in Figure \ref{fig:quine-overview}. Full details of the construction can be found in Section \ref{sec:quine}. Essentially, the tileset $Q$ of $\mathcal{Q}$ consists of a subset of ``functional'' tile types, $T_F$, and a subset of ``seed row'' tile types, $T_S$. The seed row tile types grow into a row that has an encoding of the definitions of the tile types in $T_F$, and then the tiles of $T_F$ attach to that row and grow into an assembly that computes the definitions of the tiles of $T_S$, so that the resulting assembly presents a full definition of $Q$. It then grows into a square macrotile that represents its own seed tile, having the glues of that tile represented on its sides along with the definition of $Q$. 

\paragraph{Embedded Circuits and self-description}

\emph{Self-describing circuits} are a novel device for presenting aTAM systems. These are generally useful for rigorous yet readable proofs of the behaviour of aTAM systems, which are crucial in the context of fractals.

\begin{figure}
    \centering
    \tikz{\matrix (d) at (0, -10) [every node/.style={draw, circle}, matrix of nodes, column sep=3em, row sep = 5ex, a/.style={rectangle, draw=none, inner sep=1mm}] {
            &        & |[a]| $I_2$ &            \\
            & $i_1$  & $i_1$       &            \\
|[a]|$I_0$  & $i_2$  & $m$         & |[a]| $O_0$\\
|[a]|$I_I$  & $m$    & $a$         & |[a]| $O_1$\\
            & $i_1$  & |[red]|$a$         & |[a]| $O_2$\\
};

\begin{scope}[every node/.style={font=\footnotesize}, entry/.style={pos=.9,auto}, exit/.style={pos=.1,auto,swap}, value/.style={pos=.5, fill=gray!20}]
    \draw[->] (d-1-3) edge node[entry] {$0$} node[value] {$7$} (d-2-3);
    \draw[->] (d-2-2) edge[bend right] node[exit] {$0$} node[entry] {$1$} node[value] {$7$} (d-3-2);  % (c-2-3);
    \draw[->] (d-2-3) edge[bend right] node [exit, swap] {$0$} node[entry] {$1$} node[value] {$7$} (d-3-3);

    \draw[->] (d-2-3) edge node [exit] {$0$} node[entry] {$0$} node[value] {$7$} (d-2-2);
    \draw[->] (d-3-1) edge[bend left] node[entry, swap] {$0$} node[value] {$2$} (d-3-2);
    \draw[->] (d-3-2) edge[bend left] node [exit] {$0$} node[entry, swap] {$0$} node[value] {$2$} (d-3-3);
    \draw[->] (d-3-3) edge node[exit] {$0$} node[value] {$4$} (d-3-4);

    \draw[->] (d-4-1) edge[bend left] node[swap, entry] {$0$} node[value] {$3$} (d-4-2);
    \draw[->] (d-4-2) edge[bend left] node [exit] {$0$} node[swap, entry] {$0$} node[value] {$1$} (d-4-3);
    \draw[->] (d-4-3) edge node[exit] {$0$} node[value] {$2$} (d-4-4);

    \draw[->] (d-4-2) edge node[exit] {1} node[value] {$2$} node[entry] {0} (d-5-2);
    \draw[->] (d-3-3) edge[bend right] node[exit, swap] {$1$} node[entry] {$1$} node[value] {$1$} (d-4-3);
    \draw[->] (d-3-2) edge[bend right] node[exit] {$1$} node[entry] {$1$} node[value] {$7$} (d-4-2);

    \draw[->] (d-5-2) edge[bend left] node[exit] {0} node[entry, swap] {1} node[value] {2} (d-5-3);
    \draw[->] (d-5-3) edge node[exit] {0} node[value] {$2$} (d-5-4);
%    \draw[->] (d-4-2) edge node[exit] {1} node[value] {$2$} node[entry] {0} (d-5-2);
    \draw[->] (d-4-3) edge[bend right] node[exit] {$1$} node[value] {0} node[entry] {0} (d-5-3);
\end{scope}

\tikzset{
  inputN/.pic={
    \node[draw, circle, inner sep=2pt] (gate) at (0,.5) {};
    \draw (gate) edge[bend right, ->] node[value, right=3pt] {$0$} +(0,-1.2);
  }
}
\tikzset{
  inputW/.pic={
    \node[draw, circle, inner sep=2pt] (gate) at (-.5, 0) {};
    \draw (gate) edge[bend left, ->] node[value,below=3pt] {$2$} +(1.2, 0);
  }
}

\tikzset{value/.style={pos=.5, fill=gray!20}}

\matrix[row sep=1ex, every node/.style={anchor=center}, right=of d] {
 %\node[draw, circle] (gate) at (0,0) {$a$};
 %\draw (-1.5,.2) edge[bend left, ->] node[value] {$2$} (gate);
 %\draw (-.2, 1.5) edge[bend right, ->] node[value] {$0$} (gate);
 %\draw (gate) edge[->]  node[value] {$2$} +(1.5, 0);\\
 \begin{scope}[scale=2, shift={(-.5, -.5)}]
     \draw[red] (0,0) rectangle (1,1) node[pos=.5, yshift=-1em, align=center] {$a:$\\$(0,2) \mapsto 2$};
     \draw[xshift=-2pt] (1, 0) -- (1, .4) (1, .6) -- (1,1);
     \draw (.7, .5) edge[->] node [value, above=1pt] {$2$} (1, .5);
     \draw (.65, 1) edge[->, bend right] node [value, left=1pt] {$0$} (.65, .7);
     \draw (0, .5) edge[->, bend left] node [value, above=1pt] {$2$} (.3, .5);
     \node[rectangle callout, draw, callout absolute pointer={(.3, 1)}] at (.5, 1.3) {strength $1$};
     \node[rectangle callout, draw, callout absolute pointer={(0, .3)}, anchor=west, fill=white] at (-1.2, -.2) {strength $\tau - 1$};
     \node[rectangle callout, draw, callout absolute pointer={(.95, 5pt)}, anchor=north east, fill=white] at (1, -.4) {strength $\tau$};
  \end{scope}\\
  \scoped[shift={(0,-1.5)}, scale=.8]{
     \neighcrosspics{inputN}{}{}{inputW}
  }
  \\
};}
    \caption{Left, a $2$-digit by $1$-digit multiplier computing $32$ (left, written bottom-to-top) times $7$ (top) is $224$ (right, bottom-to-top). When a gate has two input wires, each of them bends towards the other wire. Right, the tile type corresponding to the $a$ gate at the bottom-right of the circuit, and a neighborhood where it can attach. The temperature $\tau \leq$ is the maximum in-degree of a gate in the circuit (here, 2).}
    \label{fig:mod3-adder-tour}
\end{figure}

A circuit is made of gates positioned on a subset $D$ of $\ZZ$, as in \cref{fig:mod3-adder-tour}. Each of these gates has a \emph{function}, computing the outputs of the gate from the inputs. The gate is embedded on a unit square according to its \emph{wiring}, which states where its inputs come from and where its outputs go. The wiring and the function must be compatible: the wiring must have as many inputs as the function. A circuit is \emph{evaluable} if it is ready to be evaluated, that is, if it has no input wires, loops or infinite backwards paths.

These circuits are self-descriptive when there is a decoding function $\decgate$ which determines the gate at any position $p$ of the circuit from the inputs it receives. If this is the case, the circuit can be self-assembled in the aTAM. In particular, the set $D \subset \ZZ$ of positions where gates appear can be strictly self-assembled in the aTAM.

\begin{restatable}{theorem}{circuitToAtam}

  \label{thm:circuit-to-atam}
  Let $C$ be a self-describing evaluable circuit with only one gate without inputs. Then there is a standard aTAM system $S_C$ which strictly self-assembles $\operatorname{dom}(C)$.
\end{restatable}

The second column of \cref{fig:mod3-adder-tour} shows the principle of the reconstruction of the gates by the aTAM. Each tile corresponds to a gate with a set of input/output values, and the set of glues is the set of wires with values on them. Each wire encodes the set of directions with incoming wires in the gate, so the strength of each glue can be chosen so that each tile attaches when all its inputs are present.

Each of the glues on the input sides of the gate then determines the set of input sides and by self-description, the combination of inputs on that set of sides is unique.
%In \cref{}

\paragraph{The Tree Pump Theorem}

This pumping principle in the style of \cite{meunier_intrinsic_2014} deals with skinny productions, that is those which do not encircle any square larger than $m$ for some fixed $m$. The crucial property which makes them simple(-ish) is their bounded treewidth.

\begin{restatable}[Tree Pump]{theorem}{treepump}
%\begin{theorem}[Tree Pump]
  \label{thm:tree_pump}
  For any aTAM system $\calT = (S, \sigma, \tau)$ with $\sigma$ finite and connected, define the following sets of assemblies:
  \begin{itemize}[topsep=0pt,itemsep=-1ex,partopsep=1ex,parsep=1ex]
  \item for any integer $m$, $C_{m}[\calT]$ is the set of assemblies of $\calT$ which encircle an $m \times m$ square;
  \item for any real $k$ and vector $\vec{d}$, $B_{k,\vec{d}}[\calT]$ is the set of assemblies of $\calT$ which do not cover any position $\vec{p}$ such that $\vec{p} \cdot \vec{d} > k + |\dom \sigma|$
  \item for any vector $\vec{d}$, $P_{\vec{d}}[\calT]$ is the set of \emph{ultimately periodic} assemblies $A$ of $\calT$ such that there is a vector $\vec{p}$ with $\vec{p} \cdot \vec{d} > 0$ and a non-empty sub-assembly $a \sqsubseteq A$ such that $a + \vec{p} \sqsubseteq a$.
  \end{itemize}

  Then, there is a function $F: \mathbb{N} \times \mathbb{N} \to \mathbb{N}$ such that for aTAM system $\calT$ with $n$ tile types and a 1-tile seed, integer $m$ and unit vector $\vec{d}$ of $\mathbb{R}^2$,
  \[ \TerminalProds{\calT} \cap (C_m[\calT] \cup B_{F(n,m),\vec{d}}[\calT] \cup P_{\vec{d}}[\calT]) \neq \emptyset. \]
%\end{theorem}
\end{restatable}

\begin{figure}[h]
  \centering
    \tikzset{
      chemin/.pic={
        \begin{scope}[scale=.1]
          \foreach \x/\y in {0/0, 1/0, 2/0, 2/1, 2/2, 3/0, 1/2, 4/0, 4/1, 4/1}
          \path[draw, fill=gray!50!white] (\x, \y) -- ++(1,0) -- ++(0, 1) -- ++(-1,0) -- cycle;
          \coordinate (-bend) at (4, 0);
        \end{scope}
      }
    }

    \tikzset{
      u/.pic={
      \begin{scope}[scale=.8]
          
        \filldraw[fill=blue!50!white] (0,0) rectangle (.1, .1);
        \filldraw[fill=blue!50!white] (0,-.1) rectangle (.1, 0);
        \filldraw[fill=blue!50!white] (0,-.2) rectangle (.1, -.1);
        \filldraw[fill=blue!50!white] (.1,-.2) rectangle (.2, -.1);
        \filldraw[fill=blue!50!white] (.2,-.2) rectangle (.3, -.1);
        \filldraw[fill=blue!50!white] (.2,-.2) rectangle (.3, -.1);
        \filldraw[fill=blue!50!white] (.2,-.1) rectangle (.3, 0);
        \filldraw[fill=blue!50!white] (.2,0) rectangle (.3, .1);
        \filldraw[fill=blue!50!white] (.3,0) rectangle (.4, .1);
        \filldraw[fill=blue!50!white] (.4,0) rectangle (.5, .1);
      \end{scope}
      }
    }
    
    \tikzset{
      unbounded/.pic={
        \begin{scope}[scale=.5, shift={(-3, -3)}]
          \clip[draw] (-1, -1) rectangle (10, 4.5);
          \node[fill=red, circle, inner sep=1pt] (origin) at (0, 0) {};
          \node[left=.1em of origin] {$\vec{0}$};
          \coordinate (path_dest) at (5.4, 2);
          \coordinate (foot) at (1.2, 0);
          \pic[fill=gray!50!white] (invader) at (0, 0) {spaceinvader};
          \pic[fill=gray!50!white, yscale=-1, rotate=-90] (invader2) at (invader-right-foot) {spaceinvader};
          \pic[fill=gray!50!white, yscale=-1] (invader3) at (invader2-right-antenna) {spaceinvader};
          \pic[fill=gray!50!white] (invader4) at (invader3-right-foot) {spaceinvader};
          \pic[fill=gray!50!white, rotate=90] (invader5) at (invader4-right-antenna) {spaceinvader};
          \pic[fill=gray!50!white, rotate=90] (invader6) at (invader2-right-foot) {spaceinvader};
          
          % \path (foot) -- (path_dest) foreach \t [count=\ti] in {0, .2, ..., 1} {pic[pos=\t]  (chemin-\ti) {chemin}};
          \path (invader4-right-foot) -- +(10, 0) foreach \t in {0, .08, ..., 1} { pic[pos=\t] {u}};
          \node[blue, inner sep=0, fill=white, below=10pt of invader4-right-foot, anchor=north west] {periodic};

          \coordinate[shift={(1, .1)}] (a) at (invader4-right-foot);
          \coordinate[shift={(35:2)}] (b) at (a);
          \coordinate[shift={(1,0)}] (c) at (a);
          \path[purple, ->] (a) edge node[pos=.8, auto] {$\vec{d}$} (b); % ++(35:5);
          % \path[blue, ->] (a) edge node[pos=.8, auto] {$\vec{p}$} ++(0:4);
          \pic[pic text={$< \frac{\pi}{2}$}, draw=purple, thick, angle radius=4em] {angle = c--a--b};
        \end{scope}
      }      
    }

    \tikzset{
      bigCycle/.pic={
        \begin{scope}[scale=.5, shift={(-4, -4)}]
          \clip[draw] (-1, -.5) rectangle (6,5);
          \pic[fill=gray!50!white] (invader) at (0, 0) {spaceinvader};
          \path [draw=none, fill=yellow] (2, 1.5) rectangle +(2,2);
          \path[<->, dashed] (2,1.5) edge node[midway, above] {$m$} ++(2,0) edge node[midway, left] {$m$} +(0,2);
          \path (invader-right-antenna) -- ++(4, 0) 
             foreach \t in {0, .05, ..., 1} { pic[pos=\t] {code={\filldraw[thin,fill=gray!50!white] (0,0) rectangle (.1, .1);}}}
             -- ++(0, 3)
             foreach \t in {0, .05, ..., 1} { pic[pos=\t] {code={\filldraw[thin,fill=gray!50!white] (0,0) rectangle (.1, .1);}}}
             -- ++(-4, 0)
             foreach \t in {0, .05, ..., 1} { pic[pos=\t] {code={\filldraw[thin,fill=gray!50!white] (0,0) rectangle (.1, .1);}}}
             -- cycle
             foreach \t in {0, .05, ..., 1} { pic[pos=\t] {code={\filldraw[thin,fill=gray!50!white] (0,0) rectangle (.1, .1);}}};
        \end{scope}
      }
    }

    \tikzset{
      bounded/.pic={
        \begin{scope}[scale=.5]%, shift={(-1, -1)}]
          \clip[draw] (-2, -.5) rectangle (5.5, 5);
          \coordinate (origin) at (0, 0);
          \coordinate (d_circle) at (35:3);
          \coordinate (d_oppcircle) at (35:-2.5);
          \path[->] (origin) edge node[pos=.8, auto] {$F(n, m) \vec{d}$} (d_circle);
          \pic[fill=gray!50!white] (invader) at (0, 0) {spaceinvader};
          \node[fill=red, circle, inner sep=1pt] at (origin) {};
          \node[left=.1em of origin] {$\vec{0}$};
          \path[draw, gray, dashed] (d_circle) -- +(125:10);
          \path[draw, gray, dashed] (d_circle) -- +(125:-10);
          \fill[pattern=checkerboard, pattern color=red!50, path fading=east, rotate around={35:(d_circle)}] (d_circle) +(0, -10) rectangle +(10, 10);
          \path (d_circle) -- +(1.5,1.5) node [fill=white, rounded corners=2pt] {no tile};
        \end{scope}
      }
    }

    \[
        \begin{matrix}
            \left\{\tikz[baseline=(current bounding box.center)]{\pic{bigCycle};}\right\} & \left\{\tikz[baseline=(current bounding box.center)]{\pic{bounded};}\right\} & \left\{\tikz[baseline=(current bounding box.center)]{\pic{unbounded};}\right\}\\
            C_m[\calT]&B_{F(n,m),\vec{d}}[\calT]&P_{\vec{d}}[\calT]\\
        \end{matrix}
    \]
    \caption{The three sets defined by~\cref{thm:tree_pump}: $C_{m}[\calT]$ is the assemblies which encircle an $m \times m$ square, $B_{F(n, m), \vec{d}}[\calT]$ is the set of assemblies which do not reach further than $F(n,m)$ in direction $\vec{d}$, and $P_{\vec{d}}[\calT]$ is the assemblies which contain a periodic path with period $\vec{p}$ such that $\vec{p}\cdot\vec{d} > 0$.}
\label{fig:tree_pump_tour}
\end{figure}

The cases of \cref{thm:tree_pump} are detailed in \cref{fig:tree_pump_tour}. In order to do a meaningful amount of computation, a self-assembling system with $n$ tiles needs to encircle large squares ---say, of size $m$, thus hitting $C_{m}[\Ss]$. If it does not, then its productions look very much like trees drawn on $\mathbb{Z}^2$ with a $m$-cell wide brush. A branch of that tree then behaves like a finite automaton. If that automaton stops, the assembly goes no further than $F(n, m)$ in any given direction $\vec{d}$, which gives a final production in $B_{F(n,m), \vec{d}}[\Ss]$. If not, these long branches must have an ultimately periodic behavior which gives a final production in $P_{\vec{d}}[\Ss]$.
The proof proceeds in two parts, all along assuming that there are no $m \times m$ squares encircled by a production of $\calT$: first the tree-like nature of the productions of $\calT$ is formalized through a \emph{tree decomposition}. Then, the proof considers generalized assembly sequences, with ordinal time and the freedom to escape $\ZZ$. In this generalized context, repeatedly using an adaptation of the Window Movie Lemma~\cite{meunier_intrinsic_2014} on a long enough branch of the tree decomposition yields an ultimately periodic path.
The full proof is given in~\cref{sec:tree_pump}, as it involves a fair few ancillary definitions.

% \begin{proof}
%   Let $n, m \in \mathbb{N}^2$. Let $N = m^4 \cdot n$, $N$ is greater than the number of sequences of additions of elements of an $n$ element set into distinct points of an $m \times m$ square.
% \end{proof}

\subsection{From an aTAM Quine to a self-assembled DSSF}

Our first construction capable of strictly self-assembling DSSFs in the aTAM uses a smaller number of tile types (combining the tileset that is IU for standard aTAM systems and the quine tileset), but a very large scale factor.

\begin{theorem}\label{thm:dssf}
    For $z, \epsilon \in \mathbb{Q}$, where $1 < z < 2$ and $\epsilon < 1$, there exists an aTAM system $\mathcal{F}_{z,\epsilon}$ such that $\mathcal{F}_{z,\epsilon}$ strictly self-assembles a DSSF whose $\zeta$-dimension is $z \pm \epsilon$.
\end{theorem}

We prove Theorem \ref{thm:dssf} by construction. Given an arbitrary  $z, \epsilon \in \mathbb{Q}$, where $1 < z < 2$ and $\epsilon < 1$, we demonstrate an aTAM system $\mathcal{F}_{z,\epsilon} = (T,\sigma,2)$ that strictly self-assembles a DSSF whose $\zeta$-dimension is $z \pm \epsilon$. By including the IU tileset $U$ in the quine construction, the macrotile resulting from the quine's growth will then enable tiles in $U$ to simulate additional macrotiles. These macrotiles will grow in the same fashion as the individual tiles that attached to grow the quine in the first place. This process will continue at larger scales indefinitely. Because intrinsic simulation is a transitive relation (see Section~\ref{sec:DSSFs}) and because our IU system uses macrotiles with the same shape as our squared quine and does not place tiles in any macrotiles that will not map to tiles, this simulation at increasing scales produces a DSSF.
% This is done by expanding the tile types that make up the seed row of the quine construction to also include the definitions of tileset $U$ (the tileset that is IU for standard aTAM systems). This causes simulation of the original quine system by the tiles from $U$, and then for macrotiles of tiles from $U$ to simulate individual tiles of $U$ so that the system is simulated at a greater scale factor (i.e. with the entire assembly representing the first level of simulation to represent a single tile). Simulation continues to occur at an infinitely increasing set of scale factors, resulting in the recursive structure of a DSSF. 
Additionally, our construction enables the $\zeta$-dimension to be tuned by adjusting the amount of empty space in each macrotile as depicted in Figure~\ref{fig:squaring-quine-tour}
% in order to tune the fractal dimension of the assembly, empty space can be left in each of the four corners of every macrotile (as depicted in Figure \ref{fig:squaring-quine} for the quine's macrotile).

\subsection{A self-assembled DSSF described as an Embedded Circuit}

The second construction relies on \cref{thm:circuit-to-atam}: any shape which can be endowed with a self-describing circuit can be \emph{strictly} self-assembled. Taking inspiration from \cite{goodman1998matching} and others in this tradition, we observe that locally describing a recursive and self-similar structure such as that of a DSSF can be done by combining information about several levels in the substitution hierarchy in order to compute the address of each position. This affords strict-self-assembly of a DSSF with scale factor only 6, namely the Sierpiński Cacarpet $\cacarpet$ given on \cref{fig:cacarpet}. Alas, because the tileset is generated by a general theorem, its number of tile types ends up in the hundreds of millions.

\begin{restatable}{theorem}{cacarpetCircuit}
  \label{thm:cacarpet_circuit}
  There is an evaluable circuit $C_{\square}$ with domain $\cacarpet$ and with an empty input bus which is self describing. Moreover, $C_{\square}$ has only one gate without inputs.
\end{restatable}

\begin{corollary}
  \label{thm:cacarpet_ass}
  There is an aTAM system at $\tau=2$ which self-assembles the DSSF pattern $\cacarpet$.
\end{corollary}

 In order to construct the circuit $C_{\square}$, three things are needed. First, its alphabet $\Sigma$, that is, the set of messages which are going to be circulating on its wires. Then the functions of its gates, and last the circuit itself, that is,  the position of the gates and wires within $K^{\infty}$. Because $K^{\infty}$ is obtained as the fixed point of a substitution, $C^\square$ itself is going to be defined as the fixed point of a substitution $\kappa$. This substitution associates to each gate $g$ a circuit, or macro-gate $\kappa(g)$ such that:
 \begin{itemize}[topsep=0pt,itemsep=-1ex,partopsep=1ex,parsep=1ex]
     \item each gate of $\kappa(g)$ announces its address within $\kappa(g)$ in its output values,
     \item each gate of $\kappa(g)$ announces a description of $g$ in its output values, and
     \item as a circuit, $\kappa(g)$ simulates $g$.
 \end{itemize}
 In order for $C_{\square}$ to be well-defined, there is a special "seed" gate $s$ at $(0, 0)$ which has no inputs and which appears in $\kappa(s)$ at position $(0, 0)$.

This is almost sufficient for self-description: in the fixed point of $\kappa$, every gate which has a predecessor within the same macro-gate receives the value of $g$ as well as the position of a neighbor within $\kappa(g)$, from which it computes its own position. In each $\kappa(g)$ there is a gate $p$ whose inputs all are from outside $\kappa(g)$; ensuring $p$ only depends on where the inputs of $g$ are, and that it does receive that information from its predecessor gives self-description.

The full details of the circuit are in \cref{sec:cacarpet_circuit}; the idea is on \cref{fig:normal-cut-message}: each message in $C_\square$ contains two instances of pairs of a gate description and a position within $K$. The first one corresponds to the parent of the tile and the position of the tile within its parent, while the second one is used to simulate the parent.

\begin{figure}
    \centering
    \tikz{\input{tikz/empty_gate_E_to_N_nW_Se2}}
    \tikz[rotate=180]{\input{tikz/local_normal12}}\\

    \caption{The principle of $C_{\square}$. Each gate $G$ (top-left) is cut by $\kappa$ into a circuit (right). The messages are composed of two parts. On the border of $\kappa(G)$, their second part corresponds to the messages of $G$. When a gate is cut, some information is lost: the type of its input wires. This information is reconstructed from the \emph{output} wires of the predecessor, which are given in the local part of the messages from the neighboring $\kappa(G')$. The seed gate's message (bottom-left) declares "I am the tile at $(0, 0)$ within a seed gate, at all scales".} 
    \label{fig:normal-cut-message}
\end{figure}

 % Let us call the part of the circuit within an instance of $K$ a \emph{macro-gate}. Each gate is tasked with computing its position within its macro-gate, as well as doing some work for that parent, so that each macro-gate encompassing the positions of $\kappa(z)$ for some position $z$ ends up 

\subsection{A Characterization of Self-Assemblable DSSFs}

\paragraph{Statement of the characterization}

The previous construction can be adapted to any discrete self-similar fractal within which the communication pattern used by $C_\square$ can be embedded. Whether a particular fractal is amenable to hosting such a communication pattern depends on the ability of its generator $G$ to transport information to copies of itself around it, as captured by its cardinal \emph{bandwidth}. The bandwidth $G[d \leftrightarrow d']$ between two directions $d, d' \in \Dir$ is the number of disjoint paths through G from its $d$ neighbor to its $d'$ neighbor.

\begin{theorem}
    \label{thm:ptime_charac}
  There is a polynomial time algorithm which, on input $G$, decides whether:
  \begin{enumerate}[topsep=0pt,itemsep=-1ex,partopsep=1ex,parsep=1ex]
  \item there is a $k$ such that $G^k[N \leftrightarrow S] \geq 2$ and $G^k[E \leftrightarrow W] \geq 2$ and $G^\infty$ can be strictly self-assembled in the aTAM, or
  \item for all $k$, $G^k[N \leftrightarrow S] < 2$ and $G^\infty$ cannot be strictly self-assembled in the aTAM, or
  \item for all $k$, $G^k[E \leftrightarrow W] < 2$ and $G^\infty$ cannot be strictly self-assembled in the aTAM.
  \end{enumerate}
\end{theorem}

The fact that distinguishing between the cases can be done in polynomial time is quite plain and is proven in \cref{sec:limits}. The interesting part is showing that this polynomial-time dichotomy is an actual characterization.

\begin{wrapfigure}{r}{0.4\textwidth}
    \centering
    \includegraphics{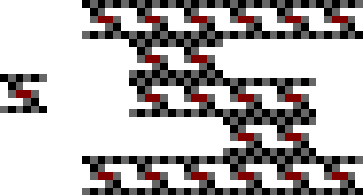}
    \caption{A generator where only the second iteration satisfies $G^2[N \leftrightarrow S] \geq 2$. Hence, $G[N \leftrightarrow S] = 1$, yet $G^\infty$ can be strictly self-assembled.}
    \label{fig:h_bizarre}    
\end{wrapfigure}

\paragraph{Bandwidth $2$ is sufficient for strict self-assembly}
%In order to compare generators, the classical notions of \emph{subgraph} and \emph{graph subdivision} are useful, accounting for marked vertices. 
Let $G, H$ be finite, connected shapes of $\mathbb{N}^2$, with $(0, 0) \in G \cap H$. Then $H$ is a \emph{subconnector} of $G$, written $H \preceq G$ if $G$ has a subgraph which is an edge subdivision of $H$, including the edges between adjacent copies of $H$ when placing  copies of $G$ on a grid.  

%if $G^+$\todo{Do $G^+$ and $H^+$ need to be defined?} has a (pointed) subgraph which is a (pointed) edge subdivision of $H^+$.
Now observe that whenever a graph $G$ has $K$ as a subconnector, the circuit of \cref{thm:cacarpet_circuit} can be embedded into $G^{\infty}$. Then changing the messages to carry addresses in $G$ rather than in $K$ yields a self-descriptive circuit with domain $G^\infty$. It actually suffices that some finite iterate $G^k$ admits $K$ as a subconnector in order to conclude, since $(G^k)^{\infty} = G^\infty$.

\begin{restatable}{lemma}{buildThroughSubconnector}
  \label{lem:build_through_subconnector}
  Let $G \ni (0,0)$ a finite, connected subgraph of $\mathbb{N}^2$ such that for some finite $k$, $K \preceq G^k$. Then there is a self-descriptive circuit with domain $G^\infty$.
\end{restatable}

When does $K \preceq G^k$? When $G[N \leftrightarrow S] \geq 2$ and $G[E \to W] \geq 2$: then  $K \preceq G^3$ as the 2 disjoint paths in each direction turn into 8, which is enough to get $K$ as a subconnector. This concludes the proof that condition 1 is sufficient to be able to strictly self-assemble $G^\infty$. 

\paragraph{Bandwidth $2$ is necessary for strict self-assembly}

In cases 2 and 3 of \cref{thm:ptime_charac}, strict self-assembly of $G^\infty$ is impossible by a generalization of the impossibility result of Hendricks et al.~\cite{hendricks_hierarchical_2020}.

\begin{restatable}{theorem}{killllama}
      \label{thm:kill_llama}
  Let $G \ni (0, 0)$ a finite shape of $\mathbb{N}$, and for $k \in \mathbb{N}$, $G^k = \sigma_G^k(\{(0, 0)\})$.

  If $\sup_{k}(G^k[N \leftrightarrow S]) = 1$ or $\sup_k(G^k[E \leftrightarrow W]) = 1$, then $G_\infty$ cannot be strictly self-assembled in the aTAM model unless $G^\infty = \mathbb{N}^2$.
\end{restatable}

Consider the case where for all $k$, $G^k[E \leftrightarrow W] = 1$. In $G^\infty$, consider all edges which are  East-West bridge of $G^k$ for some $k$. Because $\calT$ is finite, there is an infinite subset of those with the same glue on the bridge edge. From at least one of these glues $g$ which appear infinitely often, it is possible to grow unboundedly far to the north. Let $\vec{p}$ be the vector joining two instances of $g$. By \cref{thm:tree_pump}, from $g$ can grow either a periodic path going up to the north, or productions circling arbitrarily large squares. In both cases, this entails that all the $G^k$ admit $\vec{p}$ as a period. Hence $G^\infty$ is trivial.

\begin{figure}
    \centering
    \colorlet{lightgray}{gray!25}
\colorlet{conceptcol}{blue!25}

\usetikzlibrary{shapes.multipart}
\usetikzlibrary{shapes.arrows}
\tikzset{
  literature/.style={
    draw, fill=lightgray
  }
}

\tikzset{
  ours/.style={
    fill=white
    % draw, ultra thick, green!75!black, text=black
  }
}

\tikzset{
  resultbox/.style={rectangle split, rectangle split parts=2, text width=4cm, align=center, draw}
}

\tikzset{
  concept/.style={
    draw, fill=conceptcol
  }
}

\begin{tikzpicture}
  \node[literature, text width=7cm] (bigQ) {
      \textbf{Question.}
        When can Discrete Self-Similar Fractal Shapes be strictly self-assembled in the aTAM?\cite{jSSADST,jSADSSF,ScaledPierFractals,HFractals,TreeFractals}
  };

  \node (conj) [draw, literature, below=1cm of bigQ]{
      \textbf{Conjecture.} \textit{Never}\cite{InfiniteStructures,hader2023fractalNACO,InfiniteStructuresCSP}(circa 2008)
  };

  \node[resultbox, right=1cm of conj, ours, text width=4.5cm] (mainThm)
  {\textbf{\cref{thm:ptime_charac}} \nodepart{two} \textit{Iff there is enough bandwidth after iterating.}};

  \draw[->] (bigQ) -- (conj);
  \draw[->] (bigQ) -- (mainThm);

  \node[resultbox, below left=2cm of mainThm, ours] (positivePart)
  {\textbf{\cref{lem:build_through_subconnector}} \nodepart{two} \textit{Can assemble \emph{one} DSSF $\Rightarrow$ can do all shapes with enough bandwidth.}};
  \draw[->] (mainThm) -- node[pos=.5, fill=white, anchor=east] {positive direction} (positivePart);

  \node[resultbox, right=1cm of positivePart, ours] (negativePart)
  {\textbf{\cref{thm:kill_llama}} \nodepart{two} \textit{Not enough bandwidth $\Rightarrow$ large domains with a fixed periodicity}};

  \draw[->] (mainThm) -- node[pos=.5, fill=white] {negative direction} (negativePart);

  \node[resultbox, below=of negativePart, ours] (treePump)
  {\textbf{\cref{thm:tree_pump}} \nodepart{two} \textbf{Tree pump:} \textit{Tall assemblies of \emph{any TAS} either contain a big square or a periodic raising path}};

  \draw[->] (negativePart) -- node[pos=.5, fill=white] (neglim) {relies on} (treePump);

  \node[resultbox, below left=of positivePart, ours, xshift=8mm] (quineFractal)
  {\textbf{\cref{thm:dssf}} \nodepart{two} \textit{Infinite self-simulation with holes; fewer tiles, large scale factor}};
  
  \node[resultbox, right=5mm of quineFractal, ours] (circuitFractal)
  {\textbf{\cref{thm:cacarpet_circuit}} \nodepart{two} \textit{Self-describing circuit on a small-generator DSSF}};

  \draw[->] (quineFractal) -- node[pos=.5, fill=white, xshift=-1em] {implements} (positivePart);
  \draw[->] (circuitFractal) -- node[pos=.5, fill=white] {compiles into one} (positivePart);

  \node[resultbox, below=of quineFractal, ours, concept, xshift=1em] (quine)
  {\textbf{\cref{def:quine}, \cref{thm:quine}} \nodepart{two} \textbf{Quine} \textit{TAS outputting its own description}};

  \draw[->] (quineFractal) -- node[pos=.5, fill=white] (quinefraclim) {builds upon} (quine);

  \node[resultbox, below=of quine, ours] (standardIu)
  {\textbf{\cref{thm:standardIU}} \nodepart{two} \textit{Small and exact IU system for standard aTAM systems}};

  % \draw[->] (quine) -- node[pos=.5, fill=yellow] {builds Upon} (standardIu);

  \node[resultbox, below=of circuitFractal, concept, xshift=2mm] (defCircuits)
  {\textbf{\cref{def:self_describing}, \cref{thm:circuit-to-atam}} \nodepart{two} \textbf{Self-describing Circuits:}\textit{succint description for standard aTAM systems}};

  \draw[->] (quine) -- node[pos=.5, fill=yellow!25] {builds upon} (standardIu);

  \node[resultbox, below right=of standardIu, concept] (standard)
  {\textbf{\cref{def:standard}} \nodepart{two} \textbf{Standard aTAM:}  \textit{Well-behaved systems}};

  \draw[->] (standardIu) --  node[fill=yellow!25] {simulate} (standard);
  \draw[->] (defCircuits) -- node[fill=yellow!25] {describe} (standard);
  \draw[->] (circuitFractal) --  node[pos=.5] (circlim) {} (defCircuits);

  \node[resultbox, below=of treePump, concept, anchor=north, text width=4.3cm] (defFree)
  {\textbf{\cref{sec:freeSeqs}} \nodepart{two} \textbf{Free} assembly sequences};

  \node[resultbox, below =of defFree, concept, text width=5cm, xshift=4mm] (defOrdinals)
  {\textbf{\cref{def:ordinalSeqs}} \nodepart{two} \textbf{Ordinal} assembly sequences};

  \draw[->] (treePump.east) to[bend left] node[pos=.8, anchor=east] {proof auxiliary}(defFree.north east);
  \draw[->] (treePump.east) to[bend left] node[pos=.9, anchor=east] {proof auxiliary}(defOrdinals.north east);

  \node[literature, below=of defOrdinals] (wml)
  {\textbf{Window Movie Lemma} \cite{meunier_intrinsic_2014}};

  \node[literature, below=of wml, xshift=4mm] (connectedTreewidth)
  {Connected \textbf{Treewidth} \cite{DiestelGraphTheory, DBLP:journals/combinatorica/DiestelM18}};

  \draw[->] (treePump.east) to[bend left] node[pos=.9, anchor=east] {extends} (wml.north east);
  \draw[->] (treePump.east) to[bend left] node[pos=.92, fill=yellow!25, anchor=east, xshift=4em] {assumes bounded} (connectedTreewidth.north east);
  \node[literature, below=of standardIu] (wml)
  {General \textbf{IU systems} \cite{WoodsIU2013}};
  
  \draw (standardIu) -- (wml);

  \node [fill=red!50, rotate=90, single arrow, above=3cm of quineFractal] {Reading Order};

  \node[rotate=30, red, draw, thick] at (conj.north west) {\textsc{refuted}};

  \coordinate (a) at (current bounding box.south west);
  \coordinate (f) at (current bounding box.south east);
  \coordinate (b) at (quinefraclim -| a);
  \coordinate(c) at (circlim);
  \coordinate (d) at (neglim);
  \coordinate (e) at (neglim -| f);

  \begin{scope}[on background layer]
    \filldraw[yellow!25] (a) -- (b) -- (c) to[out=0, in=180, looseness=2.3] (d) -- (e) -- (f) -- cycle;
  \end{scope}

\end{tikzpicture}
    \caption{The articulation between the concepts and results of this paper. The gray boxes are from the literature, the blue boxes are new concepts. The results in the yellow part are of general interest outside the scope of fractals.}
    \label{fig:concept_map}
\end{figure}

\tableofcontents

\section{Definitions and Notations}
\label{sec:extended-defs}

%\section{Extended Definitions}

In this section we provide mathematical definitions (extending the overview provided in Section \ref{sec:definitions}) for the concepts and model used throughout the paper.

\subsection{Generic Notations}

We define the four cardinal directions in $\mathbb{Z}^2$ as $D = \{N,E,S,W\}$ and identify them with the corresponding offsets from a location $\vec{l} \in \mathbb{Z}^2$: $O = \{(0,1),(1,0),(0,-1),(-1,0)\}$.
A variable with an arrow such as $\vec{a}$ denotes a vector of values indexed by some set $S$. Given $s \in S$, $a_s$ is the element of $\vec{a}$ associated with $s$. For any sets $A, B$ and an indexing set $S$, given $\vec{a} \in A^S$ and $\vec{b} \in B^S$, the Cartesian product $\vec{a} \otimes \vec{b}$ is the vector $\vec{v} \in (A \times B)^S$ defined by $\forall s \in S, v_s = (a_s, b_s)$. Likewise, given $f: A^I \mapsto A^O$ and $g: B^I \mapsto B^O$, the function $f \otimes g: (A \times B)^I \mapsto (A \times B)^O$ is defined by $(f \otimes g)(\vec{a} \otimes \vec{b}) = f(\vec{a}) \otimes g(\vec{b})$.

The constructions in \cref{sec:abstract_models}, \cref{sec:cacarpet_circuit} and \cref{sec:tree_pump} make heavy use of sets built as Cartesian products. In an attempt to make their exposition more readable and distinguish the function of each component of the product, they are given in a ``record'' or ``object'' like syntax using '$\{\}$' for record construction and ``$x \cdot \operatorname{field}$'' for field access. That is, given an integer $k$, $k$ base sets $X_0, \ldots, X_{k-1}$ and a sequence of $k$ labels such as $L = (\operatorname{zeroth}, \operatorname{first}, \operatorname{second}, \ldots, \operatorname{not-quite-k-th})$, the element $x = (x_0, x_1, x_2, \ldots, x_{k-1}) \in \Pi_i X_i$ is written $x = \{ \operatorname{zeroth} = x_0, \operatorname{first} = x_1, \ldots, \operatorname{not-quite-k-th} = x_{k-1}\}$. Symmetrically, $x_0$ can be extracted from $x$ by writing $\FBattribute[x]{zeroth}$. A judicious set of labels can makes this notation actually useful in those sections.

\subsection{Discrete Self-Similar Fractal Shapes}

% Fractals are usually defined in some continuous space such as $\mathbb{R}^2$ or $\mathbb{C}^2$. In the context of self-assembly, one needs to work with \emph{discrete} fractals, that is, subsets of $\Z^2$ which are self-similar. 
In this paper, discrete self-similar fractal shapes are defined as fixed points of some 2 dimensional substitution. In~\cite{jSADSSF}, self-similar fractal shapes are defined as the union of an infinite hierarchy of shapes. The following definition is identical, as long as the generator contains $(0,0)$. Explicitly defining a geometrical substitution will help, as that substitution can guide constructions happening on the fractal. If the generator does not contain $(0, 0)$, the classical definition can be recovered by iterating the substitution associated with $G \cup (0, 0)$ to reach the fix-point, then iterating once the substitution associated with $G$. The substitution $\sigma_G$ based on a finite pattern $G$ replaces each pixel in a pattern $X$ with a copy of $G$.

\begin{definition}[Rectangular substitution]
  Let $G$ be a finite subset of $\mathbb{N}^2$. Let $w = \max(\{x | (x,y) \in G\})$ and $h = \max(\{y | (x,y) \in G\})$. The substitution $\sigma_G: \mathcal{P}(\mathbb{N}^2) \to \mathcal{P}(\mathbb{N}^2)$ associated with $G$ is defined by:
  \[ \forall X \subset \mathbb{N}^2, \sigma_G(X) = \{ p \in \mathbb{N}^2 | \quot{p}{G} \in X \wedge p \bmod G \in G \} \]
\end{definition}

Using substitution allows manipulating the recursive definition of DSSF with richer data than one bit per position in the following way. 

\begin{definition}[Colored substitution]
  Let $X$ be an alphabet and $G$ a finite subset of $\mathbb{N}^2$; given for each $x \in X$ a coloring $C(x) \in X^G$, the substitution $\sigma_C$ associated with $C$ is defined for each partial coloring $Y: \mathbb{N}^2 \dashrightarrow X$ by:
  \[
  \begin{cases}
    \sigma_C(Y): \mathbb{N}^2 \dashrightarrow X\\
    \sigma_C(Y): \vec{z} \mapsto C(Y(\quot{z}{G}))(z \bmod G),\\
  \end{cases}
  \]

  Where $\sigma_C(Y)(\vec{z})$ is defined whenever $Y(\quot{z}{G})$ is defined and $z \bmod G \in G$.
\end{definition}

\begin{figure}
  \centering
  \begin{tikzpicture}
    \input{tikz/colored_subst}
  \end{tikzpicture}
  \caption[Example of colored substitution]{A colored substitution on a 3-colored alphabet $\Sigma = \{ \tealdot, \pinkdot, \golddot \}$ is defined from a shape $G = \gammashape$ and a coloring of $G$ for each color of $\Sigma$. It can then be applied to any colored shape (right).} 
  \label{fig:colored_subst}
\end{figure}

\begin{definition}[Self-Similar Fractal Shape]
  Let $G$ be a finite subset of $\mathbb{N}^2$ with $(0, 0) \in G$. The discrete self-similar fractal shape with generator $G$ is the following subset of $\mathbb{N}^2$:
  \[ G^{\infty} = \bigcup_{i \in \mathbb{N}} \sigma_G^i(\{(0,0)\}). \]
  The $i$-th step of the fractal is $G^i = \sigma_G^i(\{(0, 0)\}$.
\end{definition}

One self-similar fractal shape will be of particular interest to us in this paper, wich we name \emph{Sierpinski's Cacarpet}. It is the self-similar fractal shape $\cacarpet$ with generator \(K = \{0,\ldots, 5\}^2 \setminus \{2, 3\}^2\), represented on figure~\ref{fig:cacarpet}. We note $\kappa = \sigma_K$ the associated substitution.

\begin{figure}
  \centering
  \begin{tikzpicture}
    \input{tikz/cacarpet}
  \end{tikzpicture}
  \caption{The substitution $\kappa$ of the Sierpinski's Cacarpet.}
  \label{fig:cacarpet}
\end{figure}

\subsection{The Abstract Tile Assembly Model and Standard Systems}
\label{sec:aTAM-extended}

%This section provides an expanded set of definitions related to the aTAM.

A complete definition of the aTAM is given here. This section also introduces IO markings and the concept of a \emph{standard TAS system}, which capture the common practices used in the literature to obtain intelligible aTAM systems.

%Let $\mathbb{N}$ be the set of non-negative integers, and for $n \in \mathbb{N}$, let $[n] = \{0, 1, ..., n-2, n-1\}$.
Let $\Sigma$ to be some alphabet with $\Sigma^*$ its finite strings. A \emph{glue} $g\in\Sigma^*\times\mathbb{N}$ consists of a finite string \emph{label} and non-negative integer \emph{strength}. There is a single glue of strength $0$, referred to as the \emph{null} glue. A \emph{tile type} is a tuple $t\in(\Sigma^*\times\mathbb{N})^{4}$, thought of as a unit square with a glue on each side. A \emph{tileset} is a finite set of tile types. We always assume a finite set of tile types, but allow an infinite number of copies of each tile type to occupy locations in the $\mathbb{Z}^2$ lattice, each called a \emph{tile}.
Given a tileset $T$, a \emph{configuration} is an arrangement (possibly empty) of tiles in the lattice $\mathbb{Z}^2$, i.e.\ a partial function $\alpha:\mathbb{Z}^2\dashrightarrow T$. Two adjacent tiles in a configuration \emph{interact}, or are \emph{bound} or \emph{attached}, if the glues on their abutting sides are equal (in both label and strength) and have positive strength. Each configuration $\alpha$ induces a \emph{binding graph} $B_\alpha$ whose vertices are those points occupied by tiles, with an edge of weight $s$ between two vertices if the corresponding tiles interact with strength $s$.
An \emph{assembly} is a configuration whose domain (as a graph) is connected and non-empty. The domain $\dom(\alpha) \subseteq \mathbb{Z}^2$ of an assembly $\alpha$ is known as its \emph{shape}.
We use $\Assemblies{T}$ to denote the set of all assemblies of tiles in tile set $T$.

For some $\tau\in\mathbb{Z}^+$, an assembly $\alpha$ is \emph{$\tau$-stable} if every cut of $B_\alpha$ has weight at least $\tau$, i.e.\ a $\tau$-stable assembly cannot be split into two pieces without separating bound tiles whose shared glues have cumulative strength $\tau$. %Given two assemblies $\alpha,\beta$, we say $\alpha$ is a \emph{subassembly} of $\beta$ (denoted $\alpha \sqsubseteq \beta$) if $S_\alpha \subseteq S_\beta$ and for all $p\in S_\alpha$, $\alpha(p)=\beta(p)$ (i.e., they have tiles of the same types in all locations of $\alpha$). 
Given a tile assembly system $\calT$ and an assembly $\alpha \in \Assemblies{\calT}$, an \emph{attachment candidate} is a pair $t@z$, with $t \in T$ and $z \in \mathbb{Z}^2$. It is \emph{valid} if $z \notin \dom \alpha$. It is \emph{stable} if $\alpha' = \alpha \cup \{z \mapsto t\}$ is stable. If it is both valid and stable, it is an \emph{attachment}, noted $\alpha \xrightarrow{t@z} \alpha'$.
In that case,  we say that $\alpha$ \emph{$\calT$-produces} $\alpha'$ \emph{in one step}. % written $\alpha \to^{\calT}_1 \alpha'$.
That is, $\alpha \xrightarrow{t@z} \beta$ if $\beta$ differs from $\alpha$ by the addition of a single tile of type $t$ at position $z$.
The \emph{$\calT$-frontier} is the set $\partial^{\calT}\alpha = \left\{ z | \exists (t, \beta), \alpha \xrightarrow{t@z} \beta \right\}$ of locations in which a tile could $\tau$-stably attach to $\alpha$.
%When $\calT$ is clear from context we simply refer to these as the \emph{frontier} locations.

Given a TAS $\calT=(T, \sigma, \tau)$, a sequence of $k\in\mathbb{Z}^+ \cup \{\infty\}$ assemblies $\alpha_0, \alpha_1, \ldots$ over $\mathcal{A}^T$ is called a \emph{$\calT$-assembly sequence} if, for all $i < k$, there is an attachment $t@z$ such that $\alpha_{i} \xrightarrow{t@z} \alpha_{i+1}$. The \emph{result} of an assembly sequence is the unique limiting assembly of the sequence. For finite assembly sequences, this is the final assembly; whereas for infinite assembly sequences, this is the assembly consisting of all tiles from any assembly in the sequence. We say that \emph{$\alpha$ $\calT$-produces $\beta$} (denoted $\alpha\to^{\calT} \beta$) if there is a $\calT$-assembly sequence starting with $\alpha$ whose result is $\beta$. We say $\alpha$ is \emph{$\calT$-producible} if $\sigma\to^{\calT}\alpha$ and write $\prodasm{\calT}$ to denote the set of $\calT$-producible assemblies. We say $\alpha$ is \emph{$\calT$-terminal} if $\alpha$ is $\tau$-stable and there exists no assembly that is $\calT$-producible from $\alpha$. We denote the set of $\calT$-producible and $\calT$-terminal assemblies by $\termasm{\calT}$. If $|\termasm{\calT}| = 1$, i.e., there is exactly one terminal assembly, we say that $\calT$ is \emph{directed}.

We say $\calT$ \emph{strictly self-assembles shape $P$} if for any $A \in \TerminalProds{\calT}$, $\dom A = P$.
Weak assembly of a target shape is defined as follows.
Given a TAS $\calT = (T,\sigma,\tau)$, we allow each tile type to be assigned exactly one \emph{color} from some set of colors $C$.
Let $C_P \subseteq C$ be a subset of those colors, and $T_{C_P} \subseteq T$ be the subset of tiles of $T$ whose colors are in $C_P$.
Given an assembly $\alpha \in \prodasm{\calT}$, we use $\dom_{C_p}(\alpha)$ to denote the set of all locations of tiles in $\alpha$ with colors in $C_P$.

\begin{definition}[Input and Output sides]
Let $\calT = (T,\sigma,\tau)$ be a TAS in the aTAM, and let $\alpha \in \prodasm{\calT}$ be a producible assembly in $\calT$. If $\vec{t}$ is a tile of type $t \in T$ that attaches to $\alpha$, we call the subset of sides on which $\vec{t}$ has non-null glues that form bonds with $\alpha$ at the time that $\vec{t}$ attaches its \emph{input sides} and denote them as $\texttt{IN}(\vec{t}) \subseteq D$. Let $\texttt{G}(\vec{t}) \subseteq D$ be the set of sides of $\vec{t}$ containing non-null glues. We refer to the set $\texttt{G}(\vec{t}) \setminus \texttt{IN}(\vec{t})$ as the tile's \emph{output sides}, and denote them as $\texttt{OUT}(\vec{t})$.
\end{definition}

\begin{definition}[IO marked tile set]
Let $\calT = (T, \sigma, \tau)$ be a TAS in the aTAM. We say that $T$ is an \emph{IO marked tile set} if there exists an ordered set of $4$ unique symbols $\texttt{IO}_s = \{s_N, s_E, s_S, s_W\}$ such that the following conditions hold for every tile type $t \in T$ where a tile of type $t$ is not contained in $\sigma$:
 \begin{enumerate}
    \item For some subset of sides of $t$ whose glues are non-null such that the strengths of those glues sum to exactly $\tau$, the glue labels of the glues on those sides end with the symbols from $\texttt{IO}_s$ corresponding to the directions of those sides. We say that those sides are \emph{input marked}.

    \item For all sides of $t$ that contain non-null glues which are not input marked, the glue labels of the glues of glues on those sides ends with the symbols from $\texttt{IO}_s$ corresponding to the opposite directions of those sides. We say that those sides are \emph{output marked}. 
\end{enumerate}

For each $t \in T$ such that a tile of type $t$ is contained in $\sigma$, it may have no sides that are input marked, or input marked sides whose glue strengths sum to $\le \tau$.
\end{definition}

\begin{lemma}\label{lem:IO}
Let $\calT = (T,\sigma,\tau)$ be an aTAM TAS where $T$ is IO marked. If all exposed glues on the perimeter of $\sigma$ are output marked, then for all $\alpha \in \prodasm{\calT}$, all exposed non-null glues on the perimeter of $\alpha$ are output marked. 
\end{lemma}

\begin{proof}
We prove Lemma \ref{lem:IO} by induction. Our base case is $\sigma$ which by definition is in $\prodasm{\calT}$ and is given to have all exposed non-null glues on its perimeter to be output marked. Our induction hypothesis is that, given producible assemblies $\alpha, \beta \in \prodasm{\calT}$ such that $\alpha \rightarrow^\calT_1 \beta$, if all non-null glues on the perimeter of $\alpha$ are output marked, then the same holds for $\beta$. This must be true because, since $T$ is IO marked, whatever tile $\vec{t}$ of whichever type $t \in T$ attaches to $\alpha$ to form $\beta$ must only have glues whose strengths sum to exactly $\tau$, the minimum possible that allow it to bind, that are input marked. Since all glues on the perimeter of $\alpha$ are output marked, and the only way for glues on $t$ and the perimeter of $\alpha$ to match and thus form bonds is for them to be on tile sides of opposite directions and one to be input marked and the other output marked (since those are the same symbols for the opposite sides), every input marked glue of $\vec{t}$ must bind during its attachment. This leaves only null and output marked glues remaining on $\vec{t}$ to possibly be exposed and added to the perimeter of $\beta$. Therefore, all exposed non-null glues on the perimeter of $\beta$ must also be output marked and our induction hypothesis holds and Lemma \ref{lem:IO} is proven.
\end{proof}

\begin{corollary}\label{cor:IO}
Given aTAM TAS $\calT = (T,\sigma,\tau)$ where $T$ is IO marked and all exposed glues on the perimeter of $\sigma$ are output marked, then for every tile type $t \in T$, all tiles of type $t$ that attach to any producible assemblies in $\calT$ must have the exact same input sides and output sides (noting that output sides are not required to ever form bonds with other tiles).
\end{corollary}

Corollary \ref{cor:IO} follows immediately from Lemma \ref{lem:IO} since the tiles of $T$ are IO marked and therefore can only ever use (exactly) their input marked sides as their input sides when initially binding, and any remaining non-null glues are output marked and also available as output sides. Therefore, for such systems we will also use the notation $\texttt{IN}(t)$ and $\texttt{OUT}(t)$, referring to tile \emph{type} $t$, in addition to $\texttt{IN}(\vec{t})$ and $\texttt{OUT}(\vec{t})$ referring to individual tiles of type $t$.

Throughout this paper, we will use the following two sets of symbols to IO mark tile sets: $\texttt{IO}_{s1} = \{$V$, <, \wedge, >\}$ and  $\texttt{IO}_{s2} = \{$VV$, <<, \wedge\wedge, >>\}$. Intuitively, they can be thought of as ``pointing into the tile'' when on input sides, and out of the tile on output sides.

\begin{definition}[Standard TAS]
\label{def:standard}
Let $\calT = (T, \sigma, 2)$ be a TAS in the aTAM. We say that $\calT$ is \emph{standard} if and only if:
    \begin{enumerate}
        \item $\calT$ is directed,

        \item $T$ is IO marked,

        \item For every $t \in T$, the sides that have input markings are either exactly (1) a single glue of strength-2, or (2) two diagonally adjacent strength-1 glues (i.e. not on opposite sides),

        \item All glues on the exterior of $\sigma$ are output marked, and

        \item There are no mismatches in the terminal assembly (i.e. all adjacent pairs of tile sides in $\alpha \in \termasm{\calT}$ have the same glue label and strength on both sides)

        \item In any tile location between two diagonally adjacent non-null glues or next to a strength-2 glue, exactly 1 tile may attach.
    \end{enumerate} 
\end{definition}

Note that throughout the literature of the aTAM, most constructions consist of standard aTAM systems or systems that could trivially be turned into standard systems by adding IO markings (e.g. \cite{jCCSA,jSADS,SolWin07,RotWin00}).

\subsection{Encodings of aTAM tile types and systems}

Here we provide definitions related to the ways in which aTAM tile types and systems can be encoded for use in systems that simulate their behaviors.

\begin{definition}[Glue encoding]\label{def:glue_enc}
Let $G$ be a set of glue labels, $\Sigma$ an alphabet, and $f_G: G \rightarrow \Sigma^*$ be an injective (i.e. one-to-one) function mapping glue labels to strings. For $g \in G$, we say that $f_G(g)$ is a \emph{glue encoding of $g$ over $\Sigma$}. That is, $f_G(g)$ is a unique representation of $g$ among all glues in $G$ using the fixed alphabet $\Sigma$.
\end{definition}

\begin{definition}[Glue lookup entry]\label{def:gle}
Given an IO marked tile set $T$, some $t \in T$, and alphabets $\Sigma$ and $\Sigma'$ with $\Sigma \subseteq \Sigma'$, we define a \emph{glue lookup entry of $t$ over $\Sigma'$} as a string $s \in \Sigma'^*$ that consists of glue encodings of the input sides of $t$ under $\Sigma$ followed by glue encodings of the output sides of $t$ under $\Sigma$, possibly separated and/or surrounded by additional characters in $\Sigma' - \Sigma$.
\end{definition}

\begin{definition}[Glue lookup table]\label{def:glt}
Given an IO marked tile set $T$ and alphabets $\Sigma$ and $\Sigma'$ with $\Sigma \subseteq \Sigma'$, we define a \emph{glue lookup table of $T$ over $\Sigma'$} as a string $s \in \Sigma'^*$ that consists of a glue lookup entry of each $t \in T$ over $\Sigma$, possibly separated and/or surrounded by additional characters in $\Sigma' - \Sigma$.
\end{definition}

\subsection{Intrinsic Simulation in the aTAM}\label{sec:aTAM-sim}

This section describes what it means for a TAS to intrinsically simulate another TAS. 
Intuitively, intrinsic simulation of a system $\mathcal{T}$ by another system $\mathcal{S}$ is done with respect to some scale factor $c \in \mathbb{Z}^+$ such that $c \times c$ squares of tiles in $\mathcal{S}$, called \emph{macrotiles}, represent individual tiles in $\mathcal{T}$, and there is a representation function that is able to map the macrotiles in $\mathcal{S}$ to tiles in $\calT$ (or empty space) and thus interpret the assemblies of $\mathcal{S}$ as assemblies in $\mathcal{T}$. Furthermore, the progression of the mapped macrotiles from $\mathcal{S}$ faithfully mimics the addition of tiles in $\calT$.

In the following definitions, it will be assumed that $\calT = (T, \sigma_\calT, \tau_\calT )$ is a TAS being simulated by another TAS $\calS = (S, \sigma_\calS, \tau_\calS)$. Furthermore, let $\mathbb{Z}_n$ be the set $\{0, 1, \ldots, n-1\}$ equipped with the typical notions of modular arithmetic.
Given a positive integer $c$ called the \emph{scale factor}, a \emph{$c$-block macrotile} over $S$ is a partial function $\mu:\mathbb{Z}_c \dashrightarrow S$. That is, $\mu$ assigns tiles from $S$ to some subset of the locations in a $c\times c$ block of locations. If the domain of $\mu$ is empty, then the macrotile is called \emph{empty}. During a simulation, it is assumed that the lattice $\mathbb{Z}^2$ is divided regularly into $c$-block macrotiles so that the origin occupies the south-westernmost location in the corresponding macrotile block. Given a general assembly $\alpha \in \mathcal{A}^\calS$ and some coordinates $(x,y)\in \mathbb{Z}^2$, we can recover the macrotiles from $\alpha$ by letting $\mathcal{M}^c_{x,y}[\alpha]$ refer to the $c$-block macrotile in $\alpha$ whose southwest corner occupies location $(cx,cy)$. In other words $\mathcal{M}^c_{x,y}[\alpha](i,j) = \alpha(cx+i, cy+j)$, when defined, for $(i,j)\in\mathbb{Z}^2_c$.

A partial function $R:\mathscr{B}^S_c \dashrightarrow T$ is called a \emph{macrotile representation function} from $S$ to $T$ if for any $\mu,\nu \in \mathscr{B}^S_c$ where $\mu \sqsubseteq \nu$ and $\mu \in \text{dom } R$, then $R(\mu) = R(\nu)$. In other words, if $R$ maps a macrotile $\mu$ to a tile type in $T$, then any additional tile attachments to $\mu$ should not change how it maps under $R$. This is clearly required since tiles in the aTAM cannot detach or change to other tile types so macrotiles should not be able to change their representation once assigned. With respect to an assembly sequence $\alpha_1, \alpha_2, \ldots$ in $\calS$, a macrotile at location $(x,y)$ is said to \emph{resolve} into tile type $t\in T$ at step $i$ if $\mathscr{M}^c_{x,y}[\alpha_{i-1}]$ is not in the domain of $R$ but $R(\mathscr{M}^c_{x,y}[\alpha_i]) = t$.

Given a $c$-block macrotile representation function $R$, The corresponding \emph{assembly representation function} is denoted $R^*:\mathscr{A}^S \to \mathscr{A}^T$ and is defined so that $R^*(\alpha') = \alpha$ exactly when $\alpha(x,y) = R(\mathscr{M}^c_{x,y}[\alpha'])$ for all $(x,y)\in\Z^2$. In other words, the assembly representation function is just the result of applying the macrotile representation function to each macrotile in $\calS$. Additionally, the notation ${R^*}^{-1}(\alpha)$ is used to refer to the producible preimage $\{ \alpha' \in \prodasm{\calS} \mid R^*(\alpha') = \alpha \}$ of the assembly representation function $R^*$ on $\alpha$. In other words ${R^*}^{-1}(\alpha)$ is the set of all $\calS$-producible assemblies that map to $\alpha$ under $R^*$. 

In order for intrinsic simulation to be a distinctly unique and meaningful definition compared to traditional notions of simulation, a restriction on $R^*$ is necessary. Given an assembly $\alpha'\in\prodasm{\calS}$ let $\alpha=R^*(\alpha')$. It is then said that $\alpha'$ \emph{maps cleanly} to $\alpha$ under $R^*$ if all tiles in $\alpha'$ appear in the macrotile blocks which have either themselves resolved to tiles in $\alpha$ or which are adjacent (not diagonally) to macrotiles which have resolved. This means that tiles may not exist in macrotile blocks of $\alpha'$ mapping to locations of $\alpha$ that could not possibly admit a tile attachment due to their distance from all tiles in $\alpha$. Tiles are allowed to attach in macrotile blocks of $\alpha'$ adjacent to resolved macrotiles to determine if the macrotile block might eventually resolve to a tile. These unresolved macrotiles adjacent to resolved macrotiles are called \emph{fuzz} macrotiles since at scale the tiles attaching in them resemble fuzzy hairs on the surface of a resolved assembly.

\begin{definition}
	Formally, $\alpha'$ maps cleanly to $\alpha'$ when for every non-empty block $\mathscr{M}^c_{x,y}[\alpha']$, there exists a vector $(u,v)\in\Z^2$ with length less than or equal to 1, such that $\mathscr{M}^c_{x+u,y+v}[\alpha'] \in \text{dom}\;R$.	 
\end{definition}

\begin{definition} \label{def:atam-sim-equiv-prod}
    We say that $\calS$ and $\calT$ have \emph{equivalent productions} (under $R$), written $\calS \Leftrightarrow_R \calT$, if the following conditions hold:
    \begin{itemize}
        \item $\{R^*(\alpha') | \alpha' \in \prodasm{\calS}\} = \prodasm{\calT}$,
        \item $\{R^*(\alpha') | \alpha' \in \termasm{\calS}\} = \termasm{\calT}$, and
        \item For all $\alpha'\in\prodasm{\calS}$, $\alpha'$ maps cleanly to $R^*(\alpha')$.
    \end{itemize}
\end{definition}

\begin{definition} \label{def:atam-sim-follows}
    We say that $\calT$ \emph{follows} $\calS$ (under $R$), written $\calT \dashv_R \calS$, if for all $\alpha', \beta' \in \prodasm{\calS}$, $\alpha' \to^\calS \beta'$ implies $R^*(\alpha') \to^\calT R^*(\beta')$.
\end{definition}

In the following definition let ${R^*}^{-1}(\alpha)$ refer to the producible pre-image of the assembly representation function $R^*$ on $\alpha$. That is ${R^*}^{-1}(\alpha) = \{\alpha' \in \prodasm{\calS} | R^*(\alpha') = \alpha \}$, so that ${R^*}^{-1}(\alpha)$ consists of all $\calS$-producible assemblies that represent $\alpha$ under $R$.

\begin{definition} \label{def:atam-sim-models}
    We say that $\calS$ \emph{models} $\calT$ (under $R$), written $\calS \vDash_R \calT$, if for every $\alpha \in \prodasm{\calT}$, there exists a non-empty subset $\Pi_\alpha \subseteq {R^*}^{-1}(\alpha)$, such that for all $\beta \in \prodasm{\calT}$ where $\alpha \to^\calT \beta$, the following conditions are satisfied:
    \begin{enumerate}
        \item for every $\alpha' \in \Pi_\alpha$, there exists $\beta' \in {R^*}^{-1}(\beta)$ such that $\alpha' \to^\calS \beta'$
        \item for every $\alpha''\in {R^*}^{-1}(\alpha)$ and $\beta''\in {R^*}^{-1}(\beta)$ where $\alpha'' \to^\calS \beta''$, there exists $\alpha'\in \Pi_\alpha$ such that $\alpha' \to^\calS \alpha''$.
    \end{enumerate}

    For each $\alpha \in \prodasm{\calT}$, we call the corresponding set $\Pi_\alpha$ the \emph{stem set} of $\alpha$.
\end{definition}

In Definition~\ref{def:atam-sim-models} above, the stem set $\Pi_\alpha$ of $\alpha$ is defined to be a set of assemblies representing $\alpha$ from which it is still possible to produce assemblies representing all possible $\beta$ producible from $\alpha$. Informally, the first condition specifies that all assemblies in $\Pi_\alpha$ can produce some assembly representing any $\beta$ producible from $\alpha$, while the second condition specifies that any assembly $\alpha''$ representing $\alpha$ that may produce an assembly representing $\beta$ is producible from an assembly in $\Pi_\alpha$. In this way, $\Pi_\alpha$ represents a set of the earliest possible representations of $\alpha$ where no commitment has yet been made regarding the next simulated assembly. Requiring the existence of such a set $\Pi_\alpha$ for every producible $\alpha$ ensures that non-determinism is faithfully simulated. That is, the simulation cannot simply ``decide in advance'' which tile attachments will occur.

\subsection{Intrinsic Universality}
\label{sec:iu_def}

\newcommand{\REPL}{\mathsf{REPR}}
\newcommand{\frakC}{\mathfrak{C}}

Now that we have a formal definition of what it means for one tile system to simulate another, we can proceed to formally define the concept of intrinsic universality, i.e., when there is one general-purpose tile set that can be appropriately programmed to simulate any other tile system from a specified class of tile systems.

Let $\REPL$ denote the set of all supertile representation functions (i.e., $c$-block supertile representation functions for some $c\in\Z^+$).
Define $\frakC$ to be a class of tile assembly systems, and let $U$ be a tile set.
Note that each element of $\frakC$, $\REPL$, and $\mathcal{A}^U_{< \infty}$ is a finite object, hence encoding and decoding of simulated and simulating assemblies can be defined to be computable via standard models such as Turing machines and Boolean circuits.

\begin{definition}\label{def:iu}
We say a tile set $U$ is \emph{intrinsically universal} (IU) for $\frakC$ if there are computable functions $\mathcal{R}:\frakC \to \REPL$ and $S:\frakC \to \mathcal{A}^U_{< \infty}$ such that, for each $\mathcal{T} \in \frakC$, there is a constant $c\in\Z^+$ such that, letting $R = \mathcal{R}(\mathcal{T})$, $\sigma_\mathcal{T}=S(\mathcal{T})$, and $\mathcal{U}_\mathcal{T} = (U,\sigma_\mathcal{T},\tau)$, $\mathcal{U}_\mathcal{T}$ simulates $\mathcal{T}$ at scale $c$ and using supertile representation function~$R$.
\end{definition}
That is, $\mathcal{R}(\mathcal{T})$ outputs a representation function that interprets assemblies of $\mathcal{U}_\mathcal{T}$ as assemblies of $\mathcal{T}$, and $S(\mathcal{T})$ outputs the seed assembly used to program tiles from $U$ to represent the seed assembly of $\mathcal{T}$. We refer to $\mathcal{R}$ as a \emph{representation function generator} and $S$ as a \emph{seed generator}.

% \section{Standard aTAM systems are Intrinsically Universal and admit Quines}
\section{Standard aTAM Systems are Intrinsically Universal}\label{sec:standard-IU}

We now prove that the class of standard aTAM systems is intrinsically universal by providing a construction of a tileset that is IU for it. This universal tileset will be combined with the quine system presented in the next section to make systems that are capable of self-simulation.

\subsection{Outline of IU construction}

Here we describe in broad strokes how our construction is designed. Technical details regarding individual tile gadgets are deferred to Section~\ref{sec:IU-technical}. It's important to emphasize here that there is no single best way to implement such a tileset; many of the choices we made are entirely arbitrary. This section therefore does not describe a fundamental construction, but rather one of infinitely many possible ways to implement a tileset IU for standard systems. It should also be noted that the restriction to standard systems makes designing an IU tileset significantly simpler than if the entire class of aTAM systems were to be simulated. This allows our IU tileset to consist of considerably fewer than 10,000 tile types while more general IU tilesets can easily contain orders of magnitude more \cite{DDDIU}.
% \todo{Adding a note from Florent that I still need to handle: In section 3.1, it should be emphasized that standard systems allow to have simulations with no fuzz in terminal productions, which is critical in order to get a fractal shape in the end (otherwise, higher order holes would not remain entirely empty). In particular, it would make section 6 cleaner if section 3 stated the following lemma (Lemma XYZ): for any IU system without terminal fuzz, if $m$ is the shape of the macro-tiles of the IU system, then the terminal productions of the simulations of a system $S$ are $\sigma_m(P)$, where $P$ is the set of terminal productions of $S$. Then the third point of theorem 19 would say "modification of the IU tileset to add spacing while preserving the hypothesis of lemma XYZ".}

In this section, it will be assumed that $U$ is refer to the IU tileset and we will describe the various parts of our construction under the pretext that a standard system $\calT=(T, \sigma_T, \tau_T\le 2)$ is being intrinsically simulated by a standard system $\calU = (U, \sigma_U, \tau_U=2)$.

\begin{definition}
    We say that the intrinsic simulation of a system $\calT$ by another $\calU$ is \emph{pristine} if no terminal assembly in $\calU$ has any fuzz tiles. That is, any macrotile of $\calU$ containing a tile maps to a tile type in $\calT$ under the representation function.
\end{definition}

In addition to being universal for all standard systems, our IU tileset $U$ also has the property that if the seed macrotile $\sigma_U$ encodes all of the tiles in $T$ according to the convention described in Section~\ref{sec:IU-encodings}, then the simulation of $\calT$ by $\calU$ will be pristine. This property is essential for our DSSF construction since otherwise the resulting shape would contain tiles in what should be empty locations.

\subsubsection{Macrotile structure}

In this construction, we insist that every tile attachment occurring within a macrotile block is dedicated solely to either determining how that macrotile block will resolve or to presenting information along the edges of that macrotile block to indicate to neighboring macrotiles how it has resolved. We also ensure that each macrotile block will only contain tiles in $\calU$ if the corresponding tile location in $\calT$ will contain a tile in the terminal assembly. This can be guaranteed since standard systems are locally deterministic.

The 4 sides of a completed macrotile in $\calU$ each encode 2 pieces of information: the tileset $T$ in the form of a data structure called a \emph{glue table}, and the glue to be presented along the corresponding side of the $T$-tile type $t$ being represented. Note that this encoding is predicated on the fact that $t$ has a glue with positive strength on the corresponding side. If any of the sides of $t$ contain the null glue, then the corresponding side of the macrotile will entirely consist of null glues. 

Each side of a macrotile may be logically divided into 7 sections each separated by dedicated glues. The layout of these sections is symmetric along the side of the macrotile, though the data contained in these sections is decidedly not symmetric. The center section encodes the glue table for $T$. It is surrounded by two equally sized padding sections consisting only of strength-1 blank glues that encode no information. These are surrounded by identical copies of glue encodings called \emph{glue signatures} whose purpose is to indicate which glue from $\calT$ the macrotile side represents. Finally, these are surrounded by two additional, equally sized padding sections. Note that the outer and inner padding sections need not be the same size and only serve to accommodate space inside the macrotile for passing information around.

\begin{figure}
	\centering
	\begin{subfigure}[b]{0.49\textwidth}
		\centering
		\includegraphics[width=0.9\textwidth]{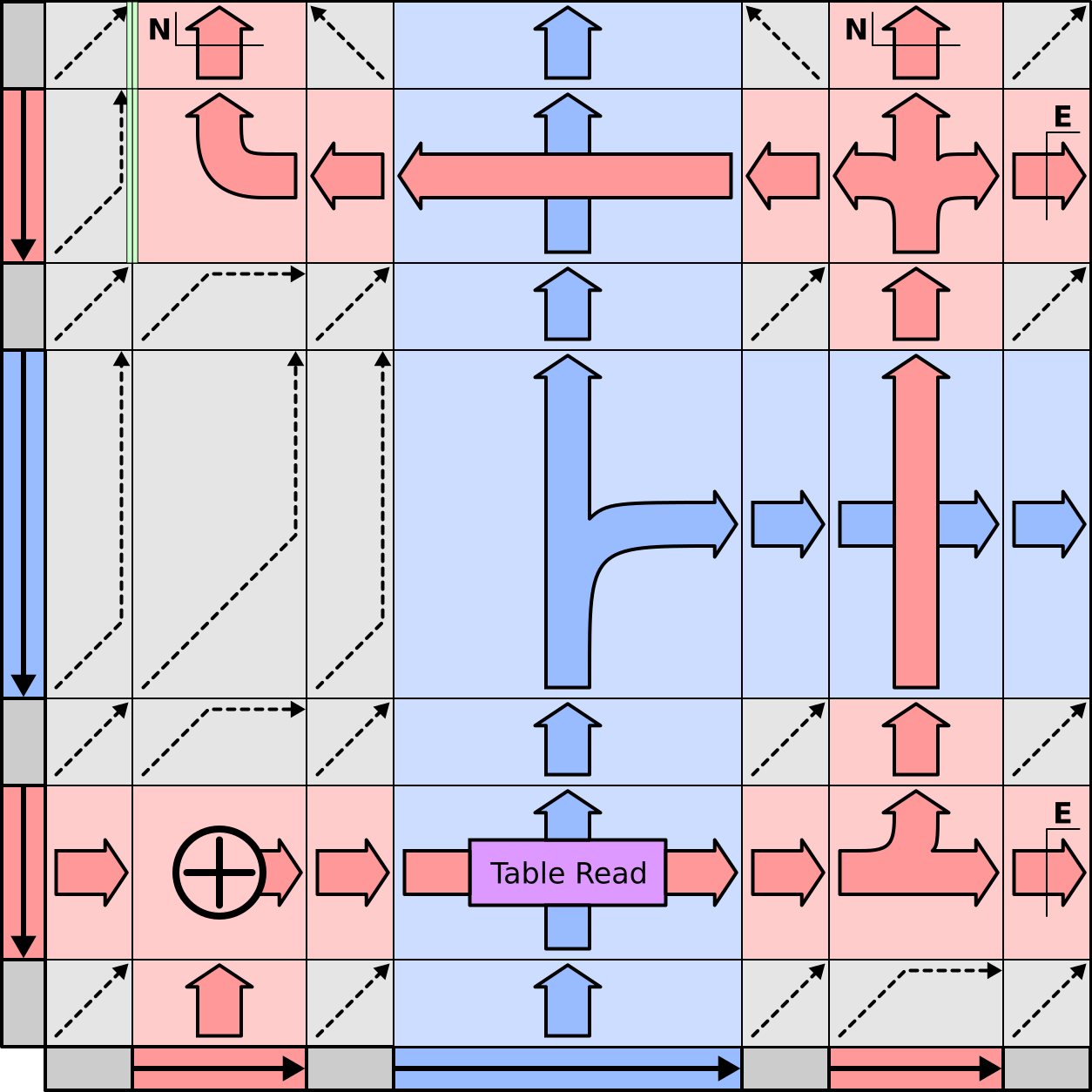}
		\caption{The layout of macrotile sections when simulating a cooperative attachment}
		\label{fig:IU-regions-cooperative}
	\end{subfigure}
	\hfill
	\begin{subfigure}[b]{0.49\textwidth}
		\centering
		\includegraphics[width=0.864\textwidth]{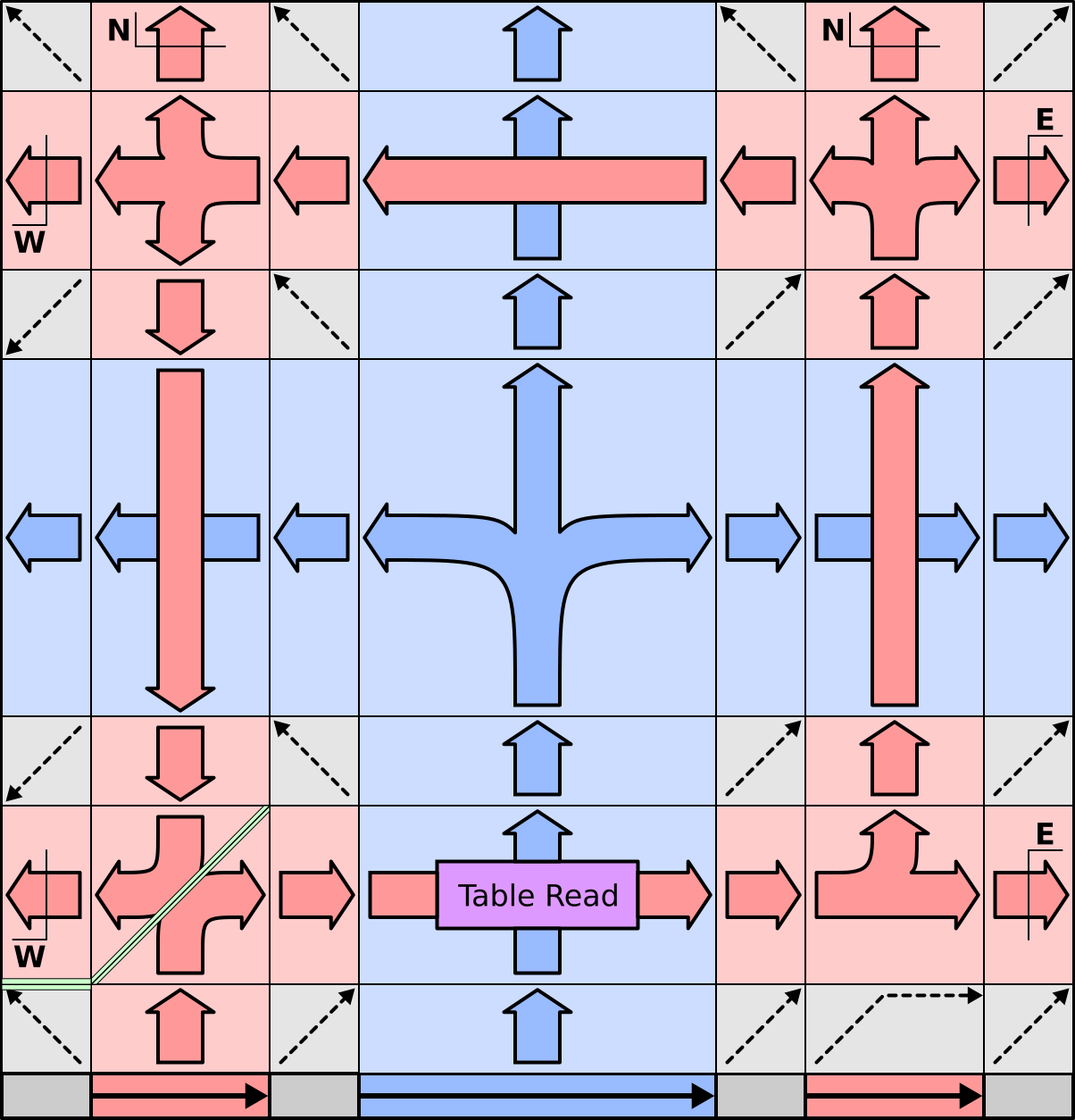}
		\caption{The layout of macrotile sections when simulating a strength-2 attachment}
		\label{fig:IU-regions-tau-strength}
	\end{subfigure}
	\caption{
        Blue arrows indicate the propagation of the glue table while red arrows indicate the propagation of glue signatures. Gray boxes indicate sections which are filled with generic tiles. Thin green lines indicate places where tiles growing from opposite directions place null glues between themselves so that mismatches are avoided.
    }
	\label{fig:IU-regions-filled}
\end{figure}

The boundaries that divide a macrotile side into 7 sections extend into the macrotile so that the entire block is divided into 49 logical sections as illustrated in Figure~\ref{fig:IU-regions-filled}. Note that these sections may differ in their functionality depending on whether a cooperative or strength-2 attachment is being simulated in the current macrotile, though there are only a small number of distinct functional tasks that a section may perform. These tasks include the following:
\begin{enumerate}
    \item \textbf{Filler sections} Filla region with generic tiles up to the boundaries (indicated by gray boxes with thin, dotted arrows),
    \item \textbf{Propagation sections} propagate and/or rotate a sequence of symbols (encoded by the glues of tiles) from one side of the section to one or more other sides (indicated by thick colored arrows in Figure~\ref{fig:IU-regions-filled}),
    \item \textbf{Propagation intersection sections} pass two sequences of symbols through each other to the opposite sides of the section ( indicated by thick arrows of different colors passing over each other),
    \item \textbf{Glue combination sections} combine glue signature information from a pair of neighboring strength-1 glues (indicated by the symbol $\oplus$ in Figure~\ref{fig:IU-regions-filled}),
    \item \textbf{Glue table read sections} read from the glue table to convert an input glue signature into an output glue signature, and
    \item \textbf{Glue output sections} ``clean up'' an output glue signature which may contain several output glue encodings (one for each output side) so that only the one for a specific output side is presented (indicated by a thick colored arrow struck-through by a thin line with a cardinal direction symbol).
\end{enumerate}

\subsubsection{Section boundary tiles}

Each section of a macrotile is separated by special \emph{section boundary} tiles. These grow along the sides of adjacent macrotiles and completed sections within the current macrotile to drive the growth of each consecutive section. These tiles keep track of the current section within a macrotile using a local coordinate system of rows and columns. Each macrotile section exists between a pair of rows and a pair of columns. For instance, when simulating a cooperative attachment, the first tile to attach between two diagonally adjacent macrotiles is a section boundary tile representing the corner between row 0 and column 0. Additional section boundary tiles may then attach to form the entirety of row 0 and column 0 of the macrotile and the sections may begin to grow by attaching to these boundary tiles.

Which function a section of a macrotile performs is dependent on the local coordinates of the section within the macrotile, so the section boundary tiles enforce that the correct function is performed in each section. Additionally, the boundary tiles also serve to pass information between macrotile sections when necessary and keep track of whether the current macrotile represents a strength-2 attachment or a cooperative one. 

\begin{figure}
    \centering
    \includegraphics[width=\textwidth]{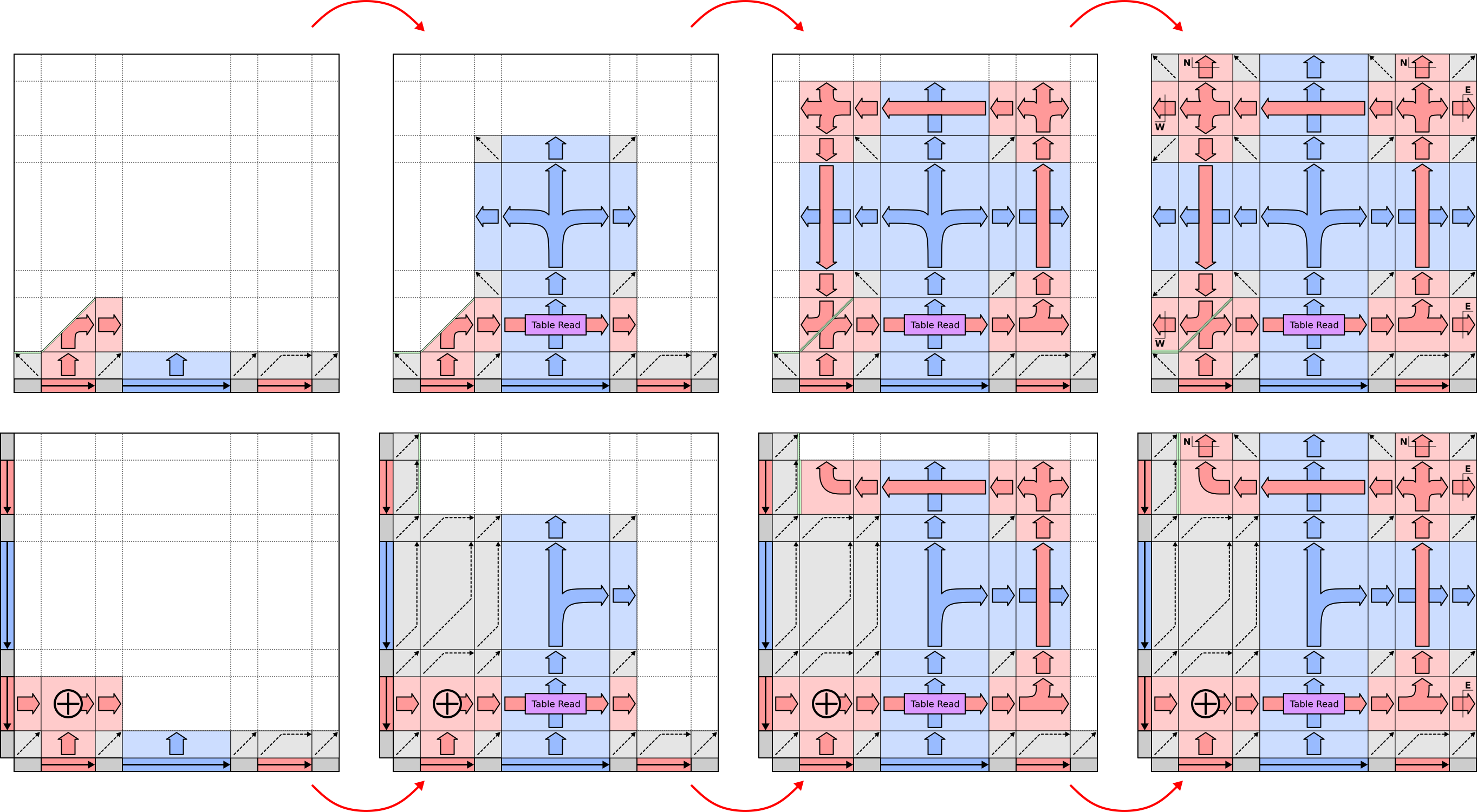}
    \caption{Different phases of IU macrotile growth when simulating strength-2 attachment to an existing macrotile on the south (top) and cooperative attachment to existing macrotiles from the south and west (bottom). For strength-2 attachments, the encoding of the $\calT$ glue (red) is simply rotated towards where the glue table (blue) read will take place. For cooperative attachments, the two glue encodings are combined into one before being propagated towards the glue table. Reading from the glue table is then performed while also propagating the glue table information towards the center of the macrotile. The resulting glue signature encoding all output glues is propagated around the center of the macrotile and finally presented along the relevant edges of the macrotile.}
    \label{fig:macrotile-growth}
\end{figure}

\subsubsection{Initiating the growth of macrotiles}

In standard systems, there are 2 distinct ways tiles may attach: using a single strength-2 glue or by the cooperation of two strength-1 glues from diagonally adjacent tiles. Growth of a macrotile simulating a cooperative attachment is initiated by a cooperative attachment between the diagonally adjacent macrotiles, while growth of a macrotile simulating a strength-2 attachment is initiated by a single strength-2 glue which is guaranteed to exist by convention on the counter-clockwise-most glue signature on the face of a completed macrotile. In this latter case, the strength-2 glue always exists in the position of the most-significant bit of the binary encoding of the glue from $\calT$, though this convention represents a completely arbitrary choice.

When simulating a cooperative attachment, the current macrotile receives input glue encodings from 2 adjacent macrotiles. These encodings are combined into a single glue signature representing the combined information from both glues. For macrotiles simulating a strength-2 attachment, this step is skipped since there is only 1 input glue for the simulated tile. Instead, the glue signature representing this single input glue is simply rotated so that it is propagated towards the glue table. In either case, our convention for glue signatures ensures that glue signatures representing a single glue are the same width as signatures representing pairs of glues and that the direction of the glues is indicated implicitly in the encoding. See Section~\ref{sec:IU-encodings} for specific details on our glue signature conventions. 

\subsubsection{Reading from the glue table}

Whether it be from a pair of input glues or a single input glue, the next step of the macrotile growth process involves using the input glue signature to read from the glue table. The glue table encodes all of the tile types in $T$ as pairs of input and output glue signatures. Reading from the glue table is done in a specific section that receives the input glue signature along one axis and the glue table encoding along the other. The input glue signature is propagated along the length of the glue table encoding using tiles that perform symbol matching logic to determine if the input glue signature exactly matches any of the input glue signatures in the glue table. If a match is found, the corresponding output glue signature is rotated in place of the input glue signature and additional matching is skipped. On the side opposite of the input glue signature is then the corresponding output glue signature, ready to be propagated to the output sides of the macrotile. Additionally, the encoding of the glue table is propagated towards the center of the macrotile unaltered so that it may be distributed to the output sides for use in adjacent macrotiles.

The representation function for our IU simulation may be defined by the output of the glue table reading gadget. Since the glue table contains information regarding all tiles to be simulated in a fixed order, the location where the glue table read gadget succeeds in finding a match uniquely corresponds to the tile type into which the macrotile should resolve. This macrotile resolution can occur as soon as the match is detected, but the exact tile attachment used to determine this is immaterial so long as resolving the macrotile occurs before any output information is presented to neighboring macrotiles.

Our convention for initiating macrotile growth ensures that the only time any tiles may attach inside of a macrotile is when two diagonally adjacent macrotiles present non-zero-strength glues to common neighbor or when a macrotile presents a strength-2 glue to a neighbor. Consequently, assuming the glue table contains encodings of all tiles to be simulated, by our requirements on standard systems, our IU simulation of standard systems will never contain fuzz tiles in unresolved macrotiles in any terminal assembly. In other words, because of the way macrotiles are initiated, assuming all tiles to be simulated are encoded in the glue table and the system to be simulated is standard, the simulation will be pristine.

\subsubsection{Propagating output information}

The definition of standard systems ensures that the glue table will always have an entry corresponding to any input glue signature that it may see. After reading from the glue table, all that remains is to propagate the resulting glue signature (which represents the output glues of the tile from $\calT$ represented by the macrotile) to the output sides of the macrotile. This is mostly done using standard tile gadgets for rotating and propagating sequences of symbols, though care needs to be taken to ensure that each side of the macrotile only has the part of the glue signature dedicated to that side. For instance, an output glue signature might contain information for glues to be presented on the north and east faces of a macrotile, but the north face should only present the north part of the signature and the east face the east part. This is handled by the \emph{glue output sections} which, using two passes over the signature, removes any part of the signature corresponding to other directions and replaces the most significant bit of the counter-clockwise-most signature on a face with a strength-2 glue in the case that the encoded glue is itself strength-2.

\subsection{Technical details} \label{sec:IU-technical}

\subsubsection{Conventions and Encodings} \label{sec:IU-encodings}

Throughout this construction, we choose and stick with a few conventions for encoding glues and tilesets. These conventions are essentially completely arbitrary and the only motivation for our choice was ease of implementation. There are certainly other choices that could easily work, and ours is in no way canonical.

\paragraph*{Glue Signatures}
For glues $g_N, g_E, g_S, g_W \in \mathscr{G}[T]$, let the notation $\langle g_N, g_E, g_S, g_W \rangle$ be called a \emph{glue signature} and denote a mapping from the cardinal directions (north, east, south, and west in that order) to the respective glues. In any situation where a glue in a glue signature is the strength-$0$ glue, the symbol $0$ can be used in the notation. For instance, the glue signature $\langle 0, g_1, g_2, 0 \rangle$ would assign the glue $g_1$ to the direction ``east'', the glue $g_2$ to ``south'', and the strength-$0$ glue to both the north and west directions. Keep in mind that glue signatures are not required to refer to a tile type with the corresponding glues on its respective sides. Instead, they are used as abstract data types in the construction, simply for when it is more convenient to keep track, for whatever reason, of multiple glues from different directions rather than individual glues. A glue signature is nothing more than the information contained in assigning a glue from $\mathscr{G}[T]$ to each cardinal direction.

Now let $\phi:\mathscr{G}[T]\times\{N,E,S,W\} \rightarrowtail \{0,1\}^m$ be an injective mapping, called the \emph{glue encoding function}, from glues of $\mathscr{G}[T]$ and the corresponding side of the tile on which the glue appears to binary strings of a fixed length $m$. Furthermore, it is assumed that there is only a single strength-$0$ glue in $\mathscr{G}[T]$ which is mapped under $\phi$ to the string $0^m$. In the case that $T$ contains tile types using multiple distinct strength-$0$ glues, it can easily be modified to use only one without altering its behavior by choosing one of the strength-$0$ glues arbitrarily and using that in place of the others. Additionally, it is assumed that all strength-$1$ glues map under $\phi$ to binary strings whose left (most significant) bit is $0$ and all strength-$2$ glues map to binary strings whose left bit is $1$. Since $\mathscr{G}[T]$ is finite, it is not difficult to devise an implementation of $\phi$ and length $m$ which satisfy these constraints.

Given a glue signature $\langle g_1, g_2, g_3, g_4 \rangle$, the notation $\phi\langle g_1, g_2, g_3, g_4 \rangle$ is abused to refer to the concatenation of the binary strings $\phi(g_1)$, $\phi(g_2)$, $\phi(g_3)$, and $\phi(g_4)$, each prefixed by a special symbol ``$\#$''. In other words, $\phi\langle g_1, g_2, g_3, g_4 \rangle$ is a string over the alphabet $\{0, 1, \#\}$ which matches the following regular expression. 
$$\#(0|1)^m\#(0|1)^m\#(0|1)^m\#(0|1)^m$$
Since $\phi$ is injective, this assignment of strings to glue signatures is invertible so that a glue signature may be recovered from its assigned string representation.

\paragraph*{Glue Tables}

During our intrinsically  simulation of $\calT$, the tiles of $\calU$ will need to somehow encode all relevant details regarding the dynamics of $\calT$. Particularly, this includes information about which tile types of $\calT$ can attach along different parts of a growing assembly. To accomplish this, each macrotile in $\calU$ keeps track of a data structure called the \emph{glue table} which encodes the full tile set $T$ as a mapping between glue signatures. 

\begin{figure}
	\centering
	\includegraphics[width=0.5\textwidth]{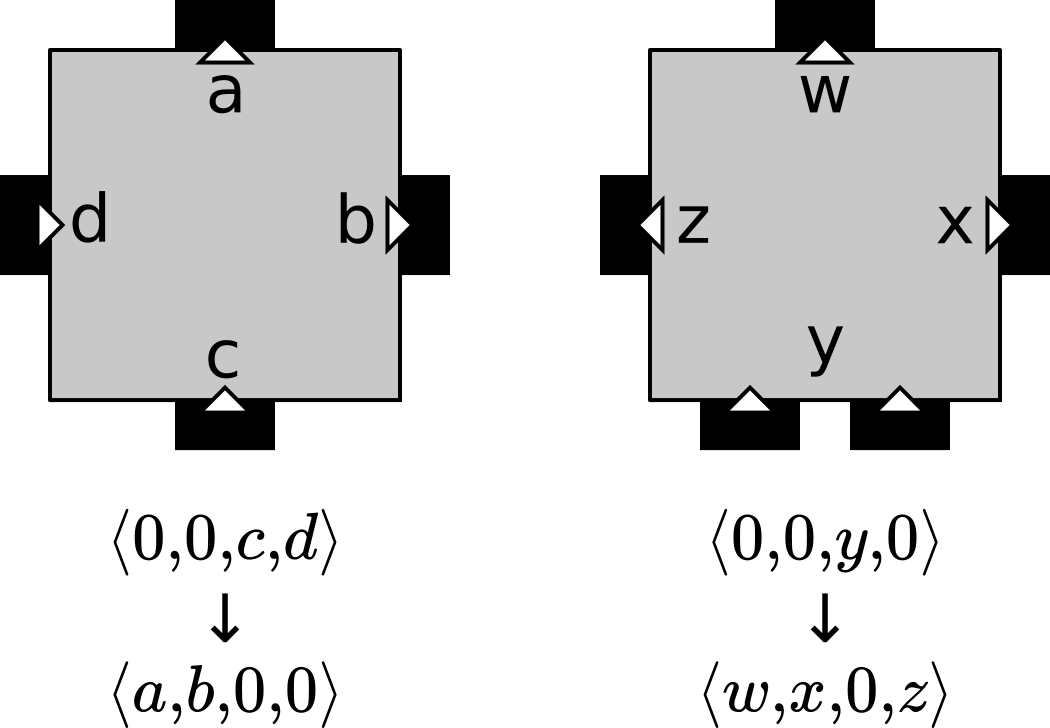}
	\caption{Two example tile types (top) and their corresponding glue table entry signatures (bottom). Glue IO-marks are indicated by small triangles, pointing into the tile type for input glues and out of the tile type for output glues. The left tile type has $2$ strength-$1$ glues as inputs on its south and west sides. The right tile type has a single strength-$2$ glue as input on its south side.}
	\label{fig:glue-table-entry-signatures}
\end{figure}

A glue table $\Gamma$ is a finite sequence of \emph{entries}, ordered pairs of glue signatures representing an input and output. Recall that glues in $T$ are IO-marked. Each tile type $t$ in $T$ is assigned an entry in the glue table where its input glues are assigned to the input signature of the glue table, and its output glues are assigned to the output signature. Output glues are assigned directly to the output signature so that the direction of the glue in the signature is the direction of the face on which the glue appears. For instance, if an output glue of type $g_o$ appears on the north face of tile type $t$, then the output signature corresponding to tile type $t$ will have $g_o$ as its north component. Input glues on the other hand, are assigned to the input glue signature with the opposite direction of the face on which they appear. For example, if an input glue $g_i$ appears on the east face of $t$, then it will be assigned to the west component of the corresponding input signature. All other components of the input and output glue signatures are assigned $0$. Figure~\ref{fig:glue-table-entry-signatures} illustrates what these signatures would look like for $2$ example tile types.

Given a tile set $T$ of size $n$ to be simulated, a glue table $\Gamma$ may be constructed by iterating over each tile type $t\in T$. For each tile type, an entry is constructed using the glues of $t$ as described above. Specifically, the input glues of each tile are assigned the signature $s^t_\text{in}$ and the output glues to the signature $s^t_\text{out}$. A glue table therefore is a sequence of pairs of glue signatures: 
$$\Gamma = (s^{t_1}_\text{in}, s^{t_1}_\text{out}), (s^{t_2}_\text{in}, s^{t_2}_\text{out}), \ldots, (s^{t_n}_\text{in}, s^{t_n}_\text{out})$$
The order of entries in the glue table is immaterial since this construction will parse glue tables using string matching logic rather than index counting logic. Given a glue table $\Gamma$ and a glue signature encoding function $\phi$, the notation $\phi\Gamma$ is abused to refer to the concatenation of glue signature encodings using the special separator symbols ``\$'', ``|$_1$'', and ``|$_2$'' in the following way.
For each tile type $t$ encoded in the glue table with input signature $s^{t}_\text{in}$ and output signature $s^{t}_\text{out}$, the corresponding glue table entry will be encoded as the string 
$$\phi s^{t}_\text{in}\;  |_1 \; |_2 \; \phi s^{t}_\text{out}$$ 
In other words, the encoding of each entry is the encoding of the input and output glue signatures separated by ``$|_1$'' and ``$|_2$''. Each of these entries is then prefixed by the separating symbol ``\$'' and concatenated together so that a glue table encoding may look like:
$$\$\; \phi s^{t_1}_\text{in}\;  |_1 \; |_2 \; \phi s^{t_1}_\text{out} \$\; \phi s^{t_2}_\text{in}\;  |_1 \; |_2 \; \phi s^{t_2}_\text{out} \cdots \$\; \phi s^{t_n}_\text{in}\;  |_1 \; |_2 \; \phi s^{t_n}_\text{out}$$ 
This encoding is summarized by a grammar whose terminal characters are $\{0,1,\#,\$,|_1,|_2\}$, non-terminals are $T$, $E$, and $S$ representing the encodings for a glue table, glue table entry, and glue signatures respectively, and production rules are:
\begin{align*}
    T & \to \$ E T \\
    T & \to \$ E \\
    E & \to S |_1 |_2 S \\
    S & \to \#(0|1)^m\#(0|1)^m\#(0|1)^m\#(0|1)^m \\
\end{align*}

% \begin{figure}
% 	\centering
% 	\includegraphics[width=0.8\textwidth]{images/IU-glue-table.png}
%     \caption{A schematic of a glue table encoded by the glues of tiles. A glue table consists of a sequence of \emph{entries} each separated by a pair of separating symbols. Each entry is made up from two glue signatures, one an input and the other an output. Finally, each signature is represented by 4 equally sized binary strings encoding glues from each of the 4 directions.}
%     \label{fig:IU-glue-table-layout}
% \end{figure}

\subsubsection{Common gadgets}

The bulk of our IU construction relies on the array of common gadgets we present here. When simulating a tile attachment, most sections within a macrotile will contain some variation of these gadgets. The gadgets presented here all exhibit rectilinear growth meaning that within each gadget, all tiles have the same pair of adjacent input directions. Rectilinear growth begins with the cooperative attachment of a tile in one corner of a section and ends with the attachment of a tile in the diagonally opposite corner.

\begin{figure}[!ht]
    \centering
    \includegraphics[width=0.4\textwidth]{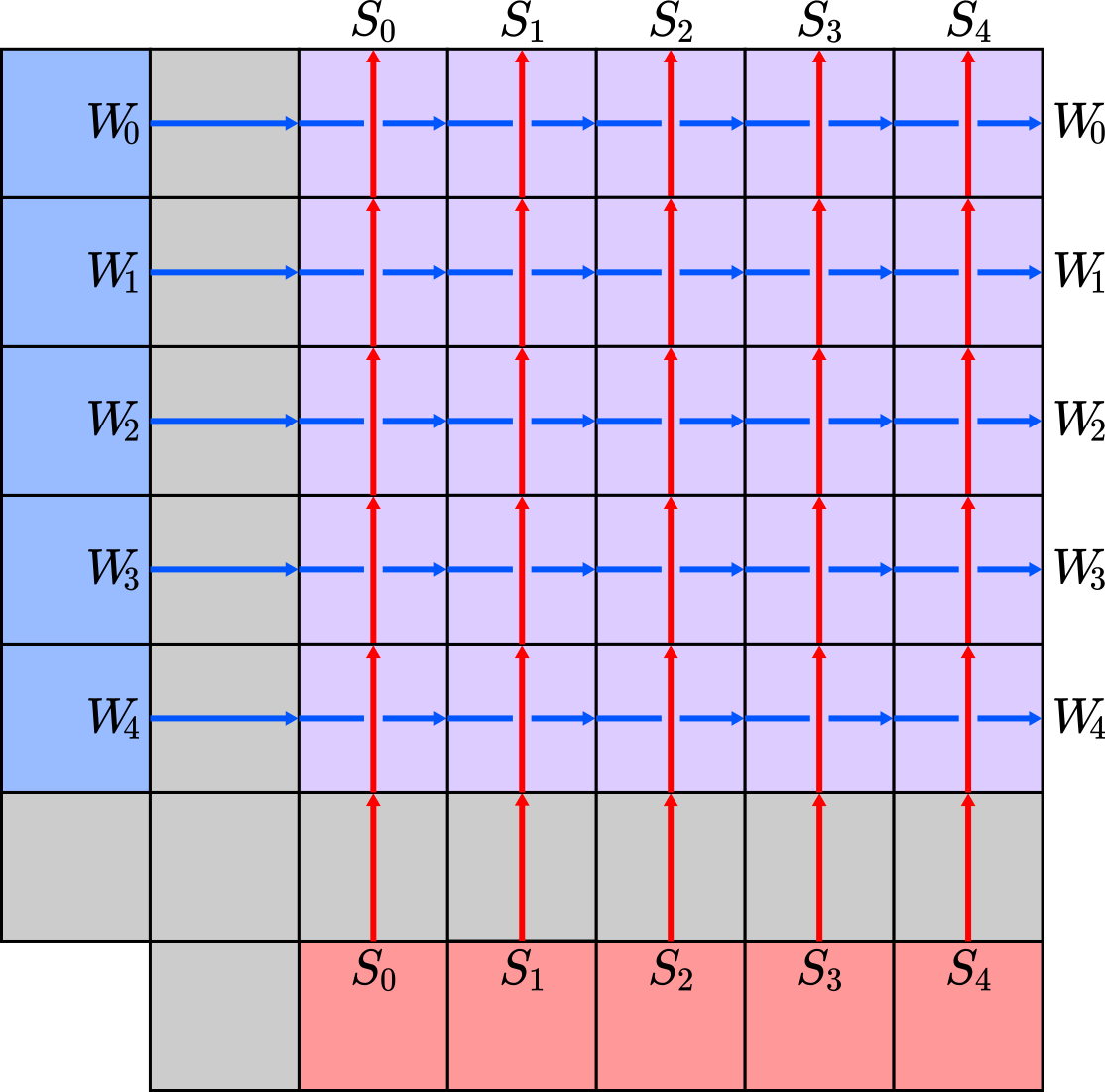}
    \caption{A schematic of tiles used to propagate two bit strings, one from south to north and the other from west to east. Arrows indicate the direction in which information is being propagated. Note that each tile in this illustration represents a schema of tile types. For instance, if the string of symbols being propagated contain only the binary bits $0$ and $1$, then there would need to be 4 distinct tiles to allow the information to propagate, one for each pair of bits (one from the west and one from the south).}
    \label{fig:IU-info-pass}
\end{figure}

Note also that it's possible to combine these gadgets so their functionality operates simultaneously in the same section. This enables, for instance, tiles to rotate a sequence of symbols in multiple directions at the same time. This does come at the cost of tile complexity however; combining gadgets requires tiles with glue labels that systematically concatenate the relevant information from the corresponding glue labels of the gadgets to be combined. The specific method of concatenating glue labels to form new ones is essentially a free choice, but effectively, this results in the tile complexity of the combined gadget being proportional to the product of tile complexity of each component gadget.

\begin{figure}[!ht]
    \centering
    \begin{subfigure}[T]{0.45\textwidth}
        \includegraphics[width=\textwidth]{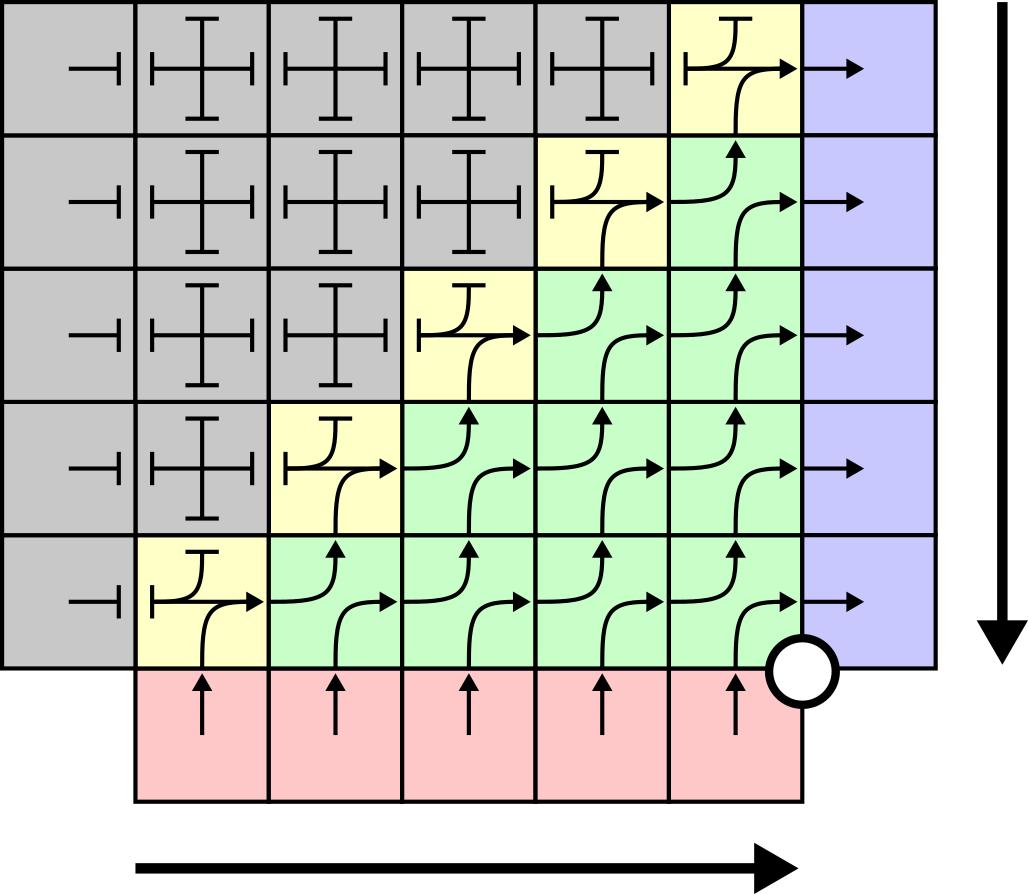}
        \caption{\emph{Cis}-rotation results in the output bits being aligned in the same direction as the input bits. Arrows to the side of the tiles indicate the orientation of the bits with the arrow pointing from most to least significant. Note that both the input and output bits have the MSB nearest the center of rotation.}
        \label{fig:gadget-rotation-cis}
    \end{subfigure}
    \hfill
    \begin{subfigure}[T]{0.45\textwidth}
        \includegraphics[width=\textwidth]{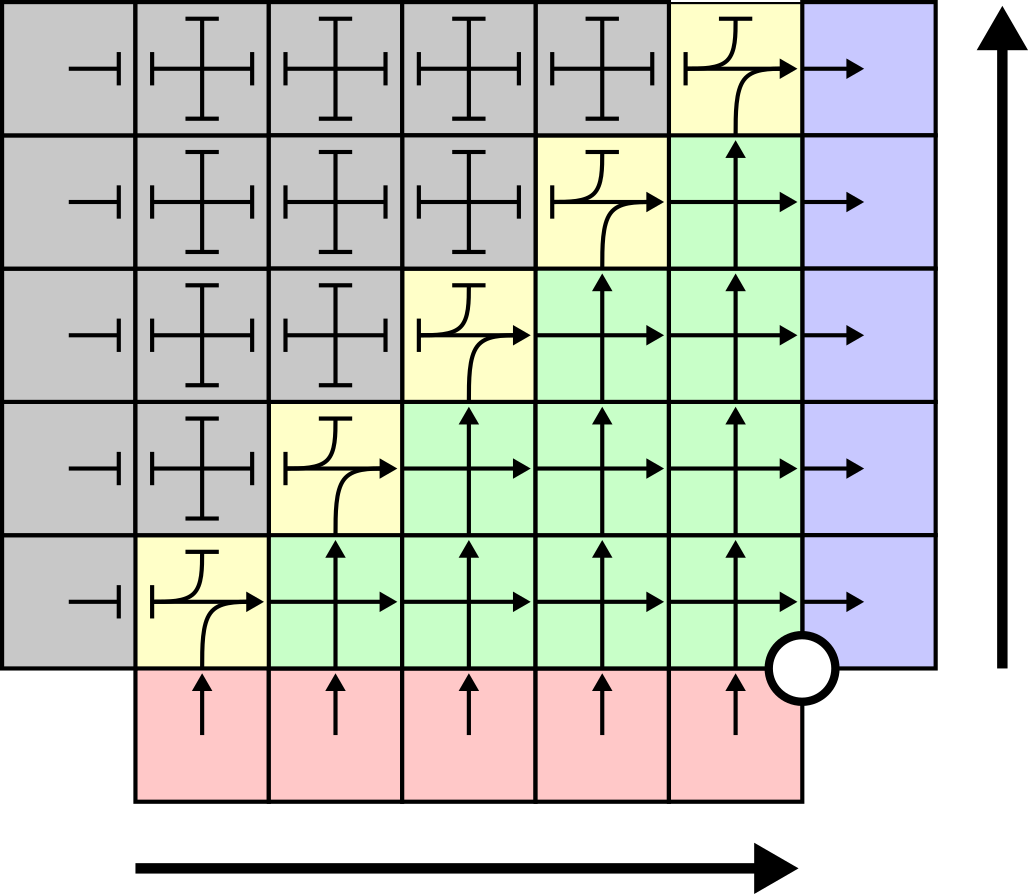}
        \caption{\emph{Trans}-rotation results in the output bits being oriented in the opposite direction of the input bits. Notice that the input bits have their MSB nearest the center of rotation, while the output bits are the opposite. Both types of rotations can be implemented using the same number of tiles.}
        \label{fig:gadget-rotation-trans}
    \end{subfigure}
    \caption{Gadgets can be made to rotate bit information encoded on the glues of tiles. After rotation, the resulting bit information will be propagating orthogonally to it's initial direction of propagation. Input bits are represented by red tiles and output bits by blue tiles. Gray tiles indicate tiles with no bit information, but which present strength-1 glues along which the rotation tiles can attach. Lines ending in a $T$-shape rather than an arrow indicate that a strength-1 attachment is occurring, but there is not bit information being propagated. Arrows indicate the propagation of a bit. The white circle indicates the center of rotation.}
    \label{fig:gadgets-rotation}
\end{figure}

The first gadget, illustrated in Figure~\ref{fig:IU-info-pass} propagates information from each of the input sides of a section to the opposite sides. This same gadget can be used to propagate information in just one direction by simply ensuring that the orthogonal direction propagates blank symbols that don't encode any information. The tile complexity for this gadget is proportional to the product of the number of distinct symbols that need to be propagated in both directions.

The next gadgets, illustrated in Figure~\ref{fig:gadgets-rotation}, rotate symbol information from one direction to an orthogonal one.

\subsubsection{Combining glue signatures}

\begin{figure}[!ht]
    \centering
    \includegraphics[width=0.8\textwidth]{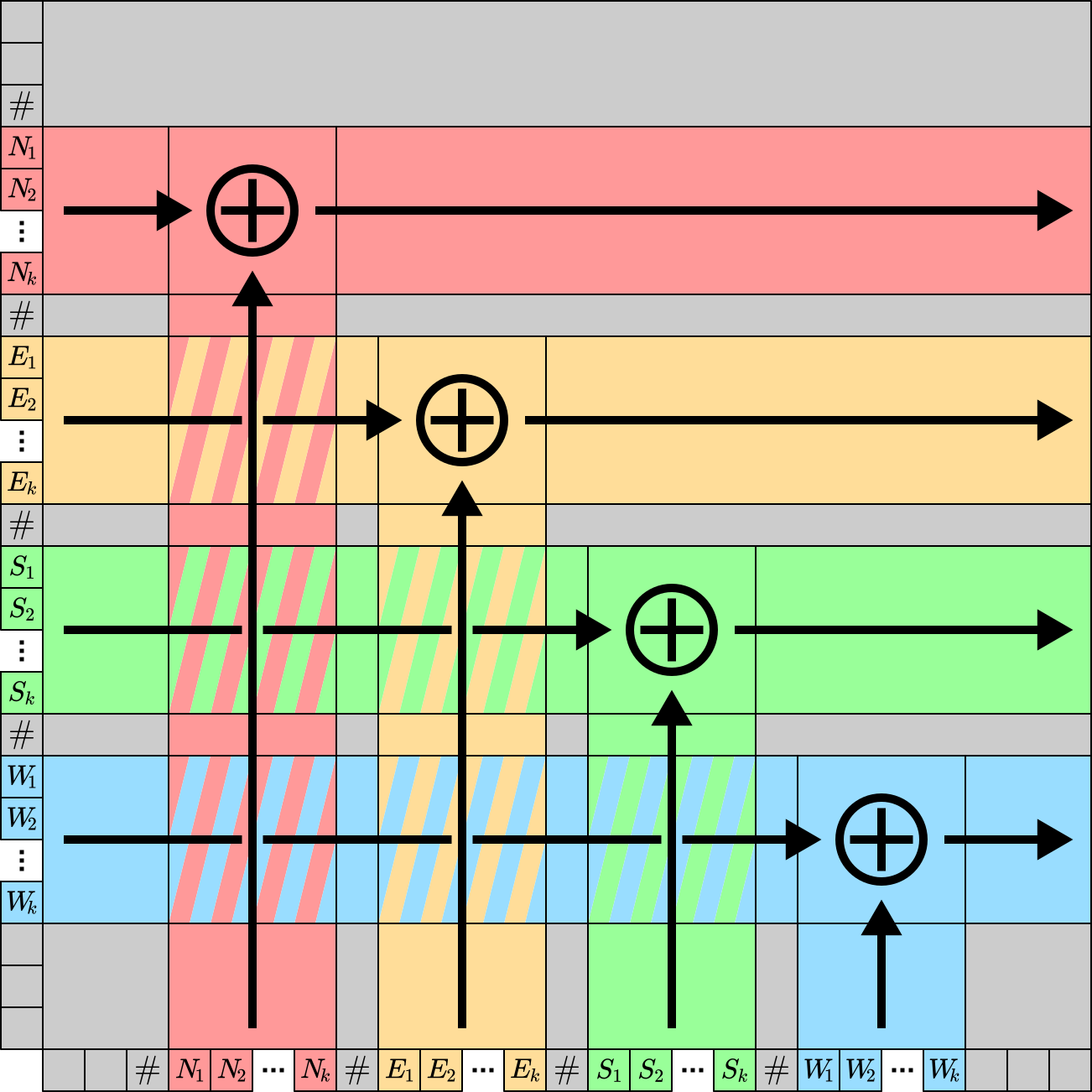}
    \caption{A schematic for the process of combining glue representations in the SW corner of a newly-forming macrotile. The adjacent macrotiles from the west and south present their glue signatures. The macrotile on the west presents an eastward facing glue so its signature only contains non-zero bits after the second glue separator (``\#''). Likewise for the south macrotile, it presents a north glue after the first glue separator. These glues are combined into a glue representation encoding both the south and west inputs. This is done using rectilinear growth, initiated from the point of cooperation between the padding regions between the diagonally adjacent tiles. Each of the binary representations in the south signature (or more generically, the counter-clockwise-most) is propagated upwards until it meets with the corresponding binary representations from the east. The binary representations are then XOR'ed together. Binary representations corresponding to different glue directions will be propagated through each other. To determine which direction a binary string corresponds to, the tiles that perform the propagation and combination also keep count of how many separators they have passed. In the end, the result is a new glue signature propagated to the east (more generally, the opposite direction of the clockwise-most cooperating macrotile) with non-zero strings for 2 directions.}
    \label{fig:IU-combine-glue-signatures}
\end{figure}

When simulating cooperative attachments between strength-1 glues, our IU tileset must combine the glue signatures presented from both of the cooperating macrotiles. Figure~\ref{fig:IU-combine-glue-signatures} depicts how information is propagated and combined as tiles attach. 

\subsubsection{Reading a glue table}

\begin{figure}
    \centering
    \includegraphics[width=0.5\textwidth]{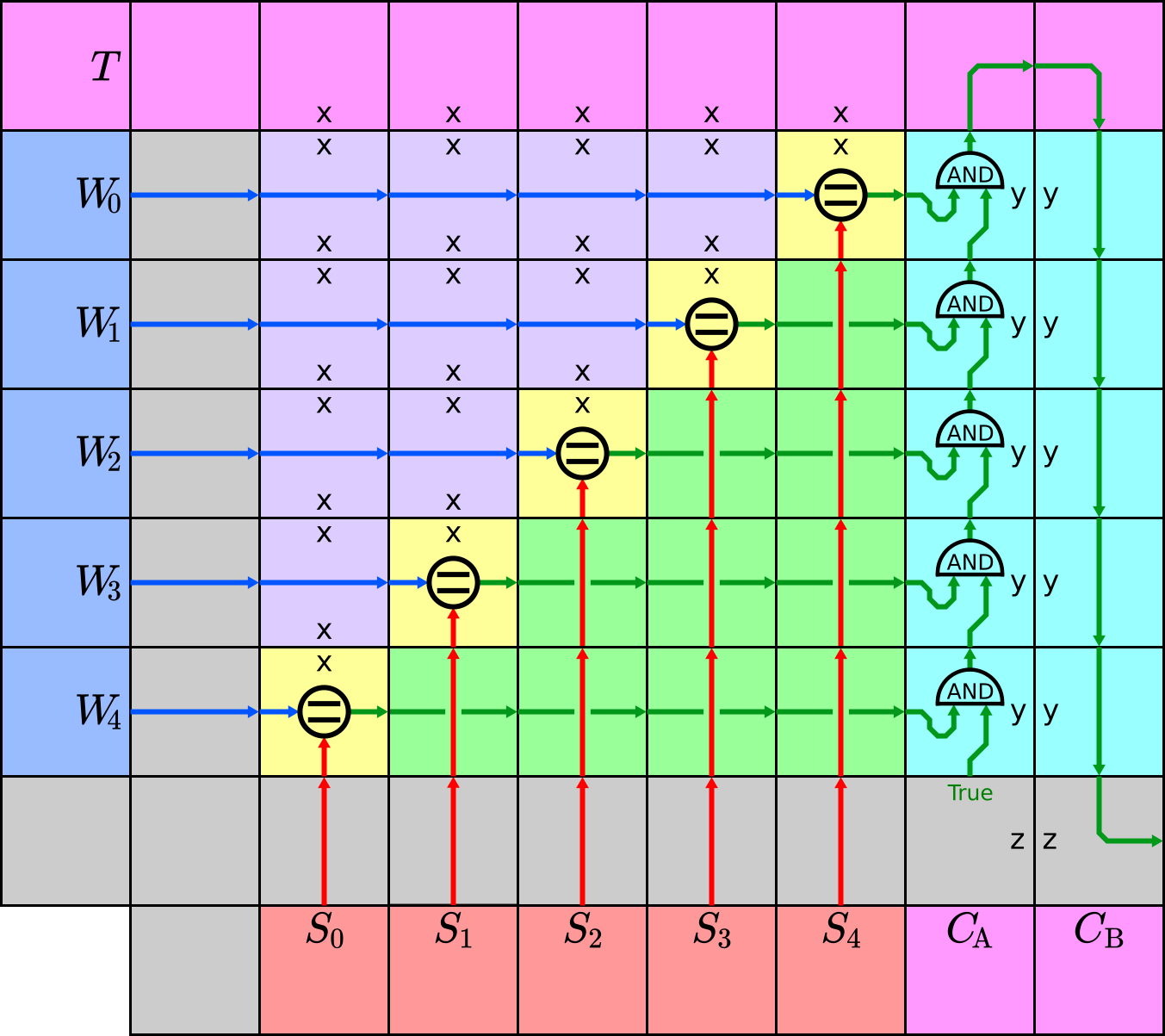}
    \caption{A schematic for a gadget which compares two strings of symbols and determines if they are equal. The string from the west $W_0\ldots W_4$ is compared with the string from the south $S_0\ldots S_4$ by means of tiles which attach in the rectangle spanned between them. These tiles propagate the information from both the west and south strings until they meet in the diagonal. If a symbol of one string matches the corresponding symbol of the other, a boolean signal (either true or false) will be propagated to the east indicating the match. Once the tiles have attached, an upward growing column of tiles will attach along the column of boolean signals and effectively AND all of them together (cyan tiles). This final signal is propagated back down to the south east of the gadget where it can be used to determine how the next gadget behaves. This behavior can be used to read from the glue table by performing a string comparison between each input entry of the glue table with the glue signature representing the current inputs to the macrotile. If the output boolean signal is true, indicating that the glue signature matches the input entry of the glue table, then it is ``known'' by the macrotile that the corresponding output entry encodes the tile to which the macrotile should resolve.}
    \label{fig:IU-glue-table-read}
\end{figure}

Throughout the IU construction, glue tables will be encoded by rows of tiles along the sides of each macrotile in the simulation. The purpose of a glue table is to help determine which simulated tile the macrotile should resolve into. To facilitate this, we introduce a tile gadget capable of reading an input glue signature from an entry of a glue table and comparing it to another glue signature representing the glues presented to the macrotile. This gadget is described in Figure~\ref{fig:IU-glue-table-read} and is capable of comparing two bit strings of the same length for equality. The result of the comparison is a column of tiles which each encode a whether the corresponding symbols in each string match. These boolean values are then reduces via the AND function to result in a single boolean value which is true if and only if all symbols match. The result can then be used to control the behavior of the tiles on the corresponding output glue signature of the glue table entry. If true, then the output glue signature may be rotated to replace the input glue signature, otherwise the input glue signature will be propagated to the next entry of the glue table for comparison. 

\section{A Self-Reproducing aTAM System: Self-Assembly of a Quine}\label{sec:quine}

We prove Theorem \ref{thm:quine} by construction, demonstrating an aTAM quine $\mathcal{Q} = (Q, \sigma, 2)$ for the tileset $U$ that is IU for standard aTAM systems (and its corresponding representation and seed generating functions $R$ and $S$) given in the proof of Theorem \ref{thm:standardIU}. Section \ref{sec:quine-blurb} contained a brief overview.

\thmQuine*

\begin{proof}
The remainder of this section consists of the full details of the construction of $\mathcal{Q}$.

\subsection{Overview of \texorpdfstring{$\mathcal{Q}$}{Q}}

A high-level, schematic depiction of $\mathcal{Q}$ can be seen in Figure \ref{fig:quine-overview}. Here we briefly describe its main components and growth process.

Overview of $\mathcal{Q}$:
\begin{enumerate}
    \item The seed $\sigma$ consists of a single tile.

    \item $Q$ is an IO marked tile set.

    \item $Q = T_S \cup T_F$, where $T_S$ is a subset of tile types we refer to as \emph{seed row} tiles, and $T_F$ is a subset of tile types we refer to as \emph{functional} tiles.
    
    \item The tile type of $\sigma$ is in $T_S$, and one copy of each tile of the types in $T_S$ attaches to the right of $\sigma$ to form a hard-coded line of length $|T_S|$ that we refer to as the \emph{seed row}.

    \item The glues on the north side of the seed row present a preliminary compressed version of the glue lookup table for $Q$ that contains glue lookup entries for only the tile types in $T_F$, as well as
    %encodings of the north and east glues of $\sigma$, plus
    additional information necessary to complete the tile lookup table and correctly format the sides of the macrotile that forms.

    \item The primary functions of the functional tiles that attach to the seed row and grow the macrotile are:
        \begin{enumerate}
            \item For each tile in the seed row, compute its compressed glue lookup entry and append that to the glue lookup table encoding so that the table eventually has entries for all tiles in $Q = T_F \cup T_S$.

            \item Decompress the glue encodings to ensure that all encodings are of the same width.
            
            \item Turn the assembly into a square while positioning and formatting the information encoded its perimeter to be consistent with the format utilized by the IU tile set. That is, turn it into a macrotile representing its own seed, with a full glue lookup table of the entire system encoded on the perimeter.
        \end{enumerate}
\end{enumerate}

In order to understand how the tiles of $T_F$ perform their work, we first describe several sets, values, and encodings that will be used in the construction.

\subsection{Glue lookup table encoding}

% Seed row symbols $\Sigma_\sigma = \{`B0!', `B1', `B2', `B3', `B4', `\$', `\mid1', `\mid2', `\#', `n', `p', `0', `1'\}$

% Glue encoding symbols $\Sigma_{GE} = \{`n', `p', `0', `1'\}$

% Glue lookup entry symbols $\Sigma_{GE} = \{`\mid1', `\mid2', `\#', `n', `p', `0', `1'\}$

% Glue lookup table symbols $\Sigma_\sigma = \{`\$', `\mid1', `\mid2', `\#', `n', `p', `0', `1'\}$

% Macrotile side symbols $\Sigma_{MT} = \{`B0', `B1', `B2', `B3', `B4', `B5', `B6', `B7', `\$', `\mid1', `\mid2', `\#', `0', `1'\}$

Let $G_V$ be the set of all glue labels on any north or south (i.e. vertically-binding) sides of tiles in $T_F$, and $G_H$ be the set of all glue labels on any east or west (i.e. horizontally-binding) sides of tiles in $T_F$. Let $g_m = \texttt{max}(|G_V|,|G_H|)$ be the size of the largest set, $G_V$ or $G_H$. Let function $\texttt{BIN}: \mathbb{N} \rightarrow \{0,1\}^*$ be defined such that $\texttt{BIN}(n)$ is the (standard) binary representation of natural number $n$. Thus, $l_g = |\texttt{BIN}(g_m)|$ is the number of bits needed for the binary representation of the size of the largest set, $G_V$ or $G_H$.

Let $l_s = |T_S|$ be the number of seed row tiles, which is also the length of the seed row. Each adjacent pair of seed row tiles has a strength-2 glue on their abutting west and east sides with a label that is unique to that pair, among all glues in $T_F \cup T_S$. (These will be the only glue labels unique to the seed row tiles, as all other glue labels will also be found on tiles of $T_F$, allowing them to bind to the seed row tiles.) This requires $l_s - 1$ unique glue labels. If we let $G_H'$ be $G_H$ unioned with these glues, then if we let $g_m' = g_m + l_s - 1$, since $l_s > 0$ we know $g_m' \ge |G_H'|$ and $g_m' \ge |G_V|$, meaning that $g_m'$ is at least as large as the size of the largest set of unique glue labels, the vertical or horizontal glue labels. We define $l_{pg} = |\texttt{BIN}(g_m')|$, the length of the binary representation of $g_m'$, and call it the \emph{padded glue length}.

\newcommand{\tttIN}{\texttt{IN}}
\newcommand{\tttOUT}{\texttt{OUT}}

Given a set of glue labels $G$, let $\texttt{LEX}(G)$ be a lexicographic ordering of the glue labels in $G$. To create the glue encodings for our construction, we use the alphabet $\Sigma_{GE} = \{0,1\}$ and define the function $\texttt{PAD\_BIN}: \mathbb{N} \times \mathbb{N} \rightarrow \{0,1\}^*$  such that $\texttt{PAD\_BIN}(n,l)$ is the string $s \in \{0,1\}^*$ consisting of the binary representation of $n$ padded on the left with the number of $0$s needed to make $|s| = l$, with the requirement that $|\texttt{BIN}(n)| \le l$.
For each $0 \le i < |\texttt{LEX}(G_V)|$, let $g_i$ be $\texttt{PAD\_BIN}(i,p)$. To create the glue encoding for a vertical glue (i.e. one on a north or south glue face) we use the index $i$ of its label in $\texttt{LEX}(G_V)$ to get $\texttt{PAD\_BIN}(i,l_{pg})$.
Then, if the glue has strength 1, we prepend a $0$ to $\texttt{PAD\_BIN}(i,l_{pg})$. 
If it has strength 2, we instead prepend a $1$. We define the glue encodings for $G_H'$ similarly, noting that the glue encodings of any two glues in the same set, $G_V$ or $G_H'$, will be unique, but two glues in different sets may have the same encoding.
Furthermore, since $l_{pg}$ is based on $g_m'$ we know that $l_{pg}$ is guaranteed to be long enough to contain the bits of any glue encoding in $G_V$ or $G_H'$ and thus any glue in $Q = T_F \cup T_S$, with possible leading $0$s for padding.
Therefore, all glue encodings will be of the same length, $l_{pg} + 1$.
Given a tile type $t$, direction $d \in \{N,E,S,W\}$, and $io \in \{\tttIN,\tttOUT\}$, we define $\texttt{GLUE\_ENC}(t,d,io)$ as a function such that, if $io = \tttIN$ and the glue on side $d$ of $t$ is an input glue, or $io = \tttOUT$ and the glue on side $d$ of $t$ is an output glue, returns the glue encoding of that glue in the specified format. Otherwise, if the value of $io$ does not match the glue's input/output status, a string of $l_{pg} + 1$ zeros is returned. One additional value that we will define for use later is $l_p = l_{pg} - l_g$.
That is, $l_p$ is the length of padded glues minus the length of the longest encoding needed for a glue in $G_V$ or $G_H$ (whichever requires the longest). Note the use of $G_H$ and not $G_H'$, so this does not include the glues specific the $T_S$, which contain the (overwhelmingly) largest proportion of all unique glue labels in $Q$.

For the glue lookup entries of our construction, we use the alphabet $\Sigma_{GLE} = \{0, 1, \#, \mid1, \mid2\}$. For each $t \in Q$, the glue lookup entry for $t$ consists of the concatenation of the following three strings (which consist of the concatenation of the encodings of input glues, two separator symbols, then concatenation of the encodings of the output glues):

`$\#\texttt{GLUE\_ENC}(t,N,\tttIN)\#\texttt{GLUE\_ENC}(t,E,\tttIN)\#\texttt{GLUE\_ENC}(t,S,\tttIN)\#\texttt{GLUE\_ENC}(t,W,\tttIN)$'

`$\mid1\mid2$'

`$\#\texttt{GLUE\_ENC}(t,N,\tttOUT)\#\texttt{GLUE\_ENC}(t,E,\tttOUT)\#\texttt{GLUE\_ENC}(t,S,\tttOUT)\#\texttt{GLUE\_ENC}(t,W,\tttOUT)$'

Given a tile type $t \in Q$, we define $\texttt{GLE}(t)$ as the function that returns the glue lookup entry for $t$ using that format. 

For the glue lookup table of our construction, we use the alphabet $\Sigma_{GLT} = \{0, 1, \#, \mid1, \mid2, \$\}$, and the table is the string  `$\$\texttt{GLE}(t_0)\$\texttt{GLE}(t_0)\ldots\$\texttt{GLE}(t_{i-1})$ for $0 \le i < |Q|$ and $t_i$ the $i$th tile type of $Q$ (i.e. the glue lookup entry of each tile type, with a $\$$ symbol prepended to each).

%Given a tile set $T$ that is IO marked, we define its glue lookup table to be the string `$T\texttt{GLE}(t_0)\#\texttt{GLE}(t_1)\#\ldots\#\texttt{GLE}(t_{i-1})T$' for $0 \le i < |T|$ and $t_i$ the $i$th tile type of $T$.

\subsection{Macrotile side encoding}

\newcommand{\B}[1]{\texttt{B#1}}
\newcommand{\vmark}[1]{\texttt{|$_#1$}}

\begin{figure}
    \centering
    \includegraphics[width=0.9\textwidth]{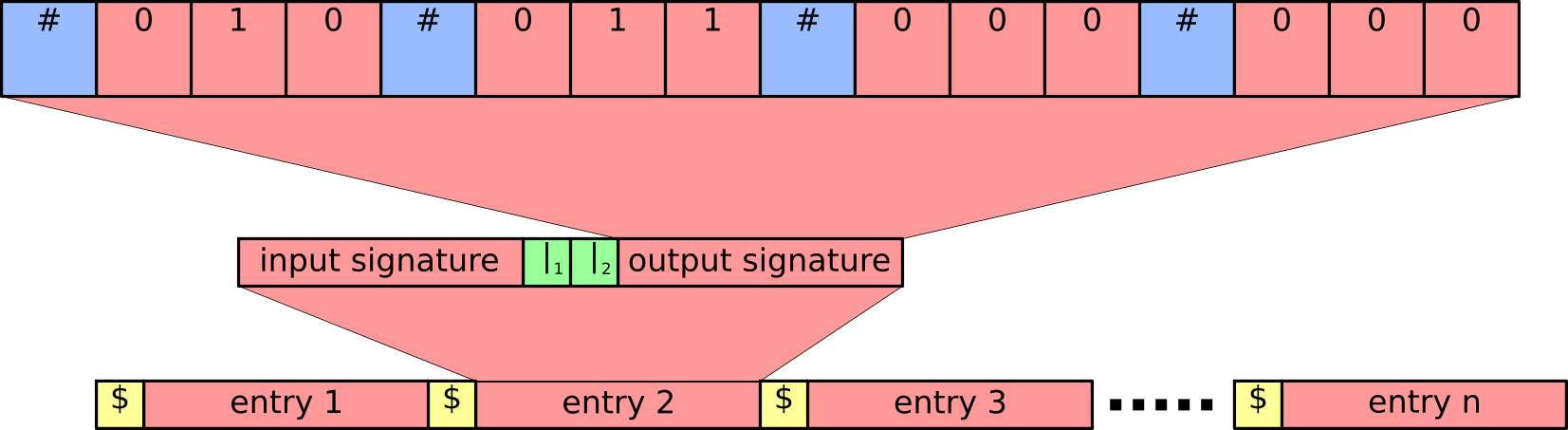}
    \caption{Example depicting the format of the glue lookup table. The glue lookup table (bottom) is composed of a list of glue lookup entries separated by $\$$ characters. Each glue lookup entry (middle) is composed of the input and output signatures of the corresponding tile separated by the two characters $|\mid1$ and $\mid2$. The input and output signatures have the same format, and an example output signature (top) is shown with output glues on the N and E sides (since the W and S glues are all $0$s, or \emph{null}). Since each of the N and E glue representations begin with a $0$, they each represent a strength-1 glue, and their labels are encoded as $10$ and $11$, respectively.}
    \label{fig:macrotile_side}
\end{figure}

Since only the north and east sides of the seed tile type, $t_\sigma$, have non-null glues, only the north and east sides of the macrotile representing $\sigma$ will encode glues and the glue lookup table. The glues on those sides of the macrotile represent $7$ distinct regions, each contained between a pair of lexicographically adjacent pairs of \emph{delimiter symbols} taken from the set $\{\B{0},\B{1},\B{2},\B{3},\B{4},\B{5},\B{6},\B{7}\}$. We will refer to the layout of the symbols and regions on the north side of the macrotile as follows (and for the east side, $g_N$ is replaced with $g_E$):

$\B{0} \langle \texttt{blank}_{out} \rangle \B{1} \langle g_N \rangle \B{2} \langle \texttt{blank}_{in} \rangle \B{3} \langle T \rangle \B{4} \langle \texttt{blank}_{in} \rangle \B{5} \langle g_N \rangle \B{6} \langle \texttt{blank}_{out} \rangle \B{7}$
\\
The strings between the pairs of delimiters are defined as follows:
\begin{enumerate}
    \item[] $\langle \texttt{blank}_{out} \rangle =$ a string of blank symbols (i.e. `\_') of length $l_{out}$, where $l_{out}$ will be defined later. 

    \item[] $\langle g_N \rangle =$ `$\#\texttt{GLUE\_ENC}(t_\sigma,N,\tttOUT)\#0^{l_{pg}+1}\#0^{l_{pg}+1}\#0^{l_{pg}+1}$'. That is, a string consisting of the glue encoding of the north glue of $t_\sigma$ followed by ``empty'' glue encodings for the other three directions (i.e. each is a string of $l_{pg}+1$ zeroes).

    \item[] $\langle \texttt{blank}_{in} \rangle =$ a string of blank symbols (i.e. `\_') of length $l_{in}$, where $l_{in}$ will be defined later. 

    \item[] $\langle T \rangle =$ the glue lookup table for $Q$.
\end{enumerate}

On the east side of the macrotile will be the same information, presented from top to bottom, with the slight difference that in the regions between $\B{1}$ and $\B{2}$, and between $\B{5}$ and $\B{6}$, the east glue of $t_\sigma$ will be encoded instead of the north, i.e. `$\#0^{l_{pg}+1}\#\texttt{GLUE\_ENC}(t_\sigma,E,\tttOUT)\#0^{l_{pg}+1}\#0^{l_{pg}+1}$'.

\subsection{Seed row encoding}

The labels on the glues on north sides of the tiles of the seed row encode a truncated, compressed, and slightly modified version of the information that will eventually be presented on the north and east sides of the macrotile.
We will refer to the layout of the symbols and regions on the north side of the seed row as follows:

$\B{0}! \langle \texttt{ctrpad} \rangle \B{1} \langle g_N,g_E \rangle \B{2} \langle \texttt{gluepad} \rangle \B{3} \langle T \rangle \B{4}$
\\
The strings between the pairs of delimiters are defined as follows:
\begin{enumerate}
    \item[] $\langle \texttt{ctrpad} \rangle =$ the binary string representing $g_m + 1$ padded to length $l_{pg}$ with $0$s on the left, then reversed in direction so that the least significant bit is on the left.  

    \item[] $\langle g_N,g_E \rangle =$ `$\#\texttt{GLUE\_ENC}(t_\sigma,N,\tttOUT)\#\texttt{GLUE\_ENC}(t_\sigma,N,\tttOUT)\#0^{l_{pg}+1}\#0^{l_{pg}+1}$'. That is, a string consisting of the glue encoding of the north glue of $t_\sigma$ followed by glue encoding of the east glue of $t_\sigma$, then 2 ``empty'' glue encodings for the other two directions (i.e. each is a string of $l_{pg}+1$ zeroes).

    \item[] $\langle \texttt{gluepad} \rangle =$ a string of $0$s of length $l_p$. 

    \item[] $\langle T \rangle =$ the compressed glue lookup table for the tiles of $T_F$.
\end{enumerate}

The ``$\B{0!}$'' symbol is a special symbol used to initiate growth and later replaced by the ``$\B{0}$'' delimiter symbol.

The northern glues of the tiles of $T_S$ in the $\langle T \rangle$ region present encodings of all of the tiles of $T_F$, and since there are $g_m$ symbols in the largest set, $G_V$ or $G_H$, which contain all of the vertical and horizontal glue labels of $T_F$, respectively, then the largest value of a glue encoding used to encode the tiles of $T_F$ is $g_m$ (since the value $0$ is reserved for the null glue). As previously mentioned, none of the vertical glue labels of the seed tiles are unique to tiles of $T_S$ (i.e. one or more tiles of $T_F$ also use each). Additionally, the east glue label of the easternmost seed row tile is $\texttt{FILL\_BOTT}$, which is shared by another tile in $T_F$ (and therefore already in $G_H$), and the south sides of the seed row tiles have no glues.
Therefore the only new glue labels that need to be encoded in order to create the glue lookup entries for the tiles of $T_S$ are those between pairs of seed row tiles.
As stated, there are $l_s - 1$ of these (where $l_s = |T_S|$). In order to allow the tiles of $T_F$ to create the glue lookup entries of the tiles of $T_S$, the value $g_m + 1$ is encoded into the $\langle \texttt{ctrpad} \rangle$ region between $\B{0}$ and $\B{1}$. This will be the value of the glue encoding of the first new glue representation.

In order to make the seed row more compact, requiring many fewer tiles for $T_S$, we compress the glue encodings contained in the glue lookup table which encodes $T_F$. This is done in two ways: (1) any glue encoding which needs to represent an empty glue location is represented by a single `$n$' character (rather than the string of $0$s of length $l_{pg}$ that will later be necessary), and (2) for the glue encoding of each non-empty glue, starting from the second leftmost bit (since the leftmost bit represents the strength of the glue), $l_p$ bits are replaced by a single `$p$' character. Those bits are guaranteed to all be $0$s by the definition of $l_p$, since it $l_p = l_{pg} - l_g$, meaning that it is the length of a fully padded binary representation of a glue minus the length of the longest encoding in $G_V$ and $G_H$, which includes any glue being encoded for $T_F$. So, the additional bits (which are only needed for the encoding of horizontal glues of seed row tiles) can be replaced by the `$p$' character and since $\langle \texttt{gluepad} \rangle$ contains a string of $l_p$ $0$s, each such `$p$' can later be replaced by $l_p$ $0$s. (Note that the only glue encodings that are not compressed in this way are those in the $\langle g_N,g_E \rangle$ region due to the physical layout of the information later used to decompress glue encodings.) The ``decompression'' is necessary because the IU tileset $U$ requires that all fields representing glues on the sides of a macrotile to be of the same width.

\subsection{Building the macrotile from the seed row}

In order to grow and transform the information encoded along the north of the seed row into the information needed for the sides of the macrotile, the following phases of assembly occur:

\begin{enumerate}
    \item Phase 1: For each tile in the seed row, the northernmost row is extended by a fixed number of tiles that consist of a preliminary representation of that seed tile, called a \emph{compressed glue lookup entry}.

    \item Phase 2: Every `$p$' symbol is replaced by a string of $l_p$ $0$s.

    \item Phase 3: Every `$n$' symbol is replaced by a string of $l_{pg}+1$ $0$s.

    \item Phase 4: The $\texttt{ctr}$ symbols in the seed row tile templates are replaced by appropriate values of the counter encoded in $\langle \texttt{ctrpad} \rangle$.

    \item Phase 5: The symbols in the regions between $\B{0}$ and $\B{1}$, and between $\B{2}$ and $\B{3}$, are each replaced by a `$\_$' symbol.

    \item Phase 6: The contents of the regions between (a) $\B{2}$ and $\B{3}$, (b) $\B{1}$ and $\B{2}$, and (c) $\B{0}$ and $\B{1}$ are copied to extend the right side of the row into regions between (a) $\B{4}$ and $\B{5}$, (b) $\B{5}$ and $\B{6}$, and (c) $\B{6}$ and $\B{7}$, respectively.

    \item Phase 7: The rectangular assembly created by the previous phases is turned into a square macrotile with correct representations on the N and E sides of the glues of $t_\sigma$ and all information rotated and spaced appropriately for simulation of the system by the IU tile set.
\end{enumerate}

We now give a brief overview of each phase. Note that for Phases 1-6, all growth is done in a zig-zag manner after the seed row. The seed row grows from left to right, and so does row 1 directly across its north. Then, row 2 and all subsequent even numbered rows grow from right to left. Row 3 and all odd rows grow from from left to right. The right-to-left growing rows always start immediately north of the furthest rightmost tile(s) and stop in the column immediately north of the seed tile (at $x$-coordinate $0$). The left-to-right growing rows always start at $x$-coordinate $0$ and either stop above the furthest rightmost tile(s) or extend the row by some constant number of tiles (depending on the phase) beyond the previously rightmost tile(s).

\subsubsection{Phase 1: adding compressed glue lookup entries for the tiles of \texorpdfstring{$T_S$}{TS}}

In this phase, the compressed glue lookup entry for each tile of the seed row is appended to the right end of the seed row. Figure \ref{fig:seed-row-ext} shows an example of the compressed lookup entry added for the seed tile.

\begin{figure}[ht]
    \centering
    \includegraphics[width=1.0\textwidth]{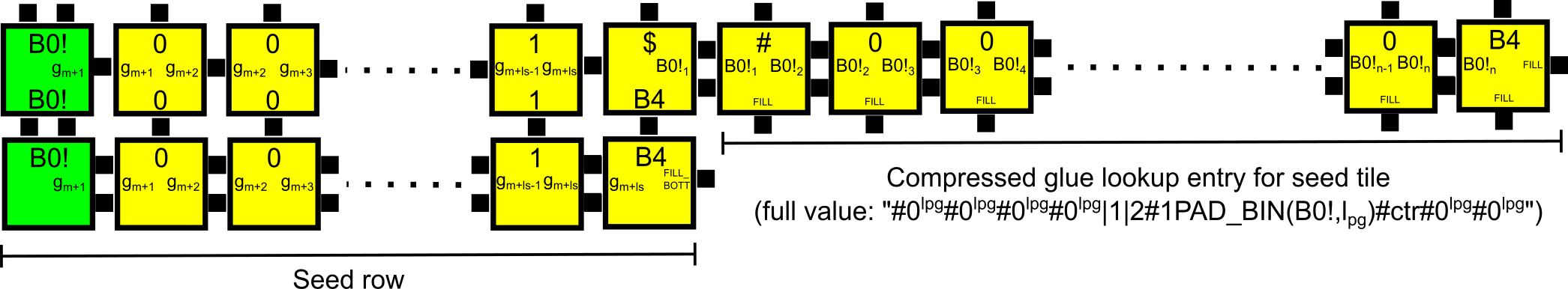}
    \caption{Example depiction of the seed row (bottom) and the compressed glue lookup table entry added for the leftmost (i.e. seed) tile (top). (Not all glue markings are shown.)\label{fig:seed-row-ext}}
\end{figure}

    \begin{enumerate}
        \item For $0 < i < |T_S| - 1$, let $t_{s_i} \in T_S$ be the $i$th tile of $T_S$ and also the $i$th tile from the left of the seed row once it has completed growth.

        \item During the growth of this phase, which is performed by the tiles of $T_F$, the full definition of each $t_{s_i}$ is inferred from its location in the seed row. For each, its northern glue can be read directly by a tile from $T_F$ attaching to its north since the seed row tiles are carefully designed so that their northern glues only consist of the following symbols:
        
        $\{\B{0!}, \B{1}, \B{2}, \B{3}, \B{4}, \$, \vmark{1}, \vmark{2}, \#, n, p, 0, 1\}$.
        
        Therefore, the tiles of $T_F$ can be created before the tiles of $T_S$ but still be guaranteed to be able to correctly bind to them and infer their definitions. The tile type of $t_{s_0}$ only has that northern glue and a strength-2 glue on its east side that allows a tile of type $t_{s_1}$ to bind, and its south and east sides contain the null glue. All other tiles in the seed row except for the rightmost, i.e. $t_{s_i}$, are similar except that they each also have a western glue that allow them to bind to $t_{s_{i-1}}$. The rightmost tile of the seed row, of type $t_{s_{|T_S|-1}}$, only has such a western glue, but its eastern glue is of strength-1 and has a label of `$\texttt{FILL\_BOTT}$' (which will be explained later).

        \item The strength-2 glues between each pair $t_{s_i}$ and $t_{s_{i+1}}$ are assigned consecutive glue encoding values beginning from $g_{m+1}$, which is the value encoded in $\langle \texttt{ctrpad} \rangle$.

        \item For example, the tile type of $t_{s_0}$ is known to be a special case with a strength-2 glue on its north whose label is $\B{0!}$ and it has one other non-null glue, on its east side, that is of strength-2 and whose label is assigned the glue encoding $g_{m+1}$. Since it is the seed tile and thus has no input, its north glue is strength-2 with the label $\B{0!}$ (which gets mapped to index number 1), and its eastern glue is assigned the current counter value (denoted by the $\texttt{ctr}$ symbol) its compressed glue lookup entry is the string 
        %`$\#n\#n\#n\#n\vmark{1}\vmark{2}\#1p00001\#\texttt{ctr}\#n\#n$' 
'$\#0^{l_{pg}}\#0^{l_{pg}}\#0^{l_{pg}}\#0^{l_{pg}}\vmark{1}\vmark{2}\#1\texttt{PAD\_BIN}(\B{0!},l_{pg})\#\texttt{ctr}\#0^{l_{pg}}\#0^{l_{pg}}$'.
        For another example, we'll describe the entry for a tile $t_{s_i}$, for $0 < i < |T_S| - 1$, assuming it has a north glue of strength-1 encoding a symbol $x \in \{\B{1}, \B{2}, \B{3}, \B{4}, \$, \vmark{1}, \vmark{2}, \#, n, p, 0, 1\}$ (which is the full set of labels of the northern glues of the tiles of $T_S$ other than the leftmost and rightmost, which are $\B{0!}$ and $\B{4}$, respectively).
        Then, the glue lookup entry for $t_{s_i}$ is `$\#0^{l_{pg}}\#0^{l_{pg}}\#0^{l_{pg}}\#\texttt{ctr}\vmark{1}\vmark{2}\#0\texttt{PAD\_BIN}(x,l_{pg})\#\texttt{ctr}\#0^{l_{pg}}\#0^{l_{pg}}$'.

        \item Since the north glue of any tile in $T_S$ is one of those in the given set and the rest of the tile's definition is fixed with the caveat that the value used to encode the horizontal glues increases by one for each glue moving to the right and a single generic variable symbol $\texttt{ctr}$ can be used to represent all of these values (to be explained in a later phase), there are a fixed number of possible compressed glue lookup entries of fixed width. Thus, the hard-coded set of tiles needed for each possible northern symbol of a seed row tile are created as part of $T_F$, before the tiles of $T_S$ are generated (therefore allowing the tiles of $T_S$ to encode them in their northern glues).

        \item Each row of Phase 1 that grows from left to right reads and stores the northern glue label $x$ of the tile that currently has the marking symbol $\texttt{ext}$ (which starts on the leftmost tile), moves the $\texttt{ext}$ symbol one tile to the right, then grows to the far right end of the row below it, whose rightmost tile has the glue label $\B{4}$. At that location, the tiles that encode the compressed glue lookup entry for symbol $x$ to attach and extend the row further to the right. The location with the $\B{4}$ symbol is changed to have a $\$$ symbol, and the $\B{4}$ symbol is placed at the rightmost end of the newly extended row.

        \item Having completed growth of that row, adding the encoding of one more tile of $T_S$, the next row grows all the way back to the left and initiates growth of the next left-to-right now that continues the process. Once the marker $\texttt{ext}$ passes the column in which the seed row ended (whose location is preserved via another special marker symbol), Phase 1 is complete since every tile of $T_S$ will now have a compressed glue lookup entry for it added to the tile lookup table.
    \end{enumerate}

    At the end of this phase, there is a (compressed) glue lookup entry for every tile in $Q$.

%The glue labels presented on the sides of the macrotile are the 15 following:
%$\{\B{0}, \B{1}, \B{2}, \B{3}, \B{4}, \B{5}, \B{6}, \B{7}, \$, \vmark{1}, \vmark{2}, \#, 0, 1, \_\}$.

%To reserve the number 0 for the null glue and assign a unique binary number to each seed row label requires 14 values and therefore 4 bits. 

\subsubsection{Phase 2: replacing \texorpdfstring{$p$}{p} symbols with strings of \texorpdfstring{$l_p$ $0$}{lp 0}s}

Phase 2 executes a relatively simple procedure. It begins by marking the location of the leftmost $p$. Then, for each $0$ in the region between $\B{2}$ and $\B{3}$, except for the final $0$, it ``inserts'' a $0$ after the marked $p$. It does this by growing a row to one position right of the marked $p$, setting the north glue of the tile at that location to $0$ while encoding the symbol currently at that location in its east glue. The tile to the east places uses that symbol for its northern glue and propagates the current value for that location to the right. This occurs until the $\B{4}$ symbol is encountered at the end of the row, at which point it is most right to a tile that extends the row rightward by one tile. For the final $0$, instead of inserting a $0$, it simply replaces the marked $p$ by $0$. In this way, the location of the marked $p$ has a string of $l_p$ $0$s inserted in its place, and by moving the mark to the next $p$ to the right after completing each such loop, that happens for all $p$ symbols. This results in all glues with the ``padding'' marker being decompressed to the full, fixed length required for all glue encodings. Figure \ref{fig:replacing-ps} shows a high-level example of one such loop iteration.

\begin{figure}
    \centering
    \includegraphics[width=1.0\textwidth]{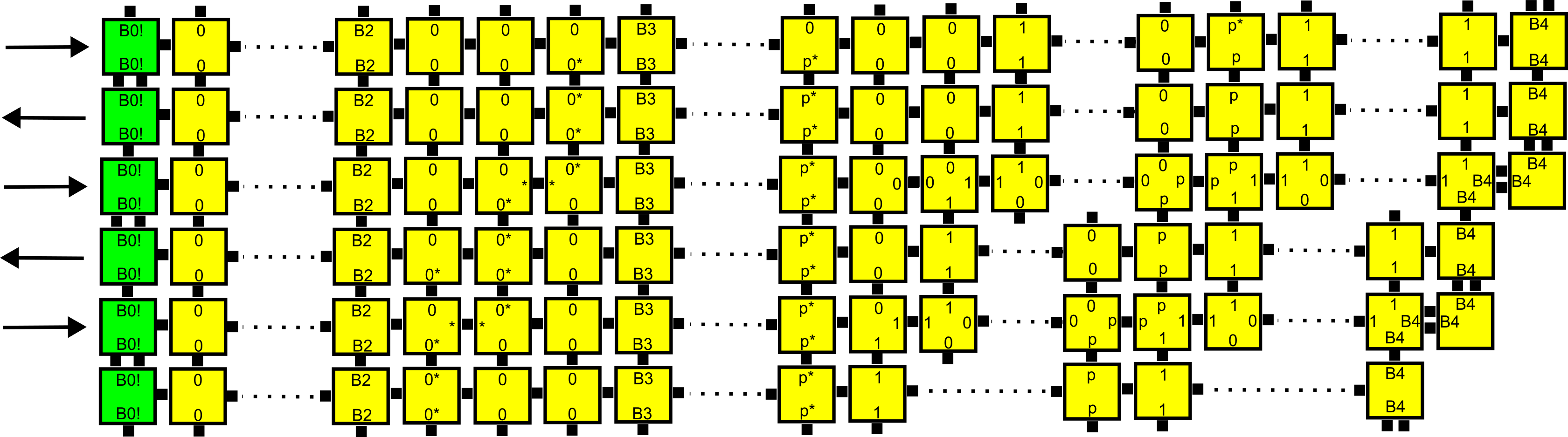}
    \caption{High-level depiction of an iteration of the main loop of Phase 2, i.e. inserting a string of $0$s (whose length is equivalent to the number of $0$s between the $\B{2}$ and $\B{3}$ delimiters) into the location of the leftmost $p$ symbol. (Note that, to highlight the main logic, not all glues are fully shown.) At the beginning of an iteration (the bottom row), the leftmost $0$ following $\B{2}$ is marked (here with a ``*'' symbol), and the leftmost $p$ is also marked. The first row of the iteration grows above it, from left to right, and moves the marker on the $0$ one position to the right, then inserts a tile representing a $0$ immediately following the marked $p$. This requires the inserted location and the rest of the tiles of the row to pass the original value of each column to the right and to change the value of the column to the value received from their left, growing the row by one tile. Right-to-left growing rows simply copy values upward. Subsequent left-to-right growing rows continue the process until the final $0$ before $\B{3}$ is encountered, at which point the marker for the $0$s is removed and that row replaces the marked $p$ with a $0$, then marks the next $p$ encountered to the right. In this example, three $0$s are between $\B{2}$ and $\B{3}$, so the marked $p$ is replaced by three $0$s. To begin the next iteration, the first $0$ after $\B{2}$ is marked. Iterations continue until no $p$ symbols remain in the row.\label{fig:replacing-ps}}
\end{figure}

\subsubsection{Phase 3: replacing \texorpdfstring{$n$}{n} symbols with strings of \texorpdfstring{$l_{pg} + 1$ $0$}{lpg + 1 0}s}

Phase 3 is performed in almost the exact same manner as Phase 2, but with the number of positions between $\B{0}$ and $\B{1}$ being used to determine how many $0$s to insert for each $n$ symbol. Irrespective of the actual bit values between $\B{0}$ and $\B{1}$, for each a $0$ is inserted, and for the final bit two $0$s are inserted so that the $n$ symbol will be replaced by a string of $0$ symbols of length $l_{pg} + 1$, since each glue label is encoded with $l_{pg}$ bits and the extra $0$ bit is in the location denoting the strength of the glue. Recall that this glue encoding, composed of all $0$s, represents the special case of the null glue.

\subsubsection{Phase 4: replacing the \texorpdfstring{$\texttt{ctr}$}{ctr} symbols with counter values}

Phase 4 is also performed in a manner very similar to the previous two phases, with just a few notable differences. For $0 \le i < |T_S|$, let $g_{s_i}$ be the label of the horizontal glue shared by two adjacent seed row tiles such that it is the $i$th such glue from the left (e.g. $g_{s_0}$ is the label of the glue shared by leftmost seed row tile and its neighbor immediately to the right). The compressed glue lookup entries for the seed row tiles (added in Phase 1) are located from left to right in the same order as the seed row tiles themselves, and within those the glues used as input are listed to the left of those used as output. The leftmost seed row tile has no input since it is $t_\sigma$, the seed of the system, and every tile to the right of that tile has a single input glue on its west side. Except for the rightmost seed row tile, each has an output to its east.
Since compressed glue lookup entries encode all such glues simply using the symbol $\texttt{ctr}$, and each pair of adjacent edges between seed row tiles share the same glue, the pattern exists where the leftmost two instances of $\texttt{ctr}$ in the table refer to the same glue as each other, the next two instances to the same glue as each other, etc.
This pattern can be seen in Figure \ref{fig:ctr_subs}.
Since the glues are encoded using values that are each one larger than the previous, from left to right, in this phase all that is required is that the current value of the counter bits is inserted into the locations of both of the two leftmost $\texttt{ctr}$ symbols, and then the counter is incremented by one and the process repeated until all copies of $\texttt{ctr}$ have been replaced.

\begin{figure}
    \centering
    \includegraphics[width=1.0\textwidth]{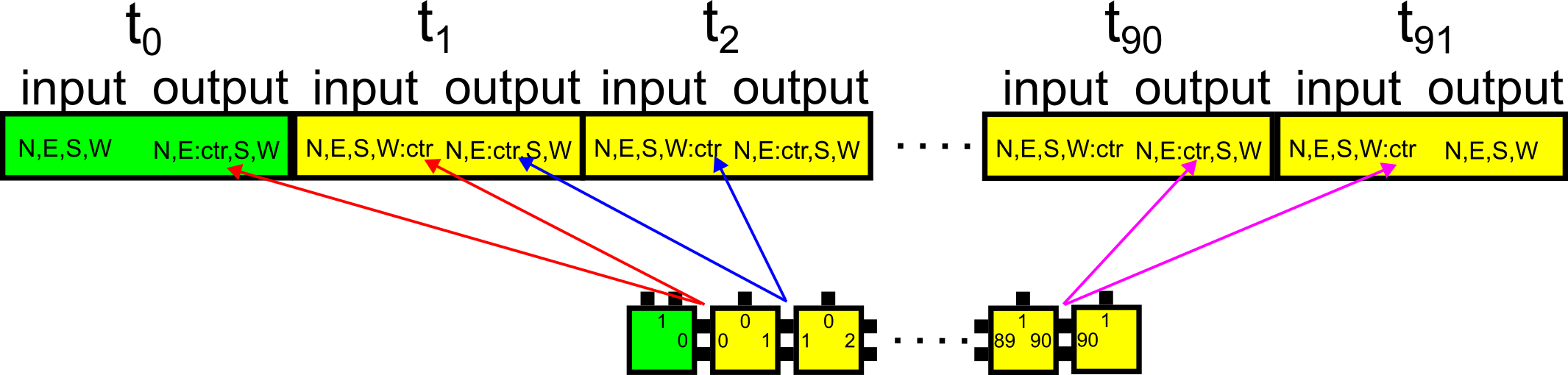}
    \caption{Schematic example of seed row tiles (bottom) and their corresponding compressed glue lookup entries (top). The only instance of the $\texttt{ctr}$ symbol in the entry for $t_0$ will be as its output glue to the east. The entry for $t_1$ has two instances of $\texttt{ctr}$ with the left representing its input glue to the west and the second its output glue to the east. Note that the leftmost two instances of $\texttt{ctr}$ in the table refer to the same glue, shared by $t_0$ and $t_1$. Therefore, the same value, which is the current value of the counter whose bits are represented in the region between $\B{0}$ and $\B{1}$, is put into both locations. Then the next two instances of $\texttt{ctr}$ also refer to the same glue, so once the counter is incremented to create a new unique encoding value for that glue, the counter's value can then be put into both locations. This pattern continues until the final, rightmost glue between seed row tiles.\label{fig:ctr_subs}}
\end{figure}

Recall that the symbols representing the bits of the counter in the seed row are in reverse order, with the least significant bit being on the left.
Since the bits are copied from left to right, and each inserted immediately following a $\texttt{ctr}$ symbol, the copied values are again reversed, to be in the correct order.
Finally, once all counter bits have been copied to the right of a $\texttt{ctr}$ symbol, the $\texttt{ctr}$ symbol is replaced by a $1$ symbol. This is in the position of the bit that specifies the glue's strength, and is a $1$ because all such glues are of strength-2.

\subsubsection{Phase 5: Blanking out spacer regions}

Phase 5 is the simplest phase. Its purpose is simply to turn all symbols in the current ``spacer'' regions ($\texttt{blank}_{out}$ between $\B{0}$ and $\B{1}$, and $\texttt{blank}_{in}$ between $\B{2}$ and $\B{3}$), into blank, i.e. ``\_'', symbols. This is done by a single row that grows left to right that detects when it is growing over one of those regions and presents only ``\_'' symbols to the north until leaving the region. Once completed, a row grows back right-to-left to reset and allow for the next phase to begin.

\subsubsection{Phase 6: Copying regions to the right}

Phase 6 also performs a relatively simple task. Namely, it copies the contents of the $\texttt{blank}_{out}$, $g_N,g_E$, and $\texttt{blank}_{in}$ regions so that they each have a copy on the right side of the glue lookup table. The ordering and a schematic depiction can be seen in Figure \ref{fig:region-copy}.

\begin{figure}
    \centering
    \includegraphics[width=1.0\textwidth]{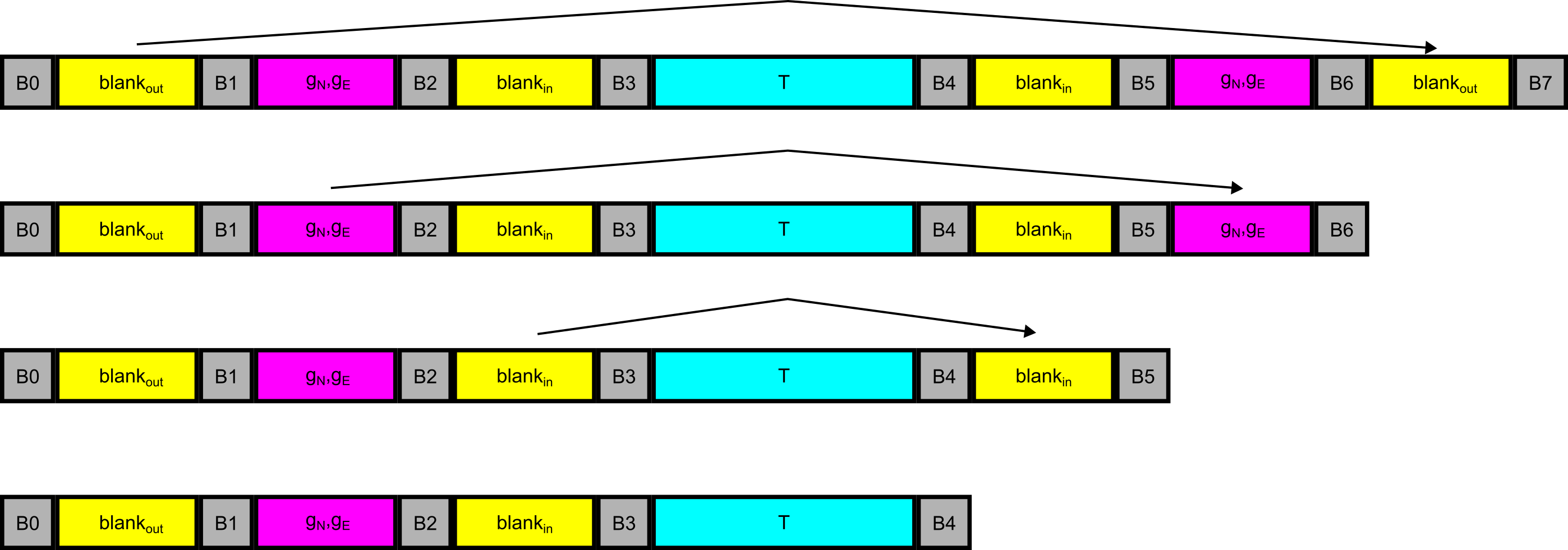}
    \caption{Schematic depiction of Phase 6 of the quine creation, the copying of 3 regions to the right. From bottom to top, the three leftmost regions are copied to the right side. Although the ordering in which the regions are copied is from right to left, the contents of each region are copied from left to right so that they end up in the same ordering on both sides.\label{fig:region-copy}}
\end{figure}

\subsubsection{Phase 7: Squaring the Quine (chmod +x quine.atam)}

In the previous stages, the rightmost tiles of each row are given a strength-1 glue labeled $\texttt{FILL}$, except for those of the first row, which are given a strength-1 glue labeled $\texttt{FILL\_BOTT}$. Additionally, every tile that attaches via a strength-2 glue on its west, and thus has no input glue on its south, has a strength-1 glues labeled $\texttt{FILL}$ on its south.
This allows two tile types, the ``filler'' type with strength-1 glues labeled $\texttt{FILL}$ on all four sides, and the ``filler bottom'' type with a strength-1 glue labeled $\texttt{FILL}$ on its north and strength-1 glues labeled $\texttt{FILL\_BOTT}$ on its east and west sides, to cause the otherwise diagonally slanted assembly to fill out into a rectangle. %(These filler tiles make up the grey portion in Figure \ref{fig:macrotile}.)

\begin{figure}
    \centering
    \includegraphics[width=0.99\textwidth]{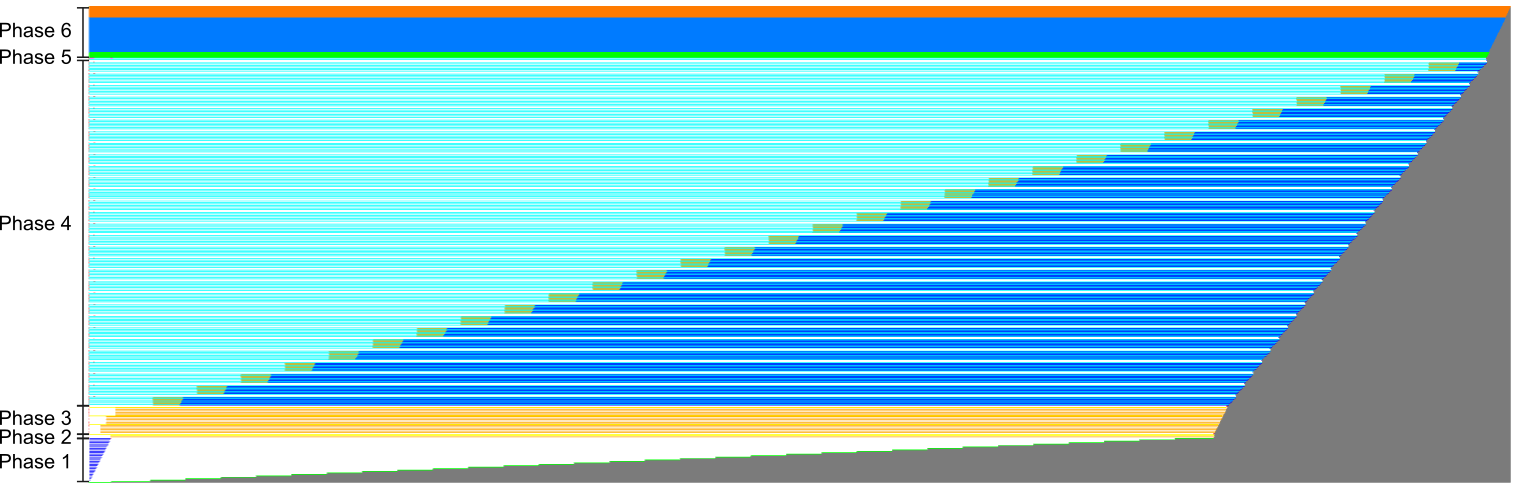}
    \caption{The rectangular assembly of the quine construction after the first 6 phases of growth (with a reduced set of tiles from $T_F$ encoded by the seed row tiles due to the size of the full assembly, so phases aren't exactly to relative scale). Growth is from the bottom upward, and phases are marked. Phase 1: the compressed glue lookup entry of each tile of the seed row (the relatively very short portion to the left of the grey portion at the bottom) is appended. Phase 2: Each `$p$' symbol is replaced. (Note that the example seed row used has few $p$ symbols, but there would actually be many more.) Phase 3: Each `$n$' symbol is replaced. Phase 4: The `$\texttt{ctr}$' symbols are replaced. Phase 5: The ``spacer'' regions are replaced with blanks. (Phase 5 is a single pair of rows, which is not very visible at this scale.) Phase 6: The regions to the left of the glue lookup table are copied to the right. The grey portion is composed of tiles of the ``filler'' type, except for the bottom row which is composed of the ``filler bottom'' type.
        \label{fig:macrotile}}
\end{figure}

An example assembly resulting after the first 6 phases, produced in the WebTAS simulator \cite{WebTAS} (with a seed row of reduced size due to the large scale factor), can be seen in Figure \ref{fig:macrotile}. In order for this assembly to become a quine and act as a macrotile seed for tileset $U$ (which is the IU tileset for standard aTAM systems), it needs to be in the shape of a square and have its perimeter glues in the format of a macrotile utilized by $U$. Note that, as this quine construction is a component our first fractal construction (i.e. Theorem \ref{thm:dssf}, the strict self-assembly of a DSSF), the way in which we create the square macrotile is slightly more complex than would be necessary solely for the proof of Theorem \ref{thm:quine}. However, this is done to simplify the final construction for Theorem \ref{thm:dssf}. Therefore, in this section we will describe the basic features needed for the current proof, and the additional complexities will be further explained in the proof of Theorem \ref{thm:dssf}.

\begin{figure}
    \centering
    \includegraphics[width=0.5\textwidth]{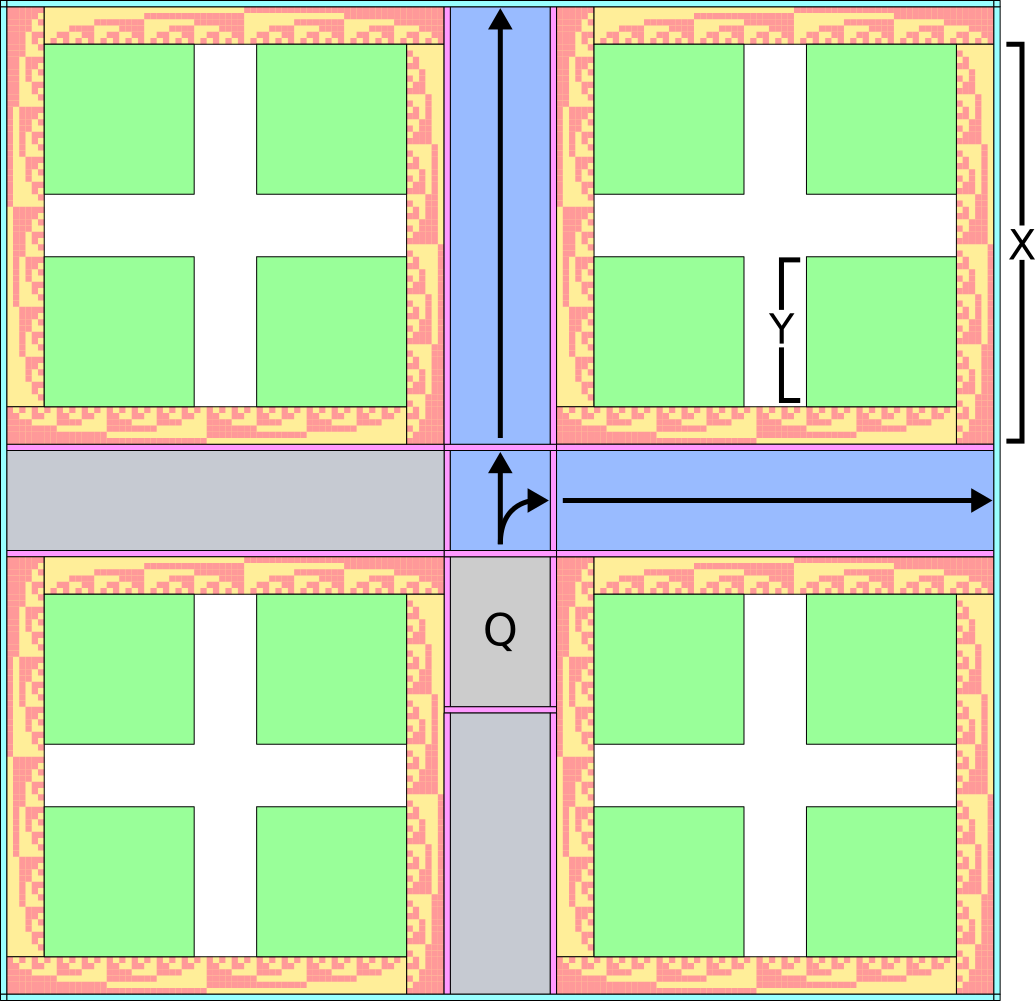}
    \caption{Schematic depiction of the formation of a macrotile square from the rectangular assembly formed after the first 6 phases (shown in grey). The dimension $X$, counted by binary counters, can be set to an arbitrary value as long as it is greater than the height of the grey rectangle, and the dimension $Y$ can be set to be any value between $0$ and $X/2$. The settings of those values will be important for the construction of Theorem \ref{thm:dssf}.}
    \label{fig:squaring-quine}
\end{figure}

Figure \ref{fig:squaring-quine} shows an overview of the process that completes the macrotile formation. The rectangular assembly formed by the first 6 phases has the information required for a macrotile used by $U$ on its north, including both the glue information that will ultimately need to be on the north, $g_N$, and that for the east, $g_E$ (since those are the two sides of the seed tile type $t_\sigma$ that are non-null). This information is first rotated using a standard string rotation gadget so that it's presented to both the north and the east. After that, 4 square frames are grown using loops of binary counter gadgets that act as the 4 corners of the macrotile (red and yellow). The size of each frame is dictated by an initial counter value and may be chosen arbitrarily (to be discussed more for the proof of Theorem \ref{thm:dssf}) as long as it is greater than the height of the rectangular assembly.
Referring to that size as $X$, $O(\log(X))$ tile types are created for the counters.
The space inside of each of the frames is empty except for solid squares of tiles that grow from each interior corner (green). The side length of these green squares, which must be $\le X/2$, is controlled by an additional binary counter made from hard-coded tiles (also discussed more later).
Referring to that size as $Y$, $O(\log(Y))$ tile types are created for these regions.
The information for the north and east sides, rotated from the rectangle, is then propagated along the square frames so that it is centered along the north and east faces of the resulting square macrotile. Additionally, generic filler tiles grow outward to the west side, and from the south of the grey rectangle to the south side.

Upon reaching the ends of the counters, the growths to each side initiate growth of a final row of tiles that forms the outer later of the macrotile. On the south and west sides, the tiles of these rows expose no glues to the south and west, respectively, making the south and west sides of the macrotile blank, representing no output glues for those sides. On the north and east sides, the following occurs during the growth of that final row:

\begin{enumerate}
    \item During the growth of the row along the north, the entries for the east glue, $g_E$, in the regions between the $\B{1}$ and $\B{2}$ delimiters and $\B{5}$ and $\B{6}$ delimiters, are changed to be all zeroes, while the entries for $g_N$ are left intact. During the growth of the row along the east, the corresponding entries for $g_N$ are changed to be all zeroes, while the entries for $g_E$ are left intact.

    \item The outward facing glues at the locations of the $\B{0}$ and $\B{7}$ delimiter symbols are changed to blank symbols (i.e. `\_') of strength 1, and as growth proceeds outward from the middle regions (blue in Figure \ref{fig:squaring-quine}) along the counters (orange and yellow in Figure \ref{fig:squaring-quine}), all outward facing glues are blank symbols of strength 1.

    \item Other than the tiles placed in these outer rows, the tiles of tileset $Q$ use IO markings that make them incompatible with the tiles of $U$. Specifically, $Q$ uses the set $\texttt{IO}_{s2} = \{\vee\vee, <<, \wedge\wedge, >>\}$, while $U$ uses the set $\texttt{IO}_{s1} = \{\vee, <, \wedge, >\}$. (This is done to make the constructions modular and ensure that there is no possibility for unintended tile attachments.) In order to make the terminal assembly of $\mathcal{Q}$ compatible with the tiles of $U$, the IO markings of the outward facing glues of the tiles of these rows use  $\texttt{IO}_{s1}$. 

    \item Once a row has completed growth along an (orange and yellow) counter, it waits until the row of the side with which it shares the corner to complete. At that point, the final two tiles of those sides cooperate to place the corner tiles.
    \begin{enumerate}
        \item On the corner tile of the northwest corner, the northern glue is a strength-2 glue with label $\B{0}$, and its western glue is null.

        \item On the corner tile of the northeast corner, its northern glue is a strength-1 glue with label $\B{7}$, and its eastern glue is a strength-2 glue with label $\B{0}$. 

        \item On the corner tile of the southeast corner, its eastern glue is a strength-1 glue with label $\B{7}$, and its southern glue is null.

        \item On the corner tile of the southwest corner, both its southern and western glues are null.
    \end{enumerate}
\end{enumerate}

Upon completion of the outer row, the assembly is terminal. The outer row essentially causes the $\texttt{blank}_{out}$ regions to be expanded to encompass the dimensions of the counters, presents the full and correct information needed by a macrotile of system using the tiles of $U$ to simulate $\mathcal{Q}$ (especially the glue encodings for the corresponding sides and the glue lookup table), has the IO markings that are compatible with the tiles of $U$, and has strength-2 glues exposed along the north and east sides that represent strength-2 glues of $t_\sigma$. As such, the terminal assembly correctly seeds itself with respect to $U$ (and representation function $R$ and seed generation function $S$).

The final point that must be shown is that $\mathcal{Q}$ is a standard aTAM system (which is the class of systems that $U$ is IU for). Recalling that all of the tiles of $Q$ are IO-marked, both facts follow immediately for the growth of phases 1 through 6, which consist solely of (1) zig-zag growth, and (2) additions of the ``filler'' and ``filler bottom'' tiles. For the zig-zag growth, all tile attachments are via a single strength-2 input glue or two diagonally adjacent strength-1 input glues, and every tile side with a non-null glue is used as either an input or output glue, meaning that no mismatches occur. Furthermore, careful design of the subsets of tiles to be specific for each phase easily ensures that no two tiles have the same set of input glues. From these facts, it follows that the zig-zag growth is also directed. There are only two filler tile types, ``filler'' and ``filler bottom'', and both have two strength-1 input glues, one on each of their north and west sides. The ``filler'' type has strength-1 output glues on its east and south, and the ``filler bottom'' only on its east. This means that all attachments are via diagonally adjacent input glues, there can be no mismatches, and all growth is directed.
Finally, the growth of phase 7, in which the rectangular assembly grows into a square macrotile also follows our requirements for a standard TAS. The tiles attach in a directed fashion.
The square frames of side length $X + \log_2(X)$ are made from standard binary counter gadgets that were easily made to be directed and without mismatches. The interior corner squares also use binary counter gadgets to control their size and are seeded from a single glue presented on the interior corners by the row of tiles that seeds the frame counters. This is the only glue that is shared between the interior squares and the frame (all others being the null glue) to avoid mismatches. The tiles that fill up the space between the frames and propagate the simulation information from the quine also have the null glue on the clockwise-most side to avoid mismatches when interfacing with the opposite side frame.

Thus $\mathcal{Q}$ is a quine and Theorem \ref{thm:quine} is proven.

\end{proof}

\subsection{Breaking circular dependencies in tile type creation}

Here we discuss a few technical details about the creation of the tile sets $T_F$ and $T_S$, which have slight circular dependencies, and how those are handled.

The high-level algorithm for generating $T_F$ and $T_S$ is the following:

\begin{enumerate}
    \item Initialize the value for the full, final width of glue representations $l_{pg} = 1$ (which will also be the width of the field needed for the counter value encoded in the seed row).
    
    \item Initialize the width of compressed glue representations $l_g = 1$.
    
    \item These will yield the width of padding $l_p = l_{pg} - l_g$ (which is a field encoded in the seed row).\label{step:pw}
    
    \item Using $l_{pg}$, generate the functional tile set $T_F$ (i.e. the tiles that perform phases 1 through 7 of the quine construction, following the descriptions previously provided for those phases). Note that $l_{pg}$ will slightly impact the size of $T_F$ by determining how many tiles must be hard-coded for the set of glue lookup entries. This is because for each symbol $x$ in the constant-sized set of symbols that can be northern glue labels for the seed row tiles, there is a set of hard-coded tiles to represent $\texttt{PAD\_BIN}(x,l_{pg})$, which is as long as $l_{pg}$.
    
    \item Get the count of unique horizontal glue labels, $g_h = |G_H|$, and unique vertical glue labels, $g_v = |G_V|$, in $T_F$.
    
    \item If $l_g < \log(g_h)$ or $l_g < \log(g_v)$:
        \begin{enumerate}
            \item Set $l_g$ equal to the maximum of $\log(g_h)$ and $\log(g_v)$.
            \item If $l_{pg} < l_g$, set $l_{pg} = l_g$.
            \item Return to Step \ref{step:pw} so that the process is redone and a (currently) accurate value for $l_g$ can be determined.
        \end{enumerate}
    
    \item Eventually $l_g$ (and $l_{pg}$) can accommodate both $g_h$ and $g_v$. This must happen since each increase of $l_g$ allows twice as many glues to be encoded but only causes a constant number of additional tiles to be generated, namely 13. That is, for the encoding of each of the 13 possible northern glue symbols in the seed row tiles, one extra bit is added and thus one extra tile is needed for each.
    
    \item Using the current $l_{pg}$, $l_g$, $l_p = l_{pg} - l_g$, and $T_F$, generate the tile set for the seed row, $T_S$, which includes the counter field padded to width $l_{pg}$, the encodings of the northern and eastern glues of the seed tile (both of length $l_{pg}$), the glue padding field of length $l_p$, and the glue lookup table containing the compressed glue lookup entries for the tiles of $T_F$ (whose glue representations depend upon $l_g$).\label{step:tS}

    \item Let $g_h' = |T_S| - 1$ (i.e. the number of unique horizontal glues between the tiles of the seed row). Note that the tiles of $T_S$ add no new vertical glues since the tiles of $T_F$ have tiles with glues that can bind to each of them.
        
    \item If $l_{pg} < \log(g_h' + g_h)$ or $l_{pg} < \log(g_v)$, increase $l_{pg}$ by 1, let $l_p = l_{pg} - l_g$ for this new value of $l_{pg}$ and return to Step \ref{step:pw}, since $l_{pg}$ was not enough bits to encode all of the unique glue labels.
        
    \item Eventually, $l_{pg}$ will be large enough (this must happen since any increase only causes one additional tile for each of the counter field of width $l_{pg}$ and the padding field of width $l_p$, so a constant number of added tiles and new glues, but a doubling of the number that can be represented.

    \item Let the final tile set $Q = T_F \cup T_S$.
\end{enumerate}

\section{An Abstract Model of Self-assembly: Self-Describing Embedded Circuits}
\label{sec:abstract_models}
\cref{sec:standard-IU,sec:quine} demonstrate that completely analyzing the behavior of an aTAM system, even when it is standard, involves a lot of details. This analysis is  made harder by the asynchronous nature of the aTAM. It is necessary in the case of intrinsically universal systems (and thus of their quines) since intrinsic universality characterizes the behavior of the universal system. On the other hand, when one is merely interested in designing a standard system with a given set of productions, it is possible to use a so-called \emph{self-descipting circuit} as a blueprint, which can then be compiled into a standard aTAM system. This device will be of use in section \cref{sec:cacarpet_circuit} to prove the existence of a standard aTAM system which strictly assembles the Sierpiński Cacarpet $\cacarpet$.

\subsection{Embedded circuits}

As a way to design and analyze complex standard aTAM systems, these can be viewed as circuits drawn on \ZZ. When defined that way, one does not need to consider the effects of concurrency between different parts of the assembly, nor how the attachments are ordered. In exchange, a condition of \emph{self-description} is needed for the circuit to translate into a TAS system. The following definitions can be skipped if one admits that a circuit is "like \cref{fig:mod3-adder}", where wires coming into the same gate bend towards each other, so that the type of each gate $g$ determines the set of inputs which will be received by its successors in addition to $g$'s output (if any).

A circuit is made of gates. Each of these gates has a \emph{function}, computing the outputs of the gates from the inputs. The gate is embedded on a unit square according to its \emph{wiring}, which states where its inputs come from and where its outputs go to. The wiring and the function must be compatible: the wiring must have as many inputs as the function.

\begin{definition}[Wire]
  A wire is a pair of a list of distinct directions $\vec{D} \in \Dir^*$ and a distinguished element $d \in D$. It is noted as a word on the alphabet $\{n, e, s, w\}$ containing the elements of $D$, with $d$ capitalized. For instance, the wire $((N, E), E)$ is noted $nE$, while $((N, E), N)$ is noted $Ne$.

  Given a wire $w = (\vec{D}, d)$, its opposite $-w$ is $(\vec{(-D_i)}, -d)$, i.e. $-Ne = Sw$.
  The set of wires is noted $\Wires$.
\end{definition}

\begin{definition}[Wiring]
  Given two integers $i, o$ such that $i + o \leq 4$, a \emph{$(i,o)$-wiring} $w$ (i.e. a wiring with \emph{in-degree} $i$ and \emph{out-degree} $o$) is given by
  \begin{itemize}
  \item a partition of $\Dir$ into three sets, $\winputset[w]$, $\woutputset[w]$ and $\winertset[w]$, where $\winputset[w]$ is ordered
  \item a map $\woutputmap[w]: \woutputset[w] \to \{0, \ldots, o - 1\}$;
  \item a map $\woutputwires[w]: \woutputset[w] \to \Wires$ such that $-d$ is the distinguished element of  $\woutputwires[w](d)$.
  \end{itemize}

  The set of all wirings is noted $\Wirings$.
\end{definition}

\begin{definition}[Input / Output / Inert Sides, Degree]
  Given a wiring $w \in \Wirings$,
  the elements of $\winputset[w]$ are the \emph{input sides} of $w$, the elements of $\woutputset[w]$ are its \emph{output sides}, and the elements of $\winertset[w]$ are its \emph{inert sides}.
  The in-degree of $w$ is the number of its input sides, its out-degree of $w$ is its number of output sides.
\end{definition}

%An element $d \in \woutputset[w]$ together with the value of $\wnextinputmap[w](d)$ is an output \emph{wire}. The set of output wires of $w$ is $\woutputwires[w]$. Thus, $Ne \in \woutputwires[w]$ denotes that $N \in \woutputset[w]$ and $\wnextinputmap[w](N) = (N, E)$, while $En \in \woutputwires[w]$ denotes that $E \in \woutputset[w]$ and $\wnextinputmap[w](E) = (N, E)$. For a given $w$, either, both or none can be true.

Additionally, each output side also determines the \emph{input} sides on the following gate and their ordering. That is, a wiring $w$ with $nE \in \woutputwires[w]$ can only have a wiring $w'$ with $\winputset[w'] = (S, W)$  to its right.

The graphical representation of a wiring is given on \cref{fig:wiring_example}. Each direction in $\winputset[w]$ has an incoming arrow, and each direction in $\woutputset[w]$ has an outgoing arrow. If $\winputset[w]$ has only one element, the corresponding arrow is straight and simple. Likewise for the outgoing arrow when $\woutputwires[w](d)$ is a singleton. When $\woutputwires[w](d)$ is made of two adjacent directions, the corresponding arrows bend to come near each other; likewise, modulo rotation, when $\woutputwires[w](N) \in \{Ne, eN\}$, the corresponding arrow bends left, and right if it is $Nw$ or $wN$. For pairs of opposite directions, an arrow with a double tip is used. Triple and quadruple inputs are not used in this paper. Inputs are numbered according to the order of $\winputset[w]$ and output wires are numbered according to $\woutputmap[w]$.

\begin{figure}
  \centering
  \begin{tikzpicture}
    \matrix (b) [matrix of nodes, column sep=10em, row sep=10ex, every node/.style={draw, circle}]{
  $w_1$ & $w_2$ & $w_3$ \\
};

\begin{scope}[every node/.style={font=\footnotesize}, entry/.style={pos=.9,auto}, exit/.style={pos=.1,auto,swap}]
\draw (b-1-1)+(-1,0) edge[-{Stealth}] node[entry] {$0$} (b-1-1);
\draw[-{Stealth}] (b-1-1) edge node[exit] {$0$} node[entry] {$0$} ++(1,0);
\draw[-{Stealth}] (b-1-1) edge[bend right] node[exit] {$1$} node[entry] {$0$} ++(0,1);
\draw[-{Stealth}] (b-1-1) edge[bend left] node[exit] {$1$} node[entry] {$1$} ++(0,-1);

\draw (b-1-2)+(-1,0) edge[-{Stealth}, bend right] node[entry] {$0$} (b-1-2);
\draw (b-1-2)+(0,-1) edge[-{Stealth}, bend left] node[entry, pos=.8] {$1$} (b-1-2);
\draw (b-1-2) edge[-{Stealth}{Stealth}] node[exit] {$0$} node[entry] {$0$} ++(0,1);

\draw (b-1-3)+(-1,0) edge[-{Stealth}{Stealth}] node[entry] {$0$} (b-1-3);
\draw (b-1-3)+(1,0)  edge[-{Stealth}{Stealth}] node[entry] {$1$} (b-1-3);
\draw (b-1-3) edge[-{Stealth}, bend left] node[exit] {$0$} node[entry] {$1$} ++(0,-1);
\end{scope}

%%% Local Variables:
%%% mode: latex
%%% TeX-master: "../main.stacs"
%%% End:
  \end{tikzpicture}
  \caption{The graphical representation of three wirings: \( w_1 = \{ \winputset = (W), \woutputset = \{ N, S, E \}, \woutputmap = \{ N \to 1, E \to 0, S \to 1 \}, \woutputwires = \{ Se, W, wN \} \} \), \( w_2 = \{ \winputset = (W, S), \woutputset = \{N\}, \woutputmap = \{ N \to 0 \}, \woutputwires = \{ Sn \} \} \) and \( w_3 = \{ \winputset = (W, E), \woutputset = \{ S \}, \woutputmap = \{ S \to 0 \}, \woutputwires = \{ wN \}  \} \). Neighboring input wires bend to come together into the wiring. Double arrows indicate a pair of opposite input arrows. The output wire in direction $d$ bends to mirror \(\woutputwires[w](d)\).}
  \label{fig:wiring_example}
\end{figure}

\begin{definition}[Gate]
  Given an alphabet $\Sigma$ two integers $i, o$, a gate $g$ is a pair $\{\gatefun: f, \gatewiring: w\}$, with $f$ a function $f: \Sigma^i \to \Sigma^o$ and $w$ an $(i, o)$-wiring.

  The set of gates on alphabet $\Sigma$ is noted $\Gates{\Sigma}$.
\end{definition}

These gates may be placed onto a subset of the grid $\Z^2$ to make \emph{circuits}. An example of a circuit, a multiplier built from a handful of adders and auxiliary gates is given on \cref{fig:mod3-adder}.

%
% TODO: inputs = {N, E} quand les inputs sont à gauche et en bas???
%
\begin{definition}[Circuit]
   Let $\Sigma$ be an alphabet. Let $D$ be an oriented subgraph of $\Z^2$, $\vec{I}$ be a sequence of arcs from $\Z^2 \setminus D$ to $D$ and $\vec{O}$ a sequence of arcs from $D$ to $\Z^2 \setminus D$.

   A circuit $C$ with dependency graph $D$, input bus $\vec{I}$ and output bus $\vec{0}$ is a map $D \to \Gates{\Sigma}$ such that:
   \begin{itemize}
   \item for any arc $E$ in direction $d \in \Dir$ between two positions $a, b \in D$, let $w_a = \gatewiring[C(a)]$ and $w_b = \gatewiring[C(b)]$; then we have \( d \in \woutputset[w_a] \), \( -d \in \winputset[w_b] \), and \(\woutputset[w_a](d) = (-d, \winputset[w_b])\),
   \item for any adjacent positions $a, b$ in $\mathbb{Z}^2$ with no arc between $a$ and $b$ in $D$, $d \in \winertset[ { \gatewiring[C(a)]  } ]$ and $-d \in \winertset[ {\gatewiring[ {C(b)} ]} ]$
   \item and for any arc $e$ of $\mathbb{Z}^2$ in direction $d \in \Dir$ between two positions $a \in D$ and $b \in  \Z^2 \setminus D$, either:
    \begin{itemize}
    \item $d \in \winertset[ { \gatewiring[ {C(a)} ] } ]$ and $e$ is neither an element of $\vec{I}$ nor of $\vec{O}$,
    \item $d \in \woutputset [{ \gatewiring[ {C(a)} ] } ]$ and $e$ is an element of $\vec{O}$,
    \item $-d \in \winputset[ { \gatewiring[ {C(a)} ] } ]$ and $-e$ is an element of $\vec{I}$.
    \end{itemize}
   \end{itemize}

   The support of $C$ is the vertex-set of $D$.
\end{definition}

\begin{figure}
  \centering
  \begin{tikzpicture}
    \matrix (c) [every node/.style={draw, circle}, matrix of nodes, column sep=3em, row sep = 7ex, a/.style={rectangle, draw=none, inner sep=1mm}] {
  &       & |[a]| $I_0$ & |[a]| $I_1$ &  & \\
  & $i_1$	& $i_2$       & $m$         & $i_1$ & \\
  |[a]| $I_2$	& $i_1$ & $m$         & $a$ & $a$        &  \\
  &       & |[a]| $O_0$ & |[a]| $O_1$ & |[a]| $O_2$ & \\
};

\begin{scope}[every node/.style={font=\footnotesize}, entry/.style={pos=.9,auto}, exit/.style={pos=.1,auto,swap}, value/.style={pos=.5, fill=gray!20}]
\draw[->] (c-1-3) edge[bend right] node[entry] {$0$} node[value] {$2$} (c-2-3);
\draw[->] (c-1-4) edge[bend right] node[entry] {$0$} node[value] {$3$} (c-2-4);
\draw[->] (c-2-3) edge[bend right] node [exit] {$0$} node[entry] {$0$} node[value] {$2$} (c-3-3);
\draw[->] (c-2-4) edge[bend right] node [exit] {$0$} node[entry] {$0$} node[value] {$1$} (c-3-4);
\draw[->] (c-3-2) edge node [exit] {$0$} node[entry] {$0$} node[value] {$7$} (c-2-2);
\draw[->] (c-3-3) edge node[exit] {$0$} node[value] {$4$} (c-4-3);
\draw[->] (c-3-4) edge node[exit] {$0$} node[value] {$2$} (c-4-4);
\draw[->] (c-2-4) edge node[exit] {1} node[value] {$2$} node[entry] {0} (c-2-5);
\draw[->] (c-3-5) edge node[exit] {0} node[value] {$2$} (c-4-5);

\draw[->] (c-2-2) edge[bend left] node[exit] {$0$} node[entry, swap] {$1$} node[value] {$7$} (c-2-3);  % (c-2-3);
\draw[->] (c-2-3) edge[bend left] node[exit] {$1$} node[entry, swap] {$1$} node[value] {$7$} (c-2-4);

\draw[->] (c-3-1) edge node[entry] {$0$} node[value] {$7$} (c-3-2);
\draw[->] (c-3-2) edge[bend left] node[exit] {$0$} node[entry, swap] {$1$} node[value] {$7$} (c-3-3);
\draw[->] (c-3-3)  edge[bend left] node[exit] {$1$} node[entry, swap] {$1$} node[value] {$1$} (c-3-4);
\draw[->] (c-3-4) edge[bend left] node[exit] {$1$} node[value] {0} node[entry] {0} (c-3-5);
\draw[->] (c-2-5) edge[bend right] node[exit] {0} node[entry, swap] {1} node[value] {2} (c-3-5);
\end{scope}

%%% Local Variables:
%%% mode: latex
%%% TeX-master: "../main"
%%% End:
  \end{tikzpicture}
  \[
    \begin{cases}
      i_1 : \Sigma \to \Sigma \\
      i_1(x) = x\\
      i_2 : \Sigma^2 \to \Sigma^2 \\
      i_2(x,y) = (x,y)\\
    \end{cases}\quad%
    \begin{cases}
      m : \Sigma^2 \to \Sigma^2 \\
      m(x,y) = (xy \bmod 10, \quot{xy}{10})\\
      a : \Sigma^2 \to \Sigma^2 \\
      a(x,y) = ((x + y) \bmod 10, \quot{(x + y)}{10})\\
    \end{cases}
  \]
  
  \caption{A multiplier circuit $M_2^1$ and the definition of the functions of its gates.  The alphabet is the set of digits $\mathbb{Z} / 10 \mathbb{Z}$. The values in the grey squares are an example of evaluation: $32 \times 7 = 224$}
  \label{fig:mod3-adder}
\end{figure}

% Given a circuit $C$, and an arc $e = (a \to b)$ in its input bus, the \emph{input wire} associated with $e$ is the pair $v@b$, with $v$ the wire $v = (\direction(e), \winputset[ {\gatewiring[C(b)]} ])$. For instance, if, as depicted \ref{fig:mod3-adder}, the gate at $(0, 3)$ in $C$ has a wiring $w$ with $\winputset[ { \gatewiring[w] } ] = (W, S)$ and $\vec{I}_1 = ((-1, 3) \to (0, 3))$, then $Ws@(0,3)$ is the input wire number $1$ of $C$.
% % In the remainder of the section, $\vec{I}$ will generally be given as a sequence of wires rather than mere arcs.
% Symmetrically, the \emph{output wire} associated with an element $e = \vec{O}_i = (a \to b)$ is  the pair $v@a$ with $v = (\direction(e), \winputset[ {\gatewiring[C(a)]} ])$. On figure~\ref{fig:kappa_normal1}, the output wire number $2$ of the circuit is $En@(5, 1)$.\todo{Est-ce que ce paragraphe est utile?}

Let $C$ be a circuit, and $A$ the arc-set of its dependency graph $D$. A function $v: A \mapsto \Sigma$ \emph{conforms} to $C$ at a position $\vec{z}$ in the support of $C$, if for any \(d \in \woutputset[ {\gatewiring[ C(\vec{z}) ]} ] \), \( v(o) = (\gatefun[C(v)])(v(i_0), \ldots, v(i_m))_k \) with $i_0, \ldots, i_m$ the input arcs of $C(\vec{z})$, $o$ its output arc in direction $d$ and $k = \woutputmap[ {\gatewiring[ C(\vec{z}) ]} ](d)$.

A circuit is \emph{well-founded} when its dependency graph has no cycle or infinite backwards paths. For a well-founded circuit, given a vector of inputs $\vec{ \imath }$ indexed by $\vec{I}$, there is a \emph{unique} function $\innerevalfunc{C}(\vec{ \imath }, -)$ which conforms to $C$ and such that the values on $\vec{I}$ are $\vec{ \imath }$. The circuit function $\evalfunc{C}: \Sigma^{ \vec{I}} \to \Sigma^{\vec{O}}$ is defined by $\evalfunc{C}(\vec{ \imath }) = (\innerevalfunc{C}(\vec{ \imath }, O_k ))_k$.

A circuit is \emph{finitely rooted} when its input bus is finite and its number of gates with in-degree $0$ is finite. It is \emph{evaluable} if it is well-founded and finitely rooted.

The circuit $M^1_2$ on figure~\ref{fig:mod3-adder} is a multiplier. Take two natural integers $a < 10, b < 100$, pose $b = b_0 + 10 b_1$ with $\forall i, 0 \leq b_i < 10$. Let $\vec{\imath}_{a,b} = (a_0, b_0, b_1)$. Then $\evalfunc{C}(\vec{\imath}_{a,b}) = \vec{o}$, with $ab = \sum o_i 10^i$. Figure~\ref{fig:mod3-adder} gives an example of the evaluation of the circuit on inputs $32$ and $7$; the values of $\innerevalfunc{C}(\vec{i}, -)$ on each arc are given on the figure.

When the alphabet $\Sigma$ is structured in two layers, i.e. $\Sigma = A \times B$, the function $f$ of a gate $g$ may act independently on layer $A$ and layer $B$, i.e. $f = f_A \otimes f_B$. Then, taking $g_A = \{ \gatewiring : w, \gatefun: f_A \} \in \Gates{A}$ and $g_B = \{ \gatewiring : w, \gatefun: f_B \} \in \Gates{B}$, $g$ can be written as $g_A \otimes g_B$. In other words, the tensor product $\otimes$ applies to pairs of gates \emph{with the same wiring}.

% % When the alphabet $\Sigma = A \times B$ is structured in two layers $A$ and $B$, it is possible to create a gate in $g_A \otimes g_B \in \Gates{\Sigma}$ by combining a gate $g_A \in \Gates{A}$  and a gate $g_B \in \Gates{B}$, each acting on one layer. This product of gates only makes sense if the wirings of $g_A$ and $g_B$ are the same. %compatible.

% % \begin{definition}[Compatible wirings]
% %   Let $w_0$, $w_1$ be two $(i, o)$-wirings, $w_0$ and $w_1$ are compatible with input permutation $\sigma_i$ if:
% %   \begin{eqnarray*}
% %     \winputset[w_0] &=& \winputset[w_1]\\
% %     % \winputmap[w_1] &=& \sigma_i \circ \winputmap[w_1]\\
% %     \woutputset[w_0] &=& \woutputset[w_1]\\
% %     \wnextinputmap[w_0] &=& \wnextinputmap[w_1].\\
% %   \end{eqnarray*}
% % \end{definition}

% % \begin{definition}[Gate superposition]
% %   Let $g_0 \in \Gates{A}, g_1 \in \Gates{B}$ such that $\gatewiring{g_0}$ and $\gatewiring{g_1}$ are compatible.
% %   Their superposition is the gate $g_0 \otimes g_1 \in \Gates(A \times B)$ which collects the inputs of $g_0$ and $g_1$ from their input sides, feeds them to their functions, then distributes the outputs to the output sides.
% % \end{definition}

\subsection{Self-description}

The process of self-assembly in the aTAM and the evaluation of an evaluable embedded circuit are somewhat alike, in that information propagates from an initial region outwards. In both processes, there is no global synchronisation, but each step can take place as soon as its inputs are ready. In the aTAM, the initial region is the seed, while in a circuit, it is the input bus together with the gates of in-degree \(0\).

There are three differences between these processes. First, the input bus of a circuit does not have an analog in the aTAM. Hence, aTAM systems are akin to \emph{closed} circuits. Secondly, there can be competition between attachments in the aTAM, as well as mismatches, while in circuits the input and output directions of each gate are fixed. This entails that an aTAM derived from a circuit by the compilation process detailed below will be standard. The last and most important difference is that in a circuit, the outputs of each gate depends not only on its input, but also on the function of the gate. For an aTAM system to simulate a circuit, this information needs to be derived from the inputs of the gate. Self-description captures the possibility to do so.

\begin{definition}[Self-describing circuit]
  \label{def:self_describing}
  An evaluable circuit $C$ on alphabet $\Sigma$ is \emph{self-describing} on input vector $\vec{ \imath }$ if there is a decoding function \(\decgate: (\Sigma \times \Dir)^{<4} \to \Gates{\Sigma}\) such that for any position $p$ with incoming arcs \((e_0, \ldots, e_{k-1}) = p + (\winputset[ {\gatewiring[C(p)] } ])\):
  \[\decgate((\innerevalfunc{C}(\vec{\imath}, e_0), d_0), \ldots, (\innerevalfunc{C}(\vec{\imath}, e_{k-1}), d_{k-1})) = C(p),\]
  where $d_j$ is the direction of $e_j$.
\end{definition}

A closed self-describing circuit really is a circuit with only one function of each in-degree. Having one function per in-degree is a harmless technicality: in a circuit, successive gates have compatible wirings, so the in-degree of each gate is known from the wiring of any of its predecessor gates, whether or not the circuit is self-describing.

\begin{lemma}
  \label{lem:one-gate}

  Let $C$ be a closed self-describing circuit. There is a closed circuit $C_1$ with only only one function of each in-degree such that $\innerevalfunc{C} = \innerevalfunc{C_1}$.
\end{lemma}

\begin{proof}
  $C_1$ has the same wirings as $C$, and for each $i$, all of its gates of in-degree $i$ have as function \(f_i: x_0 \ldots x_{i-1} \mapsto (\gatefun[\delta_C(x_0, \ldots x_{i-1})])(x_0, \ldots x_{i-1})\), where $\delta_C$ is the decoding function of $C$.
\end{proof}

\subsection{Locally deterministic patterns}
\label{sec:loc_det_pat}

If $C$ is a self-describing closed circuit, then from $\innerevalfunc{C}$, one gets a coloring of the arcs of its dependency graph with the property of \emph{local determinism}.

\begin{definition}[Locally deterministic arc coloring]
  \label{def:loc_det_pat}
  Let $P$ be a coloring of the arcs of some acyclic oriented subgraph $G$ of $\Z^2$. For each vertex $v \in G$ with in-degree $\delta$, the input vector of $v$ is \(\vec{\imath}_v = ((d_0, P(e_0)), \ldots, (d_{\delta-1}, P(e_{\delta-1})))\), where $e_i$ is the $i$-th incoming edge of \(v\) and $d_i$ its direction

  $P$ is locally deterministic if there  are two prediction functions $\pinputs: \Dir \times \Sigma \to \PDir$ and $\psymb: (\Dir \times \Sigma)^{< 4} \times \Dir \to \Sigma$ such that for each arc $e$ from vertex $a$ to vertex $b$ in direction $d \in \Dir$,
  \begin{itemize}
  \item \(P(e) = \psymb(\vec{\imath}_a, d)\)
  \item \(\pinputs(d, P(e))\) is the set of directions of the incoming arcs of $b$,
  \end{itemize}
\end{definition}

\begin{figure}
  \centering
  \def\SierpData{%
    1 1 1 1 1
    1 0 1 0 1
    1 1 0 0 1
    1 0 0 0 1
    1 1 1 1 0
  }
  \readarray\SierpData\Sierp[-,5]

  \def\z{\Sierp[\x,\y]}

  \begin{tikzpicture}[xscale=.48, yscale=.6, font=\footnotesize]
    \input{tikz/locally_det_pat_ok}
  \end{tikzpicture}\qquad%
  \begin{tikzpicture}[xscale=.48, yscale=.6, font=\footnotesize]
    \input{tikz/locally_det_pat_no_zero}
  \end{tikzpicture}\qquad%
  \begin{tikzpicture}[xscale=.48, yscale=.6, font=\footnotesize]
    \input{tikz/locally_nondet_pat}
  \end{tikzpicture}
  \caption{Three arc colorings on the alphabet $\Sigma = \{ A, \textcolor{violet}{0}, \textcolor{green!50!black}{1} \}$. All arcs go either to the right or up. The coloring $C_1$, in the upper left is locally deterministic. In the upper-right, $C_2$ is not because the two circled arcs have value $1$ and direction $E$, but the vertices they point into have different incoming directions ($\{E\}$ versus $\{N, E\}$). At the bottom, $C_3$ is not locally deterministic, because the two triangle vertices get the same incoming values, but have different outputs: $(0, 0)$ versus $(1, 1)$.}
  \label{fig:ex-locally-deterministic}
\end{figure}

\begin{lemma}
  \label{lem:self-descr-local-det}
  Let $C$ be a closed self-describing circuit on alphabet $\Sigma$, with dependency graph $D_C = (V, A)$.
  Let $C'$ be the self-describing circuit on alphabet $\Sigma \times \Wirings$ where each gate outputs its wiring in addition to its normal output.
  Then the function $\innerevalfunc{C'}$ is a locally deterministic coloring of $D_C$.
\end{lemma}

\begin{proof}
%  By lemma~\ref{lem:one-gate}, without loss of generality all the gates of $C$ with in-degree $i$ have the same  function $f_i$.

  A candidate function for $\pinputs$ takes as input the direction $d$ of an arc $e: a \to b$, and the value of $\innerevalfunc{C'}(e)$, that is, an element of $\Sigma$ corresponding to $\innerevalfunc{C}(e)$ as well as the wiring $w$ of the gate at $a$. Thus, the function $(d, x, w) \mapsto \woutputwires[w](d)$ must, by the constraints on neighboring gates in a circuit, output the set of directions of the incoming arcs at $b$.

  Likewise, a candidate function for $\psymb$ takes as input the direction of the arc $e: a \to b$, and a vector $\vec{m} \otimes \vec{d}$ of the inputs coming into $a$ with their directions. Reorder $\vec{m} \otimes \vec{d}$ so that $\vec{d}$ is in the order of the inputs of $C(a)$. Let $\{ \gatewiring: w, \gatefun : f \} = \decgate(\vec{m} \otimes \vec{d})$. Then taking $\psymb(e) = (f(\vec{m}), w)$ works.
\end{proof}

\paragraph*{From locally deterministic patterns to aTAM systems}

\begin{definition}[vertex type, atlas]
  Let $P$ be a coloring of the arcs of some acyclic oriented subgraph $G$ of $\Z^2$. For a vertex $v$ of $G$, its \emph{vertex type} is the partial function $\bar{v}: d \in \Dir \mapsto (o, x)$, where $o = -d$ if the edge $e$ in direction $d$ from $v$ is an incoming edge, $o = d$ if it is an outgoing edge, and $x \in \Sigma$ is $P(e)$. If $v$ has no edge in direction $d$, then $\bar{v}(d)$ is undefined.

  The \emph{atlas} of $P$ is the set of its vertex types.
\end{definition}

A locally deterministic coloring can be reconstructed from its atlas and the positions of its in-degree $0$ vertices.

\begin{lemma}
\label{lem:same-atlas-same-pattern}
  Let $P_1, P_2$ be locally deterministic colorings of the arcs of some oriented subgraphs of $\Z^2$, $G_1$ and $G_2$ respectively. If
  \begin{enumerate}
  \item $G_1$ and $G_2$ are acyclic and have no infinite backwards paths,
  \item $P_1$ and $P_2$ have the same atlas,
  \item $G_1$ and $G_2$ have the same vertices with in-degree $0$,
  \end{enumerate}
  then $G_1 = G_2$ and $P_1 = P_2$.
\end{lemma}

\begin{proof}
  By induction on the longest path to each position in $G_1$.
\end{proof}

Finally, a locally deterministic pattern with a unique in-degree 0 vertex can be self-assembled in the aTAM.

\begin{lemma}\label{lem:det-pat-atam}
  Let $\Sigma$ be some finite alphabet, and let $P$ be a locally deterministic coloring of the arcs of some acyclic graph $G$ with no infinite backwards path. Assume the maximal in-degree of a vertex in $G$ is $\delta$ and that $G$ has only one vertex with in-degree $0$.

  Then there is an aTAM system $S_P$ with temperature $\delta$ with $P$ as its only final production.
\end{lemma}

\begin{proof}
  Let $A_P$ be the atlas of $P$, $\pinputs$ and $\psymb$ the prediction functions of $P$. By convention, suppose $\pinputs$ returns an input vectors which is ordered clockwise, starting from direction $N$.

  First define the glues of $S_p$ and their strength. Each glue is a pair $(s, d) \in \Sigma \times \Dir$. For any symbol $s \in \Sigma$ and direction $d \in \Dir$, let $\vec{\imath} = \pinputs(s, d)$. If $d$ is the first element of $\vec{\imath}$, then its strength is \(\delta + 1 - |\vec{\imath}|\), otherwise it is $1$. Hence, for any vertex type $v \in A_P$ with input vector $\vec{\imath}_v$, the sum of the strengths on the input sides is \(\delta\).

  Then define the set of tile types $S_p$ to be $A_p$, with the glue on the $d$ side of the tile type $\bar{v}$ being $\bar{v}(d)$ if defined, and the null glue $\null$ otherwise. The seed tile of $S_p$ is the only vertex type of $A_p$ with in-degree $0$.

  Indeed, any production of $S_P$ is an initial subset of $P$, as can be seen by induction on the attachments.

  Then, consider a production $p$ of $S_P$. If \(G\) contains some position not covered by $p$, then because $G$ contains neither loops nor infinite paths, it must contain some $v$ with all its predecessors within \(\operatorname{dom}(p)\). But then the total strength of glues into $v$ is $\delta$, and the tile corresponding to $P(v)$ must be attachable there. Hence $p$ is not terminal.

  Finally, $S_P$ assembles $P$.
\end{proof}

Putting all this together, it is possible to compile a self-describing circuit into a self-assembling system.

\circuitToAtam*

\begin{proof}
  By \cref{lem:self-descr-local-det,lem:det-pat-atam}.
\end{proof}

\section{A Geometric limitation of aTAM computation: The Tree Pump Theorem}
\label{sec:tree_pump}

The dynamics of the aTAM are limited by pumping principles in the style of \cite{meunier_intrinsic_2014}, whereby under some conditions, their behavior must become periodic. The one presented here deals with skinny productions, that is those who do not encircle any square larger than $N$ for some fixed $N$. The crucial property which makes them simple(-ish) is their bounded treewidth~\cite{DiestelGraphTheory}. One remarkable feature of this pumping theorem is that it allows pumping \emph{in a chosen direction}. In \cref{sec:limits}, this pumping principle will be the final argument for ruling out the least connected of shapes from generating DSSFs which can be self-assembled.

\treepump*

\begin{figure}[ht]
  \centering
    \tikzset{
      chemin/.pic={
        \begin{scope}[scale=.1]
          \foreach \x/\y in {0/0, 1/0, 2/0, 2/1, 2/2, 3/0, 1/2, 4/0, 4/1, 4/1}
          \path[draw, fill=gray!50!white] (\x, \y) -- ++(1,0) -- ++(0, 1) -- ++(-1,0) -- cycle;
          \coordinate (-bend) at (4, 0);
        \end{scope}
      }
    }

    \tikzset{
      u/.pic={
        \filldraw[fill=blue!50!white] (0,0) rectangle (.1, .1);
        \filldraw[fill=blue!50!white] (0,-.1) rectangle (.1, 0);
        \filldraw[fill=blue!50!white] (0,-.2) rectangle (.1, -.1);
        \filldraw[fill=blue!50!white] (.1,-.2) rectangle (.2, -.1);
        \filldraw[fill=blue!50!white] (.2,-.2) rectangle (.3, -.1);
        \filldraw[fill=blue!50!white] (.2,-.2) rectangle (.3, -.1);
        \filldraw[fill=blue!50!white] (.2,-.1) rectangle (.3, 0);
        \filldraw[fill=blue!50!white] (.2,0) rectangle (.3, .1);
        \filldraw[fill=blue!50!white] (.3,0) rectangle (.4, .1);
        \filldraw[fill=blue!50!white] (.4,0) rectangle (.5, .1);
      }
    }
    
    \tikzset{
      unbounded/.pic={
        \begin{scope}[scale=.5, shift={(-3, -3)}]
          \clip[draw] (-1, -1) rectangle (10.5,7);
          \node[fill=red, circle, inner sep=1pt] (origin) at (0, 0) {};
          \node[left=.1em of origin] {$\vec{0}$};
          \coordinate (path_dest) at (5.4, 2);
          \coordinate (foot) at (1.4, 0);
          \pic[fill=gray!50!white] (invader) at (0, 0) {spaceinvader};
          \pic[fill=gray!50!white, yscale=-1, rotate=-90] (invader2) at (foot) {spaceinvader};
          \pic[fill=gray!50!white, yscale=-1] (invader3) at (invader2-right-antenna) {spaceinvader};
          \pic[fill=gray!50!white] (invader4) at (invader3-right-foot) {spaceinvader};
          \pic[fill=gray!50!white, rotate=90] (invader5) at (invader4-right-antenna) {spaceinvader};
          \pic[fill=gray!50!white, rotate=90] (invader6) at (invader2-right-foot) {spaceinvader};
          
          % \path (foot) -- (path_dest) foreach \t [count=\ti] in {0, .2, ..., 1} {pic[pos=\t]  (chemin-\ti) {chemin}};
          \path (invader4-right-foot) -- +(10, 0) foreach \t in {0, .1, ..., 1} { pic[pos=\t] {u}};
          \node[blue, inner sep=0, fill=white, below=10pt of invader4-right-foot, anchor=north west] {periodic};

          \coordinate[shift={(1, .1)}] (a) at (invader4-right-foot);
          \coordinate[shift={(35:2)}] (b) at (a);
          \coordinate[shift={(1,0)}] (c) at (a);
          \path[purple, ->] (a) edge node[pos=.8, auto] {$\vec{d}$} (b); % ++(35:5);
          % \path[blue, ->] (a) edge node[pos=.8, auto] {$\vec{p}$} ++(0:4);
          \pic[pic text={$< \frac{\pi}{2}$}, draw=purple, thick, angle radius=4em] {angle = c--a--b};
        \end{scope}
      }      
    }

    \tikzset{
      bigCycle/.pic={
        \begin{scope}[scale=.5, shift={(-4, -4)}]
          \clip[draw] (-1, -1) rectangle (7,7);
          \pic[fill=gray!50!white] (invader) at (0, 0) {spaceinvader};
          \path [draw=none, fill=yellow] (2,2) rectangle (4,4);
          \path[<->, dashed] (2,2) edge node[midway, above] {$m$} (4,2);
          \path[<->, dashed] (4,2) edge node[midway, left] {$m$} (4,4);
          \path (invader-right-antenna) -- ++(4, 0) 
             foreach \t in {0, .05, ..., 1} { pic[pos=\t] {code={\filldraw[thin,fill=gray!50!white] (0,0) rectangle (.1, .1);}}}
             -- ++(0, 4)
             foreach \t in {0, .05, ..., 1} { pic[pos=\t] {code={\filldraw[thin,fill=gray!50!white] (0,0) rectangle (.1, .1);}}}
             -- ++(-4, 0)
             foreach \t in {0, .05, ..., 1} { pic[pos=\t] {code={\filldraw[thin,fill=gray!50!white] (0,0) rectangle (.1, .1);}}}
             -- cycle
             foreach \t in {0, .05, ..., 1} { pic[pos=\t] {code={\filldraw[thin,fill=gray!50!white] (0,0) rectangle (.1, .1);}}};
        \end{scope}
      }
    }

    \tikzset{
      bounded/.pic={
        \begin{scope}[scale=.5]%, shift={(-1, -1)}]
          \clip[draw] (-2, -2) rectangle (6, 6);
          \coordinate (origin) at (0, 0);
          \coordinate (d_circle) at (35:3);
          \coordinate (d_oppcircle) at (35:-2.5);
          \path[->] (origin) edge node[pos=.8, auto] {$F(n, m) \vec{d}$} (d_circle);
          \pic[fill=gray!50!white] (invader) at (0, 0) {spaceinvader};
          \node[fill=red, circle, inner sep=1pt] at (origin) {};
          \node[left=.1em of origin] {$\vec{0}$};
          \path[draw, gray, dashed] (d_circle) -- +(125:10);
          \path[draw, gray, dashed] (d_circle) -- +(125:-10);
          \fill[pattern=checkerboard, pattern color=red!50, path fading=east, rotate around={35:(d_circle)}] (d_circle) +(0, -10) rectangle +(10, 10);
          \path (d_circle) -- +(1.5,1.5) node [fill=white, rounded corners=2pt] {no tile};
        \end{scope}
      }
    }

    \begin{eqnarray*}
      &C_{m}[\mathcal{S}] = \left\{\tikz[baseline=(current bounding box.center)]{\pic{bigCycle};}\right\},\quad
      B_{F(n,m),\vec{d}}[\mathcal{S}] = \left\{\tikz[baseline=(current bounding box.center)]{\pic{bounded};}\right\}&\\
      &P_{\vec{d}}[\mathcal{S}] = \left\{\tikz[baseline=(current bounding box.center)]{\pic{unbounded};}\right\}&\\
      %&\TerminalProds{S} \cap (C_{n,m} \cup B_{n,m} \cup P_{n,m}) \neq \emptyset&\\
    \end{eqnarray*}
    \caption{The three sets defined by~\cref{thm:tree_pump}: $C_{m}[\mathcal{S}]$ is the assemblies which encircle an $m \times m$ square, $B_{F(n, m), \vec{d}}[\mathcal{S}]$ is the set of assemblies which do not reach further than $F(n,m)$ in direction $\vec{d}$, and $P_{\vec{d}}[\mathcal{S}]$ is the assemblies which contain a periodic path with period $\vec{p}$ such that $\vec{p}\cdot\vec{d} > 0$.}
\label{fig:tree_pump}
\end{figure}

The Tree Pump Theorem states that in order to do a meaningful amount of computation, a self-assembling system with $n$ tiles needs to encircle large squares ---say, of size $m$, thus hitting $C_{m}[\Ss]$. If it does not, then its productions look very much like trees drawn on $\mathbb{Z}^2$ with a $m$-cell wide brush. A branch of that tree then behaves like a finite automaton. If that automaton stops, the assembly goes no further than $F(n, m)$ in any given direction $\vec{d}$, which gives a final production in $B_{F(n,m), \vec{d}}[\Ss]$. If not, these long branches must have an ultimately periodic behavior which gives a final production in $P_{\vec{d}}[\Ss]$.

The proof of~\cref{thm:tree_pump} needs some ancillary definitions. The objects they introduce may seem overly sophisticated in view of the concreteness of the statement of \cref{thm:tree_pump}, but one needs to soldier on. The reader may find solace in knowing that they are the result of persistent failure in one of the author's search of a more down-to-earth argument. The proof plainly needs systems endowed with the ability to escape the confines of the Euclidian plane $\mathbb{Z}^2$ and those of a merely infinite time. Hence, the definition and use of \emph{Ordinal-length} sequences of \emph{Free Assemblies}.

\subsection{Free Assemblies}
\label{sec:freeSeqs}

 In this appendix, assemblies sit on a graph called an \emph{assembly support} instead of $\mathbb{Z}^2$.

\begin{definition}[Assembly support]
  An \emph{assembly support} is a directed graph $G$ with labels in $\{N, E, S, W\}$ on its arcs, such that:
\begin{itemize}
\item each vertex has at most one incoming arc with each label,
\item each vertex has at most one outgoing arc with each label,
\item for any cycle $c$ of $G$, the sum of the labels of the arcs of $c$ is equal to $\vec{0}$ as a vector of $\mathbb{Z}^2$,
\item for any path $\pi$ of $1$ or $3$ arcs from $u$ to $v$, if the sum in $\mathbb{Z}^2$ of the labels of $\pi$ is $\vec{d} \in \Dir$, then there is an arc from $v$ to $u$ with label $-\vec{d}$.
\end{itemize}
\end{definition}

An example of a correct assembly support is given on \cref{fig:ext_arcs_steps_holes}, while \cref{fig:bad-ass-support} shows some counter-examples of labeled graphs which are not assembly supports. Because of the conditions on the incoming and outgoing arcs of each vertex, it makes sense to represent the vertices as tiles, and arcs as sides the tiles share. This hopefully helps build the intuition that assembly support are ``like $\mathbb{Z}^2$, but long separated paths which should come to the same position may avoid each other''. On \cref{fig:bad-ass-support}, the direction corresponding to each side of the tile is given explicitely. With correct Assembly Supports, like the ones on \cref{fig:ext_arcs_steps_holes}, the convention is that the direction of each side corresponds roughly to its direction on the page, and small adjustments allow tiles which would otherwise be adjacent on the page not to be. The appearance of tiles avoiding each other by going in the third dimension is deliberate, as a way to build intuition.

An \emph{embedding} from an assembly support $G$ to another assembly support $G'$ is a graph embedding which preserves the label of the arcs. An \emph{isomorphism} between $G$ and $G'$ is a graph isomorphism preserving the arc labels.

\begin{definition}[Perspective difference, Translation]
  For any two vertices $(z, z')$ of a connected component of an assembly support $G$, their \emph{perspective difference} $z' \pd z$ is $\sum_{a \in p} a$, where $p \in \Dir^*$ is the sequence of labels of a path from $z$ to $z'$. For any embedding $e: G \to \mathbb{Z}^2$, $e(z') - e(z) = z' \pd z$. By convention, $z, z'$ are in different connected components, $z' \pd z = \infty$.

  A translation between two subgraphs $A \subset G$ and $B \subset G'$ of two assembly support is an isomorphism from $A$ to $B$ which preserves perspective differences. Any isomorphism between connected subgraphs is a translation.
\end{definition}

For any Assembly Support $G$ and $z \in G$, $\ezEmbed[z, G]: G \to \mathbb{Z}^2$ is the embedding $z' \mapsto (z' \pd z)$. Since changing $z$ in $\ezEmbed[z, G]$ only amounts to a translation, it will generally be omitted. The index $G$ will be omitted as well when this causes no ambiguity.

\begin{figure}
  \centering
  \newcommand{\crookedTile}[5]{
  \begin{scope}
    \filldraw[fill=gray!25!white] #1 -- node (s) [above] {S} #2 -- node (e) [left] {E} #3 -- node (n) [below] {N} #4 -- node (w) [right]{W} cycle;
    \node at (barycentric cs:s=1,e=1,n=1,w=1) {#5};
  \end{scope}
}

\begin{tikzpicture}
  \begin{scope}[scale=1.3]
    \crookedTile{(0,0)}{(1,0)}{(1,1)}{(0,1)}{$z$}
    \filldraw[fill=gray!25!white] (1,0) -- node[above] {W} (2,0) -- node[left] {S} (2,1) -- node[below] {E} (1,1) -- node[right] {N} cycle;
    \node at (1.5,.5) {$u$};
    \crookedTile{(1,-1)}{(2,-1)}{(2,0)}{(1,0)}{$t$}

    \begin{scope}[scale=1.2,shift={(0,2)}]

      \crookedTile{(-1,0)}{(0,0)}{++(108:1)}{++(-1,0)}{$o$}
      \begin{scope}[shift={(72:-1)}]
        \crookedTile{(-1,0)}{(0,0)}{++(72:1)}{++(-1,0)}{$n$}
      \end{scope}
      \crookedTile{(0,0)}{++(36:1)}{++(108:1)}{++(36:-1)}{$p$}
      \begin{scope}[shift={(-36:1)}]
        \crookedTile{(0,0)}{++(36:1)}{++(-36:-1)}{++(36:-1)}{$e$}
      \end{scope}

      % \crookedTile{(0,0)}{++(1,0)}{++(108:1)}{++(-1,0)}
      % \crookedTile{(1,0)}{++(36:1)}{++(108:1)}{++(36:-1)}
      % \crookedTile{(1,0)}{++(-36:1)}{++(36:1)}{++(-36:-1)}
      % \crookedTile{(72:-1)}{++(0:1)}{(1,0)}{(0,0)}
    \end{scope}

    % peut aussi s'épeller aproximatif
    \begin{scope}[shift={(6,0)}]
      \crookedTile{(0,0)}{(1,0)}{(0.833,1)}{(0,1)}{$i$}
      \crookedTile{(1,0)}{(2,0)}{(1.833,1)}{(0.8333,1)}{$s$}
      \crookedTile{(-1,0)}{(0,0)}{(0,1)}{(-0.8333,1)}{$m$}
      \crookedTile{(-2,0)}{(-1,0)}{(-0.8333,1)}{(-1.8333,1)}{$\epsilon$}
      \crookedTile{(0.8333,1)}{(1.8333,1)}{(1.666,2)}{(0.666,2)}{$c$}
      \crookedTile{(-1.8333,1)}{(-0.8333,1)}{(-0.666,2)}{(-1.666,2)}{$d$}
      \crookedTile{(0.666,2)}{(1.666,2)}{(1.5,3)}{(0.5,3)}{$o$}
      \crookedTile{(-1.666,2)}{(-0.666,2)}{(-.5,3)}{(-1.5,3)}{$e$}
      \crookedTile{(0.5,3)}{(1.5,3)}{(1.333,4)}{(0.333,4)}{$u$}
      \crookedTile{(-1.5,3)}{(-0.5,3)}{(-0.333,4)}{(-1.333,4)}{$t$}
      \crookedTile{(-0.5,3)}{(0.5,3)}{(0.333,4)}{(-0.333,4)}{$n$}
    \end{scope}
  \end{scope}

\end{tikzpicture}
  \caption{\emph{Ceci n'est pas un Assembly Support}: three labelled graphs which fail to be assembly supports. Each vertex is represented by a tile, the labels of its sides are the labels of its outgoing arcs. A label on a side with no neighbor corresponds to an exterior arc. The bottom-left graph has a path of length $1$ with label $E$ from $z$ to $u$ but the $-E = W$ neighbor of $u$ is $t$. Also, the $N$ neighbor of $t$ is $u$, which has no $S$ neighbor. The one in the top right has a $3$ arc path ``$nope$'' whose labels sum to $N + E + S = E$, but $n$ is not the $W$ neighbor of $e$ (actually, there is none). The one on the right has a cycle ``$\epsilon\, miscounted$'' for which the sum of the labels is $3E + 3N + 2W + 3S = E$, which is not $\vec{0}$.}
  \label{fig:bad-ass-support}
\end{figure}

The analysis in the Tree Pump Theorem relies on an examination of the \emph{holes} of the productions of $\Ss$. These holes need to be suitably defined now that the assemblies are no longer necessarily planar.

\begin{definition}[Exterior Arc]
  An \emph{exterior arc} of an assembly support $G$ is a pair of a vertex $v$ of $G$ and a direction $d$ such that $v$ has no arc labelled $d$ in $G$.
\end{definition}

\begin{definition}[Converging Exterior Arcs, Growth]
  Two exterior arcs $(v, d)$, $(v', d')$ are \emph{converging} when $v' \pd v = d' - d$ --they virtually point to the same empty position.

  Given an assembly support $G$ and a set $Z$ of converging arcs, the assembly support $G + Z$ ($G$ grown by $Z$) is $G \cup \{\zeta\}$, where there is an arc $z \to \zeta$ (for $z \in G$) in direction $d$ whenever $(z, d) \in Z$.
\end{definition}

\begin{definition}[Step, Exterior Path, Hole]
  There is a \emph{step} between two arcs or exterior arcs $(v, d)$ and $(v', d')$ if:
  \begin{itemize}
  \item $G = G'$ and $d, d'$ form a $\pm \frac{\pi}{2}$ angle,
  \item $d = d'$ and there is an (non-exterior) arc from $v$ to $v'$ with label $d''$, with $d, d''$ forming a $\pm \frac{\pi}{2}$ angle
  \item $d, d'$ form a $\pm \frac{\pi}{2}$ angle and there is a cell $v''$ with $v'' + d = v$ and $v'' + d' = v'$.
  \end{itemize}

  A (simple) \emph{exterior path} is a sequence $P = (p_0, p_1, \ldots, p_{k})$ of arcs, such that:
  \begin{itemize}
  \item for each $i$, there is a step between $p_i$ and $p_{i+1}$, and
  \item for $0 < i < k$, $p_i$ is an exterior arc.
  \end{itemize}

  A \emph{hole} is a simple exterior path which loops back to its starting exterior arc.

  The \emph{perimeter} of a hole $H = (h_0, \ldots h_k = h_0)$ is the set of vertices appearing in the exterior vertices $h_i$.
\end{definition}

These definitions are illustrated on~\cref{fig:ext_arcs_steps_holes}, which features an assembly support with two holes of perimeter $20$. These holes are ``morally infinite'', so they resist attempts at a definition of area ---at least the author's naive ones. For an assembly which embeds into $\mathbb{Z}^2$, the free definition of hole corresponds to the intuition of a hole in an assembly, up to the detail that the exterior of the assembly forms a hole, so that all finite assemblies have at least one hole. Note that exterior paths are oriented: they turn in the positive direction around each tile.

\begin{figure}
  \centering
  \begin{tikzpicture}
\matrix[left=3em of {(0,0)}, fill=blue!25!white]{
    \begin{scope}
      \draw[->] (0, 0, 0) -- +(xyz cs:x=1) node[right] {$E$};
      \draw[->] (0, 0, 0) -- +(xyz cs:y=1) node[above] {$N$};
      \draw[->] (0, 0, 0) -- +(xyz cs:z=1) node[below left, align=center] {Artistic\\ license};
    \end{scope}\\
};   

  \foreach \x / \y / \lab in { 0 / 0 / a, 1 / 0 / b, 2 / 0 / c, 3 / 0 / d }
  \filldraw[fill=gray!25!white] (\x,\y) rectangle +(1,1) node [pos=.5] {\lab};

  \foreach \x / \y / \lab in { 3 / 4 / h}
  \filldraw[fill=gray!25!white] (\x,\y) rectangle +(1,1)  node[pos=.5] {\lab};

  \foreach \x / \y / \lab in { 4 / 4 / i, 5 / 4 / j, 6 / 4 / k}
  \filldraw[fill=gray!25!white] (\x,\y) rectangle +(1,1)  node[pos=.5] {\lab};

  \foreach \x / \y / \lab in { 4 / 4 / i, 5 / 4 / j, 6 / 4 / k}
  \filldraw[fill=gray!25!white] (\x,\y) rectangle +(1,1)  node[pos=.5] {\lab};

  \foreach \x / \y / \lab in { 6 / 3 / l, 6 / 2 / m}
  \filldraw[fill=gray!25!white] (\x,\y) rectangle +(1,1)  node[pos=.5] {\lab};

  \filldraw[fill=gray!30!white] (5, 2, -.5) -- ++(1, 0, .5) -- ++(0, 1, 0) -- ++(-1, 0, -.5) -- cycle;
  \node at (5.5, 2.5, -.25) {n};
  
  \foreach \x / \y / \lab / \labshift in { 4 / 2 / o / 0, 3 / 2 / p / .3, 2 / 2 / q / -.3}
  \filldraw[fill=gray!35!white] (\x,\y, -.5) rectangle +(1,1, 0)  node[pos=.5, shift={(\labshift, 0)}] {\lab};

  \filldraw[fill=gray!30!white] (1, 2, 0) -- ++(1, 0, -.5) -- ++(0, 1, 0) -- ++(-1, 0, .5) -- cycle;
  \node at (1.5, 2.5, -.25) {r};

  \foreach \x / \y / \lab in {0 / 2 / s, 0 / 1 / t}
  \filldraw[fill=gray!25!white] (\x,\y, 0) rectangle +(1,1, 0)  node[pos=.5] {\lab};

  \foreach \x / \y / \z / \dir [count=\n] in { %
    1.5 / 1 / 0 / 90, 2.5 / 1 / 0 / 90, 3 / 1.5 / .25 / 180, 3 / 2.5 / .5 / 180, 3 / 3.5 / .25 / 180, %
    3 / 4.5 / 0 / 180, 3.5 / 5 / 0 / 90, 4.5 / 5 / 0 / 90, 5.5 / 5 / 0 / 90, 6.5 / 5 / 0 / 90, %
    7 / 4.5 / 0 / 0, 7 / 3.5 / 0 / 0, 7 / 2.5 / 0 / 0, 6.5 / 2 / 0 / -90, 5.5 / 2 / -.25 / -90, %
    4.5 / 2 / -.5 / -90,  3.8 / 2 / -.5 / -90, 2.2 / 2 / -.5 / -90, 1.5 / 2 / -.25 / -90, 1 / 1.5 / 0 / 0 %
  }
    \draw[-{Circle}, red] (\x, \y, \z) -- +(\dir:.3) node[name=circle\n, inner sep=0] {};
    %\draw[-{Circle[open]}, red] (\x, \y, \z) -- +(\dir:.3); % node[name=circle\n, inner sep=0] {};

  \foreach \x / \y / \z / \dir [count=\n] in { %
    .5 / 0 / 0 / -90, 1.5 / 0 / 0 / -90, 2.5 / 0 / 0 / -90, 3.5 / 0 / 0 / -90,
    4 / .5 / 0 / 0, 4 / 1.5 / .25 / 0, 4 / 2.5 / .5 / 0, 4 / 3.8 / .1 / 0, %
    4.7 / 4 / 0 / -90, 5.3 / 4 / 0 / -90, %
    6 / 3.5 / 0 / 180, 5.2 / 3 / -.4 / 90, %
    4.5 / 3 / -.5 / 90,  3.8 / 3 / -.5 / 90, 2.2 / 3 / -.5 / 90, 1.5 / 3 / -.25 / 90, .5 / 3 / 0 / 90, %
    0 / 2.5 / 0 / 180, 0 / 1.5 / 0 / 180, 0 / .5 / 0 / 180 % %
  }
  {
    \draw[-{Square[open]}, blue] (\x, \y, \z) -- +(\dir:.3) node[name=square\n, inner sep=0] {};
  }

  \foreach \k [remember=\k as \lastk (initially 20)] in {1, ..., 20}
  \draw[thick, densely dotted] (circle\lastk) -- (circle\k);

  \filldraw[fill=gray!20!white] (3, 1, 0) -- (3, 2, .5) -- (4, 2, .5) -- (4, 1, 0) -- cycle;
  \node at (3.5, 1.5, .25) {e};

  \filldraw[fill=gray!15!white] (3, 2, .5) rectangle +(1,1, 0)  node[pos=.5] {f};

  \filldraw[fill=gray!20!white] (3, 3, .5) -- (3, 4, 0) -- (4, 4, 0) -- (4, 3, .5) -- cycle;
  \node at (3.5, 3.5, .25) {g};

\end{tikzpicture}
  \caption{An example of an assembly support, the shifts in the 3rd dimension between $e$ and $g$ as well as between $r$ and $n$ are artistic license to show the absence of adjacencies between those two branches. The sum of the vectors between neighbors on the cycle $abcdefghijklmnopqrsta$ is $3E + 4N + 3E + 2S + 6E + 2S = \vec{0}$. The sum of vectors between neighbors on the path $fedcbatsrqp$ is $2S + 3W + 2N + 3E = \vec{0}$, yet the vertices $f$ and $p$ are distinct. The small pending vertices next to the tiles are the exterior arcs; they form two holes, one for the square exterior arcs, and one for the circles. The dotted lines shows the steps between the exterior arcs of the ``circle'' hole. The step between the exterior arcs $(m, S)$ and $(m, E)$ stems from the first case of the definition, the one between $(b, N)$ and $(c, N)$ from the second case, and the one between $(t, E)$ and $(r, S)$ because of the third. In this last case, $s$ acts as the pivot between $r$ and $s$. Both holes have perimeter 20.}
  \label{fig:ext_arcs_steps_holes}
\end{figure}

\begin{definition}[Free Assembly]
  A free assembly $A$ of the TAS $\calT = (T, \sigma, \tau)$ is composed of an assembly support $\support[A]$ and a total function $\tiles[A]: \support[A] \rightarrow T$.

  For a TAS $\calT$, the set of free assemblies of $\calT$ is written $\FreeAssemblies{\calT}$.

  % An assembly $A$ is \emph{exact} if the function $\tiles[A]$ is total.
\end{definition}

When there is an injective embedding $e: \support[A] \to G$, it naturally defines a free assembly $e(A)$ with $\support[e(A)] = e(\support[A])$ and for $z \in e(\support[A])$, $e(A)(e(z)) =  A(z)$. An assembly $A$ \emph{embeds} into $\mathbb{Z}^2$ if $\ezEmbed$ is a one-to-one embedding $\support[A] \to \mathbb{Z}^2$; as above, this embedding naturally defines an assembly $\ezEmbed(A)$.

\begin{definition}[Free Attachment, Free Assembly Sequence]
  Let $A, A'$ be free assemblies of some TAS system $\calT = (T, \sigma, s)$, and $\eta: A \to A'$ an embedding. The tuple $(A, A', \eta)$ is an assembly candidate if:
  \begin{itemize}
  \item there is a $z \in \support[A']$ such that $\eta: \support[A] \to \support[A'] \setminus \{ z \}$ is a translation,
  \item for any $z' \in \support[A]$, $\tiles[A](z') = \tiles[A](\eta(z'))$.
  \end{itemize}

  The definitions of attachment, assembly sequence, production and terminal production extend to free assemblies. The $\mathbb{Z}^2$ version of each definition is simply the particular case where every assembly embeds into $\mathbb{Z}^2$.

  For a TAS $\calT$, the set of free assembly sequences of $\calT$ is written $\FreeSeqs{\calT}$.  
\end{definition}

The definition of free attachment is illustrated on~\cref{fig:free-attach}. The fact that $\eta$ is an isomorohpism ensures that parts of the assembly which do not touch the position $z$ are unaffected by the attachment. This preserves the locality of the aTAM process, in contrast to what happens in the FTAM~\cite{DBLP:journals/nc/Durand-LoseHPPS20} for instance. Hence, while the above definition of attachment is given from the point of view of $A'$, ``after the fact'', from the point of view of $A$, an attachment candidate can be defined from a tile type $t$ and a set $Z$ of convergent exterior arcs: $t@Z$ is the attachment candidate $(A, A', \eta)$ where:
\begin{itemize}
\item $\support[A'] = \support[A] + Z$, and $\zeta$ is the vertex of $\support[A']$ which is not in $\support[A]$,
\item $\eta$ is the identity on $\support[A]$,
\item $\tiles[A'](\zeta) = t$,
\item $\tiles[A'](z) = \tiles[A](z)$ whenever $z \neq \zeta$.
\end{itemize}

\begin{figure}
  \centering
  \begin{tikzpicture}

  \tikzset{
    tileA/.pic={
      \clip (0, 0) rectangle (1,1);
      \filldraw[fill=gray!25!white] (0, 0) rectangle (1, 1) node[anchor=center, pos=.5] {A};
      \pic at (.5, 1) {doubleGlue};
      \pic[rotate=-90] at (1, .5) {doubleGlue};
    }
  }
  
  \tikzset{
    tileB/.pic={
      \clip (0, 0) rectangle (1,1);
      \filldraw[fill=gray!25!white] (0, 0) rectangle (1, 1) node[anchor=center, pos=.5] {B};
      \pic at (.5, 1) {doubleGlue};
      \pic[rotate=180] at (.5, 0) {doubleGlue};
    }
  }

  \tikzset{
    tileBUp/.pic={
      \clip (0, 0, 0) -- (0, 1, 0.5) -- (1, 1, 0.5) -- (1, 0, 0) -- cycle;
      \filldraw[fill=gray!25!white] (0, 0)  (0, 0, 0) -- (0, 1, 0.5) -- (1, 1, 0.5) -- (1, 0, 0) -- cycle;
      \node[xslant=-.3, anchor=center] at (.5, .5, .25) {B};
      \pic[xslant=-.3] at (.5, 1, .5) {doubleGlue};
      \pic[rotate=180, xslant=-.3] at (.5, 0, 0) {doubleGlue};
    }
  }

  \tikzset{
    tileC/.pic={
      \clip (0, 0) rectangle (1,1);
      \filldraw[fill=gray!25!white] (0, 0) rectangle (1, 1) node[anchor=center, pos=.5] {C};
      \pic[rotate=-90] at (1, .5) {doubleGlue};
      \pic[rotate=90] at (0, .5) {doubleGlue};
    }
  }

  \tikzset{
    tileCDown/.pic={
      \clip (0, 0, 0) -- (1, 0, -.5) -- (1, 1, -.5) -- (0, 1, 0) -- cycle;
      \filldraw[fill=gray!25!white] (0, 0, 0) -- (1, 0, -.5) -- (1, 1, -.5) -- (0, 1, 0) -- cycle;
      \node[anchor=center, yslant=.2] at (.5, .5, -.25) {C};
      \pic[yslant=.2, rotate=-90] at (1, .5, -.5) {doubleGlue};
      \pic[yslant=.2, rotate=90] at (0, .5, 0) {doubleGlue};
    }
  }

  \tikzset{
    tileD/.pic={
      \clip (0, 0) rectangle (1,1);
      \filldraw[fill=gray!25!white] (0, 0) rectangle (1, 1) node[anchor=center, pos=.5] {D};
      \pic[rotate=-90] at (1, .5) {doubleGlue};
      \pic[rotate=180] at (.5, 0) {doubleGlue};
      \pic[rotate=-90] at (0, .5) {singleGlue};
    }
  }

  \tikzset{
    tileDnew/.pic={
      \clip (0, 0) rectangle (1,1);
      \filldraw[fill=gray] (0, 0) rectangle (1, 1) node[anchor=center, pos=.5] {D};
      \pic[rotate=-90] at (1, .5) {doubleGlue};
      \pic[rotate=180] at (.5, 0) {doubleGlue};
      \pic[rotate=-90] at (0, .5) {singleGlue};
    }
  }

  \tikzset{
    tileE/.pic={
      \clip (0, 0) rectangle (1,1);
      \filldraw[fill=gray!25!white] (0, 0) rectangle (1, 1) node[anchor=center, pos=.5] {E};
      \pic[rotate=90] at (0, .5) {doubleGlue};
      \pic at (.5, 1) {doubleGlue};
      \pic[rotate=90] at (1, .5) {singleGlue};
    }
  }

  \tikzset{
    tileEDown/.pic={
      \clip (0, 0, 0) -- (1, 0, -.5) -- (1, 1, -.5) -- (0, 1, 0) -- cycle;
      \filldraw[fill=gray!25!white] (0, 0, 0) -- (1, 0, -.5) -- (1, 1, -.5) -- (0, 1, 0) -- cycle;
      \node[anchor=center, yslant=.2] at (.5, .5, -.25) {E};
      \pic[yslant=.2, rotate=90] at (0, .5, 0) {doubleGlue};
      \pic[yslant=.2] at (.5, 1, -.25) {doubleGlue};
      \pic[yslant=.2, rotate=90] at (1, .5, -.5) {singleGlue};
    }
  }

  \tikzset{
    tileF/.pic={
      \clip (0, 0) rectangle (1,1);
      \filldraw[fill=blue!25!white] (0, 0) rectangle (1, 1)  node[anchor=center, pos=.5] {F};
      \pic[rotate=180] at (.5, 0) {doubleGlue};
      %\pic at (.5, 1) {doubleGlue};
      \pic[rotate=180] at (.5, 1) {singleGlue};
    }
  }

  \tikzset{
    startAss/.pic={
      \pic {tileA};
      \pic at (0, 1, 0) {tileBUp};
      \pic at (0, 2, .5) {tileD};
      \pic at (1, 2, .5) {tileE};
      \pic at (1, 3, .5) {tileB};
      \pic at (1, 4, .5) {tileD};
      \pic at (2, 4, .5) {tileC};
      \pic at (1, 0, 0) {tileCDown};
      \pic at (2, 0, -.5) {tileE};
      \pic at (2, 1, -.5) {tileB};
      \pic at (2, 2, -.5) {tileD};
      \pic at (3, 2, -.5) {tileE};
      \pic at (3, 3, -.5) {tileB};
      \draw[densely dotted, gray] (2, 2, .5) -- (2, 2, -.5);
      \draw[densely dotted, gray] (2, 3, .5) -- (2, 3, -.5);
      \draw[densely dotted, gray] (3, 4, .5) -- (3, 4, -.5);
    }
  }

  \tikzset{
    attachDown/.pic={
      \pic {tileA};
      \pic at (0, 1, 0) {tileBUp};
      \pic at (0, 2, .5) {tileD};
      \pic at (1, 2, .5) {tileE};
      \pic at (1, 3, .5) {tileB};
      \pic at (1, 4, .5) {tileD};
      \pic at (2, 4, .5) {tileE};
      \pic at (1, 0, 0) {tileCDown};
      \pic at (2, 0, -.5) {tileE};
      \pic at (2, 1, -.5) {tileB};
      \pic at (2, 2, -.5) {tileD};
      \pic at (3, 2, -.5) {tileE};
      \pic at (3, 3, -.5) {tileB};
      \pic at (3, 4, -.5) {tileDnew};
      \draw[densely dotted, gray] (2, 2, .5) -- (2, 2, -.5);
      \draw[densely dotted, gray] (2, 3, .5) -- (2, 3, -.5);
      \draw[densely dotted, gray] (3, 4, .5) -- (3, 4, -.5);
      \draw[densely dotted, gray] (3, 5, .5) -- (3, 5, -.5);
    }
  }

  \tikzset{
    attachUp/.pic={
      \pic {tileA};
      \pic at (0, 1, 0) {tileBUp};
      \pic at (0, 2, .5) {tileD};
      \pic at (1, 2, .5) {tileE};
      \pic at (1, 3, .5) {tileB};
      \pic at (1, 4, .5) {tileD};
      \pic at (2, 4, .5) {tileE};
      \pic at (1, 0, 0) {tileCDown};
      \pic at (2, 0, -.5) {tileE};
      \pic at (2, 1, -.5) {tileB};
      \pic at (2, 2, -.5) {tileD};
      \pic at (3, 2, -.5) {tileE};
      \pic at (3, 3, -.5) {tileB};
      \pic at (3, 4, .5) {tileDnew};
      \draw[densely dotted, gray] (2, 2, .5) -- (2, 2, -.5);
      \draw[densely dotted, gray] (2, 3, .5) -- (2, 3, -.5);
      \draw[densely dotted, gray] (4, 4, .5) -- (4, 4, -.5);
    }
  }

    \tikzset{
    attachBoth/.pic={
      \pic {tileA};
      \pic at (0, 1, 0) {tileBUp};
      \pic at (0, 2, .5) {tileD};
      \pic at (1, 2, .5) {tileE};
      \pic at (1, 3, .5) {tileB};
      \pic at (1, 4, .5) {tileD};
      \pic at (2, 4, .5) {tileEDown};
      \pic at (1, 0, 0) {tileCDown};
      \pic at (2, 0, -.5) {tileE};
      \pic at (2, 1, -.5) {tileB};
      \pic at (2, 2, -.5) {tileD};
      \pic at (3, 2, -.5) {tileE};
      \pic at (3, 3, -.5) {tileBUp};
      \pic at (3, 4, 0) {tileDnew};
      \draw[densely dotted, gray] (2, 2, .5) -- (2, 2, -.5);
      \draw[densely dotted, gray] (2, 3, .5) -- (2, 3, -.5);
    }
  }

    \tikzset{
    attachUpSolder/.pic={
      \pic {tileA};
      \pic at (0, 1, 0) {tileB};
      \pic at (0, 2, 0) {tileD};
      \pic at (1, 2, 0) {tileE};
      \pic at (1, 3, 0) {tileB};
      \pic at (1, 4, 0) {tileD};
      \pic at (2, 4, 0) {tileEDown};
      \pic at (1, 0, 0) {tileC};
      \pic at (2, 0, 0) {tileE};
      \pic at (2, 1, 0) {tileB};
      \pic at (2, 2, 0) {tileD};
      \pic at (3, 2, 0) {tileE};
      \pic at (3, 3, 0) {tileBUp};
      \pic at (3, 4, .5) {tileDnew};
      \draw[densely dotted, gray] (3, 5, .5) -- (3, 5, -.5);
    }
  }

  \node[matrix, draw] (start) at (0, 0) {\pic {startAss};\\};
  \node[matrix, draw] (both) at (15:7) {\pic {attachBoth};\\};
  \node[matrix, fill=red!50!white, draw] (up) at (-30:7) {\pic {attachUp};\\};
  \node[matrix, fill=red!50!white, draw] (solder) at (-80:7) {\pic {attachUpSolder};\\};
  \node[matrix, draw] (down) at (-120:7) {\pic {attachDown};\\};
  \node[left=1em of start] {$S$};

  \matrix[left=3em of start, fill=blue!25!white]{
    \begin{scope}
      \draw[->] (0, 0, 0) -- +(xyz cs:x=1) node[right] {$E$};
      \draw[->] (0, 0, 0) -- +(xyz cs:y=1) node[above] {$N$};
      \draw[->] (0, 0, 0) -- +(xyz cs:z=1) node[below left, align=center] {Artistic\\ license};
    \end{scope}\\

    \node {$\tau = 2$};\\
  };

  \draw[-{Stealth}] (start) -- (both) node[near end, above] {$A$};
  \draw[-{Rays[]}, red] (start) -- (up) node[near end, above] {$B$};
  \draw[-{Rays[]}, red] (start) -- (solder) node[near end, right] {$C$};
  \draw[-{Stealth}] (start) -- (down) node[near end, right] {$D$};
%  \pic at (60:7) {attachUpSolder};
%  \pic at (90:7) {attachDown};
  
  % \node[matrix, draw] {\pic {startAss};\\}
  % [grow=south east, level distance=7cm, sibling distance=6cm] child {
  %   node[matrix, draw] {\pic {attachBoth};\\}
  %   edge from parent[-{Stealth[]}]
  % }
  % child {
  %   node[matrix, draw=red] {\pic {attachUp};\\}
  %   edge from parent[red, -{Stealth[]}] node {\emoji{no-entry}}
  % }
  % child {
  %   node[matrix, draw=red] {\pic {attachUpSolder};\\}
  %   edge from parent[red, -{Stealth[]}] node {\emoji{no-entry}}
  % }
  % child {
  %   node[matrix, draw] {\pic {attachDown};\\}
  %   edge from parent[-{Stealth[]}]
  % };

\end{tikzpicture}

%%% Local Variables:
%%% mode: latex
%%% TeX-master: "../main"
%%% TeX-command-extra-options: "-shell-escape"
%%% TeX-engine: luatex
%%% End:
  \caption{Some possible and impossible free attachments from a starting free assembly $S$, at temperature $\tau$. Again, the third coordinate is a visual aid to show non-adjacencies; small dashed lines link points which embed in the same position in $\mathbb{Z}^2$. Assemblies $A$ and $D$ are reachable through one attachment $t_D@z$, where in each case, $z$ is the position in dark gray. The assembly $B$ is an attachment candidate, but it is not stable, since the new tile $t_D$ only binds through a strength-$1$ glue. The assembly $C$ can not even form an attachment candidate from $S$, as $\support[C]$ does not contain an isomorphic copy of $\support[S]$}
  \label{fig:free-attach}
\end{figure}

For clarity, a ``normal'' assembly will be referred to as a $\mathbb{Z}^2$-assembly, likewise for the other concepts which were just endowed with a ``free'' variant.

Embeddings act not only on assembly supports and assemblies, but also on assembly sequences. In doing so, they may break the correctness of attachments if they are not one-to-one; when that happens, the sequence is cut short.

\begin{definition}[Assembly Sequence Embedding]
  Let $\alpha = (A_0 \to A_1 \to \ldots) \in \FreeSeqs{\Ss}$, let $G = \support[(\lim \alpha)]$ and $G'$ an assembly support. Let $e: G \to G'$ be an embedding; since up to translation, $\support[A_0] \subset \support[A_1] \subset \ldots \subset \support[(\lim \alpha)]$, for each $i$, $e$ induces an embedding from $\support[A_i]$ into $G'$.

  Let $k$ be the largest $i$ such that $e$ is an isomorphism on $\support[A_i]$, then \(e(\alpha)\) is the assembly sequence $(e(A_0), \ldots, e(A_k))$.
\end{definition}

\subsection{Ordinal Assembly Sequences}

It is often practical to consider what happens in an assembly sequence $\alpha$ after it has placed an infinite number of tiles. For this, \emph{ordinal} assembly sequences become useful.

\begin{definition}[Ordinal Assembly Sequence]
\label{def:ordinalSeqs}
  Let $o \in \omega_1$ be a countable ordinal, an $o$-Assembly Sequence starting from an assembly $A_0$ is a sequence of length $o$ of free assemblies such that for any $i < j \leq o$, there is an injective embedding $\eta_{i \to j}: A_i \to A_j$ with:
  \begin{itemize}
  \item for $i < o$, $(A_i, A_{i+1}, \eta_{i \to i+1})$ is a free attachment,
  \item for $i < j < k \leq o$, $\eta_{j \to k} \circ \eta_{i \to j}  = \eta_{i \to k}$
  \end{itemize}

  The $i$-th production of $\alpha$ is $A_i$, and its $i$-th attachment is $\eta_{i \to i+1}$.
  
  For an $o$-Assembly Sequence $\alpha$, the notation $\lim \alpha = A_o$ is consistent with the case of usual Assembly Sequences, i.e. $\omega$-Assembly Sequences.
\end{definition}

These sequences correspond to the intuitive generalization of attachment sequences to ``times larger than infinity''. In particular, any tile attached through an ordinal attachment sequence reaches the starting assembly of the sequence through a finite number of attachments. Indeed, for a tile to have been added through an ordinal assembly sequence $\alpha$ at time $t$, the neighbors to which it attaches must have been attached at some time $t' < t$. Since these times are ordinals, such an decreasing sequence of times reaches $0$ in a finite number of steps. In particular, these ordinal assembly sequences produce the same productions as the usual $\omega$-assembly sequences.

This remark formalizes thus:
\begin{remark}
  \label{lem:pos-finite-time}
  Let $\Ss$ be an assembly system, $o$ a countable ordinal, $\alpha$ an $o$-Assembly Sequence, $k < o$ and $t@Z$ the $k$-th attachment of $\alpha$. There is a sequence $\alpha'$ of length $o'$ and a bijection $\iota: o \to o'$ such that:
  \begin{itemize}
  \item for all $t < o'$, the attachments $\alpha_t$ and $\alpha'_{\iota(t)}$ are the same,
  \item $\iota(k)$ is finite.
  \end{itemize}
\end{remark}

\begin{lemma}
  \label{lem:ord-omega}
  Let $\Ss$ be an assembly system, $o$ a countable ordinal, and $\alpha$ an $o$-Assembly Sequence. Then if $o$ is infinite, there is a $\omega$-Assembly Sequence $\alpha'$ such that $\lim \alpha' = \lim \alpha$.
\end{lemma}

\begin{proof}
  The proof goes by transfinite induction.

  If $o = \omega$, then $\alpha$ itself satisfies the conclusion of the lemma.

  If $o = p + 1$ is a successor ordinal, by induction hypothesis, there is an $\omega$-assembly sequence $\alpha''$ such that $\lim \alpha'' = A_p$, where $A_p$ is the $p$-th production of $\alpha$. Let $t@Z$ be the $p$-th attachment of $\alpha$;  there is an integer $k$ such that at time $k$, all the origin vertices of the arcs in $Z$ are attached in $\alpha''$. The attachment sequence $\alpha' = (\alpha''_0, \ldots, \alpha''_k, t@Z, \alpha''_{k+1}, \ldots)$ satisfies the conclusion of the lemma.

  If $o$ is a limit ordinal, there is a sequence $(\alpha'^i)_{i < o}$ of $\omega$-assembly sequences such that for each $i$, $\lim \alpha'^i = A_i$. Let $A'^i_j$ be the $j$-th production of $\alpha'^i$. Since $o$ is countable, it has cofinality $\omega$; pick an increasing sequence $\epsilon: \omega \to o$ such that $\sup_{i < \omega} \epsilon(i) = o$. Let $\alpha'$ be the enumeration of the set $\{A'^{\epsilon(i)}_j | j \leq i < \omega \}$ ordered lexicographically according to $(i, j)$. This sequence $\alpha'$ is an attachment sequence, and it satisfies $\lim \alpha' = \bigcup_{i < o} (A'^{\epsilon(i)}_i) = \lim \alpha$.
\end{proof}

% In the remainder of this appendix, all assembly sequences are ordinal assembly sequences.

\subsection{Holes and Fizziness}
\label{sec:fizziness}

The Tree Pump Theorem is about tree-like assemblies. Since trees are acyclic graphs, the \emph{holes} of the assemblies are inherently limited and thus play an important part in characterizing how these tree-like assemblies behave. Fizziness is the tendency of assembly sequences to create a lot of holes.

\begin{definition}[Fizziness]
  Let $A, A'$ be two free assemblies with $\support[A] \subset \support[A']$, the \emph{fizziness} $\fizziness(A, A')$ is the number of holes of $A'$ whose perimeter is not contained in $A$.
  
  Let $\alpha = (A_0 \rightarrow A_1 \rightarrow \ldots)$ be an Assembly Sequence in $\FreeSeqs{\Ss}$. The fizziness of $\alpha$, noted $\fizziness(\alpha)$ is the sequence $i \mapsto \fizziness(A_i, A_i+1)$.

  A sequence $\alpha$ is more fizzy than $\beta$, written $\alpha \moreFizzy \beta$ if $\fizziness(\alpha) > \fizziness(\beta)$ lexicographically.
\end{definition}

\begin{lemma}[Maximal Fizziness]
  Let $\calT = (S, \sigma, \tau)$ be a seeded assembly system. Assume $\sigma$ has a finite number of non-null glues on its external edges. Then there is an $\omega$-Assembly Sequence $\alpha_{\max{}} \in \FreeSeqs{\calT}$ with maximal fizziness among $\omega$-Assembly Sequences.
\end{lemma}

\begin{proof}
  Using K\H{o}nig's Lemma.
\end{proof}

Note that this lemma does not hold for $o$-assembly sequences with $o > \omega$. Indeed, after time $\omega$, there might be an infinity of possible attachments, thwarting K\H{o}nig's Lemma.

\begin{lemma}[Fizziness-Increase of Embeddings]
  Let $\Ss$ be a seeded TAS, and $\alpha \in \FreeSeqs{\Ss}$. Let $G = \support{(\lim \alpha)}$, $G'$ an assembly support, and $e: G \to G'$ an embedding.

  Then:
  \begin{enumerate}
  \item \label{lbl:embed_correct} $e(\alpha)$ is a free assembly sequence,
  \item \label{lbl:embed_heavier} $e(\alpha) \moreFizzyEq \alpha$
  \item \label{lbl:embed_same_weight} $e(\alpha) \moreFizzy \alpha$ if $e$ is not injective on $\dom(\lim \alpha)$.
  \end{enumerate}
\end{lemma}

\begin{proof}
  Let $\alpha = (t_i@z_i)_i$ be an assembly sequence in $\FreeSeqs{\Ss}$.

  For \cref{lbl:embed_correct}, the attachments of $e(\alpha)$ are valid by definition of $k$. They are stable since for each attachment, the edges adjacent to $z_i$ which make this attachment stable are preserved by $e$.

  For \cref{lbl:embed_heavier}, for each assembly $A$ of $\alpha$ which is mapped injectively, each hole of $\alpha$ is mapped by $e$ to a hole of $e(A)$ with the same labels on its border.

  For \cref{lbl:embed_same_weight}, if $e$ is not injective on $\dom(\lim \alpha)$, then there are $i < k$ such that $e(z_i) = e(z_k)$. Since $\alpha$ is a valid assembly sequence, $z_i \neq z_k$, there are two different paths from the seed to $e(z_i)$. From these two paths, it is possible to construct a hole in $\dom(e(\alpha))$ which is not a hole in $\dom(\alpha)$. Hence, from the first attachment where this hole appears, the assemblies of $e(\alpha)$ are more fizzy than the corresponding ones in $\alpha$.
\end{proof}

\begin{lemma}[Maximal Sequences are Flat]
  Let $\Ss$ be a seeded TAS, $S$ an assembly of $\Ss$ and $X \subset \FreeSeqs{\Ss, S}$ be a set of Free, Ordinal Assembly Sequences such that for any $\alpha \in X$, there is a $\alpha' \in X$ such that $\alpha' \moreFizzyEq \ezEmbed(\alpha)$.
  
  Assume $\alpha$ is a free assembly sequence with maximal fizziness within $X$. Then $\alpha$ embeds into $\mathbb{Z}^2$.
\end{lemma}

\begin{proof}
  If $\alpha$ does not embed into $\mathbb{Z}^2$ by $\ezEmbed$, then $\ezEmbed(\alpha)$ is fizzier than $\alpha$ and thus $\alpha$ cannot be maximal in $X$.
\end{proof}

  \begin{figure}
    \centering
    \tikzset{
  seed/.pic={
    \begin{scope}[scale=#1]
      \filldraw[fill={pink}] (0,0) rectangle +(1,1) node[pos=.5, font=\footnotesize] {\tikzpictext};
      \pic[scale=#1, rotate=180] at (0.5, 1) {singleGlue};
      \pic[scale=#1, rotate=90] at (1, 0.5) {singleGlue};
    \end{scope}
  }
}

\tikzset{
  tileA/.pic={
    \begin{scope}[scale=#1]
      \filldraw[fill={blue!50}] (0,0) rectangle +(1,1) node[pos=.5, font=\footnotesize] {\tikzpictext};
      \pic[scale=#1, rotate=-90] at (0, 0.5) {singleGlue};
      \pic[scale=#1, rotate=90] at (1, 0.5) {singleGlue};
    \end{scope}
  }
}

\tikzset{
  tileB/.pic={
    \begin{scope}[scale=#1]
      \filldraw[fill={yellow!50}] (0,0) rectangle +(1,1) node[pos=.5, font=\footnotesize] {\tikzpictext};
      \pic[scale=#1, rotate=-90] at (0, 0.5) {singleGlue};
      \pic[scale=#1, rotate=90] at (1, 0.5) {singleGlue};
      \pic[scale=#1] at (0.5, 0) {singleGlue};
    \end{scope}
  }
}

\tikzset{
  tileC/.pic={
    \begin{scope}[scale=#1]
      \filldraw[fill={green!50}] (0,0) rectangle +(1,1) node[pos=.5, font=\footnotesize] {\tikzpictext};
      \pic[scale=#1, rotate=180] at (0.5, 1) {singleGlue};
      \pic[scale=#1] at (0.5, 0) {singleGlue};
    \end{scope}
  }
}

\tikzset{
  tileD/.pic={
    \begin{scope}[scale=#1]
      \filldraw[fill={gray!50}] (0,0) rectangle +(1,1) node[pos=.5, font=\footnotesize] {\tikzpictext};
      \pic[scale=#1, rotate=-90] at (0, 0.5) {singleGlue};
      \pic[scale=#1, rotate=180] at (0.5, 1) {singleGlue};
    \end{scope}
  }
}

\tikzset{
  tileE/.pic={
    \begin{scope}[scale=#1]
      \filldraw[fill={purple!50}] (0,0) rectangle +(1,1) node[pos=.5, font=\footnotesize] {\tikzpictext};
      \pic[scale=#1, rotate=90] at (1, 0.5) {singleGlue};
      \pic[scale=#1, rotate=180] at (0.5, 1) {singleGlue};
    \end{scope}
  }
}

\begin{tikzpicture}[scale=.5]

  \draw[-{Stealth[]}, thick] (3,3) -- (3,5) node[pos=.5, right] {$\vec{d}$};
  \node at (8,4) {$\tau = 1$};            
  \tikzmath {
    int \t;
    for \t in {0,...,65} {
      coordinate \z;
      int \tmod;
      \tmod = mod(\t, 9);
      int \tdiv;
      \tdiv = \t / 9;
      let \tile = tileA;
      if \tmod == 0 then {
        let \tile = tileC;
      };
      if \tmod == 1 then {
        let \tile = tileB;
      };
      if \tmod == 2 then {
        let \tile = tileA;
      };
      if \tmod == 3 then {
        let \tile = tileB;
      };
      if \tmod == 4 then {
        let \tile = tileC;
      };
      if \tmod == 5 then {
        let \tile = tileD;
      };
      if \tmod == 7 then {
        let \tile = tileE;
      };
      if \tmod == 8 then {
        let \tile = tileC;
      };
      if \t == 0 then {
        let \tile = seed;
      };
      if \tmod == 0 then {
        int \y;
        \y = \t / 9;
        \z = (0, \y);
      } else {
        if \tmod < 4 then {
          \z = (\tmod + 3 * \tdiv,0);
        } else {
          if \tmod < 6 then {
            \z = (3 + 3 * \tdiv,-\tmod + 3);
          } else {
            if \tmod < 8 then {
              \z = (3 * \tdiv - \tmod + 8,-2);
            } else {
              \z = (3 * \tdiv + 1,-1);
            };
          };
        };
      };
      {
        \begin{scope}[shift={(\z)}]
            \pic["\t"] {\tile=.5};
        \end{scope}
      };
    };
  }
  \node[rotate=90] at (0.5, 9) {$\ldots$};
  \node at (25, 0.5) {$\ldots$};
  \node[left] at (0,0) {$\alpha' = $};    
\end{tikzpicture}\\
\begin{tikzpicture}
  \node[rotate=90] {$\moreFizzy$};
\end{tikzpicture}\\
\begin{tikzpicture}[scale=.5]
  \foreach \t in {0,...,65} {
    \tikzmath {
      coordinate \z;
      int \tmod;
      \tmod = mod(\t, 9);
      int \tdiv;
      \tdiv = \t / 9;
      int \thuit;
      if \t < 9 then {
        \thuit = \t;
      }
      else {
        if \t < 17 then {
          \thuit = \t - 1;
        } else {
          \thuit = \t - 2 * floor(\t / 9) + 1;
        };
      };
      let \tile = tileA;
      if \tmod == 0 then {
        let \tile = tileC;
      };
      if \tmod == 1 then {
        let \tile = tileB;
      };
      if \tmod == 2 then {
        let \tile = tileA;
      };
      if \tmod == 3 then {
        let \tile = tileB;
      };
      if \tmod == 4 then {
        let \tile = tileC;
      };
      if \tmod == 5 then {
        let \tile = tileD;
      };
      if \tmod == 7 then {
        let \tile = tileE;
      };
      if \tmod == 8 then {
        let \tile = tileC;
      };
      if \t == 0 then {
        let \tile = seed;
      };
      if \tmod == 0 then {
        int \y;
        \y = \t / 9;
        \z = (0, \y);
      } else {
        if \tmod < 4 then {
          \z = (\tmod + 3 * \tdiv,0);
        } else {
          if \tmod < 6 then {
             \z = (3 + 3 * \tdiv,-\tmod + 3);
           } else {
             if \tmod < 8 then {
               \z = (3 * \tdiv - \tmod + 8,-2);
             } else {
               \z = (\tmod + 3 * \tdiv,-\tmod);
             };
           };
        };
      };
      if \tmod == 8 then {} else {
        if \tmod == 0 then {} else {
          {
            \filldraw[fill={gray!50}] (\z) rectangle +(1,1) node[pos=.5, font=\footnotesize] {\thuit};
            \pic["\thuit"] at (\z) {\tile=.5};
          };
        };
      };
    }
  }
\pic["0"] at (0,0) {seed=.5};
  \pic["8"] at (1,-1) {tileC=.5};

  \node at (25, 0.5) {$\ldots$};
  \node[left] at (0,0) {$\alpha = $};
\end{tikzpicture}

\begin{tikzpicture}[scale=.5]

  \tikzmath {
    int \t;
    for \t in {0,...,65} {
      coordinate \z;
      int \tmod;
      \tmod = mod(\t, 9);
      int \tdiv;
      \tdiv = \t / 9;
      int \thuit;
      if \t < 9 then {
        \thuit = \t;
      }
      else {
        if \t < 17 then {
          \thuit = \t - 1;
        } else {
          \thuit = \t - 2 * floor(\t / 9) + 1;
        };
      };
      let \tile = tileA;
      if \tmod == 0 then {
        let \tile = tileC;
      };
      if \tmod == 1 then {
        let \tile = tileB;
      };
      if \tmod == 2 then {
        let \tile = tileA;
      };
      if \tmod == 3 then {
        let \tile = tileB;
      };
      if \tmod == 4 then {
        let \tile = tileC;
      };
      if \tmod == 5 then {
        let \tile = tileD;
      };
      if \tmod == 7 then {
        let \tile = tileE;
      };
      if \tmod == 8 then {
        let \tile = tileC;
      };
      if \t == 0 then {
        let \tile = seed;
      };
      let \name = $.\thuit$;
      if \tmod == 0 then {
        int \y;
        \y = \t / 9;
        \z = (0, \y);
        if \t > 0 then {
          \y = \y - 1;
          if \y == 0 then {
            let \name = $1.0$;
          } else {
            let \name = $1.\y$;
          };
        };
      } else {
        if \tmod < 4 then {
          \z = (\tmod + 3 * \tdiv,0);
        } else {
          if \tmod < 6 then {
            \z = (3 + 3 * \tdiv,-\tmod + 3);
          } else {
            if \tmod < 8 then {
              \z = (3 * \tdiv - \tmod + 8,-2);
            } else {
              \z = (3 * \tdiv + 1,-1);
              if \t > 8 then {
                \y = \tdiv - 1;
                let \name = $2.\y$;
              };
            };
          };
        };
      };
      {
        \begin{scope}[shift={(\z)}]
          \pic["\name"] {\tile=.5};
        \end{scope}
      };
    };
  }
  \node[rotate=90] at (0.5, 9) {$\ldots$};
  \node at (25, 0.5) {$\ldots$};
  \node[left] at (0,0) {$\alpha'' = $};    
\end{tikzpicture}\\
    \caption{An assembly sequence $\alpha'$ of some temperature 1 aTAM, which yields a terminal production $F = \lim \alpha'$. Below a sequence $\alpha \moreFizzy \alpha'$ which yields a smaller production $P$ which is bounded in direction $\vec{d}$ (vertically). The assembly sequence $\alpha$ is fizzier since by step $50$, it just closed its seventh hole while $\alpha'$ is lagging at only 5 holes. Because $\alpha$ skips any unprofitable attachment, $\lim \alpha$ is very much not terminal: its seed (tile number $0$) as well as every tile attached at a time of the form $16 + 7k$ has an attachable yet unfilled position. It can thus be extended into an ordinal assembly sequence $\alpha''$ which does reach $\lim \alpha$, but in time $3 \omega$. The attachment times of the form $i.j$ in $\alpha''$ should be read as $i \omega + j$. At time $\omega$, $\alpha''$ has assembled $\lim \alpha$; at time $2 \omega$, it has added the upwards path, and at time $3 \omega$, it has added the last tile to each hole of $\lim \alpha$. }
    \label{fig:recalcitrant}
  \end{figure}

\subsection{The Window Movie Lemma}

The Window Movie Lemma~\cite{meunier_intrinsic_2014} applies to free assembly sequences as well as to $\mathbb{Z}^2$ assembly sequences. In order to account for holes, movies need to record not only the glues appearing on either side of the window, but also the paths between edges of the window through either side. 

\begin{definition}[Window, Frame, Movie]
  Let $\alpha = (A_0, A_1, \ldots)$ be an assembly sequence of $\FreeSeqs{\Ss}$. A \emph{window} $W$ of $\alpha$ is a cut of $\support[(\lim \alpha)]$ separating it into two connected components $\operatorname{Near}(W)$ and $\operatorname{Far}(W)$. 

  The width $w(W)$ of a window is the maximum perspective difference between two vertices along the window.

  Let $\operatorname{Arcs}(W)$ be the set of arcs (either normal or external) of $\support[(\lim \alpha)]$ through $W$.

  The \emph{frame} $f$ on $W$ at time $k$ records for each arc $a = (v, d) \in \operatorname{Arcs}(W)$:
  \begin{itemize}
  \item $\present[f(a)]$: whether $v \in \support[A_k]$, and if so,
  \item $\glue[f(a)] = \tiles[A_k](v)(d)$, as well as
  \item $\paths[f(a)]$ the  arc of $\operatorname{Arcs}(W)$ which is reachable from $a$ through an exterior path which does not cross $W$, if there is one.
  \end{itemize}
  
  The \emph{movie} associated with $W$ is the ordered set $\movie(W)$ of distinct frames appearing on $W$.
\end{definition}

%Note that the last item of a frame ($\cc[f]$) can be a proper subset of $\operatorname{Near}(W)$ or $\operatorname{Far}(W)$.

\begin{figure}
  \centering
  \tikzset{
  extPath/.style={-{Stealth[scale=.75]}}
}

\tikzset{
  doubleGlue/.pic={
    \begin{scope}[scale=.2]
      \fill[fill=black] (-.7, -1) rectangle (-.2, 0);
      \fill[fill=black] (.2, -1) rectangle (.7, 0);
    \end{scope}
  }
}

\tikzset{
  theTileA/.pic={
    \clip (0, 0) rectangle (1,1);
    \filldraw[fill=gray!25!white] (0, 0) rectangle (1, 1) node[anchor=center, pos=.5] {#1};
    \pic at (.5, 1) {doubleGlue};
    \pic[rotate=-90] at (1, .5) {doubleGlue};
  }
}

\tikzset{
  theTileB/.pic={
    \clip (0, 0) rectangle (1,1);
    \filldraw[fill=gray!25!white] (0, 0) rectangle (1, 1) node[anchor=center, pos=.5] {#1};
    \pic at (.5, 1) {doubleGlue};
    \pic[rotate=180] at (.5, 0) {doubleGlue};
  }
}

\tikzset{
  theTileC/.pic={
    \clip (0, 0) rectangle (1,1);
    \filldraw[fill=gray!25!white] (0, 0) rectangle (1, 1) node[anchor=center, pos=.5] {#1};
    \pic[rotate=-90] at (1, .5) {doubleGlue};
    \pic[rotate=90] at (0, .5) {doubleGlue};
  }
}

\tikzset{
  theTileD/.pic={
    \clip (0, 0) rectangle (1,1);
    \filldraw[fill=gray!25!white] (0, 0) rectangle (1, 1) node[anchor=center, pos=.5] {#1};
    \pic[rotate=-90] at (1, .5) {doubleGlue};
    \pic[rotate=180] at (.5, 0) {doubleGlue};
  }
}

\tikzset{
  theTileE/.pic={
    \clip (0, 0) rectangle (1,1);
    \filldraw[fill=gray!25!white] (0, 0) rectangle (1, 1) node[anchor=center, pos=.5] {#1};
    \pic[rotate=90] at (0, .5) {doubleGlue};
    \pic[rotate=180] at (.5, 0) {doubleGlue};
  }
}

\tikzset{
  theTileF/.pic={
    \clip (0, 0) rectangle (1,1);
    \filldraw[fill=gray!25!white] (0, 0) rectangle (1, 1) node[anchor=center, pos=.5] {#1};
    \pic at (.5, 1) {doubleGlue};
    \pic[rotate=90] at (0, .5) {doubleGlue};
  }
}

\begin{tikzpicture}
  \foreach \pos / \tile [count=\i from 0] in {(0,0)/A, (1,0)/C, (0,1)/B, (0,2)/B, (0,3)/D, (1,3)/C, (2,3)/E, (2,2)/B, (2,1)/B, (2,0)/F}
  {
    \begin{scope}[shift={(0,0)}]
      \pic at \pos {theTile\tile=\i};
    \end{scope}
    
    \begin{scope}[shift={(-4, -1.5)}]
      \foreach \frame in {\i, ..., 9} {
        \begin{scope}[shift={(4*.30*\frame, 0)}, transform shape, scale=.30]
          \pic at \pos {theTile\tile={}};
        \end{scope}
      }
    \end{scope}
  }
  \path[draw, red, ultra thick] (-.2, 2) node[left] {$W$}  -- (1.2, 2) (1.8, 2) -- (3.2, 2);
  \path[draw, red, dashed, -{Stealth[]}] (1.2, 2) to[bend left] node[above] {$(1, 0)$}(1.8, 2);

  \node[above left] at (-4.5, -1.5) {Assemblies};
  \node[left] at (-4.5, -2) {Frames};
  \node[left] at (-4.5, -3.5) {Movie};

  \tikzset{frame0/.pic={
      \fill[yellow!50!white] (-.05, -.25) rectangle (0.8, .25);
      \fill[yellow!50!white] (-.05, -.15) rectangle (0.8, .15);
      \path[draw, red, thick] (0, 0) -- (0.25, 0) (0.5, 0) -- (0.75, 0);
    }
  }

  \tikzset{frame1/.pic={\pic {frame0};}}
  
  \tikzset{frame2/.pic={
      \fill[yellow!50!white] (-.05, -.25) rectangle (0.8, .25);
      \pic[scale=.5] at (0.125,0) {doubleGlue};
      \draw[extPath] (0,0) to[out=-135, in=180] (.125, -.15) to [out=0, in=-45] (.25,0);
      \path[draw, red, thick] (0, 0) -- (0.25, 0) (0.5, 0) -- (0.75, 0);
    }}

  \tikzset{frame3/.pic={
      \fill[yellow!50!white] (-.05, -.25) rectangle (0.8, .25);
      \draw[extPath] (0,0) to[out=-135, in=180] (.125, -.15) to [out=0, in=-45] (.25,0);
      \pic[scale=.5] at (0.125, 0) {doubleGlue};
      \pic[scale=.5, rotate=180] at (0.125, 0) {doubleGlue};
      \draw[extPath, rotate around={180:(0.125,0)}] (0,0) to[out=-135, in=180] (.125, -.15) to [out=0, in=-45] (.25,0);
      \path[draw, red, thick] (0, 0) -- (0.25, 0) (0.5, 0) -- (0.75, 0);
    }}

  \tikzset{frame4/.pic={\pic {frame3};}}
  \tikzset{frame5/.pic={\pic {frame3};}}
  \tikzset{frame6/.pic={\pic {frame3};}}

  \tikzset{frame7/.pic={
      \fill[yellow!50!white] (-.05, -.25) rectangle (0.8, .25);
    \pic[scale=.5] at (0.125, 0) {doubleGlue};
    
    \draw[extPath] (0.75, 0) to[out=90, in=90] (0, 0);
    \pic [scale=.5,rotate=180] at (0.125, 0) {doubleGlue};
    \draw[extPath] (0,0) to[out=-135, in=180] (.125, -.15) to [out=0, in=-45] (.25,0);
    \draw[extPath] (.25, 0) to [out=45, in=135] (.5, 0);
    \pic [scale=.5,rotate=180] at (0.625, 0) {doubleGlue};
    \path[draw, red, thick] (0, 0) -- (0.25, 0) (0.5, 0) -- (0.75, 0);
  }}

\tikzset{frame8/.pic={
    \fill[yellow!50!white] (-.05, -.25) rectangle (0.8, .25);

    \pic[scale=.5] at (0.125, 0) {doubleGlue};
    \pic [scale=.5,rotate=180] at (0.125, 0) {doubleGlue};
    \pic[scale=.5] at (0.625, 0) {doubleGlue};
    \pic [scale=.5,rotate=180] at (0.625, 0) {doubleGlue};
    \path[draw, red, thick] (0, 0) -- (0.25, 0) (0.5, 0) -- (0.75, 0);

    \draw[extPath] (0.75, 0) to[out=90, in=90] (0, 0);
    \draw[extPath] (0,0) to[out=-135, in=180] +(.125, -.15) to [out=0, in=-45] (.25,0);
    \draw[extPath] (.25, 0) to [out=45, in=135] (.5, 0);
    \draw[extPath] (0.5, 0) to[out=-135, in=180] +(.125, -.15) to [out=0, in=-45] (0.75,0);
  }}
  
  \tikzset{frame9/.pic={
    \fill[yellow!50!white] (-.05, -.25) rectangle (0.8, .25);
    \pic[scale=.5] at (0.125, 0) {doubleGlue};
    \pic [scale=.5,rotate=180] at (0.125, 0) {doubleGlue};
    \pic[scale=.5] at (0.625, 0) {doubleGlue};
    \pic [scale=.5,rotate=180] at (0.625, 0) {doubleGlue};
    \path[draw, red, thick] (0, 0) -- (0.25, 0) (0.5, 0) -- (0.75, 0);

    \draw[extPath] (0.75, 0) to[out=90, in=90] (0, 0);
    \draw[extPath] (.25, 0) to [out=45, in=135] (.5, 0);
    \begin{scope}[rotate around={180:(.375,0)}]
      \draw[extPath] (0.75, 0) to[out=90, in=90] (0, 0);
      \draw[extPath] (.25, 0) to [out=45, in=135] (.5, 0);
  \end{scope}
  }}
  
  \begin{scope}[shift={(-4,-2)}]
    \foreach \i in {0, ..., 9} \pic[shift={(4*.30*\i,0)}] {frame\i};
  \end{scope}

  \begin{scope}[shift={(-3.9,-3.5)}]
    \fill[black] (-0.3,-1) rectangle (11.8,1);

    \begin{scope}[shift={(-.35,0)}]
      \foreach \x in {0, .4, ..., 12} {
        \fill[fill=white] (\x, -.65) rectangle +(.2, -.2);
        \fill[fill=white] (\x, .85) rectangle +(.2, -.2);
      }
    \end{scope}
    \foreach \i [count=\c from 0] in {0, 2, 3, 7, 8, 9}
    \pic[shift={(2*\c,0)}, scale=2] {frame\i};
  \end{scope}
\end{tikzpicture}

%%% Local Variables:
%%% mode: latex
%%% TeX-master: "../main"
%%% TeX-command-extra-options: "-shell-escape"
%%% TeX-engine: luatex
%%% End:
  \caption{Example assembly sequence $\alpha$ and window $W$. On the main picture, the labels of the tiles of $\lim \alpha$ are the order in which they attach. The small pictures are the successive assemblies of $\alpha$, their associated frames and the obtained movie. The movie is made up of the sequence of the unique frames, in order of apparition.}
  \label{fig:glue_movie_example}
\end{figure}

\cref{fig:glue_movie_example} gives an example free assembly sequence $\alpha$ with a window $W$. The arrows between the edges on either side of $W$ represent the relation ``$\paths$''.

\begin{lemma}[Window Movie Lemma]
  \label{lem:wml}
  Let $\alpha \in \FreeSeqs{\Ss, \sigma_A}$ and $\beta \in \FreeSeqs{\Ss, \sigma_B}$ two assembly sequences. Assume there are two windows $A$ in $\alpha$ and $B$ in $\beta$ and a translation $\tau$ such that $\tau(A) = B$, $\movie(A)$ is the same as $\movie(B)$ up to translation by $\tau$, and lastly $\tau$ can be extended to a translation from $\far(A)$ to $\far(B)$ which maps $\sigma_B \cap \operatorname{Far}(B)$ to $\sigma_A \cap \operatorname{Far}(A)$.

  Let $\alpha^\dagger = \alpha_{|\far(A)} \in \FreeSeqs{\Ss, \lim \alpha \cap \near(A)}$ be the sequence of attachments of $\alpha$ within $\far(A)$, and likewise $\beta^\dagger = \alpha_{|\far(B)} \in \FreeSeqs{\Ss, \lim \beta \cap \near(B)}$ be the sequence of attachments of $\beta$ witihn $\far(B)$.

  Let $\alpha'$ be the candidate assembly sequence consisting of $\alpha$ where for every $k$, the attachment $\alpha^\dagger_k$ is replaced with $\beta^\dagger_k$.
  
  Then there is a window $A'$ on $\alpha'$ and two translations $\tau_N: \operatorname{Near}(A') \to \operatorname{Near}(A)$ and $\tau_F: \operatorname{Far}(A') \to \operatorname{Far}(B)$ such that:
    \begin{equation}
    \begin{cases}
      \forall z  \in \operatorname{Near}(A'),  &\tiles[(\lim \alpha')](z) = \tiles[(\lim \alpha)](\tau_N(z))\\
      \forall z  \in \operatorname{Far}(A'), &\tiles[(\lim \alpha')](z) = \tiles[(\lim \beta)](\tau_F(z))
    \end{cases}\label{eq:transl}
  \end{equation}

  Moreover, if $\beta^\dagger \moreFizzy \alpha^\dagger$, then $\alpha' \moreFizzy \alpha$.
\end{lemma}

\begin{figure}
  \centering
  \input{tikz/tree_pump.tex}
  \caption{The Window Movie Lemma: consider two assembly sequences $\iota$ (the ibex) and $\beta$ (the bunny), with their productions depicted in the top row. The sequence $\iota$, has two windows $W_t$ and $W_h$. In $W_t$, $\far(W_t)$ is the head of the ibex, $\near(W_t)$ its body. In $W_h$, $\far(W_h)$ is the tip of the horn. Each of them is associated with a translation, $\tau_t$ and $\tau_h$ respectively, which sends the window to another window with the same movie, and satisfying the hypothesis of~\cref{lem:wml}. Applying \cref{lem:wml} in each case gives two new assembly sequences represented on the bottom row. Applying it with $W_t$ yields a fearsome chimera, while its application to $W_h$ yields an ibex with a horn so long it needs to bend in the third dimension to avoid piercing its spine, i.e. avoiding the conflict that would arise were our attachment sequences not free.}
  \label{fig:tpl}
\end{figure}

\begin{proof}
  Define $\alpha'$ as $\alpha$ where for each $k$, the $k$-th attachment $t@Z$ within $\far(A)$ has been replaced by the $k$-th attachment within $\far(B)$ of $\beta$. When doing this, for an attachment $t@Z$ within $\far(B)$, any arc $a = (v, d)$ in $Z$ with $v \in \near(B)$ is replaced by the arc $\tau^{-1}(a)$.

  The translations $\tau_N$ and $\tau_F$ can be defined inductively on the attachments of $\alpha'$, such that \cref{eq:transl} holds.
  
  The proof that $\alpha'$ is a valid free assembly sequence is the same as in the $\mathbb{Z}^2$ case. Note that since $B$ is a cut of $\support[(\lim \beta)]$, any attachment of $\beta$ in $\far(B)$ can be replayed in $\alpha'$ in $\far(A')$ without conflicts since they do not create any adjacency outside $\operatorname{Far}(\beta)$.

  Observe that every attachment in $\operatorname{Far}(B)$ which affects holes in $\beta$ affects the same number of holes in $\alpha'$: holes which are wholly within $\operatorname{Far}(B)$ have had all of their tiles attached in $\alpha'$, and holes which go through $B$ become holes through $A$ because the part within $\operatorname{Near}(B)$ has the same connections in $\operatorname{Near}(A)$ at that frame in the movie.
\end{proof}

The fact that the grafting process of \cref{lem:wml} is increasing in the fizziness of the far part of the assembly sequences implies that when the original assembly sequences have maximal fizziness, so does the chimera sequence $\alpha'$.

% \begin{corollary}
%   Let $X$ be a set of assembly sequences, suppose $\alpha$ and $\beta$ satisfy the hypothesis of \cref{lem:wml} and are of maximal fizziness within $X$.

%   If the sequences $\alpha'$ and $\beta'$ obtained by applying \cref{lem:wml} in both directions are also in $X$, they are of maximal fizziness within $X$ too.
% \end{corollary}\ruggedtodo{proof?}

\cref{lem:wml} will be most useful in this paper in the case where the cuts $A$ and $B$ are on the same branch of the assembly. In this case, by iterating \cref{lem:wml}, it is possible to get a production with a periodic subassembly.

\begin{corollary}
  \label{lem:celeri_branche}
  Let $\alpha = (A_i)_{i < o}$ be an assembly sequence with two windows $A$ and $B$ satisfying the hypothesis of \cref{lem:wml} and such that $\far(B) \subset \far(A)$. Let $\tau$ be the translation between $A$ and $B$. Then, there is an assembly sequence $\alpha^{\tau}$ such that $F = \lim \alpha^{\tau} \cap \far(A)$ verifies $\tau(F) \subset F$, and within $\near(B)$, $\alpha^{\tau}$ does the same attachments in the same order as $\alpha$.
\end{corollary}

\begin{proof}
  Let $\tau$ be the translation sending $A$ to $B$.
  
  Show by induction that for any $k$, there is a sequence $\alpha^k$ such that the movie on the windows $A$ and $B$ within $\alpha^k$ are the same as in $\alpha$, and for each $z \in \near(B) \cap \far(A)$ and $j \leq k$, $(\lim \alpha^k)(\tau^j(z)) = (\lim \alpha)(z)$, and the attachments done by $\alpha^k$ and $\alpha^j$ within $\bigcup_{i \leq j} \tau^i(\near(B) \cap \far(A))$ are the same.

  For $k = 0$, it suffices to pick $\alpha^0 = \alpha$.
  Assume that $\alpha^k$ satisfies the induction hypothesis. Then \cref{lem:wml} applies, and the assembly sequence it yields, $\alpha^{k+1}$, satisfies the induction hypothesis at rank $k+1$.

  For a pair $(j, k) \in \mathbb{N}$, if for all $i \leq j$, $\alpha^k_i \in \bigcup_{l \leq k} \tau^k(\near(B) \cap \far(A))$, then the $j$ first attachments are unchanging after rank $k$: for all $m \geq k$ and $i \leq j$, $\alpha^m_i = \alpha^k_i$. Let $\alpha^{\vec{p}}$ be the sequence of such unchanging attachments. Let $m$ be the supremum of the $j \in \mathbb{N}$ such that the $j$ first attachments are unchanging after some rank $k$. Define $\alpha^{\vec{p}}$ as the sequence of length $m$ with $\alpha^{\vec{p}}_j = \alpha^k_j$, where $k$ is such that the $j$ first attachments are unchanging.

  It is easy to check that indeed, $\lim \alpha^{\vec{p}} \cap \far(A)$ has period $\vec{p}$: any attachment in $\alpha^{\vec{p}}$ within $\far(A)$ gets picked up by subsequent applications of \cref{lem:wml} which translate it by $\vec{p}$.
\end{proof}

In this situation with a branch and two cuts with the same movie, it is also possible to cut assembly sequences short, so that they can do their business in $\near(A)$ without needing to mess with $\far(B)$.

\begin{corollary}
  \label{lem:couic-couic}
  Let $\alpha$ be an assembly sequence with two windows $A$ and $B$ satisfying the hypothesis of \cref{lem:wml} and such that $\far(B) \subset \far(A)$. Then there is an assembly sequence $\alpha'$ which does the same attachments as $\alpha$ within $\near(A)$, but has no attachment in $\far(B)$.
\end{corollary}

\begin{proof}
  Let $j$ be the last time that $\alpha$ does an attachment in $\near(A)$ next to $A$, i.e. the date of the last frame in the movie of $A$ where a tile is attached on the near side of the window. By \cref{lem:pos-finite-time}, up to a reordering of $\alpha$, $j$ is finite.

    Let $\alpha^0$ be the prefix of length $j$ of $\alpha$. For each $k$, if $\alpha^k$ has any attachment in $\far(B)$, it is possible to use \cref{lem:wml} between $B$ and $A$ to obtain a sequence $\alpha^{k+1}$ with fewer attachments within $\far(B)$. Hence, there is a finite $k$ such that $\alpha^k$ has no attachments in $\far(B)$, and $\lim \alpha^k \cap \near(A) = \lim \alpha^0 \cap \near(A)$. Since by time $j$, the movie on $A$ is over, the rest of the attachments of $\alpha$ which take place in $\near(A)$ can be replayed after $\alpha^k$.
\end{proof}

\subsection{The Tree Pump Theorem}
\label{sec:tree_pump_tree_pump}

After all these considerations about the fantastic properties of Ordinal Free Assembly Sequences, it is now time to come back to Earth, or rather $\mathbb{Z}^2$. The object is now to prove \cref{thm:tree_pump}, for which an investigation of the properties of systems whose $\mathbb{Z}^2$-assembly sequences do not circle large squares is in order.

The statement of \cref{thm:tree_pump} is given again, recall that it deals with the $\mathbb{Z}^2$-productions of the aTAM system $\Ss$, hence in its statement, $\TerminalProds{\Ss}$, $C_m[\Ss]$, $B_{F(n,m),\vec{d}}[\Ss]$ and $P_{\vec{d}}[\Ss]$ are sets of $\mathbb{Z}^2$-assemblies.

\treepump*

When considering the statement of \cref{thm:tree_pump}, the holes of the $\mathbb{Z}^2$-productions of $\Ss$ might as well be filled in. Hence, the definition of their fill-in, illustrated on \cref{fig:treedec}.

\begin{definition}[Fill-in]
  For a subgraph $D \subset \mathbb{Z}^2$, define the \emph{fill-in} $\complete{D}$ as the subgraph of $\mathbb{Z}^2$ induced by the positions $p$ such that there is no infinite path from $p$ in $\mathbb{Z}^2 \setminus D$.
\end{definition}

A $\mathbb{Z}^2$ assembly $A$ which does not circle any square larger than $m \times m$ looks like a tree, as expressed by the notion of \emph{Connected Treewidth}~\cite{DBLP:journals/combinatorica/DiestelM18}. This tree is obtained by grouping the positions of $\support[A]$ in connected sets of size at most $2m$, known as \emph{bags}, organized in a tree in such a way that:
\begin{itemize}
\item any arc of $\support[A]$ is between two positions which appear in the same bag, and
\item for any position $z \in \support[A]$, the set of bags which contain $z$ forms a subtree.
\end{itemize}

This decomposition is represented on \cref{fig:treedec}.

\begin{figure}
  \centering
  \input{tikz/treedec.tex}
  \caption{The domain $D$ of a production $P$, its fill-in $\complete{D}$ and an associated connected tree decomposition of $\treedec{D}$. The blue part of each bag $b$ of $\treedec{D}$ is its intersection $W_b$ with its parent, which disconnects the vertices appearing in its subtree from the rest of $\complete{D}$. The choice of the root of $\treedec{D}$ is arbitrary and does not affect the sets $W_b$, up to renaming.}
  \label{fig:treedec}
\end{figure}

\begin{lemma}
  \label{lem:treedec}
  Let $\Ss$ an aTAM system, $m$ an integer and $A$ a $\mathbb{Z}^2$ assembly of $\Ss$ such that $A \notin C_m[\Ss]$.

  Then $\support[A]$ has Connected Tree Width $2m$.
\end{lemma}

\begin{proof}
Indeed, it has treewidth $m$: otherwise it would contain an $m \times m$ grid as a minor; that minor would have to be realized in $\mathbb{Z}^2$ as a subgraph which would encircle some $m \times m$ square. Moreover, since $\complete{P}$ does not encircle any empty position, it does not have any \emph{geodesic cycle} of length more than $4$.
\end{proof}

The point of using Connected Treewidth rather than the usual treewidth is that thanks to the connectivity of the bags of $\treedec{P}$, the distances in $\treedec{P}$ reflect those in $\complete{P}$:
\begin{itemize}
\item there is a function $r$ such that for any vertex $v \in \complete{P}$, the subtree of bags of $\treedec{P}$ containing $v$ has a size at most $r(m)$,
\item there are two increasing functions $d_m, D_m$ such that for any vertices $u, v$ at distance $\delta$ in $\complete{P}$, any bags $B_u \ni u$ and $B_v \ni v$ of $\treedec{P}$ are at distance $\Delta$ with $d_m(\delta) \leq \Delta \leq D_m(\delta)$.    
\end{itemize}

As a consequence, if two tiles are far enough in an assembly $A$ and the path between them does not go through the seed, there must be two windows cutting that path which satisfy the hypothesis of \cref{lem:celeri_branche}.

\begin{lemma}
  Let $\Ss$ be an aTAM system with $n$ tiles and $m$ an integer.
  
  Let $\alpha = (A_i)_{i \leq o} \in \FreeSeqs{\Ss, A_0}$ be such that for all $i \leq o$, $\ezEmbed(A_i) \notin C_m[\Ss]$.

  There is a constant $F(n, m)$ such that if $z, z' \in \support[\ezEmbed(\lim \alpha)] \setminus \support[A_0]$ are such that the distance between $z'$ and  $z$ is greater than $F(n, m)$, then there are two windows $A$ and $B$ with $z \in \near(A)$ and $z' \in \far(B)$ which satisfy the hypothesis of \cref{lem:wml}.
\end{lemma}
  
\begin{proof}
  let $W(n, m)$ be the number of movies with $n$ glues on a window of width $2m$. Note that $W(n, m)$ counts the number of arrangements of the edges within the window, the perspective difference between the connected components of the window, as well as the events on the frames of the movie. Let $F(n, m) = d^{-1}_m(W(n,m))$.
  
  Let $P = \support[\lim \alpha]$. By \cref{lem:treedec}, pick a tree decomposition $\treedec{P}$ with connected bags of size at most $2m$. Let $\delta$ be the distance between $z$ and $z'$. If $\delta \leq F(n, m)$, by the pigeonhole principle, there must be two windows $A$ and $B$ between $z$ and $z'$ with the same movie.
\end{proof}

In the case where the assembly embeds into $\mathbb{Z}^2$, one actually controls the direction of the vector between the two windows.

\begin{lemma}
  \label{lem:useless}
  Let $\Ss$ be an aTAM system with $n$ tiles, $m$ an integer and $\vec{d}$ a unit vector.
  
  Let $\alpha = (A_i)_{i \leq o} \in \FreeSeqs{\Ss, A_0}$ be such that for all $i \leq o$, $A_i$ embeds into $\mathbb{Z}^2$.

  There is a constant $F'(n, m)$ such that if $z, z' \in \support[\lim \alpha] \setminus \support[A_0]$ are such that the distance between $(z' - z) \cdot \vec{d} > F(n, m)$, then there are two windows $A$ and $B$ with $z \in \near(A)$ and $z' \in \far(B)$ which satisfy the hypothesis of \cref{lem:wml}, and such that the vector $\vec{p}$ of the translation between $A$ and $B$ satisfies $\vec{p} \cdot \vec{d} > 1$.
\end{lemma}

\begin{proof}
  Pick $F'(n, m) = 2 F(n,m) + 1$. If $(z' - z) \cdot \vec{d} > F'(n, m)$, then in $\treedec{\support[\lim \alpha]}$ there there are two bags, one $b$ containing $z$, the other $b' \ni z'$ such that on the path between $b_z$ and $b_{z'}$, there are $2 F(n,m) + 1$ bags $(b_i)_i$ such that for all $x \in b_i, y \in b_j, (x - y) \cdot \vec{d} \geq i - j$. Thus, there are $i, j$ such $j > i + 1$ and two windows, $A$ between $b_i$ and $b_{i+1}$, and $B$ between $b_j$ and $b_{j+1}$ which have the same movie. Because the windows $A$ and $B$ are separated by at least two elements of $(b_i)$, the translation vector $\vec{p}$ between them satisfies $\vec{p} \cdot \vec{d} > 1$.
\end{proof}

Thus, there is a simple argument to prove \cref{thm:tree_pump}: pick a sequence of maximal fizziness which produces a terminal assembly of $\Ss$ which is neither in $B_{F'(n,m),\vec{d}}[\Ss]$ nor in $C_m[\Ss]$. \cref{lem:useless} yields two windows with the same movie; by applying \cref{lem:celeri_branche} between them, a production in $P_{\vec{d}}[\Ss]$ appears. Alas, the sequence on which  this argument is founded may simply fail to exist: the $\omega$-sequences of maximal fizziness may fail to reach a terminal production. One could extend them in order to reach a terminal production, but there may not be a sequence of maximal fizziness among these (ordinal) extensions.

Thus, it is necessary to prevent the pesky tendency ordinal sequences have to try and buy time by spacing out their attachment so as to generate an infinite family of sequences with increasing fizziness. \emph{Straight} sequences are the answer: they do not have the wiggle room to misbehave.

\begin{definition}
  Let $\Ss$ be an aTAM system, and $\prec$ an arbitrary total order on its sequences. A free, ordinal assembly sequence $\alpha = (A_i)_{i < o} \in \FreeSeqs{\Ss}$ is $\prec$-\emph{straight} if for every pair of window $N, F$ such that $N$ and $F$ have the same movie and $\support[A_0] \subset \near(N)$, $\alpha$ is the smallest sequence for $\prec$ obtained by applying \cref{lem:celeri_branche} to $N$ and $F$ in $\alpha$.

  The order $\prec$ is henceforth fixed and left implicit.
\end{definition}

\begin{lemma}
  \label{lem:finite_choice}
  Let $\calT = (S, \sigma, \tau)$ be an aTAM system and $w$ an integer.

  Let $X \subset \FreeSeqs{\calT}$ be a set of assembly sequences, and $d$ an integer.
  If for any sequence $\alpha \in X$ and any position $z \in \support[{ \lim \alpha }]$ at distance more than $d$ from $\sigma$, there are two windows $A$ and $B$ such that $\support[ \sigma ] \subset \near(A)$ and $z \in \far(B)$, then there is a finite number of straight sequences in $X$.
\end{lemma}

\begin{proof}
  In this case, each straight sequence in $X$ is determined by what it does in a radius $d$ around $\sigma$.
\end{proof}

\begin{lemma}
  let $\calT = (S, \sigma, \tau)$ be an aTAM system and $\alpha$ a straight assembly sequence of $\Ss$. There is a straight assembly sequence $\alpha'$ which extends $\alpha$ such that $\lim \alpha' \in \TerminalProds{\calT}$.
\end{lemma}

\begin{proof}
  Let $P = \lim \alpha$. If $P \notin \TerminalProds{\calT}$, then there is an attachment $t@Z$ which is possible in $P$.

  Let $\alpha'$ be $\alpha$ with $t@Z$ appended at its end, and $z$ the position of that last attachment in $\support[{ \lim \alpha' }]$. Assume that $Z$ is such that the distance between $\support[ \sigma ]$ and $z$ is minimal.

  If there are no pairs of windows $(N, F)$ in $\alpha'$ such that $\far(F) \subset \far(N)$, $\movie(N) = \movie(F)$,  $\support[ \sigma ] \subset \near(N)$ and $z \in \far(N)$, then $\alpha'$ is straight.

  If there are two windows  $(N, F)$ in $\alpha'$ such that $\far(F) \subset \far(N)$, $\movie(N) = \movie(F)$,  $\support[ \sigma ] \subset \near(N)$ and $z \in \far(F)$, then by \cref{lem:couic-couic}, there is a place closer to $ \sigma $ than $z$ where an attachment was possible.

  Lastly, if there is a pair of windows  $(N, F)$ in $\alpha'$ such that $\far(F) \subset \far(N)$, $\movie(N) = \movie(F)$,  $\support[ \sigma ] \subset \near(N)$ and $z \in \far(N) \cap \near(F)$, then the sequence $\alpha''$ obtained by applying \cref{lem:celeri_branche} in $\alpha'$ is straight and has $\alpha$ as a prefix, since the new attachments (the repetitions of $t@Z$) are done last in each repetition of the movie in $\far(N) \cap \near(F)$.

  This process yields a straight assembly sequence $\alpha'$ which strictly extends $\alpha$. It can be repeated until $\lim \alpha' \in \TerminalProds{\Ss}$.
\end{proof}

\begin{lemma}
  \label{lem:crash_recovery}
  Let $\calT = (S, \sigma, \tau)$ be an aTAM system and $\alpha$ a straight assembly sequence of $\Ss$. Then $\ezEmbed(\alpha)$ is a prefix of a straight sequence $\alpha'$. Moreover, the connected treewidth of $\support[ {\lim \alpha'} ]$ is no greater than that of  $\support[{ (\lim \ezEmbed(\alpha)) }]$.
\end{lemma}

\begin{proof}
  If $\alpha$ embeds cleanly in $\mathbb{Z}^2$, there is nothing to prove since $\ezEmbed(\alpha) = \alpha$.

  Otherwise, $\ezEmbed(\alpha)$ stops because one of its attachment $t@z$ creates a hole which does not exist in $\alpha$. Because $\alpha$ is straight, there cannot be a pair of windows with the same movie separating $\sigma$ and $z$. Applying \cref{lem:celeri_branche} in $\ezEmbed(\alpha)$ on each pair of windows with the same movie which are the closest to $\sigma$ yields a straight sequence $\alpha'$ extending $\ezEmbed(\alpha)$.

  Fix a tree decomposition of $\support[(\lim \ezEmbed(\alpha))]$. Because $\alpha$ is straight, among all tree decompositions of $\support[ {\lim \alpha'} ]$, there is one where each bag which lies in the far side of a pair of windows is a translation of some bag close to the seed. This tree decomposition has the same width as the decomposition of $\support[(\lim \ezEmbed(\alpha))]$.
\end{proof}

At last, all the ingredients are there to prove \cref{thm:tree_pump}.

\begin{proof}[Proof of \cref{thm:tree_pump}]
  Define a sequence $(\alpha_i)$ of straight assembly sequences as follows:
  \begin{itemize}
  \item Fix $\alpha_0$ to be the assembly sequence with zero attachments: $\alpha_0 = (\sigma)$;
  \item for $i$ even, $\alpha_{i+1}$ is a straight extension of $\alpha_i$ which reaches a terminal production;
  \item for $i$ odd, $\alpha_{i+1}$ is obtained from $\alpha_i$ by \cref{lem:crash_recovery}.
  \end{itemize}

  For every $i$, either $\alpha_{i+1} = \alpha_i$ or $\alpha_{i+1} \moreFizzy \alpha_i$: at the even steps, if $\alpha_{i+1} \neq \alpha_i$, then it is a proper extension and is thus more fizzy; at the odd steps, if $\alpha_{i}$ does not embed into $\mathbb{Z}^2$, then $\ezEmbed(\alpha_i)$ is more fizzy, and so is its extension $\alpha_{i+1}$.

  If $\TerminalProds{\Ss} \cap C_m[\Ss] = \emptyset$, then for each $i$ odd, the Connected Tree-width of $\ezEmbed(\alpha_i)$ is at most $2m$, hence the Connected Tree-width of $\alpha_{i+1}$ is at most $2m$. But then, by \cref{lem:finite_choice}, $\alpha_{i}$ can only take a finite number of different values for $i$ even.

  Thus, the sequence $(\alpha_i)$ is eventually stationary, and its fixed point $\beta$ is a straight sequence which embeds into $\mathbb{Z}^2$ and reaches a terminal production. If $\TerminalProds{\Ss} \cap B_{F(n,m), \vec{d}} \neq \emptyset$, this terminal production must reach at least $F(n, m)$ in direction $\vec{d}$. Thus $\support[ {\lim \beta} ]$ has a branch with two windows $A, B$ such that $\movie(A) = \movie(B)$ and the vector sending $A$ to $B$ verifies $\vec{p} \cdot \vec{d} > 1$ (by \cref{lem:useless}). Because $\beta$ is straight, that branch is periodic, thus $\lim \beta \in \TerminalProds{\Ss} \cap P_{\vec{d}}$.
\end{proof}

\section{From an aTAM Quine to a Self-Assembled Discrete Self-Similar Fractal}\label{sec:DSSFs}

In this section, we present our first fractal construction, which is an aTAM system that performs an infinite series of nested simulations of itself, at greater and greater scale factors. Importantly, unlike previous IU constructions (e.g. \cite{IUSA,DirectedNotIU,DDDIU}), this construction begins with a seed consisting of a single tile. That tile first grows into an $m \times m$ macrotile whose sides have a full definition of the system and the glues of the seed tile on its output sides, using the construction of the proof for Theorem \ref{thm:quine}. Then, since that portion of the system has been designed so that it meets the requirements of a standard TAS, the tile set that is IU for the class of standard aTAM systems treats that macrotile as a scaled version of the seed tile, and simulates the growth of the original $m \times m$ macrotile but with each tile represented by an $m \times m$ macrotiles, resulting in an $m^2 \times m^2$ macrotile. By combining the tilesets of the two results, we are able to cause this process of simulation at increasing scale factor to happen for an infinite series of scale factors.

Then, to cause the resulting assembly to be a discrete self-similar fractal with $\zeta$-dimension $< 2$, we make slight modifications to the previous IU and quine constructions so that the macrotiles contain a specified amount of empty space. Then, the repeated simulation at greater and greater scales yields a DSSF.

% \begin{theorem}\label{thm:dssf}
%     For $z, \epsilon \in \mathbb{Q}$, where $1 < z < 2$ and $\epsilon < 1$, there exists an aTAM system $\calT_{z,\epsilon}$ such that $\mathcal{F}_{z,\epsilon}$ strictly self-assembles a DSSF whose $\zeta$-dimension is $z \pm \epsilon$.
% \end{theorem}

We prove Theorem \ref{thm:dssf} by construction. Given an arbitrary  $z, \epsilon \in \mathbb{Q}$, where $1 < z < 2$ and $\epsilon < 1$, we demonstrate an aTAM system $\mathcal{F}_{z,\epsilon} = (T,\sigma,2)$ that strictly self-assembles a DSSF whose $\zeta$-dimension is $z \pm \epsilon$. Our construction utilizes the constructions from the proofs of Theorems \ref{thm:standardIU} and \ref{thm:quine} with slight modifications. Those modifications and the proofs of their correctness, and thus the overall correctness of the proof of Theorem \ref{thm:dssf}, are included in subsections dedicated to each of the following:
\begin{enumerate}
    \item Modification to allow simulation to occur at greater and greater scales instead of stopping after first level.

    \item Proof that simulation is correct at each level.

    \item Modifications to the quine macrotile and the IU system's macrotiles to add spacing for targeted fractal dimensions.

    \item Derivation of the resulting fractal dimension.
\end{enumerate}

\subsection{Nested self-simulation at an infinitely increasing series of scales}

The aTAM quine $\mathcal{Q} = (Q,\sigma,2)$ presented in the proof of Theorem \ref{thm:quine} creates a finite terminal assembly, which we'll call $\alpha_Q$.
As it is a quine with respect to the tileset $U$ (the tileset that is IU for the class of standard aTAM systems) and its representation and seed generation functions $R$ and $S$, $\alpha_Q$ is a macrotile mapping to $\sigma$ under $R$ (and having the same shape and exterior glues as $S(\sigma)$).
This in turn means that, if $\alpha_Q$ is used as the seed for a system containing the tiles of $U$ (and $\tau=2$ since both $\mathcal{Q}$ and $U$ utilize $\tau=2$), which we'll call $\mathcal{Q}^+$ (i.e. $\mathcal{Q}^+ = (U,\alpha_Q,2)$), then $\mathcal{Q}^+$ would simulate $\mathcal{Q}$ under $R$ (at some scale factor $m$) and its terminal assembly, $\alpha_{Q^+}$, would be $\alpha_Q$ scaled by $m$ (as interpreted under $R$). Additionally, there would be two macrotile locations that map to empty space under $R$ but which contain fuzz (which does not compromise the correctness of the simulation).
Let $c$ be the dimensions of $\alpha_Q$, which is a square.%, and therefore $mc$ the dimensions of the $mc \times mc$ terminal assembly of $\mathcal{Q}^+$ that doesn't include the fuzz.
\begin{figure}
    \centering
    \includegraphics[width=0.8\textwidth]{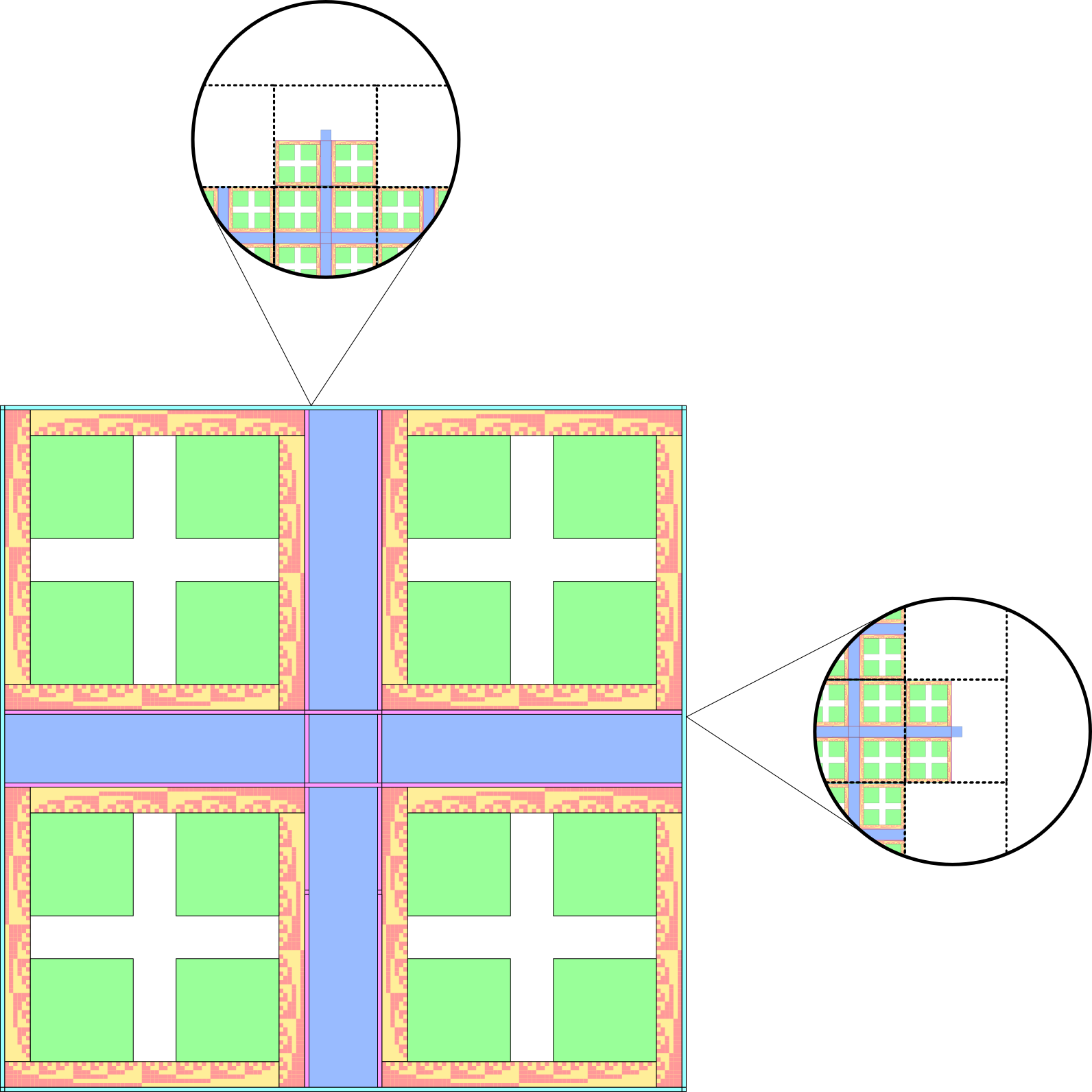}
    \caption{Without encoding the IU tileset in the glue table, $Q^+$ will grow up to
    the point where the quine has acted as the seed for a scaled copy of itself. Note that the large macrotile illustrated here is an order 2 structure made from macrotiles itself. In other words, $Q^+$ simulates $Q$ using macrotiles. The zoomed-in views on the north and east side depict the fuzz macrotile locations that will contain partial macrotiles. These correspond to the positions along the side of $Q$ where control is given to the IU tileset, however growth stops here because the glue tables on the tiles in $Q^+$ do not contain encodings for the individual tiles in the IU tileset $U$. These will not map to tiles under the representation function since the glue table failed to find a match, so $Q^+$ shows that $Q$ acts as a seed for itself.}
    \label{fig:enter-label}
\end{figure}
The macrotile locations in $\mathcal{Q}^+$ containing the fuzz would be those mapping to the locations along the counter-clockwise-most encoding of the strength-2 glue being presented by the simulated macrotile. This is purely a chosen convention though and there is no particular significance to this location over any other.
Thus, in the scaled simulation of $\mathcal{Q}^+$, macrotile growth would be initiated in those locations (because strength-2 glues are simulated there), but the growth by the tiles of $U$ that attempts to find a match in the glue lookup table would fail because the tiles of $U$ are not encoded in it.
This would result in terminal growth that is valid fuzz, mapping only to empty space.

In order to create $\mathcal{F}_{z,\epsilon}$ so that the simulation continues indefinitely at all scales, we simply modify the procedure used to create $\mathcal{Q}$ by including the tile types from $U$ in $T_F$ in the glue table. With the tile types of $U$ also encoded in the glue lookup table, the simulation will continue where $\mathcal{Q}^+$ terminated. The following section details that the relation of intrinsic simulation is transitive and, consequently, this will result in the increasingly higher order simulations of $Q$ at an infinite series of larger scales and thus a DSSF.

\subsection{Correctness of nested self-simulation at infinitely increasing scales}

Here we prove that having equivalent productions is transitive.

\begin{lemma}
    Let $\mathcal{S}$, $\mathcal{T}$, and $\mathcal{U}$ be TAS's satisfying $\mathcal{S}\Leftrightarrow_{R_1} \mathcal{T}$ and $\mathcal{T} \Leftrightarrow_{R_2} \mathcal{U}$, then $\mathcal{S}\Leftrightarrow_{R_2 \circ R_1} \mathcal{U}$.
\end{lemma}

\begin{proof}
    By hypothesis $\mathcal{A}[\mathcal{T}] = R_1^*(\mathcal{A}[\mathcal{S}])$ and $\mathcal{A}[\mathcal{U}] = R_2^*(\mathcal{A}[\mathcal{T}])$. Thus $R_2^*(R_1^*(\mathcal{A}[\mathcal{S}])) = R_2^*(\mathcal{A}[\mathcal{T}]) = \mathcal{A}[\mathcal{U}]$. The same holds true for the sets of terminal assemblies.

    It must also be the case that assemblies in $\calS$ map cleanly to those in $\calU$ since all fuzz locations in $\calT$-assemblies $R_2$-mapping to $\calU$-assemblies are fuzz locations in the $\calS$-assemblies $R_1$-mapping to the $\calT$-assemblies. 
\end{proof}

This result says that the follows relationship is transitive
\begin{lemma}
    If $\mathcal{S}$, $\mathcal{T}$, and $\mathcal{U}$ are TAS's satisfying $\mathcal{T}\dashv_{R_1} \mathcal{S}$ and $\mathcal{U} \dashv_{R_2} \mathcal{T}$, then $\mathcal{U}\dashv_{R_2 \circ R_1} \mathcal{S}$.
\end{lemma}

\begin{proof}
    Let $\alpha'', \beta'' \in \mathcal{A}[\mathcal{S}]$ such that $\alpha'' \to^{\mathcal{S}} \beta''$. We know that $R_1^*(\alpha'') \to^{\mathcal{T}} R_1^*(\beta'')$ by hypothesis. Furthermore, let $\alpha' = R_1^*(\alpha'')$ and let $\beta' = R_1^*(\beta'')$. By hypothesis, since $\alpha' \to^{\mathcal{T}} \beta'$, it must be the case that $R_2^*(\alpha') \to^{\mathcal{U}} R_2^*(\beta')$. Consequently, $R_2^*(R_1^*(\alpha'')) \to^{\mathcal{U}} R_2^*(R_1^*(\beta''))$ and the lemma is proved.
\end{proof}

\begin{lemma}
    If $\mathcal{S}$, $\mathcal{T}$, and $\mathcal{U}$ are TAS's satisfying $\mathcal{S}\vDash_{R_1} \mathcal{T}$ and $\mathcal{T} \vDash_{R_2} \mathcal{U}$, then $\mathcal{S}\vDash_{R_2 \circ R_1} \mathcal{U}$.
\end{lemma}

\begin{proof}
    Let $\alpha \in \mathcal{A}[\mathcal{U}]$ and let $\Pi^{\mathcal{T}}_\alpha \subset \mathcal{A}[\mathcal{T}]$ be the stem set of $\alpha$ under $R_2$. For each element $\alpha' \in \Pi^{\mathcal{T}}_\alpha$, let $\Pi_{\alpha'}^{\mathcal{S}}$ be the corresponding stem set of $\alpha'$ under $R_1$ in $\prodasm{\mathcal{S}}$. Now we define $\Pi^{\mathcal{S}}_\alpha = \bigcup_{\alpha' \in \Pi^{\mathcal{T}}_\alpha} \Pi_{\alpha'}^{\mathcal{S}}$ to be the union of these stem sets. We will now show that $\Pi^{\mathcal{S}}_\alpha$ is the stem set of $\alpha$ under $R_2\circ R_1$.

    By definition, if $\alpha'' \in \Pi^{\mathcal{S}}_\alpha$, then $\alpha'' \in \Pi_{\alpha'}^{\mathcal{S}}$ for some $\alpha' \in \Pi^{\mathcal{T}}_\alpha$. Consequently, $R_1^*(\alpha'') \in \Pi^{\mathcal{T}}_\alpha$ and thus $R_2^*(R_1^*(\alpha'')) = \alpha$. Now, let $\beta \in \prodasm{\mathcal{U}}$ such that $\alpha \rightarrow^{\mathcal{U}} \beta$. We will show that the two conditions in the definition of \emph{models} hold.
    
    For the first condition, let $\alpha'' \in \Pi^{\mathcal{S}}_\alpha$. By definition, $\alpha''\in\Pi^\mathcal{S}_{\alpha'}$ for some $\alpha'\in \Pi^{\mathcal{T}}_\alpha$. Furthermore, by definition, there exists some $\beta' \in {R_2^*}^{-1}(\beta)$ such that $\alpha' \to^{\mathcal{T}} \beta'$ and there exists some $\beta'' \in {R_1^*}^{-1}(\beta')$ such that $\alpha'' \to^{\mathcal{S}} \beta''$. Since $\beta'' \in {R_1^*}^{-1}(\beta')$ and $\beta' \in {R_2^*}^{-1}(\beta)$, we may conclude that $\beta'' \in {R_1^*}^{-1}({R_2^*}^{-1}(\beta))$

    For the second condition, let $\alpha'' \in {R_1^*}^{-1}({R_2^*}^{-1}(\alpha))$ and $\beta''\in {R_1^*}^{-1}({R_2^*}^{-1}(\beta))$ with $\alpha'' \to^{\mathcal{S}} \beta''$. We want to say that $\Pi_{\alpha}^\mathcal{S}$ contains an assembly which $\calS$-produces $\alpha''$. Because $\calT$ models $\calU$ we know that there exists an assembly which $\calT$-produces $R_1(\alpha'')$ in the corresponding stem set $\Pi^\calS_{\alpha'}$. The conclusion then follows by condition 2 of $\calS$ modelling $\calT$.
\end{proof}

\begin{theorem}
    If $\calS$ intrinsically simulates $\calT$ and $\calT$ intrinsically simulates $\calU$, then $\calS$ intrinsically simulates $\calU$.
\end{theorem}

This result follows from the previous lemmas. It is also clear that the scale factor of the intrinsic simulation of $\calU$ by $\calS$ is the product of the component scale factors.

To see why $\mathcal{F}_{z,\epsilon}$ strictly assembles into a DSSF, we reiterate some properties of our construction. First, our Quine tileset will grow from a single tile seed $\sigma$ into a square $\alpha^1_\sigma$ which resembles a macrotile for our IU tileset. This macrotile will have all of the information needed for our IU tileset to simulate the growth from $\sigma$ to $\alpha^1_\sigma$ using macrotiles instead of individual tiles. The result will be a larger assembly $\alpha^2_\sigma$ whose shape consists each occupied tile location in $\alpha^1_\sigma$ substituted by the shape of a macrotile. Importantly, our IU tileset is configured to simulate tiles using macrotiles so that each has a shape identical to that of $\alpha^1_\sigma$, and furthermore, without the need for fuzz so long as all tiles are included in the glue table encoding.
Consequently, since all of the IU tiles are encoded into $\mathcal{F}_{z,\epsilon}$ so that the glue table along $\alpha^1_\sigma$ contain 
the IU tileset entries, there is nothing stopping the tileset from growing from the individual seed $\sigma$ into $\alpha^2_\sigma$. Since $\mathcal{F}_{z,\epsilon}$ intrinsically simulates itself at some scale, transitivity implies that it simulates itself at all powers of that scale.

\subsection{Adding space to macrotiles}

Since the quine assembly will act as the seed macrotile for the simulation of itself by $U$, the exact shape of the square, will act as a generator for a DSSF and influence the $\zeta$-dimension of the resulting assembly. Here we present a scheme for ``squaring'' the result from our quine in such a way that the $\zeta$-dimension of the eventual DSSF may be effectively freely chosen.

Let $W$ and $H$ be the width and height respectively of the rectangle produced by the Quine tile set. Note that both of these numbers depend on the number of tile types to be encoded in the representation produced by the Quine. Specifically, if the Quine produces the representation of a tile set $T$ with $t$ distinct tile types, then both $W$ and $H$ grow as $\Theta(t \log_2 t)$.

Each square frame is made from 4 binary counter gadgets which grow into a closed loop. The length of each of these counters may be chosen arbitrarily in advance, so let $X$ be the chosen length. Therefore, each counter that makes up the frame will effectively grow into an $X$ by $\lceil \log_2X \rceil$ rectangle. The side length of each square frame is therefore $F = X + \lceil \log_2 X \rceil$ and the entire macrotile square will therefore end up being $2F + W$ tiles on each side.

Within each of the square frames, 4 solid square blocks of tiles grow from each interior corner to fill in some of the space within each frame. The side length of these squares is precisely $Y$ tiles where $Y$ is some counter value chosen in advance. The number of tiles within each of these blocks, its area, is therefore $Y^2$. Furthermore, the tiled area of each square frame is therefore $4(X\lceil \log_2 X\rceil + Y^2)$. Since the area of the Quine is $W \cdot H$ and the area of the rectangles propagating the Quine information to the north and east is $W(X + \lceil \log_2 X \rceil)$, we can compute the tiled area of the entire square macrotile to be

$$ A = 16(X\lceil \log_2 X\rceil + Y^2) + 2W(X + \lceil \log_2 X \rceil) + WH$$

It's important to note that there's a bit of an awkward dependency here; the width $W$ depends on the number of tile types to be encoded by the Quine, and in order to change the value of $X$ or $Y$, the number of tile types to be encoded must change. In other words, increasing $X$ or $Y$ will result in an increase to $W$ as well. In order to handle this, we may divide our overall tileset into two parts, one which has only those tiles that encode the values of $X$ and $Y$ which we will call $T_{X,Y}$ and all of the other tiles including those for the Quine, IU simulation, and square frame which we will call $T_{else}$. Both $X$ and $Y$ are encoded as binary numbers so the size of $T_{X,Y}$ is simply $\lceil \log_2 X \rceil\cdot\lceil \log_2 Y \rceil$. We can make a simplification here by noting that $Y$ will never be larger than half of $X$ and so it's bit representation will be no longer than that of $X$. Therefore, we can safely say that the number of tiles in $T_{X,Y}$ is $\Theta(\log_2 X)$. Since $|T_{else}|$ is simply a constant, the size of $W$ is therefore $\Theta(\log_2 X \cdot \log_2 \log_2 X)$. The same holds for $H$. Combining these results, we find that the side length of the resulting macrotile square is
$$S = 2(X + \lceil \log_2 X \rceil) + \Theta(\log_2 X \cdot \log_2 \log_2 X)$$
while the area occupied by tiles in the square macrotile will be
$$A = 16(X\lceil \log_2 X\rceil + Y^2) + \Theta(X \cdot \log_2 X \cdot \log_2 \log_2 X) $$

\subsection{Derivation of the resulting fractal dimension}

As shown in \cite{hader2023fractalNACO}, for a DSSF with a generator $G$ which has a minimum bounding box that is square with side length $s$, we know that the $\zeta$-dimension of the DSSF will be
$$\frac{\log_2|G|}{\log_2 s}$$

In our construction the Quine macrotile square acts as the generator for our DSSF, so the side length of the bounding box $s$ is just $S$ while the number of tile locations occupied in the generator $|G|=A$. Consequently, the $\zeta$-dimension of our resulting DSSF will be 
$$\frac{\log_2(A)}{\log_2(S)}$$

Attempting to plug in the previously deduced values of $A$ and $S$, we could find a messy expression for the exact $\zeta$-dimension given our values of $X$ and $Y$ and the total number of tile types in the tile set to be simulated. Instead, we note that something interesting happens when the limit is taken as $X$ grows to infinity. We show that we can make the $\zeta$-dimension converge to any desired value $d\in(1,2]$ by choosing $Y$ so that it grows according to $X^{d/2}$. This provides us with a means of choosing our $\zeta$-dimension to an arbitrary precision: simply choose $Y$ so that its size is proportional to $X^{d/2}$ and increase the value of $X$ until your $\zeta$-dimension is within the desired tolerance of $d$.

To see this, assume that $Y=X^{d/2}$ and note that this means that the dominant factor in the limit of $A$ will be $16 Y^2 = 16 X^d$ and the dominant factor in the limit of $S$ will be $2X$. Therefore
\begin{align*}
    \lim_{X\to\infty} \left(\frac{\log_2(A)}{\log_2(S)}\right) & = \lim_{X\to\infty} \left(\frac{\log_2(16X^d)}{\log_2(2X)}\right) \\
    & = \lim_{X\to\infty} \left(\frac{4+\log_2(X^d)}{1+\log_2(X)}\right) \\
    & = \lim_{X\to\infty} \left(\frac{4+d\log_2(X)}{1+\log_2(X)}\right) \\
    & = d \\
\end{align*}

\section{A Self-Describing Embedded Circuit for the Sierpinski Cacarpet}
\label{sec:cacarpet_circuit}
Our second fractal construction relies on the self-describing circuit technique of \cref{sec:abstract_models}.

\cacarpetCircuit*

The remainder of this section is the description of $C_{\square}$. This circuit is built from two sets of messages: layer 1 messages in $\Sigma_1$ and layer 2 messages in $\Sigma_2$, and three fractal structures: a wiring layer $W_{\square}: K^\infty \to \Wirings$, and two function layers, $F_1$ operating on $\Sigma_1$ and $F_2$ operating on $\Sigma_2$.

The wiring layer $C_{w}: K^{\infty} \to \Wirings$ is the fixed point of a substitution $\kappa_w: \Wirings \to \Wirings^K$ starting from a seed wiring $s_w$:
\[
  \begin{cases}
    C_{w}(\vec{0}) = s_w\\
    C_{w}(\vec{z}) = \kappa_w(C_{w}\lfloor \frac{z}{K} \rfloor)(z \bmod K)\\
  \end{cases}
\]

The definition of the function layers is a bit more indirect: there are two sets of labels $L_1$ and $L_2$ respectively, with two rules $\kappa_1: \Wirings \to L_1^K$ and $\kappa_2: \Wirings \times L_2 \to L_2^K$. The rule $\kappa_2$ takes the form of a substitution.

Their fixpoints, starting from two seed labels $\mathfrak{s}_1$ and $\mathfrak{s}_2$ define two tiling of $K^\infty$ with labels of $L_1$ and $L_2$ respectively: $\Lambda_1$ and $\Lambda_2$. On layer $1$, $\Lambda_1$ is defined for each position according to the wiring of its parent:
\[
  \Lambda_1(\vec{z}) = \kappa_1(C_{w}(\lfloor\frac{z}{K}\rfloor))(z \bmod K)
\]

On layer $2$, $\Lambda_2$ is defined for each position according to the wiring \emph{and label} of its parent in a substitutive manner:
\[
  \begin{cases}
    \Lambda_2(\vec{0}) = \mathfrak{s}_2\\
    \Lambda_2(\vec{z}) = \kappa_2(C_{w}(\lfloor\frac{z}{K}\rfloor), C_{2}(\lfloor\frac{z}{K}\rfloor))(z \bmod K)\\
  \end{cases}.
\]

Finally, the actual gate functions are defined from two functions, $\finst{1}$ on layer 1 and $\finst{2}$ on layer 2, which define a function on $\Sigma_1$ (respectively $\Sigma_2$) from four elements, two wirings and two labels in $L_1$ (respectively $L_2$), one each for the gate and its parent. Together with the substitutions $\kappa_i$, they define a circuit $\mu_i(w, l)$ with domain $K$ known as the layer-$i$ meta-gate obtained by $w$ and $l$:
\[
  \begin{cases}
    \mu_1(w, l): K \to \Gates{\Sigma_1}\\
    \mu_1(w, l): \vec{z} \mapsto \{ \gatewiring: \kappa_w(w)(\vec{z}), \gatefun: \operatorname{instantiate}_1(w, l, \kappa_w(w)(\vec{z}), \kappa_1(w)(\vec{z}))\}\\
    \mu_2(w, l): K \to \Gates{\Sigma_2}\\
    \mu_2(w, l): \vec{z} \mapsto \{ \gatewiring: \kappa_w(w)(\vec{z}), \gatefun: \operatorname{instantiate}_2(w, l, \kappa_w(w)(\vec{z}), \kappa_2(w, l)(\vec{z}))\}.\\
\end{cases}
\]

In this circuit, the wiring of each gate is given by $\kappa_w(w)$, and its function is given by instantiating the label given by $\kappa_i$. When iterating this process, a gate $g$ appearing at position $\vec{z}$ in some parent meta-gate $\mu_i(l,w)$, the child meta-gate $\mu_i(g)$ associated with $g$ is defined as $\mu_i(\gatewiring[g], \kappa_i(l,w))$. The functions $\operatorname{instantiate}_i$ are each injective in their last argument, ensuring this does not create any ambiguity.

Finally, these meta-gates beget the two layers of circuits $C_1$ and $C_2$ by:
\[
  \begin{cases}
    C_i(\vec{0}) = \{ \gatewiring: s_w, \gatefun: s_i \}\\
    C_i(\vec{z}) = \mu_i(C_i(\lfloor\frac{\vec{z}}{K}\rfloor)(\vec{z} \bmod K)\\
  \end{cases}.
\]

By this definition, the wirings of $C_1(\vec{z})$ and $C_2(\vec{z})$ are the same ---they are given by $C_w$, so $C_\square: \vec{z} \mapsto C_1(\vec{z}) \otimes C_2(\vec{z})$ defines a circuit with alphabet $\Sigma_1 \times \Sigma_2$ and domain $K^\infty$.

The next subsections are the description of $C_w$, followed by those of $C_1$, then $C_2$, and finally the proof of that $C_{\square}$ is self-describing.

\subsection{The wiring layer}

The seed wiring $s_w$ is represented on the left of \cref{fig:kappa_seed}. On the right of the same figure is its image $\kappa_w(s_z)$.

\begin{figure}
  \centering
  \begin{tikzpicture}
    \matrix[column sep=3em] {
      \input{tikz/seed_alone}&
      \input{tikz/seed_local}\\
    };
  \end{tikzpicture}
  \caption{$\kappa_1(w_s, \gcolor{seed})$, the first iteration of $\kappa_1$ on the seed gate. The wirings are given graphically; all gates have the \gcolor{normal} label, except the $s$ at $(0,0)$ which has label \gcolor{seed}. For the wiring with two inputs, the fat arrow represents the input number 0; all output wires have number 0: $\woutputmap[w](d) = 0$}
  \label{fig:kappa_seed}
\end{figure}

For any wiring $w$, $\kappa_1(w)$ will only contain wirings with at most two inputs sides, and these input sides are adjacent. Thus, all gates in $C_w$ have at most two input sides and they are adjacent. Hence, $\kappa_w$ only needs to be defined on wirings with that property.

The definition of $\kappa_w$ is isotropic: $\kappa_w$ commutes with a rotation of $\pi / 2$ and with reflections. This property will be upheld simply by giving the definition of $\kappa_w$ \emph{up to rotation and reflection}. 

Lastly, for each input wire in $w$ there are two wires in the input bus of $\kappa_w(w)$, and for each output wire in $w$ there are two wires in the output bus of $\kappa_w(w)$. The position of these wires are as follows, up to rotation:

\begin{tabular}{cc|ccc}
  \multicolumn{2}{c}{input of $w$} & \multicolumn{3}{c}{input $i$ of $\kappa_w(w)$}\\
  $\winputset[w]$ & direction & position & direction & $\winputset[i]$ \\
  \hline
  \multirow{2}{*}{$(W)$} & \multirow{2}{*}{$W$} & $(0,2)$ & $W$ & $(W)$\\
                           &                      & $(0,3)$ & $W$ & $(S, W)$\\
  \hline
  \multirow{4}{*}{$(S, W)$} & \multirow{2}{*}{$W$} & $(0,0)$ & $W$ & $(S, W)$\\
                              &                      & $(0,1)$ & $W$ & $(S, W)$\\
                              & \multirow{2}{*}{$S$} & $(0,0)$ & $S$ & $(S, W)$\\
                              &                      & $(1,0)$ & $S$ & $(S, W)$\\
  \hline
  \hline
  \multicolumn{2}{c}{output of $g$} & \multicolumn{3}{c}{output $o$ of $\kappa_w(g)$}\\
  $\woutputwires[w](d)$ & direction $d$ & position & direction $d'$ & $\woutputwires[i](d')$ \\
  \hline
  \multirow{2}{*}{$(W)$} & \multirow{2}{*}{$E$} & $(5, 2)$ & $E$            & $W$\\
                           &                      & $(5, 3)$ & $E$            & $(S, W)$\\
  \hline
  \multirow{2}{*}{$(S, W)$} & \multirow{2}{*}{$E$} & $(5, 0)$ & $E$            & $(S, W)$\\
                              &                      & $(5, 1)$ & $E$            & $(S, W)$\\
  \hline
  \multirow{2}{*}{$(S, W)$} & \multirow{2}{*}{$N$} & $(0, 5)$ & $N$            & $(S, W)$\\
                              &                      & $(1, 5)$ & $N$            & $(S, W)$\\
  \hline
\end{tabular}

Thus, if $w$ has inputs $\winputset[w] = (W)$, $\woutputset[w] = \{N, E, S\}$, outputs $\woutputwires[w](N) = Sw$, $\woutputwires[w](E) = sW$ and $\woutputwires[w](S) = N$, then $\kappa_w(w)$ has the corresponding input and output busses and wires:
\[
  \begin{cases}
    \vec{I} = \begin{pmatrix}
        \wirepos{(-1,2) \to (0, 2)}{W}\\
        \wirepos{(-1,3) \to (0, 3)}{Ws})\\
    \end{pmatrix}\\
    \vec{O} = \begin{pmatrix}
        \wirepos{(0, 5) \to (0, 6)}{Sw}\\
        \wirepos{(1, 5) \to (0, 6)}{Sw}\\
        \wirepos{(5, 0) \to (6,0)}{Ws}\\
        \wirepos{(5, 1) \to (6, 1)}{Ws}\\
        \wirepos{(2,0) \to (2, -1)}{N}\\
        \wirepos{(3,0) \to (3, -1)}{Nw})\\
    \end{pmatrix}
  \end{cases}
\]

\begin{figure}
  \centering
  \begin{tikzpicture}
    \matrix[column sep=2em, row sep=0] {
      \scoped[rotate=180, scale=.7]{\input{tikz/in_wires_E}} &%
      \scoped[rotate=90, scale=.7]{\input{tikz/in_wires_E}} &%      
      \scoped[rotate=0, scale=.7]{\input{tikz/in_wires_E}} &%
      \scoped[rotate=-90, scale=.7]{\input{tikz/in_wires_E}} &\\
      \scoped [rotate=180]{\input{tikz/empty_gate_E}}; &
      \scoped [rotate=90]{\input{tikz/empty_gate_E}};  &% 
      \scoped [rotate=0]{\input{tikz/empty_gate_E}};   &
      \scoped [rotate=-90]{\input{tikz/empty_gate_E}}; &\\
      \scoped[rotate=180, scale = .7]{\input{tikz/in_wires_NE}} &
      \scoped[rotate=90, scale=.7]{\input{tikz/in_wires_NE}} &
      \scoped[rotate=0, scale=.7]{\input{tikz/in_wires_NE}} &
      \scoped[rotate=-90, scale=.7]{\input{tikz/in_wires_NE}} &\\
      \scoped [rotate=180]{\input{tikz/empty_gate_NE}};  &
      \scoped [rotate=90]{\input{tikz/empty_gate_NE}};   &
      \scoped [rotate=0]{\input{tikz/empty_gate_NE}};    &
      \scoped [rotate=-90]{\input{tikz/empty_gate_NE}};  &\\      
    };
  \end{tikzpicture}
  \caption{Location of the input wires in $\kappa_1(g)$ according to the input sides of $w = \gatewiring{g}$. The output wires mirror the input sides of the neighboring tiles.}

  \label{fig:kappa_wires}
\end{figure}

\paragraph*{Wirings with no inputs}

All wirings without input are sent by $\kappa_w$ to the array of wirings represented on \cref{fig:kappa_seed}. The seed wiring $s_w$ is the only one to effectively appear when iterating $\kappa_w$.

\paragraph*{Wirings with one input}

A wiring with one input is cut according to \cref{fig:kappa_normal1}.

\begin{figure}
  \centering
  \begin{tikzpicture}
    \matrix[column sep=3em, row sep=5mm, cells={anchor=center}] {
      \scoped [rotate=180]{\input{tikz/empty_gate_E_to_N_nW_Se}};&
      \scoped [rotate=180]{\input{tikz/local_normal1}};\\
    };
  \end{tikzpicture}
  \caption{$\kappa_1(w, l)$ when $\winputset[w] = (W)$. The wirings are given graphically; on layer 1, all gates have the \gcolor{normal} label (black), except for the gate at $(0, 3)$ which has label \gcolor{input} (red). Dotted segments are parts of the wiring only if the corresponding element is in the outputs of $w$. For wirings with two inputs, the fat arrow represents the first input. The $\dagger$ marks the potential position for label $\gcolor{dA}$ on layer 2.}
  \label{fig:kappa_normal1}
\end{figure}

The wirings in $\kappa_w(g)$ depend on the input and outputs of $w$. Since $w$ has only one input, it is sufficient, up to rotation, to examine the case where $\winputset[w] = \{W\}$. The black arrows of \cref{fig:kappa_normal1} depict the case where $\woutputset[w] = \emptyset$. For each $d \in \woutputset[w]$, some extra wires are needed. The additional wires for an output in the $N$ direction depend on whether the directions appearing in $\woutputwires[w](N)$ are $\{S\}$, $\{S, E\}$ or $\{S, W\}$. The wires for the other output directions are derived from these by rotation. \Cref{fig:kappa_normal1} has one direction with each case, showing the complete range of possibilities for the extra wires; they are represented in dotted lines on the figure.

\paragraph*{Wirings with two (adjacent) inputs} The image of such a wiring $w$ by $\kappa_w$ is defined on \cref{fig:kappa_input}, which is rotated and reflected so that the $S$ side of the figure is mapped to the $0$ input of $w$, and the $W$ side to the $1$ input of $w$.%\ruggedtodo{\textsc{reformuler}}

\begin{figure}
  \label{fig:kappa_input}
  \centering
    \begin{tikzpicture}
    \matrix[column sep=3em, row sep=5mm, cells={anchor=center}] {
      \input{tikz/empty_gate_SW_input2}&
      \input{tikz/local_input2};\\
    };
  \end{tikzpicture}
  \caption{$\kappa_1(w, l)$ when $\winputset{w} = (S, W)$. The wirings are given graphically; all gates have the \gcolor{normal} label except for the input gate at $(0, 0)$ which has label $\gcolor{input}$. This configuration gets rotated and reflected according to the inputs of the gate, so that the arrows on the outer ring form a path from the side with input $0$ to the side with input $1$, ``the long way around''. The '*' marks the \emph{special} position on layer 2, and the $\dagger$ marks the position where label $\gcolor{dA}$ may appear.}
\end{figure}

\begin{lemma}
  \label{lem:wiring-normal}
    For each gate $w$, $\kappa_w(w)$ is the wiring of a normal circuit. Moreover, $C_w$ is the wiring of a normal, closed circuit.
\end{lemma}

\begin{proof}
  This follows from observation of the schemata describing each $\kappa_w(w)$ and observing that for two wirings $w, w'$, if an output side in direction $d$ of $w$ matches an input side in direction $-d$ of $w'$, then the corresponding sides of $\kappa_w(w)$ and $\kappa_w(w')$ also match.

  Additionally, each iteration of $\kappa_w$ on $s_w$ is without inputs, so $C_w$ is closed.
\end{proof}

\subsection{Layer 1}
\label{sec:local_layer}

The components of layer 1 are an alphabet $\Sigma_1$, a finite set of labels $L_1$, a substitution $\kappa_1: \Wirings \mapsto L_1^K$, and for each label $l \in L_1$, and an instantiation function $\operatorname{instantiate}_1$. Layer 1 can then be realized as described above into a circuit with the wirings of $C_w$ and the functions given by instantiating the fixpoint of $\kappa_1$.

The set of labels is $L_1 = \{\gcolor{s}_1, \gcolor{normal}, \gcolor{input} \}$.

The elements of $\Sigma_1$ are \emph{layer 1 messages}.

\begin{definition}[Layer 1 Message]
  A layer 1 message is a triple $\{\lmpos = \vec{z} \in K_1, \lmpwiring = w \in \Wirings, \lmpar = l \in L_1\}$. \end{definition}

The circuit $C_1$ is going to be defined by iterating $\kappa_w$ and $\kappa_1$ starting from $s_w$ and $\gcolor{s}_1$, then instantiating the labels.

\paragraph*{Labels}

The label function $\kappa_1: \Wirings \to L_1^K$ assigns the label $\gcolor{normal}$ at all positions except:
\begin{itemize}
\item $\kappa_1(s_w)(\vec{0}) = \gcolor{s}_1$
\item if $w$ has at least one input, then $\kappa_1(w)(\vec{z}) = \gcolor{input}$ for the position $\vec{z}$ which receives the input wire $0$ in $\kappa_w(w)$.
\end{itemize}

\paragraph*{The seed function}

The starting label is $\gcolor{s}_1$, it only ever appears at position $(0,0)$ in $\kappa_1(s_w, \gcolor{s}_1)$.

Since $s_1$ has no inputs, its associated function $f_s = \finst{1}(s_w, \gcolor{s}_1, s_w, \gcolor{s}_1)$ is a constant function with value $f_{s}() = \{\lmpos: (0, 0), \lmpcolor: \gcolor{s}_1, \lmpwiring: s_w\}$.

\paragraph*{Gate functions for layer 1}

There are two types of functions for the gates output by $\kappa_1(g)$ other than the seed. They have a either an \emph{increment} function $\fincr{k,D}$ with $k \in \{1, 2\}$ and $D \in \Dir$ or a \emph{reparenting} function $\frepar{k, D, w, c}$, with $k \in \{1, 2\}$, $D \in \Dir$, $w$ a wiring and $c \in C$ a color. The versions with $k=2$ take two inputs but ignore the second one. The functions $\fincr{}$ and $\frepar{}$ are defined as follows:

\begin{eqnarray*}
  \lmpar[\fincr{1, D}(m)] &=& \lmpar[m],\\
  \lmpos[\fincr{1, D}(m)] &=& \lmpos{m} + D\\
  \lmpar[\frepar{1, D, w, c}(m)] &=& (w, c)\\
  \lmpos[\frepar{1, D, w, c}(m)] &=& \lmpos[m] + D\\
  \fincr{2, D}(m, m') &=& \fincr{1, D}(m)\\
  \frepar{2, D, w, c}(m, m') &=& \frepar{1, D, w, c}(m).\\
\end{eqnarray*}

The function of each gate is fixed from its label and wiring, and those of its parent through \finst{1} as follows:
\[
  \begin{cases}
    \finst{1}(w_p, l_p, w, \gcolor{normal}) = \fincr{k, D} \\
    \finst{1}(w_p, l_p, w, \gcolor{input}) = \frepar{k, D, w_p, l_p},\\
  \end{cases}
\]
where $k$ is the number of inputs of $w$, and $D$ is the direction of its first input.

\paragraph*{Behavior and Self-Description of Layer 1}

The circuit $C_1$ obtained from $\kappa_1$ is normal, by \cref{lem:wiring-normal}. Let $e = \innerevalfunc{C_1}: K^\infty \times \Dir \to \Sigma_l$ be the evaluation function of $C_1$. That function $e$ enjoys a simple description, which reflects the fact that in $C_1$, the different meta-gates do not actually communicate. On layer 2 however, there will be some communication between meta-gates, as described in the next section.

\begin{lemma}
  \label{lem:local_eval}
  Let $w \in \Wirings$, $l \in L_1$ and $a: p \mapsto p'$ be an internal or outgoing arc in $\kappa_w(w)$. If $w$ has no inputs, assume $l = \gcolor{s}_1$.

  For any input $\vec{\imath}$ of $\mu_1(w, l)$, let $m = \innerevalfunc{\mu_1(w,l)}(\vec{\imath}, a)$ be the value of arc $a$ on input $\vec{\imath}$. Then $\lmpos[m] = p$, $\lmpwiring[m] = w$ and $\lmpcolor[m] = l$.
\end{lemma}
\begin{proof}
  
  By induction on the non-incoming arcs of the (acyclic) dependency graph $D$ of $\kappa_w(w)$.

  If $w$ is the seed wiring $s_w$, then the root of $D$ is also $s_1$ and its outputs satisfy $\lmpos[e_g(a)] = (0, 0)$, $\lmpwiring[e_g(a)] = \gatewiring[s_1]$ and $\lmpcolor[e_g(a)] = \gcolor{s}_1$.

  Otherwise, $\kappa_1(w)$ has a unique position $\vec{z_i}$ with label $\gcolor{input}$, corresponding to a gate in $\mu_1(l,w)$ with function $\frepar{k, D, w, l}$, for some $k$ and $D$. By definition of $\frepar{}$, its outputs satisfy $\lmpos[e_g(a)] = \vec{z_i}$ , $\lmpwiring[e_g(a)] = w$ and $\lmpcolor[e_g(a)] = l$.

  In both cases, each other gate $g'$ at position $\vec{z}$ of $\mu_1(l,w)$ has function $\fincr{k, D}$, where $k$ is the number of inputs of $g'$, and $D$ is the direction of its first input arc $i = \vec{z} - D \overset{D}{\mapsto} z$. By induction, $\lmpos[e_g(i)] = z - D$ and $\lmpar[e_g(i)] = (\gatewiring[g], \gatecolor[g])$. By definition of $\fincr{k, D}$, each of the output arcs $o$ of $g'$ verify $\lmpos[e_g(o)] = (\vec{z} - D) + D = \vec{z}$ and $\lmpwiring[e_g(a)] = w$ and $\lmpcolor[e_g(a)] = l$.
\end{proof}

Additionally, $C_1$ is ``mostly self-describing'': the first input of a gate $g$ is enough to recover $g$, except for the value of $w$ and $c$ in gates with a function of the form $\frepar{k, D, w, c}$.

\begin{lemma}
  \label{lem:layer1-self-descr}
  For a position  $p \in K^{\infty}$, let $ \vec{e} = \winputset{\gatewiring[C_1(p)]}$ be the input arcs of $C_1(p)$; let $d_i$ be the direction of $e_i$.
  
  There is a function \(\decgate: \Sigma_1 \times \Dir \to \Gates{\Sigma_1} \cup \bot\) such that for all $p \in K^{\infty}$,
  \begin{itemize}
  \item if \(\gatefun[C_1(p)]\) is $\fincr{k, d_0}$, then \(\decgate(\innerevalfunc{C_1}(e_0), d_0) = C_1(p)\)
  \item if \(\gatefun[C_1(p)]\) is $\frepar{k, d_0, -, -}$ then \(\decgate(\innerevalfunc{C_1}(e_0), d_0) = \bot\).
  \end{itemize}
\end{lemma}

\begin{proof}
  The function $\decgate$ is defined as follows: let $m \in \Sigma_1$ be a local message, and $d \in \Dir$. Let $p = \lmpos[m]$, if $p - d \notin \{0, \ldots, 5\}^2$, then $\decgate(m, d) = \bot$. Otherwise, $\decgate(m, d)$ is the gate at position $p$ in $\mu_1(\lmpwiring[m], \lmpar[m])$.
  
  By \cref{lem:local_eval}, $\decgate$ satisfies the lemma.
\end{proof}

Moreover, when $\decgate(e(e_O), d_0) = \bot$, $g$ itself cannot be determined, but its label and wiring can, as well as $\mu_1(g)$.

\begin{corollary}
  \label{thm:layer1-decode}
    For a position  $p \in K^{\infty}$, let $ \vec{e} = \winputset[\gatewiring[C_1(p)]]$ be the input arcs of $C_1(p)$; let $d_i$ be the direction of $e_i$,
  
    \begin{itemize}
    \item there is a function \(\decgatew: \Sigma_1 \times \Dir \to \Wirings\) such that for each position $p \in K^{\infty}$, $\decgatew(\innerevalfunc{C_1}(e_0), d_0) = \gatewiring[C_1(p)]$
    \item there is a function \(\decgatel: \Sigma_1 \times \Dir \to \Wirings\) such that for each position $p \in K^{\infty}$, $\decgatel(\innerevalfunc{C_1}(e_0), d_0) = \Lambda_1(p)$
%    \item there is a function \(\decgatecut: \Sigma_1 \times \Dir \to \Gates{\Sigma_1}^{K}\) such that for all $p \in K^{\infty}$, $\decgatecut(\innerevalfunc{C_1}(e_0), d_0) = \mu_1(C_1(p))$.\ruggedtodo{Utile?}
\end{itemize}
\end{corollary}

\begin{proof}
%  $\decgatecut$ can be built from $\decgatew$ and $\decgatel$.

  For \decgatew{} and \decgatel{}, it suffices to observe that whenever $\decgate(\innerevalfunc{C_1}(e_0), d_0) = \bot$ at some position $p$, $\innerevalfunc{C_1}(p)$ has label \gcolor{input}, and its wiring only depends on its position within its metagate.
\end{proof}

\subsection{Layer 2}

A second layer is needed in order to get full self-description of the circuit on $K^\infty$. This layer routes global information between the meta-gates so that the input gate of each meta-gate can be indentified by its incoming global message. The construction needs to ``tie the knot'', so it is not only needed to recover the identity of the input gate on the layer 1, but also on layer 2 itself.

This layer is defined by a set $L_2$ of labels, an alphabet $\Sigma_2$, the label substitution $\kappa_2$ and its instantiation function. In contrast with layer 1, $\kappa_2$ takes as input a wiring, as well as a label. This makes the definition of $\Lambda_2$ recursive. In contrast with layer 1, on layer 2, the two-step definition of gates using labels and the function $\operatorname{instantiate}$ is actually \emph{needed} because of a subtelty related to the tying of the knot. The functions of some gates make use of $\kappa_2$. Thus $\kappa_2$ shall not directly manipulate the gates or their function, lest the definition of layer 2 becomes cyclical and possibly ill-founded. %In other words, each gate function is allowed to observe which function is associated with each gate in a meta-gate, but not to actually run them.

The set of layer 2 labels is $L_2 = \{\gcolor{s}, \gcolor{i_1}, \gcolor{i_2}, \gcolor{dL}, \gcolor{dA} \}$.

\paragraph*{Layer 2 messages}

A message on layer 2 is an element of $\Sigma_2$; it is built from:
\begin{itemize}
\item $\mloc[m]\tikzmark{local} \in L_2$,
\item $\mglob[m]$\tikzmark{global}, itself consisting of:
  \begin{itemize}
  \item $\gmcolor[ {\mglob[m]} ] \in L_2$.\tikzmark{anc-layer2}
  \item $\gmanclocal[ {\mglob[m]} ] \in \Sigma_1$\tikzmark{anc-layer1}
  \end{itemize}
\end{itemize}

These messages make $C_2$ self-describing by completing the information available in layer 1 and used in \cref{lem:layer1-self-descr}.
A message $m_2$ output by a gate $g_2$ in $\kappa_2(p)$ identifies $p$ through $\mloc[m_2]$ if $g_2$ is not the input gate of $\kappa_2(p)$, as in \cref{lem:layer1-self-descr}. %\Highlight[fill=purple!30!white]{identifies $p$}
The global part $\mglob[m_2]$ identifies \emph{some ancestor} of $g_2$, on layer 1 through $\gmanclocal[ {\mglob[m]} ]$ and on layer 2 through $\gmcolor[ {\mglob[m]} ]$.
This global information will enable the determination of the entry gate of each meta-gate, on both layers.
From two messages $m_l, m_g \in \Sigma_2$, two messages $(m_1, m_2) = \extractfunc(m_l,m_g) \in \Sigma_1 \times \Sigma_2$ can be extracted by reading the local information from $m_l$, and the global information from $m_g$, as follows:
\[
  \begin{cases}
    m_1 = \gmanclocal[ {\mglob[m_l]} ]\\
    \mloc[m_2] = \gmcolor[ {\mglob[m_l]} ]\\
    \mglob[m_2] = \mglob[m_g].\\
  \end{cases}
\]

The converse operation, embedding, takes as input three messages, a payload $(p_1, p_2) \in (\Sigma_1 \times \Sigma_2)$ and a context $c \in \Sigma_2$, and yields two messages $m_l$ and $m_g$
\[
  \begin{cases}
    \mloc[m_l] = \mloc[m_g] = \mloc[c]\\
    \gmcolor[ {\mglob[m_l]} ] = \mloc[p_2].\\
    \gmanclocal[ {\mglob[m_l]} ] = p_1\\
    \mglob[m_g] = \mglob[p_2].\\
  \end{cases}
\]

Extracting and embedding are dual operations, in the sense that  \[\forall c, p_1, p_2, \extractfunc \circ \embedfunc(c, p_1, p_2) = (p_1, p_2).\]

\paragraph*{The substitution $\kappa_2$}

Given a wiring $w$ and a label $l \in L_2$, the substitution $\kappa_2$ yields a label for each position in $K$;
\begin{itemize}
\item $\gcolor{s}_2$ for the seed gate, i.e. for position $(0,0)$ if $w$ has no inputs;
\item $\gcolor{dL}$ for the gate receiving the input number 0 of the meta-gate;
\item $\gcolor{dA}$ for the gate receiving the input number 1 of the meta-gate whenever $l = \gcolor{dL}$;
\item when $w$ has two inputs, there is a \emph{special} position within the meta-gate where the value depends on $l$ as follows:
  \begin{itemize}
  \item $\gcolor{i2}$ if $l \in \{\gcolor{i2}, \gcolor{dL}\}$,
  \item $\gcolor{dA}$ if $l = \gcolor{dA}$;
  \end{itemize}
\item otherwise, $\gcolor{i_1}$ for gates with one input and $\gcolor{i_2}$ for gates with two inputs.
\end{itemize}

The special position is marked on \cref{fig:kappa_normal1,fig:kappa_input} by a star.

\paragraph*{The layer-2 seed gate}

The function $f^2_s = \finst{2}(s_w, \gcolor{s}_2, s_w, \gcolor{s}_2)$ of the seed gate is a constant function with value $f_{s}^2() = \{\mloc: \gcolor{s_2}, \mglob: \{\gmcolor: \gcolor{s_2}, \gmanclocal: f^1_s() \} \}$. Recall that $f^1_s()$ is the message output by the seed gate on layer 1.

\paragraph*{Gate functions for layer 2}

The functions of the gates depend on their label and on the direction $D$ of their first input, as dictated by the function $\operatorname{instantiate}$:%\ruggedtodo{\textsc{reformuler}}
\[
  \begin{cases}
    \operatorname{instantiate}(D, \gcolor{s}) = f^2_s\\
    \operatorname{instantiate}(D, \gcolor{i1}): x \mapsto x\\
    \operatorname{instantiate}(D, \gcolor{i2}): (x, y) \mapsto (x, y)\\
    \operatorname{instantiate}(D, \gcolor{dL}) =  \fdecode[D]{L} \circ \fgincr{G}{D}\\
    \operatorname{instantiate}(D, \gcolor{dA}) = (m_l, m_g) \mapsto \fdecode[D]{A}(\fgincr{G}{D}(m_l), m_g),\\
  \end{cases}
\]
where $D$ is the direction of the first input of $w_p$.

The gate with labels \gcolor{i1} or \gcolor{i2} are wires; their function is the identity function of arity $1$ or $2$ respectively.

Given $m \in \Sigma_2$, the increment functions $\fgincr{G}{D}$ increments $\lmpos[\gmanclocal{\mglob{m}}]$ by the unit vector of direction $D$.

The two decoding functions $\fdecode{L}$ and $\fdecode{A}$ are based on the same underlying function $\fdecode{}: \Dir \times \Sigma_1 \times \Sigma_2 \to \Sigma_1 \times \Sigma_2$ defined as follows: let $m_1 \in \Sigma_1, m_2 \in \Sigma_2$, pose
$z = \lmpos[m_1]$, $z_a = \lmpos[ {\gmanclocal[ {\mglob[m_2]} ]} ]$, $a = \gmanclocal[ {\mglob[m_2]} ] $. Let $l_p$ be defined as $\kappa_2(\lmpwiring[a], \lmpcolor[a], \gmcolor[ {\mglob[m_2]}], p')$ if $z \in K_1$, and $\gcolor{dL}$ otherwise. Then $\fdecode[D]{}(m_1, m_2)$ is the pair $(m'_1, m'_2)$ with:
\begin{eqnarray*}
  m'_1 &=& \begin{Bmatrix}
    \lmpos&:& z \operatorname{mod} K\\
    \lmpwiring&:& \decgatew(a, D)\\
    \lmpcolor&:& \decgatel(a, D)\}
  \end{Bmatrix}\\
  m'_2 &=& \begin{Bmatrix}
    \mloc &:& p\\
    \gmcolor[ { \mglob } ] &: &\gmcolor[\mglob[m]]\\
    \gmanclocal[ { \mglob } ] &: &
      \begin{Bmatrix}
        \lmpwiring &:& \lmpwiring[ { \gmanclocal[ { \mglob[m] } ] } ]\\
        \lmpcolor &:& \lmpcolor[ { \gmanclocal[ { \mglob[m] } ] } ]\\
        \lmpos &:& z_a \bmod K_1\\
      \end{Bmatrix}
    \end{Bmatrix}
\end{eqnarray*}

% The definition of $\fdecode{}$ is somewhat lengthy, the important bit about that function is its properties. First of these, $\fdecode{}$ plugs the hole in the self-description of layer 1.

% \begin{lemma}
%   Let $D \in \Dir, p \in K^\infty$, $g = C_1(p)$, $e$ the input wire number $0$ of
%   $g$ in $C_1$. Let $m = \fincr{D}(\innerevalfunc{C_1}(e))$ and $m'$ be such such that $\gmanclocal{\mglob{m'}} = m$. Pose $(m'_1, m'_2) = \fdecode{}(m')$. Then $g = \kappa_1(\lmpwiring{m'_1}, \lmpcolor{m'_1}, \lmpos{m'_1})$.
% \end{lemma}

% \begin{proof}
%   First note that by the definition of $\kappa_1$, every meta-gate in $C_1$ indeed has output wires, and they all carry the same value, so the statement of the lemma does make sense.
  
%   By construction, since $m'_1$ is built from $\decgatew$ and $\decgatel$.
% \end{proof}

For a direction $D$ and a message $m_2 \in \Sigma_2$, take an arbitrary $m_1 \in \Sigma_1$ and let $(m'_1, m'_2) = \fdecode[D]{}(m_1, m_2)$; the value of $m'_2$ does not depend on $m_1$, so $\fdecode[D]{L}(m_2)$ is defined to be the $m'_2 \in \Sigma$ returned by $\fdecode{}(m_1, m_2)$ for any $m_1$.

% An analogous property holds on layer 2, where $\fdecode{L}{D}$ yields the label of non-seed gates given 

% \begin{lemma}
%   Let $l_1 \in L_1, l_2 \in L_2, w \in \Wirings$. Let $g_1$ be a gate of layer 1 with label $l_1$ and wiring $w$, and $g_2 = \{ \gatewiring: w, \gatefun: \operatorname{instantiate}(l_2) \}$.

%   Let $p \in K$, $l'_2 = \kappa_2(w, l_2, l_2, p)$, and $g'_2$ the corresponding gate.

%   If $g'_2$ is not the seed gate, let $D$ be the direction of its input number $0$.

%   Then for any $m$ such that $\gmcolor{\mglob{m}} = l_2$, $\lmpos[\gmanclocal{\mglob{m}}] = (p - D \operatorname{mod} K)$, $\lmpcolor[\gmanclocal{\mglob{m}}] = l_1$ and $\lmpwiring[\gmanclocal{\mglob{m}}] = w$,
%   \[ \mloc{\fdecode{L}{D}(m)} = l'_2 \]
% \end{lemma}

For a direction $D$ and $m_l, m_g \in \Sigma_2$, $\fdecode[D]{A}(m_l, m_g)$ is defined as follows. Let $(m_1, m_2) = \extractfunc(m_l,m_g)$, $(m'_1, m'_2) = \fdecode[D]{}(m_1, m_2)$. Then $\fdecode[D]{A}(m_l, m_g)$ is the pair $m'_l, m'_g$ with:
\[
  \begin{cases}
    \mloc[m'_l] = \mloc[m'_g] = \mloc[m_l]\\
    \gmanclocal[ { \mglob[m'_l] } ] = m'_1\\
    \gmcolor[ { \mglob[m'_l] } ] = \mloc[m'_2]\\
    \mglob{m'_g} = \mglob[m']\\
  \end{cases}
\]

The function $\fdecode{A}$ is engineered in order to enjoy the following property, a kind of commutation between $\fdecode{}$ and $\extractfunc$.

\begin{lemma}
  \label{prop:decodeL_incr}
  For any direction $D \in \Dir$,
  \[\fdecode[D]{} \circ \extractfunc = \extractfunc \circ \fdecode[D]{A} \]
\end{lemma}

\begin{proof}
  By computation.
\end{proof}

\paragraph*{Properties of $C_2$}

The messages passing through the circuit $C_2$ built above hold all the necessary information for $C_\square$ to be self-describing: in other words, $C_2$ carries all the information needed to determine the entry gate of each meta-gate in $C_1$, as well as the information needed to determine each of its gates.

The $\gmanclocal[\mglob]$ part of the messages on input $0$ each meta-gate simulate the gates of layer 1, as long as each meta-gate simulating a gate with label $\gcolor{input}$ receives on input $1$ the message of its parent meta-gate.

\begin{lemma}
  \label{lem:eval_layer1}

  let $w \in \Wirings$ with $k$ inputs, $l_2 \in L_2$, and $M$ be the circuit $\mu_2(w, l_2)$. Note that $M$ has $2k$ inputs. Let $D$ be the direction of the first input of $w$.

  Let $\vec{\imath} \in \Sigma^{2k}$, and let $\vec{o}$ be the output of $M$ on input $\vec{\imath}$. Pose $i_a = \gmanclocal[\mglob[i_0]]$ and $i_p = \gmanclocal[\mglob[i_1]]$.

  Then, if $g_1 = \decgate(i_a, D) \neq \bot$, then $o_0$ is the output of $g_1$ on input $i_a$; otherwise, if $\decgate(i_a, D) = \bot$, then $o_0$ is the output of $\decgate_c(i_p, D)(\lmpos[i_a])$ on input $i_a$.
\end{lemma}

\begin{proof}
  By computation.
\end{proof}

The rest of the $\mglob$ part of the messages allows $C_2$ to simulate itself.

\begin{lemma}
  \label{lem:eval_subst1}

  Let $D \in \Dir$, $w \in \Wirings$ with one input in direction $D$, $l_2 \in L_2$. Let $f = \operatorname{instantiate}(D, l_2)$ and $C = \mu_2(w, l_2)$.

  Let $c \in \Sigma_1, i_1 \in \Sigma_1, i_2 \in \Sigma_2$ and $(m_l, m_g) = \embedfunc(c, i_1, i_2)$. Let $(o_0, o_1)$ be the outputs of $C$ on input $(m_l, m_g)$. Then $\extractfunc(o_0, o_1) = f(i_2)$.
\end{lemma}

\begin{proof}
  The proof proceeds by case on $l_2$, which can be either $\gcolor{i_1}$ or $\gcolor{dL}$.
  If $l_2 = \gcolor{i_1}$, then $f$ is the identity function, and it suffices to follow the wirings to check the result.

  If $l_2 = \gcolor{dL}$, then following the wirings reduces the desired equality to the definition of $\fdecode[D]{L}$.

  % If $l_2 = \gcolor{s}$, then $ $.
  % \ruggedtodo{Do the case study.}
\end{proof}

\begin{lemma}
  \label{lem:eval_subst2}
  Let $w \in \Wirings$ with two inputs, $l_2 \in L_2$. Let $f = \operatorname{instantiate}(l_2)$ and $C = \mu_2(w, l_2)$.

  For $k \in \{0, 1\}$, let $c^k \in \Sigma_1, i_1^k \in \Sigma_1, i_2^k \in \Sigma_2$ and $(m_l^k, m_g) = \embedfunc(c^k, i_1^k, i_2^k)$. Let $(o_0^0, o_1^0, o_0^1, o_1^1)$ be the outputs of $C$ on input $(m_l^0, m_g^0, m_l^1, m_g^1)$. Then for $k \in \{0, 1\} \extractfunc(o_0^k, o_1^k)$ is the $k$-th component of $f(i_2^0, i_2^1)$.
\end{lemma}

\begin{proof}
  The proof proceeds by case on $l_2$. If $l_2 \in \{\gcolor{s}, \gcolor{i2}, \gcolor{dA} \}$, following the wirings in $\kappa_2(w, l_1, l_2)$ confirms that the lemma holds.

  If $l_2 = \gcolor{dL}$, the lemma follows from \cref{prop:decodeL_incr} by again following the wirings.
\end{proof}

Together, these properties entail a substitutive structure of the messages in $C_\square$. Going up the hierarchy, the message between two gates of $C_\square$ can be extracted from the messages between the corresponding meta-gates.

\begin{lemma}
  \label{lem:eval_mg}
  Let $a = p \to p'$ be a wire between two positions $p, p'$ of $K^{\infty}$ in direction $D$. Let $a_l, a_g$ be the two wires crossing the edges between $p K$ and $p' K$ in clockwise order looking in direction $D$ (i.e., if $D$ is $E$, $a_l$ is the northernmost of the two; if $D$ is $S$, the westernmost…).

  Let $m = \innerevalfunc{C_\square}(a)$, $l = (l_1, l_2) = \innerevalfunc{C_\square}(a_l)$ and $g = (g_1, g_2) = \innerevalfunc{C_\square}(a_g)$.

  Then $\extractfunc(l_2, g_2) = m$.
\end{lemma}

\begin{proof}
  By induction on $p$, following the wires of $C_\square$.
\end{proof}

%Going down the hierarchy, the local part of the messages (layer 1 and \mloc on layer 2) describe the parent of each gate and the position of the gate within that parent's meta-gate.

\begin{lemma}
  \label{lem:info-loc}
  Let $\vec{p} \in K^\infty$. Let $g_1 = C_1(\vec{p})$, $g_2 = C_2(\vec{p})$, $g'_1 = C_1(\lfloor \frac{\vec{p}}{K} \rfloor)$, $g'_2 = C_2(\lfloor \frac{\vec{p}}{K} \rfloor)$. Let $g = g_1 \otimes g_2 = C_\square(\vec{p})$, and $g' = g'_1 \otimes g'_2 = C_\square(\lfloor \frac{\vec{p}}{K} \rfloor)$. Let $e$ be an output wire of $g$, and $m_1 \times m_2 = \innerevalfunc{C_\square}(e)$ its value in $C_\square$.

  Then $\mloc[m_2]$ is the label of $g'_2$, $\lmpcolor[m_1]$ is the label of $g'_1$, $\lmpwiring[m_1]$ is $\gatewiring[g']$, and $\lmpos[m_1]$ is $\vec{p} \bmod K$.
\end{lemma}

\begin{proof}
  By the previous lemma, input $0$ of each meta-gate $\mu_1(l_1, w) \otimes \mu_2(l_2, w)$ encodes $l_1, l_2$ and $w$. The gate after that input has label $\gcolor{dL}$, so by definition of its function $\fdecode{L}$, its output satisfies the lemma. The other gates in the meta-gate preserve the local part of the messages.
\end{proof}

This local information is just what is needed to reconstruct each gate from its \emph{output}, which is just short of self-description.

\begin{corollary}
  There are is a function $\decgate': \Sigma_1 \times \Sigma_2 \to \Wirings \times L_1 \times L_2 \times K$ such that for any position $p \in K^{\infty}$ and any wire $a: p \to p'$ of $C_\square$, \[\decgate'(\innerevalfunc{C_\square}(a)) = (\Lambda_1(\quot{p}{K}), \Lambda_2(\quot{p}{K}), C_w(\quot{p}{K}), p \bmod K).\]
\end{corollary}

With a tiny bit of extra work, each gate $g$ can be reconstructed from its input number 0, making $C_\square$ self-descriptive.

\begin{theorem}
  The circuit $C_\square$ is self-descriptive.
\end{theorem}

\begin{proof}
  Each gate $g$ in position $p$ can be reconstructed from its set of input directions and the message on its input number $0$.

  If $g$ has no inputs, then $g = s_1 \otimes s_2$.

  Otherwise, let $D$ be the direction of its first input wire, and $m = (m_1, m_2)$ the message coming into its first wire.

  Let $(w, l_1, l_2, p') = \decgate'(m)$. Note that $p' = (p - D) \bmod K$. If $p' + D$ belongs to $K$, then $\bigquot{p}{K} = \bigquot{p - D}{K}$, so $p$ belongs to the same meta-gate as its predecessor $p - D$, hence $C_\square(p)$ is $\mu_1(w, l_1)(p' + D) \otimes \mu_2(w, l_2)(p' + D)$.

  Otherwise, since $p' + D \notin K$, it must be the case that $\bigquot{p}{K} = \bigquot{p-D}{K} + D$: $p$ and its predecessor $p - D$ are in neighboring meta-gates. Let $a = \bigquot{p}{K}$ be the position of the parent gate. The position $a \bmod k$ of $a$ within its meta-gate is the position where input $0$ enters that meta-gate.

  A close examination of the values of $\mu_1(w, l_1)$ and $\mu_2(w, l_2)$ for all $w, l_1, l_2$ reveals that in each (non-seed) meta-gate the label and wiring of the gate in the position where input $0$ comes only depends on the input directions of $w$. By \cref{lem:info-loc}, those directions are exactly those of $\woutputwires[ { \lmpwiring[ { \gmanclocal[ { \mglob[m_2] } ] } ] } ](D)$.
\end{proof}

This concludes the proof of \cref{thm:cacarpet_circuit}. By \cref{thm:circuit-to-atam}, this means that there is an aTAM system $S_\square$ which strictly self-assembles $K^\infty$.

\section{Admissible Generators for DSSF Strict Self-Assembly}
\label{sec:limits}
One, two, infinity? We now want to characterize the generators $G$ for which $G^\infty$ is amenable to strict self-assembly.

\subsection{Generalizing the Circuit Construction}

The circuit-based construction of \cref{sec:cacarpet_circuit} can be adaptated to any self-similar discrete fractal within which the communication pattern used by $C_\square$ can be embedded. Whether a particular fractal is amenable to hosting such a communication pattern depends on the ability of its generator $G$ to transport information to copies of itself around it.

\begin{definition}
  Let $G$ be a finite subset of $\mathbb{N}^2$, the grid $G^{\#}$ is the subset of $\mathbb{Z}^2$ defined by:
  \[ G^{\#} = \{\ p \in \mathbb{Z}^2 | p \bmod G \in G \} \]

  % Let $\{d,  d'\} \subset \Dir$, the $d-d'$-flow in $G$ 

  The grid neighborhood graph $G^{+}$ of $G$ is the subgraph of $G^{\#}$ induced by the distance $1$ neighborhood of $G$:
  \[ G^+ = G \cup \{ p \in G^{\#} | \exists d \in \Dir, p + d \in G \}. \]

  For a direction $d \in \Dir$, the $d$-port in $G^{+}$ is \( G^{+d} = \{ p \in G^{\#} | p - d \in G \}. \)

  For $d, d' \in \Dir$, the $(d, d')$-bandwidth of $G$ is the number \( G[d \leftrightarrow d'] \) of vertex-disjoint paths from $G^{+d}$ to $G^{+d'}$ in $G^+$.
\end{definition}

\begin{figure}
  \centering
  \includegraphics[width=\textwidth]{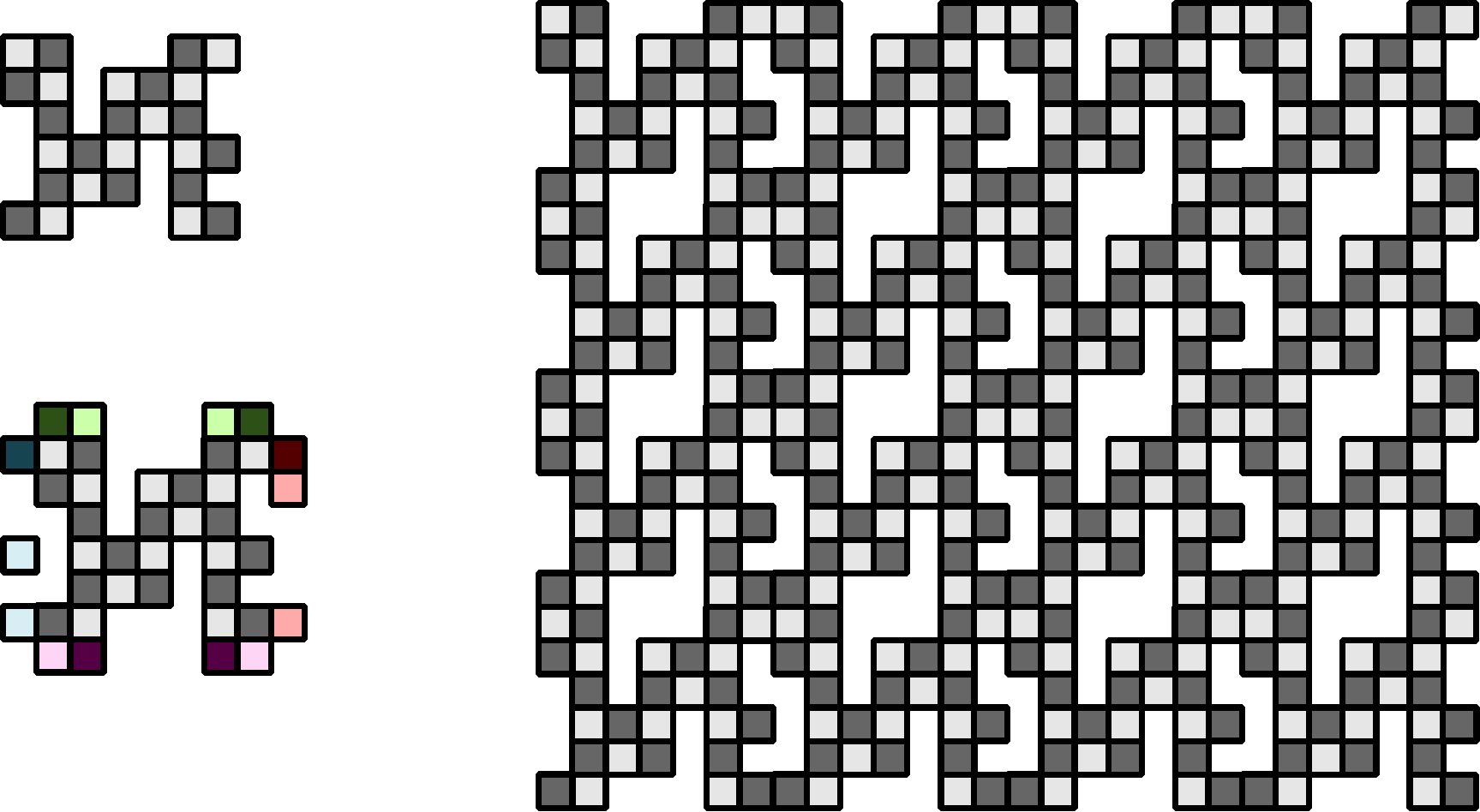}
  \caption{A finite shape $G$, its grid neighborhood $G^{+}$ and its grid graph $G^{\#}$.}
  \label{fig:grid_neigh}
\end{figure}

In order to compare generators, the classical notions of \emph{subgraph} and \emph{graph subdivision} is useful, accounting for marked vertices.

\begin{definition}[Pointed subgraph]
  Let $G, H$ be two graph, each with marked vertices. The graph $H$ is a \emph{pointed subgraph} of $G$ if $H$ is a subgraph of $G$ in such a way that any marked vertex of $H$ is mapped to a marked vertex of $G$. The graph $G$ may have extra marked vertices.
\end{definition}

\begin{definition}[Edge subdivision]
  Let $G = (V, E)$ be a graph, with marked vertices $(v_0, \ldots, v_{k-1}) \in V^k$. The edge subdivision operation for an edge $e = \{u, v\} \in E$ is the deletion of $e$ from $G$ and the addition of a new vertex $w \notin V$ and of the edges $\{u,w\}$ and $\{w,v\}$.

  This operation generates a new graph H, where the same vertices are marked as in $G$.
  \[H = (V \cup \{w\}, (E \setminus \{u,v\}) \cap \{\{u,w\}, \{w,v\}\})\]
\end{definition}

\begin{definition}[Graph Subdivision]
A graph with marked vertices which has been derived from $G$ by a sequence of edge subdivision operations is called a pointed subdivision of $G$.
\end{definition}

\begin{definition}[Subconnector]
  Let $G, H$ be finite, connected shapes of $\mathbb{N}^2$, with $(0, 0) \in G \cap H$.

  Then $H$ is a subconnector of $G$, written $H \preceq G$ if $G^+$ has a (pointed) subgraph which is a (pointed) subdivision of $H^+$.
\end{definition}

The notion of subconnector is well-suited to the study of substitutions given the following properties.

\begin{remark}
  \label{lem:preceq_sigma}
  Let $G, H, I$ be  finite, connected shapes of $\mathbb{N}^2$, with $G \preceq H$, then:
  \begin{eqnarray*}
    \sigma_I(G) \preceq \sigma_I(H)\\
    \sigma_G(I) \preceq \sigma_H(I)\\
  \end{eqnarray*}
\end{remark}

\buildThroughSubconnector*

\begin{proof}
  First, notice that $\kappa_w$ can be completed to cover the case where $w$ has two opposite input directions, as represented on \cref{fig:subst_w_extra} (modulo rotation and reflection).

  \begin{figure}
    \centering
    \begin{tikzpicture}
      \matrix {
        \begin{scope}
          \input{tikz/empty_gate_EW_to_N_Se}
        \end{scope}; &
        \begin{scope}
          \input{tikz/local_opposite}
        \end{scope}; \\
      };
    \end{tikzpicture}
    \caption{The value of $\kappa_w(w)$ when $\winputset[w] = \{E, W\}$. This case is not needed in \cref{sec:cacarpet_circuit}, but it is necessary to generalize $\kappa_w$ to $G$ when $K^+$ is a subdivision of $G^+$. No position needs to be provisionned for $\gcolor{dA}$, since both inputs must be in the same metagate.}
    \label{fig:subst_w_extra}
  \end{figure}

  Fix the vertices of $G^+$ which represent the vertices of $K^+$, the ones that sit in the middle of its edges, and the ones which are pending leaves.

  Then it is possible to define a substitution $\gamma_w: \Wirings \to \Wirings^G$ by using $\kappa_w$ to fix a subdivision of $G^+$, then adding the extra vertices to $\gamma_w(w)$. This process may create vertices in $\gamma_w(w)$ with two opposite inputs, for which the extra cases of \cref{fig:subst_w_extra} are necessary.

  \begin{figure}
    \centering
    \includegraphics[width=.5\textwidth]{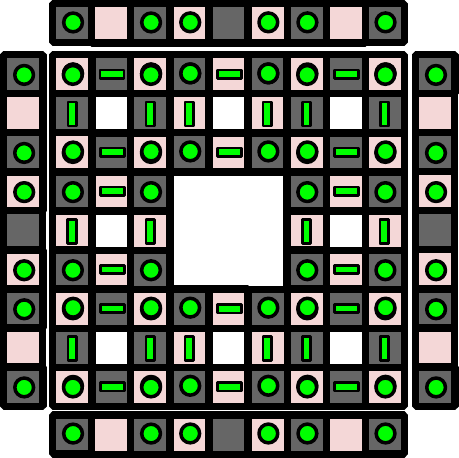}
    \caption{The second iteration $S$ of the Sierpinski Carpet (in the center) is a subconnector of $K$: $S^+$ contains a subdivision of $K^+$, in which the cells with a circle correspond to vertices of $K^+$, and those with a bar are obtained by dividing the edge of $K^+$ the bar represents. Not all edges are subdivided: adjacent circles in $G^+$ are indeed adjacent in $K^+$.}
    \label{fig:gamma_w}
  \end{figure}

  The label substitutions $\kappa_1$ and $\kappa_2$ can also be adaptated to $G^+$, defining $\gamma_1$ and $\gamma_2$. In $\gamma_1$, all vertices which do not represent a vertex of $K$ have label $\gcolor{normal}$. For layer $2$, each vertex of $G^+$ representing a vertex of $K$ keeps its label, each new vertex gets a label \gcolor{i_1}.
  
  Set $\Sigma_1^G = \Sigma_1$, except that $\lmpos$ takes values in $G$ rather than $K$. Then $\Sigma_2^G$ is $\Sigma_2$, except that $\gmanclocal$ takes values in $\Sigma_1^G$ rather than $\Sigma_1$.

  Then a circuit $C_G$ on $\Sigma_1^G \times \Sigma_2^G$ can be defined from $\gamma_w$, $\gamma_1$ and $\gamma_2$ like $C_\square$. That circuit $C_G$ has the same properties as $C_\square$ and is also self-descriptive. When accounting for the local part of the messages, gates which are upstream from all the $K$-vertices show the local message of the corresponding input meta-gate.

  Thus $C_G$ is self-descriptive.
\end{proof}

The question is now which $G$ are such that $G^{+}$ has a subdivision of $K^+$ as a subgraph. This condition seems constraining since $K$ is a rather large graph with its 32 vertices. Yet, one can make any $G$ larger by iterating the substitution generated by $G$ before trying to self-assemble $G^{\infty}$.

\begin{remark}
  Let $G \ni (0, 0)$ be a finite shape of $\mathbb{N}^2$. For any $k > 0$, $G^{\infty} = (G^k)^\infty$.
\end{remark}

% In~\cite{DBLP:journals/nc/PatitzS10}, Patitz and Summers observe that for $G^\infty$ to self-assemble strictly, it must avoid pinch-points.

By iterating $\sigma_G$, it is quite easy to get an instance of $K$ as a subconnector, as long as one starts with sufficient vertical and horizontal bandwidth.

\begin{lemma}
  Let $G \ni (0, 0)$ be a finite shape of $\mathbb{N}^2$. If $G[N \leftrightarrow S] \geq 2$ and $G[E \to W] \geq 2$, then $K \preceq G^3$
\end{lemma}

\begin{proof}
  Let $H = \{0, 1\}^2$. If $G[N \leftrightarrow S] \geq 2$ and $G[E \to W] \geq 2$, then $H \preceq G$: pick two disjoint North-South paths, two disjoint East-West paths, their four intersections can act as the vertices of $H$.

  By \cref{lem:preceq_sigma}, since $H \preceq G$, $H^3 \preceq G^3$. But $H^3$ is an $8 \times 8$ grid, so $K \preceq H^3$.

  Hence, $K \preceq G^3$.
\end{proof}

\subsection{Limits to Strict Assembly of DSSFs in the ATAM model}

This characterization of generators for which the associated fractal can be strictly self-assembled is tight, by a generalization of the impossibility result of Hendricks et al.~\cite{hendricks_hierarchical_2020}.

\begin{lemma}
\label{lem:kill_giraffe}
  Let $G \ni (0, 0)$ a finite shape of $\mathbb{N}$.

  If for all $k > 0$, there is an $x_k$ such that there is only one $y$ with $(x_k, y) \in (G^k)^+$ and $(x_k, y+1) \in (G^k)^+$, then $G^\infty$ cannot be strictly self-assembled in the aTAM model unless $G = \{\vec{0}\}$.
\end{lemma}

\begin{proof}
  Let $w$ and $h$ be the width and height of $G$. Assume that there is an aTAM $\mathcal{G}$ which self-assembles $G^{\infty}$ at temperature $\tau$.

  Let $C_k$ be the set of all glues which appear on the eastern edge of position $(x_k, y)$ for some value of $y$. Since $\mathcal{G}$ assembles $G^{\infty}$, all the $C_k$ are non-empty, so let $C_{\infty} = \bigcap_k (\bigcup_{k' > k}C_{k'})$ be the set of glues appearing in infinitely many $C_k$; $C_{\infty}$ is also non-empty. Let $F_k \subseteq C_k$ be the set of glues $g$ such that there is a position $\vec{z} = (x_k, y)$ and a production $\Pi_{k,t}$ of $\mathcal{G}$ where:
  \begin{itemize}
  \item the eastern glue of $\Pi_{k,t}(\vec{z})$ is $g$
  \item $\Pi_{k,t}$ contains no tile right of $x_k$
  \end{itemize}
  Like $C_\infty$, $F_\infty = \bigcap_k(\bigcup_{k' > k} F_{k'})$ is non-empty. The sets $C_k$ and $F_k$ are illustrated on \cref{fig:Ck_and_Fk}.

  \begin{figure}
    \centering
    \begin{tikzpicture}[scale=.48, yscale=-1]
      \input{tikz/Ck_and_Fk}
    \end{tikzpicture}
    \caption{The set $C_k$ is the set of all glues which attach across the vertical lines at coordinate $x_k$. Ignoring any tiles which are not visible in the picture, $C_0 = C_1 = \{ a \cdot E, b \cdot E, c \cdot E \}$, $C_2 = \{ a \cdot E, b \cdot E\}$ and $C_3 = \{ c \cdot E \}$. The tiles marked with a $1$ are the ones which appear in their column in some production without other tiles right of the red line; the subset $F_k$ of $C_k$ contains their east glues: , $F_0 = F_1 = \{ a \cdot E, b \cdot E \}$, $C_2 = \{ a \cdot E \}$ and $C_3 = \{ c \cdot E \}$}
    \label{fig:Ck_and_Fk}
  \end{figure}
  
  Now let $t$ be a glue of $\mathcal{G}$, and $s_t$ be a (new) tile with $t$ on its eastern side, and $\epsilon$ on all other sides. Let $\mathcal{G}_t[x \geq 1]$ be the set of productions of $\mathcal{G}$, starting from the seed configuration with $s_t$ at $\vec{0}$ and attaching no tiles at positions $x < 1$. Let $u_t = \min_{p \in \mathcal{G}_t[x>1]} \max \{y | (x,y) \in p\}$ and $d_t =  \min_{p \in \mathcal{G}_t[x>1]} \min \{y | (x,y) \in p\}$. These two quantities are illustrated on \cref{fig:ut_and_dt}.

  \begin{figure}
    \centering
    \begin{tikzpicture}
      \input{tikz/ut_and_dt}
    \end{tikzpicture}
    \caption{The definition of $u_t$ and $d_t$ given the two productions reachable from the glue $t$, without going left of the seed.}
    \label{fig:ut_and_dt}
  \end{figure}

  Since $G^\infty \subset \mathbb{N}^2$, it does not contain any infinite southward path. Therefore, for any $k$ and $t \in F_k$, $d_t > - \infty$, otherwise from $\Pi_{k,t}$ it would be possible to produce paths going arbitrarily far down, out of $\mathbb{N}^2$. Let $d_\infty = \max_{t \in F_{\infty}} \{d_t  | t \in \mathbb{G} \}$, $d_\infty$ is finite. On the other hand, $u_\infty = \max \{ u_t | t \in F_{\infty} \}$ must be infinite. Indeed, if it is finite, let $s = u_\infty -d_\infty + 1$. Consider a maximal production $\Pi^*$ obtained by only placing tiles left of the vertical line $L^*$ at $x$-coordinate $x_s$. Starting from $\Pi^*$, the only attachable positions are isolated points on $L^*$, separated by a distance at least $w^s$. From each of them, it is possible to grow an arm which reaches no higher than $s$ upwards, whence, as illustruted on \cref{fig:dt_finite_ut_infinite}, the part right of $L^*$ of that production is contained within a union of $(u_\infty - d_\infty)$-width horizontal bands, so its domain cannot be $G^\infty$.%\ruggedtodo{Could it actually, but anyway, then it is ripe for pumping}.

  \begin{figure}
    \centering
    \begin{tikzpicture}[scale=.4]
      \input{tikz/dt_finite}
    \end{tikzpicture}%
    \quad%
    \begin{tikzpicture}[scale=.4]
      \input{tikz/ut_infinite}
    \end{tikzpicture}
    \caption{Faulty terminal productions which can be reached, left if some $d_{t}$ is infinite, right if all $u_t$ are finite: $\min_t(d_t) = -1$, $u_t = 1$, $u_s = 2$.}
    \label{fig:dt_finite_ut_infinite}
  \end{figure}

  For $t$ such that $u_t = +\infty$, all terminal productions on the right half-plane reach infinitely high. Thus, by applying~\cref{thm:tree_pump} with increasing values of $m$, either one of these terminal productions contains a periodic path going up, or there are productions $P$ with arbitrarily large squares in their fill-in $\complete{P}$. In both cases, for each $k$, there is a production $F_k$ for which some $k \times k$ square is inaccessible from any point $(0, y)$ with $y > 0$ without crossing $F_k$, as illustrated on~\cref{fig:peephole}.
    
  For a large enough level of substitution, any glue $t$ for which $u_t$ is finite does not even grow one ``meta-cell'' up. Let $F_{!} = \{ g \in F_\infty | u_g = +\infty \}$, and $s_{!}$ be such that $\max \{ u_g | g \notin F_{!} \} < h^{s_{!}}$ and $|d_\infty| <  h^{s_{!}}$.

  Consider a vertical line $L_x$ at some position $x$ between copies of $G^{s_!}$ in $G^\infty$. Let $P_y$ be a maximal subproduction of $P$ which can be assembled without attaching any tile right of $L_y$. In $P_y$, let $\vec{z}_{x} = (x, y)$ be the lowest position on $L_{x}$ containing a tile $T$ of $F_{s!}$.
  %, note $t = T \cdot E$. Let $P^!$ be the production obtained by pasting some production $\Pi_! \in \mathcal{G}_t[x \geq 1]$ to the right of $\vec{z}^!$. In productions obtained from $P^!$, the positions to the right of the path formed by $\ell \Pi_!$ attach to the seed through $\vec{z}^!_n$. As seen previously, this part right of $\ell \Pi_!$ can be taken to be $F_k$, and thus contain parts of $G^\infty$ which circle arbitrarily large squares.

  \begin{figure}
    \centering
    \begin{tikzpicture}
      \input{tikz/peephole}
    \end{tikzpicture}
    \caption{The part of the production that sees the seed only through $\vec{z}_!$ contains cycles of $G^\infty$ of arbitrary size.}
    \label{fig:peephole}
  \end{figure}

  By the pigeonhole principle, there must be two vertical lines $L_n$ and $L_m$ between copies of $G^{s_{!}}$ such that $P_n(\vec{z}_n) = P_m(\vec{z}_m)$. Let $\vec{v}$ be the vector $\vec{z}_m - \vec{z}_n$. By construction of $P_n$ and $P_m$, for any production $P_n^{!}$ obtained from $P_n$, there is a production $P_m^{!}$ obtained from $P_m$ such that the part of $P_n^{!}$ above and to the right of $\vec{z}_n$ and the part of $P_m^{!}$ above and to the right of $\vec{z}_m$ are identically filled. Hence, there is a vector which preserves either arbitrarily large cycles of $G^\infty$ or its intersection with a quarter-plane, thus $G$ must be trivial.
\end{proof}

Still open is the case where in all the iterates $(G^k)^+$, there is a vertex which disconnects either the north and south or the east and west, but no straight line which crosses $(G^k)^+$ only once. A more precise variant of the previous proof takes care of that case.

\killllama*

\begin{proof}
    Assume without loss of generality that $\sup(G^k[E \leftrightarrow W]) = 1$.

    The proof is the same as that of \cref{lem:kill_giraffe}, except that instead of straight lines cutting the fractal only once every $h^k$, one needs to consider cuts which pass through the first occurence of each level of the east-west bridges. These cuts are not necessarily straight, but each of them remains within a bounded vertical band, which suffices for the remainder of the proof.
\end{proof}

\subsection{Deciding Bounded Bandwidth}

\cref{thm:kill_llama} gives a dichotomy between generators for which the associated fractal can be strictly self-assembled and those for which it cannot. There is a polynomial time algorithm for determining which of \cref{lem:build_through_subconnector} or \cref{thm:kill_llama} applies.

\begin{theorem}
  There is a polynomial time algorithm which, on input $G$ decides whether:
  \begin{itemize}
  \item there is a $k$ such that $G^k[N \leftrightarrow S] \geq 2$ and $G^k[E \leftrightarrow W] \geq 2$ and thus $G^\infty$ can be strictly self-assembled in the aTAM, or
  \item for all $k$, $G^k[N \leftrightarrow S] < 2$ and thus $G^\infty$ cannot be strictly self-assembled in the aTAM, or
  \item for all $k$, $G^k[E \leftrightarrow W] < 2$ and thus $G^\infty$ cannot be strictly self-assembled in the aTAM.
  \end{itemize}

  Notice that the second and third case are \emph{not} mutually exclusive.
\end{theorem}

\begin{proof}
  Without loss of generality, assume $G$ is connected.

  First, use a max-flow algorithm to compute $G[d \leftrightarrow d']$ for all directions $d$, $d'$.
  If both $G[E \leftrightarrow W]$ and $G[N \leftrightarrow S]$ are $2$ or more, then the first case applies.

  Let $D$ be a set of pairs of directions, $v$ is a $D$-disconnector if $v$ disconnects $G^{+d}$ from $G^{+d'}$ for any $(d, d') \in D$. If $G[E \leftrightarrow W] = 1$, then there is at least one $\{EW\}$-disconnector. For any $D$-disconnector, define $\Cause(v, D)$ as the set of pair of directions $(\delta, \delta')$ such that there are $(d, d') \in D$ and a path from $G^{+d}$ to $G^{+d'}$ which enters $v$ from direction $\delta$ and leaves it through direction $\delta'$.

  Compute the disconnect causation graph $\Delta$. It is a directed graph whose vertices are couples $(v, D)$ where $v$ is a $D$-disconnector, and there is an arc from $(v, D)$ to each vertex $(v', \Cause(v, D))$. The causation graph $\Delta$ has size at most $64 |G|$ and the existence of each of its arcs can be tested in polynomial time.

  The graph $\Delta$ contains a cycle reachable from $(v, \{EW\})$, if and only if for all $k$, $G^k[E \leftrightarrow W] < 2$, likewise for $N$ and $S$. Indeed, for $k > 0$, a vertex $v$ is a $D$-disconnector in $G^k$ if and only if:
  \begin{itemize}
  \item The vertex of $\quot{v}{G^{k-1}}$ of $G$ corresponding to the copy of $G^{k-1}$ containing $v$ is a $D$-disconnector, and
  \item the vertex $(v \bmod G^{k-1})$ of $G^{k-1}$ corresponding to the position of $v$ within that copy is a $\Cause(v,D)$-disconnector.
  \end{itemize}

  Hence, from $\Delta$, it is possible to distinguish between the three cases.
\end{proof}

\section{Open Questions}
\label{sec:questions}

The tools we have introduced here —standard systems, quines, self-describing circuits— as well as our characterization of the behavior of bounded-treewidth productions in the aTAM through~\cref{thm:tree_pump} have passed the test of strict self-assembly of fractals, which has been an open question in self-assembly for more than a decade.

We are curious to know what the landscape of fractal self-assembly is in 3D. In cellular automata, Tilings, as well as in the aTAM, it is well-known that uncomputability can appear with the third dimension. While the Tree-Pump Theorem certainly would work in 3D, as well as on any ``reasonable grid'', in $\mathbb{Z}^3$, the periodic path obtained from it does not isolate a large domain. Hence, the question of characterizing DSSFs which can be strictly self-assembled in $\mathbb{Z}^3$ remains open.

The Tree Pump Theorem links \emph{bounded} connected treewidth with having a periodic, and thus compuatble behavior. This leads to the question of quantitative bounds: can an analogue of Complexity Theory be built on treewidth for the aTAM? This also begs the question whether using connected treewidth is fundamental to this pumping principle: are there bounded treewidth, aperiodic shapes which can be strictly self-assembled?

\printbibliography

\appendix

%\section{Notations}
%\input{common_notations.tex}

\end{document}